\documentclass[aps,showpacs,showkeys,preprintnumbers]{revtex4}
\usepackage{graphicx}
\usepackage{color}
\usepackage{amsmath}
\usepackage{amssymb}
\usepackage{multirow}
\usepackage{titlesec}
\makeatletter
\renewcommand\paragraph{\@startsection{paragraph}{4}{\z@}%
            {-2.5ex\@plus -1ex \@minus -.25ex}%
            {1.25ex \@plus .25ex}%
            {\normalfont\normalsize\bfseries}}
\makeatother
\def \be  {\begin{equation}}
\def \ee  {\end{equation}}
\def \ee  {\end{equation}}
\def \bea {\begin{eqnarray}}
\def \eea {\end{eqnarray}}

\newcommand{\nn}{\nonumber}

\begin{document}

\preprint{ECTP-2019-02}
\preprint{WLCAPP-2019-02}
\hspace{0.05cm}
\title{Extensive/nonextensive statistics for $p_{\mathtt{T}}$ distributions of various charged particles produced in p$+$p and A$+$A collisions in a wide range of energies}

\author{Abdel Nasser  Tawfik}
\email{atawfik@nu.edu.eg}
\affiliation{Nile University - Egyptian Center for Theoretical Physics (ECTP), Juhayna Square off 26th-July-Corridor, 12588 Giza, Egypt}
\affiliation{Institute for Theoretical Physics, Goethe University, Max-von-Laue-Str. 1, D-60438 Frankfurt am Main, Germany}

\author{Hayam Yassin}
\email{hiam_hussien@women.asu.edu.eg}
\author{Eman R. Abo Elyazeed}
\email{eman.reda@women.asu.edu.eg}
\affiliation{Physics Department, Faculty of Women for Arts, Science and Education, Ain Shams University, 11577 Cairo, Egypt}

\date{\today}

\begin{abstract}

We present a systematic study for the statistical fits of the transverse momentum distributions of charged pions, Kaons and protons produced at energies ranging between $7.7$ and $2670~$GeV to the extensive Boltzmann-Gibbs (BG) and the nonextensive statistics (Tsallis as a special type and the generic axiomatic nonextensive approach). We also present a comprehensive review on various experimental parametrizations proposed to fit the transverse momentum distributions of these produced particles. The inconsistency that the BG approach is to be utilized in characterizing the chemical freezeout, while the Tsallis approach in determining the kinetic freezeout is elaborated. The resulting energy dependence of the different fit parameters largely varies with the particle species and the degree of (non)extensivity. This manifests that the Tsallis nonextensive approach seems to work well for p$+$p rather than for A$+$A collisions. Nevertheless, discussing deeper physical insights of nonextensive statistical approaches isn't targeted, drawing a complete picture of the utilization of Tsallis statistics in modeling the transverse momentum distributions of several charged particle produced at a wide range of energies and accordingly either disprove or though confirm the relevant works are main advantages of this review. We propose analytical expressions for the dependence of the fit parameters obtained on the size of the colliding system, the energy, as well as the types of the statistical approach applied. We conclude that the statistical dependence of the various fit parameters, especially between Boltzmann and Tsallis approaches could be understood that the statistical analysis {\it ad hoc} is biased to the corresponding degree of extensivity (Boltzmann) or nonextensivity (Tsallis). Alternatively, the empirical parameterizations, the other models, and the generic (non)extensive approach seem to relax this biasness.

\end{abstract}


\maketitle

\tableofcontents

\section{Introduction} 
\label{intro}

In high-energy collisions, large transverse momenta and particle yields are likely generated \cite{Tawfik:2014eba}. The statistical nature of such a particle production process was first proposed by Koppe \cite{Tawfik:2013tza}. In 1950, Fermi introduced a solid statistical theory assuming concentration of energy in a small spatial volume which - through multiple successive processes - decomposes into many {\it smaller} final-states \cite{Fermi1950,Fermi1950book}. Assuming a varying number of ideal particles including an aggregation of oppositely charged particles and fulfilling the conservation laws, a generalization to quantum statistics was introduced by Magalinski and Terletskii in 1957  \cite{Magalinskii1957}. Based on a statistical bootstrap approach, Fast and Hagedorn introduced in 1963 the mass spectrum function characterizing the abundances of the {\it so-far-detected} hadron resonances and introducing the concept of limiting temperature \cite{Hagedorn1963,Hagedorn1963b}. This very short overview highlights key milestones of the use of the {\it extensive} Maxwell-Boltzmann statistics in high-energy collisions. Over the last five decades, enormous numbers of papers reporting on various statistical characteristics of the particle production have been published. Interested readers are kindly advised to consult recent review articles, such as ref. \cite{Tawfik:2014eba}. There is a common consensus among particle physicists that Maxwell-Boltzmann statistics describes well the particle multiplicities and their fluctuations and correlations within the so-far explored range of beam energies.

With the introduction of Tsallis nonextensive statistics \cite{Tsallis:1987eu}, different implications even in high-energy physics were proposed \cite{Bediaga:1999hv,Parvan:2016rln}. This nonextensive approach assumes the same phase-space as in the {\it extensive} statistics, but it replaces the Boltzmann factors by the so-called $q$-exponential functions, with $q>1$. A large number of research papers has been published so far dealing with this type of nonextensivity, for instance \cite{Beck:2000nz,Wilk:1999dr,Walton:1999dy,Alberico:2000nc,Bediaga:1999hv,Zimanyi:2005nn,Trainor:2007zj,Wilk:2008ue,Biro:2008er,at14,Tripathy:2016hlg,Khuntia:2016ikm,Bhattacharyya:2015hya,Deppman:2012qt,Alberico:2005vu,Parvan:2016rln}. Recently, AT explained \cite{Tawfik:2016pwz} that if one limits Tsallis nonextensivity to the related algebraic operations one is apparently not assuring a proper implementation on the high-energy particle production. The possible interactions besides the fluctuations, the correlations, and the likely modifications in the phase space due to symmetry change, for instance, seem not being incorporated through $q$-statistics. In other words, a kind a global equilibrium seems being assumed no matter whether energy varies and/or the interacting system rapidly spatio-temporarily evolutes. To remain within the scope of the present research paper, we intentionally disregard the recent proposal that the degree of nonextensivity can be best taken into account when - instead of this types of nonextensivity - a wider class of superstatistics should be implemented, e.g. generic axiomatic nonextensive approach \cite{Tawfik:2016pwz,Thurner:2010if,Tawfik:2017bul}, where both Boltzmann-Gibbs (BG) and Tsallis approaches represent very special cases in it. 

Nuclear physics, for instance, has a long history with the nonextensive statistics. Here, the nonextensive approaches are rightfully based on clear physical arguments. Weisskopf accounted for the fact that a high-energy emission reduces the temperature of the remaining nucleus \cite{Weisskopf1937}. This idea was extended to heavy ions canonical suppression \cite{Cheng:2001dz,Tounsi:2001ck}. The Tsallis-based arguments differ in that they are mainly useful to parametrize nonextensivity, on one hand. On the other hand, they offer little physical insights. Not only absence of deeper physical insights, but also the utilization itself seems not privileged a common agreement among the particle physicists.  This research article reviews and discusses on the Tsallis approach and focuses on its utilization in modeling the transverse momentum distributions of various charged particles produced in a wide range of energies.

Although the frequently reported surprising success of Tsallis statistics in reproducing transverse momentum ($p_{\mathtt{T}}$) distributions of some charged particles in p$+$p collisions at high energies, various pieces of the puzzle are still missing \cite{Kassner:2016obj,Zheng:2015mhz,Gao:2015qsq,Zheng:2015gaa,Zheng:2015tua,Wilk:2015pva,Marques:2015mwa,Urmossy:2015kva,Cleymans:2013rfq,Rybczynski:2014cha}. For instance, the reproduction of $p_{\mathtt{T}}$ of the charged particles produced in A$+$A is mysteriously still overseen. Many colleagues believe that this is not as successful as in p$+$p collisions. In this regard, different questions should be answered, for example, \begin{itemize}
\item a) whether the system size affects the possibility of nonextensivity of such produced particles; mainly well-identified bosons and baryons representing low-lying Goldstone mesons and the most-stable fermions? 
\item b) whether the fits of the transverse momentum spectra of the various particles characterize the chemical or the kinetic freezeouts? 
\item c) whether the possible flow diminishes the proposed nonextensivity of these particles, when moving from p$+$p to A$+$A collisions? 
\item d) whether this {\it very special} type of nonextensivity is indeed able to describe the statistical nature of the particle production? 
\end{itemize}
For example, if the answer to the last question, for instance, is "yes", one would expect that other aspects of the particle production, such as the particle multiplicity (particle yields and their ratios) could (should) be reproduced by Tsallis statistics, as well. Apparently, this is not the case \cite{Tawfik:2017bul,Deppman:2012us,Tawfik:2016pwz}!

The transverse momentum ($p_{\mathtt{T}}$) quantifies the projection of the four-momentum onto the plane with a transverse (perpendicular) orientation to the collision axis ($z$-axis). Accordingly, $p_{\mathtt{T}}=p \sin(\theta)$, where $p$ is the four-momentum and $\theta$ is the initial polar angle of the particle of interest with respect to the vertex position along the collision axis $z$. If the longitudinal momentum or rapidity ($y$) is integrated in the thermal distribution with no flow, the same $p_{\mathtt{T}}$-distribution can be obtained \cite{Kataja:1990tp}. Studying the $p_{\mathtt{T}}$-distributions within thermal approaches gives - among others - an estimation for the temperature of the fireball (emission source) \cite{Turbide:2003si}. The transverse mass and energy can also be determined.

The Tsallis distribution was utilized, for instant, to fitting different particle spectra at midrapidity from central d$+$Au, Cu$+$Cu, and Au$+$Au collisions at RHIC and p$+$Pb and Pb$+$Pb collisions at LHC energies \cite{Zheng:2015gaa}. Although the strong medium effects in nucleus-nucleus collisions (for example Cu$+$Cu and Au$+$Au), it was concluded that the Tsallis approach can be used to fit most of the particle spectra in d$+$Au and p$+$Pb collisions, where the medium effects are assumed being very weak. In  A$+$A collisions, the Tsallis approach is found capable to fit very well all the particle spectra at the RHIC energies except the little deviation observed for proton and $\Lambda$ at low $p_{\mathtt{T}}$ \cite{Zheng:2015gaa}. At LHC energies, the Tsallis distribution can only fit part of the particle spectra  either in low or high $p_{\mathtt{T}}$ region. In order to reproduce the entire $p_{\mathtt{T}}$ region, new formula with an additional degree of freedom should be proposed, for instance, 
\bea
\left(E \frac{d^3 N}{dp^3}\right)_{|\eta|<a} &=& 
A \frac{\exp\left[-\frac{b}{T} \arctan\left(\frac{E_T}{b}\right) \right]}{\left[1+\left(\frac{E_T}{b}\right)^4\right]^c}, \label{eq:newprmtrs}
\eea
where $A$, $b$, $T$, and $c$ are free parameters. Expression (\ref{eq:newprmtrs}) is inspired by Fokker-Planck equation \cite{Banerjee:2010}, where the exponent $2$ in the numerator is replaced by $4$ \cite{Zheng:2015gaa}. The main idea proposed in ref. \cite{Zheng:2015gaa} is that a transition from the exponential distribution to Tsallis distribution should be conjectured to take place at intermediate $p_{\mathtt{T}}$.  This proposed transition was done {\it ad hoc} in order to get a new expression to be used in fitting the intermediate $p_{\mathtt{T}}$ spectra from A$+$A collisions by taking into account degrees of freedom greater than the ones available to the BG statistics.

Furthermore, a well-thorough study of the traverse momentum spectra of well-identified particles produced in p$+$p collisions at RHIC and LHC energies with the Tsallis distribution was conducted in refs. \cite{Khandai:2013gva,Saraswat:2017kpg}. It was found that the rapidity and energy dependence of the $p_{\mathtt{T}}$ spectra in p$+$p collisions describes well the experimental results from STAR, PHENIX, ALICE and CMS programs \cite{Zheng:2015tua}. This becomes possible when cascade particle production mechanism was included. The energy dependence of the temperature ($T$) and $q$ (where $q$ is thought similar as the parameter $n$) of the Tsallis distribution has been discussed in great detail \cite{Zheng:2015tua}. 

It should be noticed that almost all relevant experimental results are detectable differential quantities, such as  $d N/d^3p=\int d^3p (p^{\mu}/p^0) f_0(x,p)$, satisfying normalization conditions, so that the total number, for instance, reads
\bea
N &=& \int_S \frac{d N}{d^3 p} d^3 p= \int_S n(x) u^{\mu} d\sigma_{\mu},
\eea
where $f_0(x,p)$ is the phase-space distribution function, $S$ is the surface, $d\sigma_{\mu}$ is a time-like normal vector, and $u^{\mu}$ is the four velocity. Similarly, the transverse momentum and transverse momentum distribution, respectively, can be expressed as 
\bea
p_{\mathtt{T}} &=& \int_S \frac{d N}{d^3p} p^x d^3p = \int_S T^{\mu x} d\sigma_{\mu}, \\
\frac{d N}{p_{\mathtt{T}} dp_{\mathtt{T}}} &=& \gamma V \int \frac{d^3 p}{p^0} \frac{d [p^{\mu} u_{\mu} f(x,p)]}{p_{\mathtt{T}} dp_{\mathtt{T}}},
\eea
where $T^{\mu x}$ is the energy-momentum tensor.

The present papers focuses on a comprehensive characterization of the transverse momentum distributions of various charged particles produced in p$+$p and A$+$A collisions at beam energies ranging between $7.7$ and $2670~$GeV. We also present a short review on the various experimental parametrizations, where measured $p_{\mathtt{T}}$ are well fitted but not necessarily within $q$-statistics; the Tsallis-type, for instance. We shall discuss on the possible reasons that even this type of nonextensivity seems to be successful for p$+$p but not for A$+$A collisions! Also, we highlight that the resulting fit parameters are not only depending on the energy but - among others - on the particle species, themselves!

\section{Approaches}

\subsection{Statistical-thermal approaches}

\subsubsection{Transverse momentum distributions}

\paragraph{Extensive statistics}

As proposed in literature \cite{Parvan:2016rln,Cleymans:2013rfq}, the four-momentum in the nonrelativistic limit, i.e. $m\gg p$, could be replaced by the transverse momentum ($p_{\mathtt{T}}$), the transverse mass [$m_{\mathtt{T}}=(p_{\mathtt{T}}^2+m^2)^{1/2}$], and the rapidity ($y$), with the dispersion relation $E=m_{\mathtt{T}}\, \cosh(y)$, so that $d^3p=p_{\mathtt{T}}\, m_{\mathtt{T}}\, \cosh(y)\, dp_{\mathtt{T}}\, dy\, d\phi$, where $\phi$ is the azimuthal angle. At finite temperature ($T$), chemical potential ($\mu$), and volume ($V$) and assuming a full detector acceptance, i.e. $\int d\phi=2\pi$, the extensive Boltzmann-Gibbs (BG) and Fermi-Dirac and Bose-Einstein (F$|$B) statistical approaches, respectively, give
\bea
\frac{1}{2 \pi p_{\mathtt{T}}}\, \left.\frac{d^2\, N}{dp_{\mathtt{T}}\, dy}\right|_{\mathtt{BG-extensive}} &=& \frac{g V}{(2 \pi)^3}  \, m_{\mathtt{T}}\, \cosh(y) \left[\exp\left(\frac{\mu-m_{\mathtt{T}}\, \cosh(y)}{T}\right)\right], \label{MBpT}\\
\frac{1}{2 \pi p_{\mathtt{T}}}\, \left.\frac{d^2\, N}{dp_{\mathtt{T}}\, dy}\right|_{\mathtt{F|B-extensive}} &=& \pm \frac{g V}{(2 \pi)^3}  \, m_{\mathtt{T}}\, \cosh(y) \left[\exp\left(\frac{m_{\mathtt{T}}\, \cosh(y)}{T} - \mu\right)\pm 1\right]^{-1}, \label{QSpT}
\eea
which can be derived straightforwardly from the distribution function in the corresponding four-space,
\bea
\left. N\right|_{\mathtt{BG-extensive}} &=& g V \int_0^{\infty} \frac{d^3 p}{(2 \pi)^3} e^{\frac{\mu-\varepsilon}{T}}, \label{MBN}\\
\left. N\right|_{\mathtt{F|B-extensive}}  &=& \pm g V \int_0^{\infty} \frac{d^3 p}{(2 \pi)^3} \frac{e^{\frac{\mu-\varepsilon}{T}}}{1\pm e^{\frac{\mu-\varepsilon}{T}}},  \label{QSN}
\eea
where $\varepsilon=(p^2+m^2)^{1/2}$ is the extensive dispersion relation is natural units, $g$ is the degeneracy factor and $\mu$ combine all types of chemical potentials.

\paragraph{Tsallis and Tsallis-factorized nonextensive statistics}

For the seek of simplicity, the Tsallis statistics is applied to the Maxwell-Boltzmann distributions, Eq. (\ref{MBN}). Accordingly, the total number of particles can be estimated - within Tsallis nonextensive statistics - as follows.
\bea
N &=& g V \int_0^{\infty} \frac{d^3 p}{(2 \pi)^3} \left[1+(q-1) \frac{E-\mu}{T}\right]^{\frac{-q}{q-1}},
\eea
where $q>1$ is key parameter defining the degree of Tsallis-nonextensivity. The dependence of $q$ on $T$, $\mu$ and $\sqrt{s_{\rm NN}}$ shall be reviewed later on. The four-momentum distribution can be expressed as 
\bea
E \frac{d^3 N}{d^3 p} &=& E \frac{g V}{(2 \pi)^3} \left[1+(q-1) \frac{E-\mu}{T}\right]^{\frac{-q}{q-1}}.
\eea
At finite chemical potential and non-vanishing rapidity, the transverse-momentum distribution becomes
\bea
\frac{1}{2 \pi p_{\mathtt{T}}} \left.\frac{d^2\, N}{dp_{\mathtt{T}}\, dy}\right|_{\mathtt{Tsallis-nonextensive}} &=& V \frac{g}{(2 \pi)^3}  \, m_{\mathtt{T}}\, \cosh(y) \left[1+(q-1)\frac{m_{\mathtt{T}}\, \cosh(y)-\mu}{T}\right]^{-q/(q-1)}. \label{TsallispT}
\eea
When $q\rightarrow 1$, BG statistics can be resembled, straightforwardly \cite{Tsallis:1987eu,Parvan:2016rln}, Eq. (\ref{MBpT}). An upper bound on $q$ is naturally set by the given derivatives \cite{Parvan:2016rln}. Per definition, both expressions describe an ideal nonextensive gas, i.e.  one can sum over the constituents of such an additive gas. On the other hand, the given single-particle multiplicity distribution is conjectured as a powerful modeling for the transverse momentum distribution of the individual produced particle \cite{Parvan:2016rln}. It was argued that many-particle multiplicity distributions using Tsallis statistics doesn't factorize into a product of single-particle ones \cite{Parvan:2016mbv}. It was pointed out that only a factorization approximation allows the use of an explicit single-particle distribution function in Tsallis statistics \cite{Buyukkilic1995}. 
\bea
\frac{1}{2 \pi p_{\mathtt{T}}}\, \left.\frac{d^2\, N}{dp_{\mathtt{T}}\, dy}\right|_{\mathtt{factorized}} &=& V \frac{g}{(2 \pi)^3}  \, m_{\rm {\mathtt{T}}}\, \cosh(y) \sum_{N=0}^{N_0} \frac{\tilde{\omega}^N}{N!}\, h_0(N) \nn \\
& & \left[1+\frac{q-1}{q}\frac{\Lambda - m_{\mathtt{T}}\, \cosh(y)-\mu(N+1)}{T}\right]^{-q/(q-1) + 3N}, \label{TsallispTfactor}
\eea
where $\tilde{\omega}=g V T^3/\pi^2$,
\bea
h_0(N) &=& \left\{
\begin{array}{ll}
\frac{[q/(1-q)]^{3N} \Gamma[1/(1-q)-3N]}{\Gamma[1/(1-q)]}, \quad & q<1, \\
 & \\
\frac{[q/(q-1)]^{3N} \Gamma[q/(1-q)]}{\Gamma[q/(q-1)+3N]}, \quad & q>1.
\end{array}
\right.
\eea
and the norm function $\Lambda$ is to be determined  from a norm equation \cite{Parvan:2016mbv}. It was concluded that at SPS energies, where the entropic parameter gets a value very close to unity, the Tsallis factorized statistics, Eq. (\ref{TsallispTfactor}), seems to deviate from the Tsallis non-factorized statistics, Eq. (\ref{TsallispT}). At higher energies, both types of Tsallis statistics become indistinguishable \cite{Parvan:2016mbv}. This result strengthens the argumentation that the implementation of Tsallis approach on particle production at top RHIC and LHC energies should be conducted, carefully \cite{Tawfik:2017bul,Tawfik:2016pwz}.

At $y=0$ (mid-rapidity), Eqs. (\ref{TsallispT}) - (\ref{TsallispTfactor}) can be reformulated \cite{Parvan:2016rln,Parvan:2016mbv}. Also, for a given rapidity range $y_0<y<y_1$, the transverse momentum distribution can be extended to include an integration over $dy$ \cite{Parvan:2016rln,Parvan:2016mbv}. For a systematic fit of $p_{\mathtt{T}}$ distributions, we highlight that some experimental measurements are normalized to the geometrical factor $2 \pi p_{\mathtt{T}}$, while others are not. This might affect the given experimental uncertainties. We disregard this slight difference while comparing to our calculations.   

\paragraph{Generic and generic-factorized nonextensive statistics}

To introduce a generalized statistical approach to a large statistical system having various types of nonextensivity, such as the particle production at high energies, two asymptotic properties, each is associated with a scaling function, have been proposed \cite{Thurner:2010if}. Each scaling function is characterized by one critical exponent. These are $c$ for first and $d$ for second property, by which an equivalence class of entropies can be defined, uniquely, 
\bea
S_{c,d}[p] &=& \sum_{i}^{\Omega} {\cal A} \Gamma(d+1, 1 - c \ln p_i) + {\cal B} p_i, \label{eq:NewExtns1}
\eea
where $\Omega$ is the number of states, $\Gamma(a,b)=\int_{b}^{\infty}\, dt\, t^{a-1}\, \exp(-t)$ being incomplete $\Gamma$-function and ${\cal A}$ and ${\cal B}$ are arbitrary parameters.  In the limit $\Omega \rightarrow \infty$, each admissible system approaches one of these equivalence classes. It was concluded \cite{Thurner:2010if} that the universality classes $(c, d)$ not only introduce generic entropy and characterize it entirely, Eq. (\ref{eq:NewExtns1}), but also specify the particle distribution functions, 
\begin{eqnarray}
f_{c,d,r}(x) = \exp\left\{-\dfrac{d}{1-c}\left[\mathtt{W}_k\left(B\left(1-x/r\right)^{1/d}\right) - \mathtt{W}_k(B)\right]\right\}, \label{eq:ps1}
\end{eqnarray}
where $\mathtt{W}_k$ is $k$-th branch of Lambert-{$\mathtt{W}$} function, which has real solutions at $k=0$ for all classes with $d\geq 0$ and at $k=-1$ for $d<0$, as well. $B\equiv (1-c)r/[1-(1-c)r]\, \exp \left\{(1-c)r/[1-(1-c)r]\right\}$ with $r=(1-c+c d)^{-1}$. At $k=0$, the asymptotic expansion of Lambert-{$\mathtt{W}$} function reads
\bea
{\mathtt{W}}_{k=0}(x) &=& \sum_{n=1}^{\infty} \frac{(-1)^{n-1}\, n^{n-2}}{(n-1)!}\, x^n.
\eea

The properties of this new (non)extensivity entropy, Eq. (\ref{eq:NewExtns1}), lead to
\begin{eqnarray}
\dfrac{1}{1-c} &=& \lim_{N \rightarrow \infty} N\, \frac{\Omega^{\prime}}{\Omega}, \label{5} \\
d &=& \lim_{N \rightarrow \infty} \log \Omega\, \left(\dfrac{1}{N} \frac{\Omega}{\Omega^{\prime}}+c-1\right), \label{6}
\end{eqnarray}
while the number of microstates ($\Omega$) is related to the distribution function 
\begin{eqnarray}
\Omega(N) &=& \dfrac{1}{f_{c,d}(-\varphi\, c\, N)}
  \exp \left\{ \dfrac{d}{1-c} \mathtt{W}_k\left(\dfrac{(1-c)\exp [(1-c)/c\, d]}{c\, d} \left[ \dfrac{\varphi\, c\, N}{r} \right]^{1/d}\right) \right\}, \label{eq:states1}
\end{eqnarray}
where $\varphi$ is given by 
\begin{eqnarray}
\varphi = \dfrac{d}{d N}\, S_g &=& \Omega^{\prime}\, \left(g(1/\Omega)-\dfrac{1}{\Omega}\, g^{\prime} (1/\Omega)\right). \label{scd}
\end{eqnarray}

For BG statistical distribution, the probability can be expressed as
\bea
p_i &=& \frac{1}{z}\, f_{c,d,r}(x_i),
\eea
where $f_{c,d,r}(x_i)$ was given in Eq. (\ref{eq:ps1}) and the partition function can be constructed as follows.
\bea
z &=&  \sum_i f_{c,d,r}(x_i).
\eea
At finite temperature ($T$) and finite chemical potential ($\mu$), the single-particle distribution function  reads
\bea\langle n_{p\sigma}\rangle &=& \frac{1}{z} \sum_{\left\{n_{p\sigma}\right\}}  \frac{n_{p\sigma}}{\prod_{p\sigma} n_{p\sigma}!}\; f_{c,d,r}\left(\sum_{p\sigma} n_{p\sigma} X\right) = f_{c,d,r}(x_p),
\eea
where $x_p=(\epsilon_p-\mu)/T$, $\sum_{n_{p\sigma}} n_{p\sigma}\, f =f_p$, if $x\rightarrow x_p$ and 
\bea
f_{c,d,r}\left(\sum_{\left\{n_{p\sigma}\right\}} x \right) &=& \exp\left\{\frac{-d}{1-c}\left[W_k\left(B\left(1+\frac{\Lambda}{r T} - \sum_{n_{p\sigma}}\left(\frac{E_p-\mu}{T}\right)\right)^{1/d}\right)-W_k(B)\right]\right\}. \hspace*{8mm}
\eea
Then, the partition function can be given as,
\bea
z &=& \sum_{\left\{n_{p\sigma}\right\}}  \frac{1}{\prod_{p\sigma} n_{p\sigma}!}\; f_{c,d,r}\left(\sum_{p\sigma} n_{p\sigma} X\right) = \exp\left(\sum_{p\sigma} \langle n_{p\sigma}\rangle\right). \label{eq:PFcdr1}
\eea
The conversion from classical to quantum statistics is straightforward.
\bea
z &=& \pm \sum_{\left\{n_{p\sigma}\right\}}  \frac{1}{\prod_{p\sigma} n_{p\sigma}!}\; \left[f_{c,d,r}\left(1\pm \sum_{p\sigma} n_{p\sigma} X\right)^{-1} \right], \label{eq:PFcdr2}
\eea
where $\pm$ represent fermions and bosons, respectively. 

The averaged number  reads
\bea
\langle N \rangle &=& \sum_{n_{p\sigma}} \langle n_{p\sigma} \rangle.
\eea
To estimate the norm function for the proposed generic nonextensive statistics,
\bea
\sum_{\left\{n_{p\sigma}\right\}}  \frac{1}{\prod_{p\sigma} n_{p\sigma}!}\; f_{c,d,r}\left(x_i\right) &=&1,
\eea
we start with
\bea
\sum_{N=0}^{\infty} \int_0^{\infty} dE \, f_{c,d,r}\left(x\right) \, W_{N,E}=1,
\eea
where 
\bea
W_{N,E} &=& \sum_{\left\{n_{p\sigma}\right\}} \frac{1}{\prod_{p\sigma} n_{p\sigma}!} \;
\delta\left(\sum_{p\sigma} n_{p\sigma} -N\right) \;
\delta\left(\sum_{p\sigma} n_{p\sigma}\epsilon_p -E\right), \\
E &=& \sum_{n_{p\sigma}} \langle n_{p\sigma} \rangle \, \epsilon_p.
\eea
In a grand-canonical ensemble of a Maxwell-Boltzmann ideal gas, the might reexpress this as,
\bea
W_{N,E} &=& \frac{1}{N!} \left(\frac{g V}{\pi^2}T^3\right)^N\; \frac{E^{3N-1}}{\Gamma(3N)}.
\eea

Then, the transverse momentum distribution can be expressed as 
\bea
\frac{1}{2\pi p_{\mathtt{T}}} \frac{d^2 N}{dy dp_{\mathtt{T}}} &=& \frac{g V T}{8 \pi^3} m_{\mathtt{T}} \cosh y \sum_{N=0}^{N_0} \frac{\tilde{\omega}^N}{N!} h_0(0) \times \nonumber \\
&& \frac{r^{-1}\, (1-c)^{-1}\; W_0\left[B\left(1+\frac{\Lambda}{r T} - \frac{m_{\mathtt{T}} \cosh y - \mu(N+1)}{r T}\right)^{1/d}\right]}
{\left[T+\Lambda - m_{\mathtt{T}} \cosh y + \mu(N+1)\right] \left\{1+W_0\left[B\left(1+\frac{\Lambda}{r T} - \frac{m_{\mathtt{T}} \cosh y-\mu(N+1)}{r T}\right)\right]^{1/d}\right\}}. \hspace*{8mm}
\eea

\subsubsection{$p_{\mathtt{T}}$ spectra and kinetic freezeouts}

In this section, we discuss whether the $p_{\mathtt{T}}$ spectrum distributions are to be related to chemical or to kinetic freezeouts. The earlier is a late stage of the temporal evolution of high-energy collisions at which the particle abundances can be described by the equilibrium distribution functions determined by - at least - a set of three types of parameters, namely the temperatures ($T$), the baryon chemical potentials ($\mu$) and the fireball volumes ($V$). The number of produced particles is likely fixed, i.e. no further {\it chemical} processes take place or no longer change in the produced particles. At this stage, the produced particles would exist in their ground states as well as in various exited states. The kinetic freezeout is conjectured to take place at lower temperatures, e.g. the system cools down during a very late stage of the colliding system. The corresponding {\it effective} temperature describes the degree of excitation of the {\it elastically} interacting system. It is conjectured that during this stage thermal equilibrium in which the $p_{\mathtt{T}}$ spectrum distributions of the produced particles are no longer changed, is apparently reached. As mentioned, distinguishing between these two phases is a great objective to be achieved. We review recent studies aiming at proposing answers to the question which quantitatively characterizes which stage? 

From the statistical fits of the $p_{\mathtt{T}}$ spectrum distributions, various thermal parameters including the fireball geometry and the expansion velocity could be extracted. As elaborated in the previous section, extensive and nonextensive approaches were frequently utilized to characterize the particle production and throughout indirectly model the final state of the interacting system \cite{Tawfik:2014eba}.  Alternatively, this can also be achieved through {\it direct} comparison with calculations based on multisource thermal model, single freezeout scheme \cite{Stachel:2013zma}, and incoherent-multiple-collision model \cite{imcm}, for instance. It was pointed out \cite{Danielewicz:1984ww} that both Blast-wave \cite{Siemens:1978pb} and thermal fireball models \cite{Westfall:1976fu} seem not capable to model the $p_{\mathtt{T}}$ momentum spectra. Models inspired by Hydrodynamics \cite{Schnedermann:1993ws} are assumed to propose a good estimation for the kinetic freezeout parameters; kinetic temperature ($T_{\mathtt{kin}}$) and the average transverse radial flow velocity [$\langle\beta\rangle=2\beta_S/(2+n)$] with $\beta_S$ is the surface velocity and $n$ is the exponent of the flow profile and thus offer fruitful information about the collision dynamics. Such models propose that the produced particles are locally thermalized at $T_{\mathtt{kin}}$ and flow with a common $\langle\beta\rangle$,
\bea
\frac{d N}{p_{\mathtt{T}} dp_{\mathtt{T}}} &\propto& \int_0^R r\, dr\, m_{\mathtt{T}}\, I_0\left(\frac{p_{\mathtt{T}}}{T_{\mathtt{kin}}} \sinh \rho(r)\right) K_1\left(\frac{m_{\mathtt{T}}}{T_{\mathtt{kin}}} \cosh \rho(r)\right), 
\eea
where $m_{\mathtt{T}}=(p_{\mathtt{T}}^2+m^2)^{1/2}$ is the transverse mass with $m$ being mass of the particle of interest. $I_0$ and $K_1$ are modified Bessel functions. $\rho(r)=\tanh^{-1}\beta$, where $\beta=\beta_S (r/R)^n$ and $r/R$ is the relative radial position in the thermal source. It should be noted that some of these approaches assumes extensive statistics.

It is apparent that the produced particles are considerably affected by the dynamics of the interacting system. Various observations support unambiguously this conclusion. With these regards, it was noticed that the shape of the momentum distributions depend on the freezeout hypersurface \cite{Anderlik:1998cb}. From a hydrodynamic description, i.e. modeling the system as an expanding hadron fluid, both strength and duration of the expansion can be characterized and an effective equation of state could be determined \cite{Lokhtin:1996ht}. 

Another finding connected with the statistical fits of the $p_{\mathtt{T}}$ momentum spectra is the so-called nonextensive effective temperature \cite{biroBook,Tawfik:2016pwz}. Following the proposal that the effective temperature differs from the real temperature, both effective and real temperatures were extracted from mean transverse flow velocity and mean flow velocity of produced particles \cite{Wei:2016ihj}. Their dependence on rest and moving masses, centralities, and center-of-mass energies could be extracted, as well \cite{Wei:2016ihj,Lao:2015zgd,Huovinen:2006jp}. It was argued that both types of temperatures can be related to each other \cite{Tawfik:2016pwz}. Such an extrapolation shall be elaborated in section \ref{sec:extrapolateT}.

\subsection{Empirical parameterizations}

This section reviews another alternative that the transverse momentum  $p_{\mathtt{T}}$ distributions of different charged particles produced at various energies can be fitted statistically without conciliating to a concrete statistical approach. On one hand, experimentalists mostly aim at a best description of their experimental results apart from the possible complications and the constrains of some theoretical approaches. On the other hand, the resulting parametrizations for different $p_{\mathtt{T}}$  distributions become precise, so that they turn to offer excellent descriptions for the characterization of the $p_{\mathtt{T}}$ distributions. One of the advantages of these parametrizations is that the system {\it almost} alone manifest its degree of extensivity or nonextensivity. For instance, when increasing the energy, the system produced couldn't conserve the same degree of extensivity or nonextensivity as assumed when implementing Boltzmann or Tsallis approach. For a better comparison, we divide the various experimental parametrizations according the system sizes.

\subsubsection{A$+$A collisions}

It is well known that the STAR experiment - among others - runs a successful program for the $p_{\mathtt{T}}$  distributions of the well-defined charged particles ($\pi^{\pm}$, $K^{\pm}$, $p$ and $\bar{p}$) produced in Au$-$Au collisions \cite{Adamczyk:2017iwn,Abelev:2008ab}. Recently, a systematic beam energy scan at $7.7$, $11.5$, $19.6$, $27$, and $39~$GeV was reported \cite{Adamczyk:2017iwn} for $0.2<p_{\mathtt{T}}<2~$GeV. These results are depicted in Figs. $12$-$16$. The results at higher energies shall be discussed, as well. We first introduce three examples on parametrizations proposed by the STAR collaboration \cite{Adamczyk:2017iwn}. At mid-rapidity $y<0.1$ and in (GeV/c)$^2$ units, the $p_{\mathtt{T}}$ distributions have been fitted to Bose-Einstein, $m_{\mathtt{T}}$-exponential, and double-exponential, respectively,
\bea
\frac{1}{2 \pi p_{\mathtt{T}}}\frac{d^2\, N}{dp_{\mathtt{T}}\, dy} & = & c_{\rm BE} \left[\exp\left(\frac{m_{\mathtt{T}}}{T_{\rm BE}}\right)-1\right]^{-1},  \label{eq:BE1}\\
\frac{1}{2 \pi p_{\mathtt{T}}}\frac{d^2\, N}{dp_{\mathtt{T}}\, dy} & = & c_{m_{\mathtt{T}}}  \exp\left[\frac{-(m_{\mathtt{T}}-m)}{T_{m_{\mathtt{T}}}}\right], \label{eq:mT1}\\
\frac{1}{2 \pi p_{\mathtt{T}}}\frac{d^2\, N}{dp_{\mathtt{T}}\, dy} & = & c_1 \exp\left(\frac{-p^2_{\mathtt{T}}}{T_1^2}\right) + c_2 \exp\left(\frac{-p^2_{\mathtt{T}}}{T_2^2}\right). \label{eq:DE1}
\eea 
Other parametrizations such as Boltzmann $\propto m_{\mathtt{T}}\exp(-m_{\mathtt{T}}/T)$ and $p_{\mathtt{T}}$-exponential $\propto \exp(-p_{\mathtt{T}}/T)$ have been proposed, as well \cite{Adamczyk:2017iwn}. It is obvious that these expressions suggest various power-scales. Depending on the resulting parameters, which are summarized in Tab \ref{tab:1}, one would be able to favor one or another power-scale. It would be of great interest to compare between them and the ones proposed by Tsallis statistics. The latter is blindly applicable in low as well as in high $p_T$ regions and therefore was categorically criticized, for instance in ref. \cite{Bialas:2015pla}. We also compare these fits with the ones using the same parametrizations but at the energies $62.4$, $130$, and $200$ \cite{Abelev:2008ab}. The results are added to Tab. \ref{tab:1}, as well.

\begin{table}
  \begin{tabular}{|c|c|r|r|r|r|r|r|r|r|}
    \hline
     &  & $7.7$ GeV & $11.5$ GeV & $19.6$ GeV & $27$ GeV & $39$ GeV & $62.4$ GeV &$130$ GeV & $200$ GeV \\ \hline
    \multirow{2}{*}{$\pi^+$} & $c_{\rm BE}$ GeV$^{-2}$ & $331.804$ & $398.084$ & $479.707$ & $493.251$ & $474.231$ & $575.202$ & $167.267$ & $422.939$\\
    \cline{2-10} & $T_{\rm BE}$ GeV & $0.204$ & $0.211$ & $0.217$ & $0.222$ & $0.230$ & $0.25$ & $0.236$ & $0.259$\\  \hline
    \multirow{2}{*}{$\pi^-$} & $c_{\rm BE}$ GeV$^{-2}$ & $370.501$ & $433.36$ & $498.458$ & $521.712$  & $497.888$ & $741.325$ & $159.963$ & $560.647$\\
      \cline{2-10} & $T_{\rm BE}$ GeV & $0.200$ & $0.207$ & $0.216$ & $0.220$ & $0.227$ & $0.228$ & $0.23$ & $0.243$\\  \hline
     \multirow{2}{*}{$K^+$} & $c_{m_{\mathtt{T}}}$ GeV$^{-2}$ & $20.981$ & $24.475$ & $27.561$ & $28.117$ & $28.159$ & $155.855$ & $80.419$ & $192.892$ \\
         \cline{2-10} & $T_{m_{\mathtt{T}}}$ GeV & $0.221$  & $0.227$ & $0.234$ & $0.240$ & $ 0.245$ & $0.269$ & $0.25$ & $0.275$\\  \hline
       \multirow{2}{*}{$K^-$} & $c_{m_{\mathtt{T}}}$ GeV$^{-2}$ & $8.662$ & $13.099$ & $18.165$ & $21.645$ & $22.070$ & $137.548$ & $63.217$ & $178.744$\\
      \cline{2-10} & $T_{m_{\mathtt{T}}}$ GeV & $0.203$ & $0.212$ & $0.228$ & $0.230$ & $0.244$ & $0.27$ & $0.25$ & $0.275$\\  \hline
      \multirow{4}{*}{$p$} & $c_1$ GeV$^{-2}$ & $20.923$ & $17.462$ & $7.949$ & $10.947$ & $1.533$ & $5.205$ & $2.87$ & $2,957$ \\
      \cline{2-10} & $T1$ GeV &  $0.905$ & $0.899$ & $0.999$ & $0.947$ & $1.218$ & $0.822$ & $0.827$ & $0.98$ \\  
      \cline{2-10} & $c_2$ GeV$^{-2}$ &  $19.429$ & $49.943$ & $6.136$ & $ 0.009$ & $7.641$ & $2.202$ & $0.336$ & $0.046$\\  
      \cline{2-10} & $T2$ GeV &  $0.00062$ & $0.00500$ & $0.72454$ & $0.02321$ & $0.89727$ & $1.193$ & $1.249$ & $1.095$ \\  \hline
       \multirow{4}{*}{$\bar{p}$} & $c_1$ GeV$^{-2}$ & $0.154$ & $0.579$ & $1.490$ & $ 0.0026$ & $2.819$ & $2.323$ & $1.463$ & $1.523$ \\ 
       \cline{2-10} & $T_1$ GeV & $0.907$ & $0.894$ & $0.945$ & $0.0024$ & $0.981$ & $0.851$ & $0.88$ & $0.928$\\  
       \cline{2-10} & $c_2$ GeV$^{-2}$ & $0$ & $0$ & $0.000095$ & $2.070$ & $0.00011$ & $1.208$ & $0.37$ & $1.074$\\
       \cline{2-10} & $T_2$ GeV & $0.40511$ & $0.51028$ & $0.08611$& $0.96683$ & $0.09363$ & $1.179$ & $1.224$ & $1.084$ \\ \hline
     \end{tabular}
     \caption{Proportionality constants ($c$) and inverse slopes ($T$) obtained from the statistical fits of $p_{\mathtt{T}}$ spectra distributions for the well-identified pions, Kaons, protons, and their-particles measured in central Au$-$Au central collisions ($0–5\%$) at mid-rapidity ($y<0.1$) and energies $7.7$, $11.5$, $19.6$, $27$, $39$,\cite{Adamczyk:2017iwn}, $62.4$, $130$, and $200$ \cite{Abelev:2008ab} to Bose-Einstein [Eq. (\ref{eq:BE1})], $m_{\mathtt{T}}$-exponential [Eq. (\ref{eq:mT1})], and double-exponential functions [Eq. (\ref{eq:DE1})], respectively. 
     \label{tab:1} }
\end{table}

Table \ref{tab:1} summarizes the results on the proportionality constants ($c$) and the inverse slopes ($T$) deduced from statistical fits of Bose-Einstein, Eq. (\ref{eq:BE1}), $m_{\mathtt{T}}$-exponential, Eq. (\ref{eq:mT1}), and double-exponential functions, Eq. (\ref{eq:DE1}) to the $p_{\mathtt{T}}$ spectra measured in the STAR experiment in $0-5\%$ collisions for pions, Kaons, and protons and their anti-particles, respectively. It is obvious that the inverse slopes, $T$, of the $p_T$ spectra of pions are smaller than that of the Kaons, which in turn are smaller than that of the protons.  Also, it is obvious that
\begin{itemize}
\item there is a general pattern observed that $T$ increases with the increase in the center-of-mass energies,
\item the resulting $T$ obtained from $p_{\mathtt{T}}$ spectra of pions, Kaons, and protons (particles)  are slightly greater than that from their counterparts (anti-particles), respectively, and
\item especially for protons and anti-protons, the increase of $T$ with the center-of-mass energies might be monotonic except at $19.6~$GeV or at $27~$GeV, where a rapid increase or decrease is registered. With this regard, we recall that this region of the center-of-mass energies shall be precisely scanned in the forthcoming energy scan program of the STAR experiment as it might reveal interesting new physics! 
\end{itemize}

For nonextensive Tsallis and generic axiomatic statistics, we notice that
\begin{itemize}
\item the temperature $T$ increases with the increase in the center-of-mass energies $\sqrt{s_{\mathtt{NN}}}$ except (with an exception at $200~$GeV) for all particles except for pions, 
\item for pions there is a inverse proportionality between $T$ and $\sqrt{s_{\mathtt{NN}}}$, and
\item the resulting $T$ from  $p_{\mathtt{T}}$ spectra of the anti-particles are slightly greater than that from $p_{\mathtt{T}}$ spectra of their particles.
\end{itemize}
It is obvious that the values of the resulting temperatures range between $0.204$ and $0.245~$GeV for pions and Kaons. But for protons, there are two sets of resulting temperatures, first term of Eq. \ref{eq:DE1} results in $0.894<T<1.218~$GeV, while the second term gives $0.00062<T<0.96683~$GeV. 

For pions and Kaons, the temperatures determined are found greater than the freeze-out temperatures which means that both transverse momentum spectra are nearly stemming from the hadronization phase. But in case of protons, the temperatures are very large. There is a huge difference between these and the freeze-out temperatures. Accordingly, we conclude that the transverse momentum distributions for protons are likely to be stemming from the transition phase quark-hadron, i.e. earlier than the hadronization phase.

On the other hand, we notice that the proportionality constants seem to have a monotonic increase with the center-of-mass energy, especially for pions and Kaons, but not for protons, especially at $19.6~$GeV or at $27~$GeV. This would be interpreted that at these two energies both thermal and chemical freezeout temperatures likely exceed the one deduced from Blast-wave and thermal models, respectively. In other words, the overestimation becomes greater at $19.6~$GeV or at $27~$GeV pointing to new physics that both sets of statistical approaches become distinguishable. This is another phenomenological observation supporting the idea that this energy range deserves finer analysis. Future facilities such as the Facility for Antiproton and Ion Research (FAIR) at GSI, Darmstadt, Germany and  the Nuclotron-based Ion Collider fAcility (NICA) at JINR, Dubna, Russia are designed - among others - to cover such an energy range. Also, the RHIC Beam Energy Scan program in the STAR experiment  targets this energy region, as well.

Prior to the results reported in ref. \cite{Adamczyk:2017iwn}, STAR collaboration published detailed $p_{\mathtt{T}}$ spectrum distributions of $\pi^{\pm}$, $K^{\pm}$, $p$ and $\bar{p}$, at $62.4$, $130$, and $200~$GeV in Au$-$Au collisions at mid-rapidity \cite{Abelev:2008ab}. Besides parametrizations based on $p_{\mathtt{T}}$-exponential, $m_{\mathtt{T}}$-exponential, and Boltzmann expressions, other expressions have been proposed, as well, such as $p_{\mathtt{T}}$-Gaussian and $p_{\mathtt{T}}^3$-exponential, 
\bea
\frac{d N}{p_{\mathtt{T}}\, dp_{\mathtt{T}}} & = & c_{p_{\mathtt{T}}^2} \exp\left(-\frac{p^2_{\mathtt{T}}}{T_{p_{\mathtt{T}}^2}^2}\right), \label{eq:NpT1} \\
\frac{d N}{p_{\mathtt{T}}\, dp_{\mathtt{T}}} & = & c_{p_{\mathtt{T}}^3} \exp\left(-\frac{p^3_{\mathtt{T}}}{T_{p_{\mathtt{T}}^3}^3}\right).  \label{eq:NpT2}
\eea
We would like to emphasize that the parameters in Eqs. (\ref{eq:BE1}) - (\ref{eq:DE1}) aree determined and listed out in Tab. \ref{tab:1}.

\begin{table}
  \begin{tabular}{|c|c|c|c|c|}
    \hline
     &  & $62.4$ GeV & $130$ GeV & $200$ GeV \\ \hline
    \multirow{2}{*}{$\pi^+$} & $c_{p_{\mathtt{T}}}$-exponential GeV$^{-2}$ & $897.773$ & $1071.82$ & $1128.81$ \\
    \cline{2-3} & $T_{p_{\mathtt{T}}}$-exponential GeV & $0.1996$ & $0.201$ & $0.209$ \\  \hline
    \multirow{2}{*}{$\pi^-$} & $c_{p_{\mathtt{T}}}$-exponential GeV$^{-2}$ & $916.824$ & $1086.12$ & $1175.46$ \\
    \cline{2-3} & $T_{p_{\mathtt{T}}}$-exponential GeV & $0.1996$ & $0.201$ & $0.206$ \\  \hline
     \multirow{2}{*}{$\pi^+$} & $c_{\rm BE}$ GeV$^{-2}$ & $759.95$ & $938.961$ & $937.542$ \\
     \cline{2-3} & $T_{\rm BE}$ GeV & $0.21$ & $0.208$ & $0.22$ \\  \hline
    \multirow{2}{*}{$\pi^-$} & $c_{\rm BE}$ GeV$^{-2}$ & $702.611$ & $840.006$ & $856.455$ \\
    \cline{2-3} & $T_{\rm BE}$ GeV & $0.221$ & $0.2213$ & $0.233$ \\  \hline
     \multirow{2}{*}{$K^+$} & $c_{m_{\mathtt{T}}}$ GeV$^{-2}$ & $184.534$ & $201.305$ & $149.856$ \\
      \cline{2-3} & $T_{m_{\mathtt{T}}}$ GeV & $0.27$  & $0.284$ & $0.322$ \\  \hline
       \multirow{2}{*}{$K^-$} & $c_{m_{\mathtt{T}}}$ GeV$^{-2}$ & $159.003$ & $178.227$ & $124.089$ \\
      \cline{2-3} & $T_{m_{\mathtt{T}}}$ GeV & $0.27$ & $0.284$ & $0.35$ \\  \hline
      \multirow{4}{*}{$p$} & $c_1$ GeV$^{-2}$ & $3.119$ & $2.738$ & $3.055$ \\
      \cline{2-3} & $T_1$ GeV &  $0.982$ & $0.991$ & $1.107$ \\  
      \cline{2-3} & $c_2$ GeV$^{-2}$ &  $4.98$ & $4.233$ & $4.125$ \\  
      \cline{2-3} & $T_2$ GeV &  $1.19$ & $1.324$ & $1.438$ \\  \hline
       \multirow{4}{*}{$\bar{p}$} & $c_1$ GeV$^{-2}$ & $2.32$ & $1.462$ & $1.523$ \\ 
       \cline{2-3} & $T_1$ GeV & $1.122$ & $1.182$ & $1.2$ \\  
       \cline{2-3} & $c_2$ GeV$^{-2}$ & $1.33$ & $3.44$ & $4.102$ \\
       \cline{2-3} & $T_2$ GeV & $1.2$ & $1.224$ & $1.25$\\ \hline
      \multirow{2}{*}{$p$} & $c_{p_{\mathtt{T}}}$-Gaussian GeV$^{-2}$ & $8.103$ & $6.879$ & $7.079$ \\
      \cline{2-3} & $T_{p_{\mathtt{T}}}$-Gaussian GeV &  $1.101$ & $1.225$ & $1.278$ \\  
      \multirow{2}{*}{$\bar{p}$} & $c_{p_{\mathtt{T}}}$-Gaussian GeV$^{-2}$ & $3.535$ & $4.798$ & $5.503$ \\
      \cline{2-3} & $T_{p_{\mathtt{T}}}$-Gaussian GeV &  $1.188$ & $1.248$ & $1.266$ \\ \hline 
     \end{tabular}
\caption{The proportionality constants ($c$) and the inverse slopes ($T$) obtained from the statistical fits of $p_{\mathtt{T}}$ spectra distributions for the well-identified pions, Kaons, protons, and their anti-particles measured in Au$-$Au central collisions ($0-5\%$) at mid-rapidity ($y<0.1$) and energies $62.4$, $130$, and $200~$GeV \cite{Abelev:2008ab} by using different parametrizations given in Eqs. (\ref{eq:BE1}), (\ref{eq:mT1}), (\ref{eq:DE1}), (\ref{eq:NpT1}), (\ref{eq:NpT2}).
\label{tab:2}
}
\end{table}

Table \ref{tab:2} shows the results on the effective temperatures as determined from the charged particles and their anti-particles in central Au-Au collisions at $62.4$, $130$, and $200~$GeV \cite{Abelev:2008ab} by using the parametrizations given in Eqs. (\ref{eq:BE1}), (\ref{eq:mT1}), (\ref{eq:DE1}), (\ref{eq:NpT1}), (\ref{eq:NpT2}). We notice that for all such particles and anti-particles there is an increase of the effective temperatures with the increase in the energies. On the other hand, the results for pions are lower than the ones for Kaons, which in turn are lower than the ones for protons and anti-protons.

\subsubsection{p$+$p collisions}

Due to the non-Abelian energy loss of parent parton penetrating through dense medium (jet quenching), which is likely available in A$+$A collisions, the possible suppression in high $p_{\mathtt{T}}$ spectra of leading hadrons, such as pions, Kaons and protons in A$+$A- compared to p$+$p-collisions was proposed as canonical signatures for the quark-gluon plasma (QGP) formation \cite{Gyulassy:2003mc}. While it was concluded \cite{dEnterria2005} that the high $p_{\mathtt{T}}$ spectra of hadrons produced in central Au$+$Au collisions at top RHIC energies, $\sqrt{s_{NN}}=130$, and $200~$GeV, are found strongly suppressed by a factor of $4-5$ \cite{Adcox:2001jp,Adler:2002xw,Adler:2003qi,Adams:2003kv,Back:2003qr,Arsene:2003yk} comparing to the results from p$+$p-collisions at the same energies \cite{Adams:2003kv,Adler:2003pb}, it was reported that the high $p_{\mathtt{T}}$ spectra of pions produced in central A$+$A-collisions at CERN-SPS energies are enhanced relative to their p$+$p counterparts \cite{Aggarwal:2001gn,ReffInt15b,Aggarwal:1998vh,Wang:1998hs,Wang:1998ww,Wang:2001cy}.
The later would manifest the {\it ''Cronin effect''} in the hadron production, which was first observed in $1975$ \cite{Cronin:1974zm} and effectively refers to an enhancement in the high $p_{\mathtt{T}}$ spectra due to multiple interactions  produced off $pA$- and A$+$A-collisions, e.g. bound nucleons seem to accumulate leading to the high $p_{\mathtt{T}}$ spectra of the leading hadrons. On the other hand, the suppression observed at top RHIC energies can be understood due to two phenomena \cite{Kopeliovich:2002yh}. The first one is the multiple interactions within the colliding heavy-ions, {\it ''Cronin effect''}. The second one describes the final state interactions with the dense medium. Accordingly, the dense medium properties could be characterized, when the Cronin effect for such nuclear collisions can be determined, precisely.

From this short review on the results of the high $p_{\mathtt{T}}$ from p$+$p- and A$+$A-collisions, we are now able to summarize that the suppression and the enhancement of $p_{\mathtt{T}}$ spectra of the leading hadrons measured in A$+$A-collisions relative to the ones in p$+$p-collisions sheds light on the Cronin effect, the medium modification and signatures for the QGP formation, in nuclear matter. Let us first recall some proposals for the dynamics behind the Cronin effect. A parameter-free approach which describes well the available experimental data was suggested in ref. \cite{Kopeliovich:2002yh}. It is based on an assumption that the mechanism of the multiple interactions should be affected by the collision energy, on one hand. At low collision energies, high $p_{\mathtt{T}}$ of partons are incoherently produced off different nucleons, while at high collision energies their production becomes coherent, i.e. the coherence length, the distances along a projectile momentum direction, linearly increases with the energy \cite{Gribov:1967hh}
\bea
l_c &=& \frac{\sqrt{s_{\rm NN}}}{m_{\rm N}\; k_{\mathtt{T}}},
\eea 
where $k_{\mathtt{T}}$ is the  transverse momentum of the parton which is produced at midrapidity. Later on, such these partons likely undergo confinement phase transition forming hadrons that shall be detected and their $p_{\mathtt{T}}$ measured. This expression could be interpreted by means of the Heisenberg uncertainty principle and the QCD renormalizability as follows \cite{Geiger:1995ak}. If $l_c$ becomes shorter than the averaged internucleon separation, then the projectile interacts incoherently with individual nucleons similar to the p$+$p scattering, otherwise the interaction is coherent \cite{Ayala:1995kg,Geiger:1995ak}. On the other hand, if the coherence length becomes short, then the possible broadening in the $p_{\mathtt{T}}$ spectrum distribution \cite{Liang:2004ph}, which might be produced due to initial and/or final interactions, should not be interpreted as a medium effect on the parton distribution of the nucleus \cite{Kopeliovich:2002yh}. The possible broadening is momentum spectra was proposed as a signature for the QGP formation \cite{Liang:2004ph}. If $l_c$ becomes longer than the nuclear radius, it is conjectured that all amplitudes, for instance of $\lambda$'s of $x$'s, interfere, coherently \cite{Ayala:1995kg,Geiger:1995ak}. This leads to a collective parton distribution of the nucleus. The amplitudes with large parton momentum related to large $l_c\sim (m_{\rm N} x)^{-1}$ likely overlap and thus have no correlations with individual nucleons. This means that factorization can be applied, while the parton distribution becomes modified \cite{Diehl:2017wew,Kasemets:2017vyh}.

We have briefly reviewed the reproduction of the $p_{\mathtt{T}}$ spectra measured in A$+$A and p$+$p collisions at top RHIC and at SPS energies. A strong suppression is found in the earlier, while an enhancement is to be concluded in the latter \cite{dEnterria2005}. For the seek of completeness, we recall that the Cronin effect could be well confirmed in fixed target $pA$-collisions at Fermilab energies $20-40~$GeV \cite{Antreasyan:1978cw,Straub:1992xd} and in $\alpha\alpha$-collisions at ISR energies $31~$GeV \cite{Angelis:1985fk}.

Now, we can review the various parameterizations for $p_{\mathtt{T}}$ distributions:
\begin{itemize}
\item First, we recall the five-parameter functional form for the inclusive cross-section distribution at ISR energies \cite{dEnterria2005}
\bea
E\frac{d^3\, \sigma_{\mathtt{pp}\rightarrow\pi\mathtt{X}}}{dp^3} &=& A\left[\exp\left(a p_{\mathtt{T}}^2 + b p_{\mathtt{T}}\right)+\frac{p_{\mathtt{T}}}{p_{\mathtt{0}}}\right]^{-n},
\eea
where $A=265.1~$mb GeV$^{-2}$, $a=-0.0129~$GeV$^{-1}$, $b=0.049~$GeV, $p_{\mathtt{0}}=2.639~$GeV, and $n=17.95$. 

\item Second, from the different parametrizations utilized at $0.9$, $2.76$ and $7~$TeV, we start with  the modified Hagedorn function. This was utilized by the ALICE collaboration for a better parametrization for low $p_{\mathtt{T}}$ differential cross-section \cite{Abelev:2013ala}, 
\bea
\frac{1}{2 \pi p_{\mathtt{T}}} \frac{d^2 \sigma_{\mathtt{ch}}^2}{dp_{\mathtt{T}}\, d\eta} &=& A \frac{p_{\mathtt{T}}}{m_{\mathtt{T}}} \left(1+\frac{p_{\mathtt{T}}}{p_{\mathtt{T,0}}}\right)^{-n}, 
\eea 
where $n$ is a $q$-like exponent. It was concluded that this expression behaves, at high $p_{\mathtt{T}}$, as power law (asymptotic), while at low $p_{\mathtt{T}}$ the quantities within the brackets, $(1+p_{\mathtt{T}}/p_{\mathtt{T,0}})^{-n}$, represent an exponential function with an inverse slope parameter $p_{\mathtt{T,0}}/n$. The differential cross section, in left-hand side,  $d^2 \sigma_{\mathtt{ch}}^2/dp_{\mathtt{T}}\, d\eta$, was measured as $\sigma_{\mathtt{MB_{OR}}}^{\mathtt{NN}}$ multiplied by the number of charged particles per event differential yield of charged particles in minimum bias collisions, $d^2 N_{\mathtt{ch}}^{\mathtt{MB_{OR}}}/dp_{\mathtt{T}}\, d\eta$.

\item Third, the ALICE results on differential cross-sections have been confronted to NLO-pQCD calculations \cite{Sassot:2010bh}. The scaling exponent $n$ was estimated by comparing the spectra $x_{\mathtt{T}} = 2p_{\mathtt{T}}/\sqrt{s_{\mathtt{NN}}}$ at different center-of-mass energies $\sqrt{s_{\mathtt{NN}}}$ and $\sqrt{s_{\mathtt{NN}}^\prime}$ for fixed $p_{\mathtt{T}}$, 
\begin{equation}
n(x_T)=-
\frac{\ln \left[ \sigma_{\mathtt{inv}}(s_{\mathtt{NN}},x_{\mathtt{T}})/\sigma_{\mathtt{inv}}(s_{\mathtt{NN}}^{\prime},x_{\mathtt{T}}) \right]}
{\ln(\sqrt{s_{\mathtt{NN}}}/\sqrt{s_{\mathtt{NN}}^{\prime}})},
\label{eq:nexp}
\end{equation}
where the quantities with prime are the ones compared to the quantities without prime.

\item Fourth,  the LHC results on $p_{\mathtt{T}}$ spectra of well-identified and unidentified particles have been parametrized, as well  \cite{Chatrchyan:2012qb,Aamodt:2011zj,Khachatryan:2010xs,Khachatryan:2010us,Adams:2004ep}. The generic parameterization reads
\bea
\frac{d^2 N}{dy\, dp_{\mathtt{T}}} &=& p_{\mathtt{T}}\, \frac{d N}{dy}\, \frac{(n-1)(n-2)}{n T [n T+(n-2)m]}    \left[1+\frac{m_{\mathtt{T}}-m}{n T}\right]^{-n}, \label{eq:lavy}
\eea
where $n$ is an exponent (though being equivalent to Tsallis $q$-parameter) and $T$ gives an inverse slope parameter (equivalent to temperature). $d N/dy$ is the yield distribution. From experimental point-of-view, the systematic errors count possible contributions from all individual detectors, overall normalization errors, and the uncertainties in the extrapolation to large $p_{\mathtt{T}}$. For pions, which are the lowest Goldstone bosons and accordingly more abundant than other particles, at mid-rapidity, $p/E\equiv p_{\mathtt{T}}/m_{\mathtt{T}}$. Accordingly, the rhs of the previous expression should be multiplied by  $p_{\mathtt{T}}/m_{\mathtt{T}}$ for pions, especially. This type of parametrization is known as Tsallis-Pareto or Tsallis-Levy \cite{Biro:2008hz}. Through a private communication, Tsallis is rejecting that this has any thing to do with the Tsallis-type of statistics. The various fit parameters are detailed in the Tables 4 and 5 of ref. \cite{Aamodt:2011zj}. For the seek of comparison, we give in Tab. \ref{tab:3} the results reported in ref. \cite{Aamodt:2011zj}.
\end{itemize}

\subsection{Other models for $p_{\mathtt{T}}$ spectra distributions}
\label{sec:othrm}

This is another alternative to describe the experimental results. The system {\it almost} alone manifest its degree of extensivity or nonextensivity. The modeling proposed likely empowers the statistical system with such ability. This can be seen when the energy goes from few GeV to few TeV, the produced system accordingly should have different degrees of extensivity or nonextensivity and this should be reflected in the statistical approach applied.

Assuming that many emission sources are formed in the high energy collisions, a multisource thermal model with different interacting mechanisms and different detection samplings has been proposed \cite{Liu:2008am,Liu:2008ar,Liu:2014nra}. It was conjectured that the sources of one group are in a local equilibrium. This allows to apply singular distributions and to assign to the multisources one common temperature and same degrees of freedom, as well. It is apparent that the emission of the produced particles off the multisources constructs the final-state spectrum, which can be characterized statistically by a multi-component distribution law \cite{YA-QIN:2012}. In light of this, the authors of ref. \cite{Wei:2015yut} proposed that a multicomponent Erlang $p_{\mathtt{T}}$ spectra distribution estimates the mean $p_{\mathtt{T}}$ of each group and Tsallis statistics determined the effective temperature of the whole interacting system, which may have group-by-group fluctuations in different local thermal equilibrium. To this end, we recall first the  assumption that the particles generated off one emission source obeys an exponential function of $p_{\mathtt{T}}$ spectra distribution \cite{Liu:2014nra,Wei:2016ihj}
\bea
f_{i j} (p_{{\mathtt{T}}_{i j}}) &=& \frac{1}{\langle p_{{\mathtt{T}}_{i j}}\rangle}\; \exp\left[\frac{p_{{\mathtt{T}}_{i j}}}{\langle p_{{\mathtt{T}}_{i j}}\rangle}\right],
\eea
where $p_{{\mathtt{T}}_{i j}}$ is $p_{\mathtt{T}}$ spectra stemming from $i$-th source in $j$-th group and $\langle \cdots\rangle$ gives the mean value. When summing up over $N_j$ sources in $j$-th group, the single-component Erlang distribution could be constructed as
\bea
f_{j} (p_{\mathtt{T}}) &=& \frac{p_{\mathtt{T}}^{N_j-1}}{(N_j-1)!\; \langle p_{{\mathtt{T}}_{i j}}\rangle^{N_j-1}}\; \exp\left[\frac{p_{{\mathtt{T}}_{i j}}}{\langle p_{{\mathtt{T}}_{i j}}\rangle}\right],
\eea
where $p_{\mathtt{T}}$ stands for transverse momentum contributed by $N_j$ sources. When summing up over all groups, the multi-component Erlang distribution is obtained
\bea
f_E(p_{\mathtt{T}}) &=& \sum_{j=1}\,  w_j\; f_{j} (p_{\mathtt{T}}),
\eea
where $w_j$ is the relative weight of $j$-th groups. The mean transverse momentum reads
\bea
\langle p_{{\mathtt{T}}}\rangle &=& \sum_{j=1} w_j\, N_j \langle p_{{\mathtt{T}}_{i j}}\rangle.
\eea

It was assumed that the temperature deduced from multi-component Erlang and Tsallis fits to mean and transverse spectra of various produced particles at RHIC and LHC energies is identical to the kinetic one characterizing thermal freezeout \cite{Wei:2016ihj}. The flow velocity could also be extracted. Various colliding systems including $p+p$, Cu$+$Cu, Au$+$Au, Pb$+$Pb, and p$+$Pb at different centralities have been analyzed \cite{Wei:2016ihj}. To all these systems, the multisource thermal model was also utilized. The effective temperature and the real (physical) temperature could be determined. The latter is likely identical to the thermal freezeout temperature of the interacting system. The mean transverse flow velocity and the mean flow velocity of produced particles, as well as, relationships among these quantities were extracted, as well. Furthermore, this extensive study determined the dependence of the effective temperature and the mean and the transverse momentum distributions on the rest mass, the moving mass, the centrality, and the center-of-mass energy. Also, the dependence of the thermal freezeout temperature and the mean and the transverse flow velocity on centrality, center-of-mass energy, and system size was analyzed \cite{Wei:2016ihj}.

The rapidity and the energy dependences of the transverse momentum spectra for charged particles in p$+$p collisions were analyzed by using two types of Tsallis-approaches, e.g. with and without thermodynamic description, where experimental results from the STAR, PHENIX, ALICE, and CMS experiments could be well reproduced \cite{Zheng:2015mhz}. It was found that the temperatures obtained with the thermodynamic description is smaller than the ones without such description \cite{Zheng:2015mhz}. In the Tsallis distribution with thermodynamical descriptions, there is an extra term $m_T$ which is responsible for the discrepancies of the temperatures from the other type of Tsallis-approaches.

The experimental results from light flavour particles; $p$, $\pi$, $K$ and their anti-particles measured in Au$+$Au at $200~$GeV and from strange particles; $K_S^0$, $\Lambda$, $\Xi$, $\Omega$ measured in Cu$+$Cu at $200~$GeV are confronted to the multi-source thermal model. On the other hand, it was found that the results of $p$, $\pi^+$, $K^+$, $\phi$ from Pb$+$Pb at $2.76~$TeV, as well as $\pi^++\pi^-$, $K^++K^-$, $p+\bar{p}$, $\Lambda+\bar{\Lambda}$, $K_S^0$ from p$+$Pb at $5.02~$TeV, and $p$, $\pi^+$, $K^+$, $\Lambda$, $\phi$, $\Sigma^-+\bar{\Sigma^+}$ at $0.9$ and $7~$TeV from p$+$p collisions, which apparently combine different colliding systems and different energies, are well described by means of the standard Tsallis approaches, e.g. the ones compelling with Fermi-Dirac or Bose-Einstein statistics and with the two- and three-component standard distributions \cite{Wei:2015oha}. From this analysis, a dependence of the effective temperature on the rest mass of the particle $m_0$ was proposed \cite{Wei:2015oha}. The proposed relation between the effective temperature $T$ also expressed as $T_{T-S}$ or $T_T$ or $T_S$ and the particle’s rest mass $m_0$ is given as \cite{Adler:2003cb,Takeuchi:2015ana,Heiselberg:1998es,Heinz:2004qz,Russo:2015xtz}
\begin{equation}
T = T_0 + a\; m_0,\label{eq:Tm0}
\end{equation}
where $T_0$ is intercept also given as $T_{T-S_0}$ or $T_{T_0}$ or $T_{S_0}$ of the linear relation between $T$ and $m_0$ which, at $m_0= 0$. Also, $T_0$ is known as the kinetic freezeout temperature of interacting system.

The transverse momentum spectra of pions, Kaons, and proton and their anti-particles at mid-rapidity in p$+$p collisions at $7~$TeV measured by the ALICE experiment have been analyzed by using different techniques \cite{Adam:2015qaa,Andrei:2014vaa}. This allows for precise measurements within different $p_{\mathtt{T}}$-ranges including low for pions and moderated range for both Kaons and protons \cite{Adam:2015qaa}. The dependence on the particle mass of results from Pb$+$Pb collisions at $2.76~$TeV was parametrized by using BG Blastwave fits \cite{Andrei:2014vaa}. Various collective phenomena in small colliding systems such as central p$+$p and p$+$Pb and peripheral Pb$+$Pb have been investigated.  

Furthermore, the relativistic stochastic model in the three dimensional (non-Euclidean) rapidity space was used to describe the transverse momentum spectra for anti-protons from Au$+$Au collisions. The radial symmetric diffusion in Euclidean space reads \cite{Suzuki:2005zb}
\begin{equation}
\frac{\partial f}{\partial t} = \frac{D}{\sinh^2y} \frac{\partial}{\partial y} \left[\sinh^2y \frac{\partial f}{\partial y}\right],
\end{equation}
where $D$ is the diffusion constant. At initial conditions, i.e. $t = 0$, we get
\begin{equation}
f(y,0) = \frac{\delta(y-y_0)}{4 \pi \sinh^2y}.
\end{equation}
At $k_B T = m \sigma(t)^2$, where $k_B$ is the Boltzmann constant and $\sigma(t)^2 = 2 D T$, there is a coincidence between the relativistic stochastic model and the Maxwell-Boltzmann. Therefore, the temperatures could be estimated from the Maxwell-Boltzmann distribution function at low and high $p_{\mathtt{T}}$-ranges under the assumption that in the lower momentum limit the distribution function approaches the Maxwell-Boltzmann when the rapidity becomes smaller than unity ($y\ll 1$) or $p_T\ll m$ \cite{Suzuki:2005zb}.

\subsection{Extrapolation to Boltzmann temperature}
\label{sec:extrapolateT}

As discussed in Sec. \ref{sec:othrm}, the effective temperature can be expressed in terms of the particle’s rest mass, Eq. (\ref{eq:Tm0}) \cite{Wei:2015oha}. Thus, the intercept $T_0$ (known as $T_{T-S_0}$, $T_{T_0}$, or $T_{S_0}$) gives the temperature at the kinetic freezeout of the interacting system or the real temperature (source) \cite{Adler:2003cb,Takeuchi:2015ana,Heiselberg:1998es,Heinz:2004qz,Russo:2015xtz}. Alternatively, $T_0$ could be understood as the quantity related to massless particle. Furthermore, the parameter $a$ can be given as a function of the $v_0^2/2$, where $v_0$ is the transverse radial flow velocity \cite{Takeuchi:2015ana,Heiselberg:1998es}. It was concluded that $a = v^2_0/2$ is only valid within the low $p_{\mathtt{T}}$-region \cite{Takeuchi:2015ana,Heiselberg:1998es}. On the other hand, $a = v^2_0/2$ within $p_{\mathtt{T}}>2~$GeV$/$c would be only valid, if the radial radial flow velocity becomes modified $v_{T-S0}$ or $v_{T0}$ or $v_{S0}$ corresponding to Tsallis-standard and Tsallis or even to the standard distributions. The latter include standard Boltzmann, Fermi-Dirac, and Bose-Einstein distributions, which can be summarized as
\begin{eqnarray}
f_i(p_{\mathtt{T}}) &=& \frac{1}{N} \frac{d N}{d p_{\mathtt{T}}} = C_{i0} p_{\mathtt{T}} \sqrt{p^2_{\mathtt{T}} + m^2_0} \int^{y_{max}}_{y_{min}} \cosh y \left[\exp\left(\frac{\sqrt{p^2_{\mathtt{T}} + m^2_0}\cosh y}{T_i}\right)+S\right]^{-1} dy,
\label{Boltz:comp} 
\end{eqnarray}
where $C_{i0}$ is normalization constant \cite{Wei:2015oha} and $T_i$ is the effective temperature for the $i$-th component. Furthermore, it was concluded that $T_{T-S}$, $T_T$, and $T_S$ all increase with the increase in the collision centrality, where $T_{T-S} \leq T_T < T_S$ for a given set of data \cite{Wei:2015oha}.

The two-Boltzmann, where $i=2$ in Eq. (\ref{Boltz:comp}) and the Tsallis distributions have been utilized in studying the transverse momentum distributions of the final-state particles produced in high-energy collisions at LHC energies \cite{Liu:2014nsa}. It was found that the resulting two temperatures refer to fluctuations taking place in the interacting system. The temperature fluctuations have been investigated under the consideration of an interacting system of groups with different sizes. Other observables such as the transverse energy and the multiplicity have been related to these fluctuations, as well. Tsallis statistics seems to describe well the temperature fluctuations and the degree of non-equilibrium, as fas this is related to nonextensivity. In other words, the degree of non-equilibrium is merely referring to the change in the non-extnsivity parameter $q$. Thus, this study has showed that Tsallis statistics describes well the fluctuations in both $T$ and $q$. From two-Boltzmann distributions, the temperature of the interacting system $T$ can be given as \cite{Gao:2015qsq}
\begin{equation}
T = k_1 T_1 + k_2 T_2,
\end{equation}
where $k_1$ and $k_2$ are constants denoting the contributions from first and second Boltzmann distribution, respectively. Alternatively, the temperature of the interacting system can be given as $T=k_1 T_1 +(1 - k_1)T_2$ \cite{Liu:2014nsa}. 

The transverse momentum spectra of the charged particles produced in Au$+$Au collisions at RHIC and in Pb$+$Pb collisions at LHC energies with different centrality intervals were analysis by the multisource thermal model in which the Tsallis distributions, the Boltzmann distributions, (two-component) Tsallis distributions, and the (two-component) Boltzmann distributions are implemented \cite{Gao:2015qsq}. It was concluded that there is a linear correlation between the effective temperatures obtained from both Tsallis and Boltzmann distributions,  
\begin{equation}
T_T = (0.956 \pm 0.009) T_B + (-0.034 \pm 0.004).
\end{equation}
$T_T$ refers to Tsallis temperature while $T_B$ refers to Boltzmann temperature. 

In Au$+$Au collisions at RHIC energies, it was concluded that the effective temperatures $T_T$ and $T_B$ increase with the increasing in the particle masses but decrease with the increasing in the centrality. The comparison between the two types of temperatures results in $T_T<T_B$, and also helps in estimating the nonextensive parameter $q$. It was found that the latter is almost not changing in most cases \cite{Gao:2015qsq}. 

From the values obtained for the effective temperatures for charged particles measured in Pb$+$Pb collisions at $2.76~$TeV, it was noticed that both $T_T$ and $T_B$ increase with the increase in the particle masses but they are not depending on the centrality of the collisions, especially, when moving from central to semi-central collisions. Also, it was found that $T_T<T_B$ but $q$ does not depend nearly on the centrality of the collisions. On the other hand, $q$ slightly increases from semi-central to peripheral collisions\cite{Gao:2015qsq}.

The transverse momentum spectra of strange particles produced in Pb$+$Pb, p$+$Pb, and p$+$p collisions at different center of mass energies with different multiplicities which measured by the CMS experiment have been described by using both Tsallis and Boltzmann statistics \cite{Yassin:2018svv}. The effective temperatures, the Tsallis temperature $(T_{Ts})$ and the Boltzmann temperature $(T_{Boltz})$, are found increasing with the increase in both the mass and the strangeness number of the particles and also increasing with the multiplicities but $q$ decreasing with the increase in the particle's mass and also decreasing with the increase in the multiplicities. There is a linear correlation between the extracted temperatures from the two types of statistics
\begin{equation}
T_{Ts}= a T_{Boltz} + b
\end{equation}
where $a = 1.2465 \pm 0.0138$ and $b = -160.499 \pm 5.386$. So the two temperatures were found related to each other as $T_{Ts} < T_{Boltz}$. The values of the constants $a$ and $b$ change as the change in the particle mass as 
\begin{eqnarray}
&&\mathbf{For \;\;K_s^0} \;\;\;T_{Ts} = (1.3714 \pm 0.0092)\; T_{Boltz} + (−177.514 \pm 2.503) \\
&&\mathbf{For \;\;\;\Lambda} \;\;\;\;T_{Ts} = (1.3856 \pm 0.0255)\; T_{Boltz} + (−209.84 \pm 10.2) \\
&&\mathbf{For \;\;\;\Xi^-} \;\;T_{Ts} = (1.39513 \pm 0.0212)\; T_{Boltz} + (−249.213 \pm 9.963)
\end{eqnarray}

\section{Results}
\label{sec:res}

\subsection{Statistical-thermal approaches}

\subsubsection{A$+$A Collisions}

Figure \ref{fig:GenericAll} shows the fit parameters for the charged particles and anti-particles as  functions of the center-of-mass energies in A$+$A collisions obtained from three types of statistics, namely Boltzmann (top panels), Tsallis (medium panel), and generic statistics (bottom panel). These parameters are the chemical potential $\mu$ (left panel), the temperature $T$ (medium panel), and the volume $V$ (right panel). We notice that the chemical potential is inversely proportional to the center-of-mass energies for all particles. This is valid for the different types of statistics. For Boltzmann statistics, the temperature increases with the increase in the energies for all particles. For both Tsallis and generic axiomatic statistics the temperature increases as well with the increase in energies for Kaons and protons and their anti-particles but for pions it decreases. With respect to the volume, we notice that the volume increases with the increase in energies for all particles except for proton (here the volume decreases with the energies). 

Figure \ref{fig:Genericd} presents the non-extensive parameters $q$ and $d$ for Tsallis and generic statistics as functions of the center-of-mass energies for charged particles and anti-particles. We find  that $q$ decreases with the increase in energies for Kaons and protons but increases for pions, left panel. On the other hand, $d$ decreases with the increase in the center-of-mass energies for all particles and anti-particles. 

From the resulting $(c,d)$ that $c=0.9995$ and remains unchanged, while $d$ is positive but less than unity, review right panel of Fig. \ref{fig:GenericdNN}, i.e. $(c,d)\equiv(1,d>0)$, we conclude that these result is stretched exponentials and asymptotically stable classes of entropy. In this particular case, this means that 
\bea
S_{\eta}(p) &=& \sum_i \Gamma\left(\frac{\eta+1}{\eta},-\ln p_i\right) - p_i \Gamma\left(\frac{\eta+1}{\eta}\right),
\eea
where $\eta=1/d$ is characterized as stretching exponent distribution, i.e. $\eta>0$ \cite{AP1999}. At positive $d$, the branch of Lambert-$W$ functions, which are the real solutions of $x=W_k(x) \exp(W_k(x))$ is the one at $k=0$. This is defined by the solutions of $W_0(x)\sim x-x^2+\cdots$. Within the given $\eta$-region, there are three cases: 
\begin{itemize}
\item at $\eta<1$, $S_{\eta}$ is known as superadditive,
\item at $\eta>1$, $S_{\eta}$ is known as subadditive, and
\item at $\eta=1$, $S_{\eta}$ is characterized by positivity, equiprobability, concavity and irreversibility. 
\end{itemize} 
The third case means that three Shannon-Khinchin axioms, i.e. the continuity, the maximality, and the expandability, besides the extensivity are verified. This reproduces the {\it logarithmic} BG nonextensive entropy.
As an example, let us assume that $(c,d)\equiv(1,2)$, where $\eta=1/2$. Then, we get that
\bea
S_{1,2}(p) &=& 2 \left(1-\sum_i p_i \ln p_i\right) + \frac{1}{2} \sum_i p_i \left(\ln p_i\right)^2.
\eea\rm{
This is a superposition of two entropy terms. Furthermore, it is apparent that $S_{1,2}(p)$ is superadditive and its asymptotic behavior is dominated by the second term.

Coming back to the stretched exponent distributions, which are characterized by $c\rightarrow 1$, we recall that the BG extensive entropy is to be recovered at $d=1$. Also, the Tsallis nonextensive entropy is to be restored, at $d=0$. The Lambert exponential is given as
\bea
\lim_{c\rightarrow 1} \varepsilon_{c,d,r}(x) &=& \exp\left\{-dr\left[\left(1-x/r\right)^{1/d}-1\right]\right\},
\eea
where $r=(1-c+cd)^{-1}$ determining the distribution function, especially at small probabilities of microstates ($x$) but not effecting the asymptotic properties. The values obtained for the two equivalent classes $(c,d)$ make it suitable to recall their physical meaning and their relation to (non)equilibrium mechanism. For more detailes, interested readers are advised to consult ref. \cite{Tawfik:2018ahq}.

Figure \ref{fig:AllStatisticsPerParticle} depicts the fit parameters obtained from the three types of statistics as functions of center-of-mass energies. Left panel shows a comparison for pions, middle panel illustrates Kaons, while right panel presents protons. The top panel gives the dependence of $\mu$ on center-of-mass energies. An inverse proportionality between $\mu$ and the energies for all particles is obtained by using all types of statistics. Middle panel shows the temperature as a function of the energies. It is noticed that the temperature increases with increasing energies for Kaons and protons from the three types of statistics while this dependence is only obtained by using Boltzmann statistics for pions. By using non-extensive statistics, we find that the temperature of pions decreases with the increase in energies. Bottom panel presents the dependence of the volume on the center-of-mass energies. We find that the volume increases with the increase in energies for pions and Kaons by using the three types of statistics. For protons (antiprotons) such a dependence is only obtained when using Tsallis statistics. But for protons and by using Boltzmann and by using generic axiomatic statistics, the volume is found decreasing with the increase in the energies. 

We conclude that by using Boltzmann statistics there is a general behavior that the temperature increases with the increase in energies for all particles. At $200~$GeV, the temperature has values smaller than the ones at lower energies. Also, we conclude that the temperature obtained from anti-particles are slightly greater than the ones from the particles. But for nonextensive statistics, the temperature increases with the increase in the energies for all particles except for pions. Another exception for pions could be highlighted that there is a reverse proportionality between the temperature and the energies. The resulting temperature from $p_{\mathtt{T}}$ spectra of anti-pions, anti-Kaons, and anti-protons are slightly greater than that from $p_{\mathtt{T}}$ spectra of their particles.

\subsubsection{p$+$p Collisions}

Figure \ref{fig:GenericAllNN} depicts the fit parameters for the charged particles and anti-particles as functions of the center-of-mass energies in p$+$p collisions as obtained from three types of statistics, namely Boltzmann (top panels), Tsallis (medium panel), and generic axiomatic statistics (bottom panel). These parameters are the chemical potential $\mu$ (left panel), the temperature $T$ (medium panel), and the volume $V$ (right panel). It is found that the chemical potential is approximately inversely proportional to the center-of-mass energies for all particles for Boltzmann and generic axiomatic statistics. But by using Tsallis statistics there is some variations at low energies. Also, for Boltzmann and generic axiomatic statistics, the temperature increases with the increase in the energies for all particles. For Tsallis statistics, the temperature decreases with the increase in energies for all particles and their anti-particles except at low energies, where the temperature has a reverse trend. With respect to the volume, we notice that the volume increases with the increase in energies for all particles by using Boltzmann and generic axiomatic statistics. But the volume decreases with the increase in energies by using Tsallis statistics. 

Figure \ref{fig:GenericdNN} shows the non-extensive parameters $q$ and $d$ obtained from Tsallis and generic axiomatic statistics, respectively, as functions of the center-of-mass energies for charged particles and anti-particles. In left panel, we find that $q$ increases with the increase in energies for all particles. On the other hand, $d$ decreases with the increase in the center-of-mass energies for all particles and anti-particles. With this regard, we recall that the equivalent class $c=0.9995$ for all particles. 

Figure \ref{fig:CMSworld} shows a comparison of $dN/dy$ and $\langle p_{\mathtt{T}}\rangle$ measured in various experiments in a wide range of energies; UA2~\cite{Banner:1983jq},
E735~\cite{Alexopoulos:1993wt}, PHENIX~\cite{Adare:2011vy},
STAR~\cite{Abelev:2006cs}, ALICE \cite{Aamodt:2011zj} and CMS \cite{Khachatryan:2010us}. It is obvious that the energy dependence for both quantities seems to be consistent with a power-law increase \cite{Khachatryan:2010us}.

Figure \ref{fig:AllStatisticsPerParticleNN} presents the fit parameters obtained from the three types of statistics as functions of the center-of-mass energies. The left panel shows a comparison for pions, middle panel illustrates Kaons, while right panel presents protons and antiprotons. The top panel gives the dependence of $\mu$ on the center-of-mass energies. An inverse proportionality between $\mu$ and the energies is obtained for all particles by using all types of statistics. The middle panel shows the temperature as a function of the center-of-mass energies. It is to be noticed that by using Boltzmann and generic axiomatic statistics the temperature increases with increasing energies for all particles. By using Tsallis statistics, we find that the temperature decreases with the increase in energies. The bottom panel presents the dependence of the volume on the center-of-mass energies. We find that by using Boltzmann and generic axiomatic statistics the volume decreases with the increase in energies for all particles. By using Tsallis statistics, the volume is found nearly independent on the increase in the energies.

Table \ref{tab:3} presents the results obtained from the statistical fits of the momentum spectra of combined positive and negative particles measured in p$+$p collisions at $0.9~$GeV with statistical and systematic uncertainties, \cite{Aamodt:2011zj} to Eq. (\ref{eq:lavy}). We notice that the temperature $T$ and the transverse momentum $p_{\mathtt{T}}$ increase with the increase in the particle mass, while the nonextensive parameter $n$, which is related to $q$, is found nearly independent on the type of particles at $900~$GeV. It is approximately in the range between $6$ to $8$. This means that $q$ ranges between $1.14$ and $\sim 1.2$.

For p$+$p collisions, we conclude that, there is a general behavior for the resulting temperature $T$, namely $T$ decreases with the increase in energies for all studied particles by using all types of statistics (extensive and nonextensive) except for Tsallis, where the temperature increases with the increase in the energies at low energies only. For the three types of statistics, the fit parameter $\mu$ decreases with the increase in the energies for all particles and anti-particles. But by using Tsallis statistics, there are some variations at low energies. The volume of the system is found to increase with the increase in energies for all particles and their anti-particles by using Boltzmann and generic axiomatic statistics. But the obtained $V$ by using Tsallis statistics is found to have a reverse behavior. The nonextensive parameter $q$ from Tsallis statistics increases with the increase in the energies for all particles and their anti-particles. Also, the nonextensive parameter, equivalent class, $d$ which is obtained by using generic axiomatic statistics, decreases with the increase in the energies for all types of particles.

\begin{table}[h!]
  \centering
  \label{tab:3}
  \begin{tabular}{|c|c|c|c|c|c|} \hline
    Particle & $d N/dy$ & $T/$GeV & $n$ & $\langle p_{\mathtt{T}}\rangle/$GeV & $\chi^2/$ndf\\
    \hline
    $\pi^++\pi^-$ & $2.977\pm0.007\pm0.15$ & $0.126\pm0.0005\pm0.001$ & $7.82\pm0.06\pm0.1$ & $0.404\pm0.001\pm0.02$ & $19.69/30$ \\ \hline
    $K^++K^-$ & $0.366\pm0.006\pm0.03$ & $0.160\pm0.003\pm0.005$ & $6.08\pm0.2\pm0.4$ & $0.651\pm0.004\pm0.05$ & $8.46/24$ \\ \hline
    $p+\bar{p}$& $0.162\pm0.003\pm0.012$ & $0.184\pm0.005\pm0.007$ & $7.5\pm0.7\pm0.9$ & $0.764\pm0.005\pm0.07$ & $15.70/21$ \\ \hline
\end{tabular}
\caption{Results of the statistical fits for the experimental results on the combined positive and negative particles spectra measured in p$+$p collisions at $0.9~$GeV to Eq. (\ref{eq:lavy}) \cite{Aamodt:2011zj}. The statistical and systematic uncertainties are indicated. } 
\end{table}


\begin{figure}[htb]
\centering{
\includegraphics[width=5cm,angle=-0]{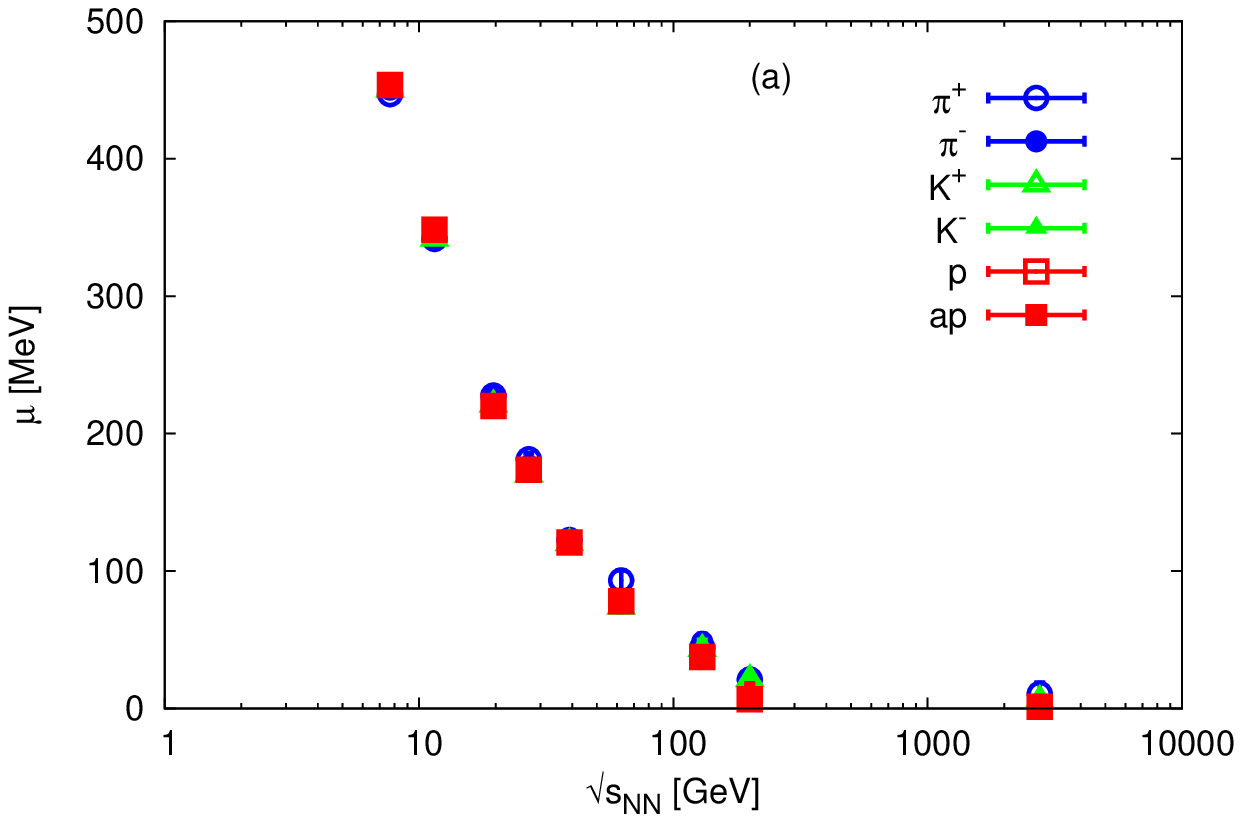}
\includegraphics[width=5cm,angle=-0]{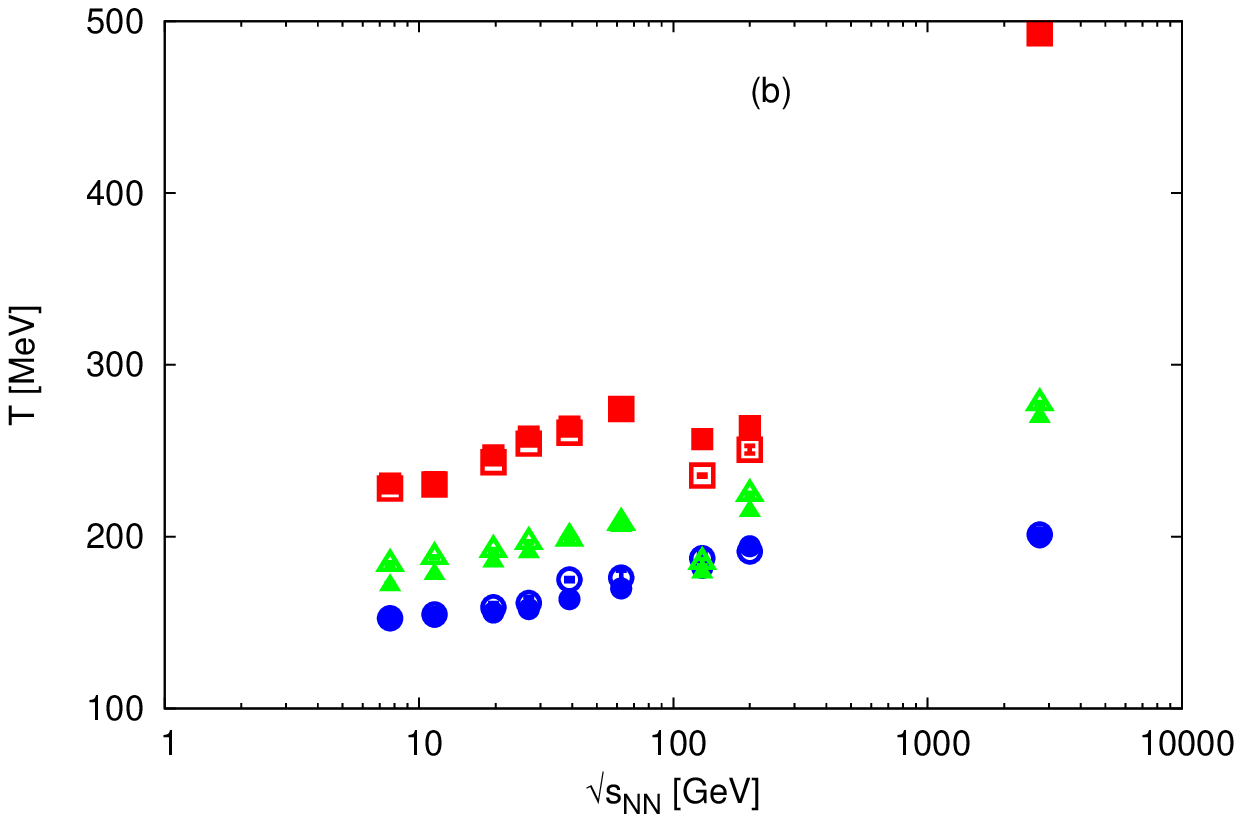}
\includegraphics[width=5cm,angle=-0]{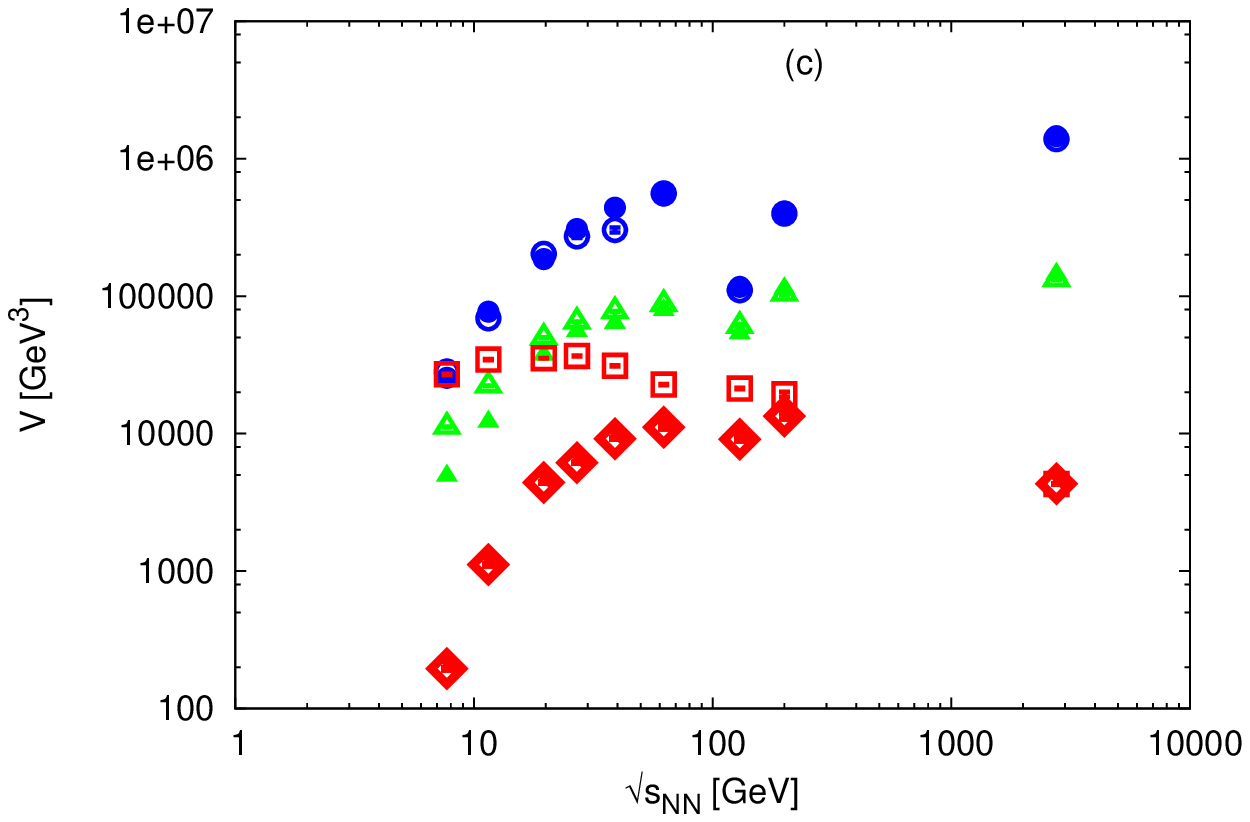}\\
\includegraphics[width=5cm,angle=-0]{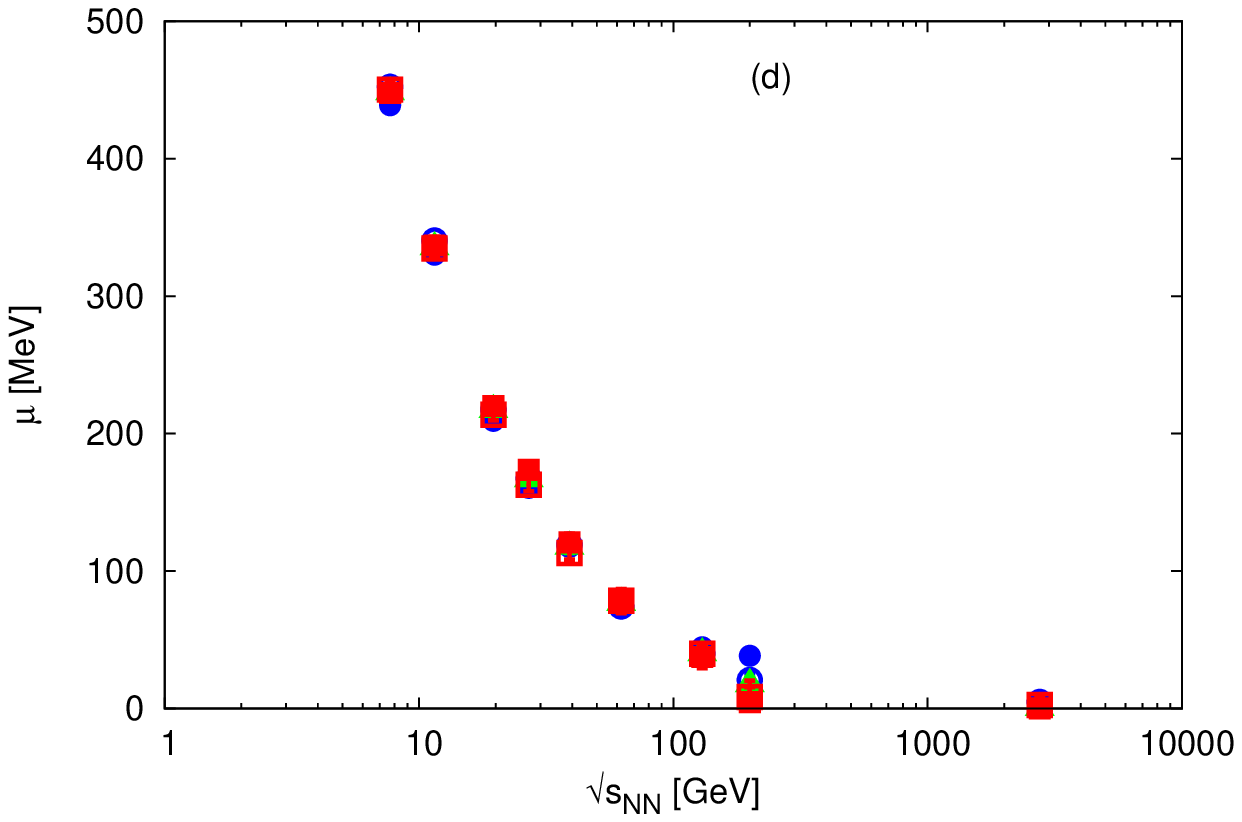}
\includegraphics[width=5cm,angle=-0]{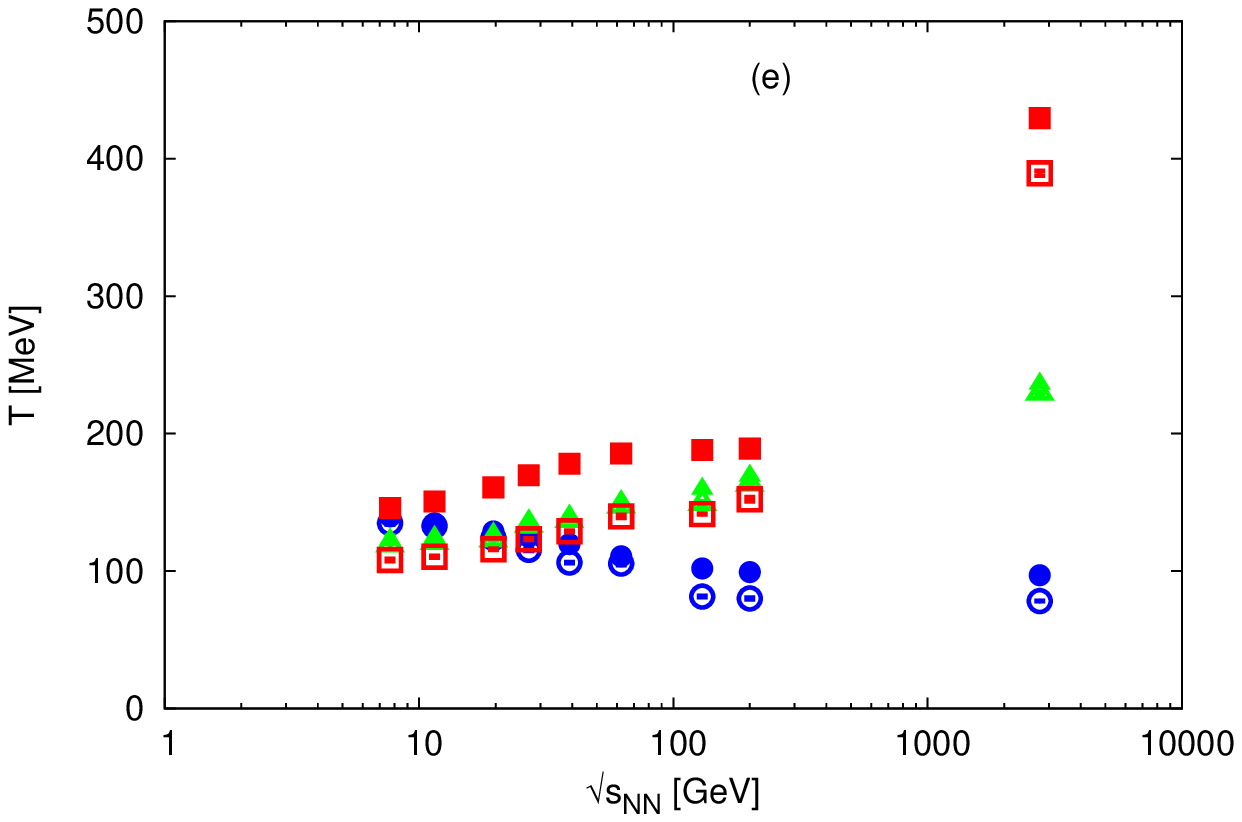}
\includegraphics[width=5cm,angle=-0]{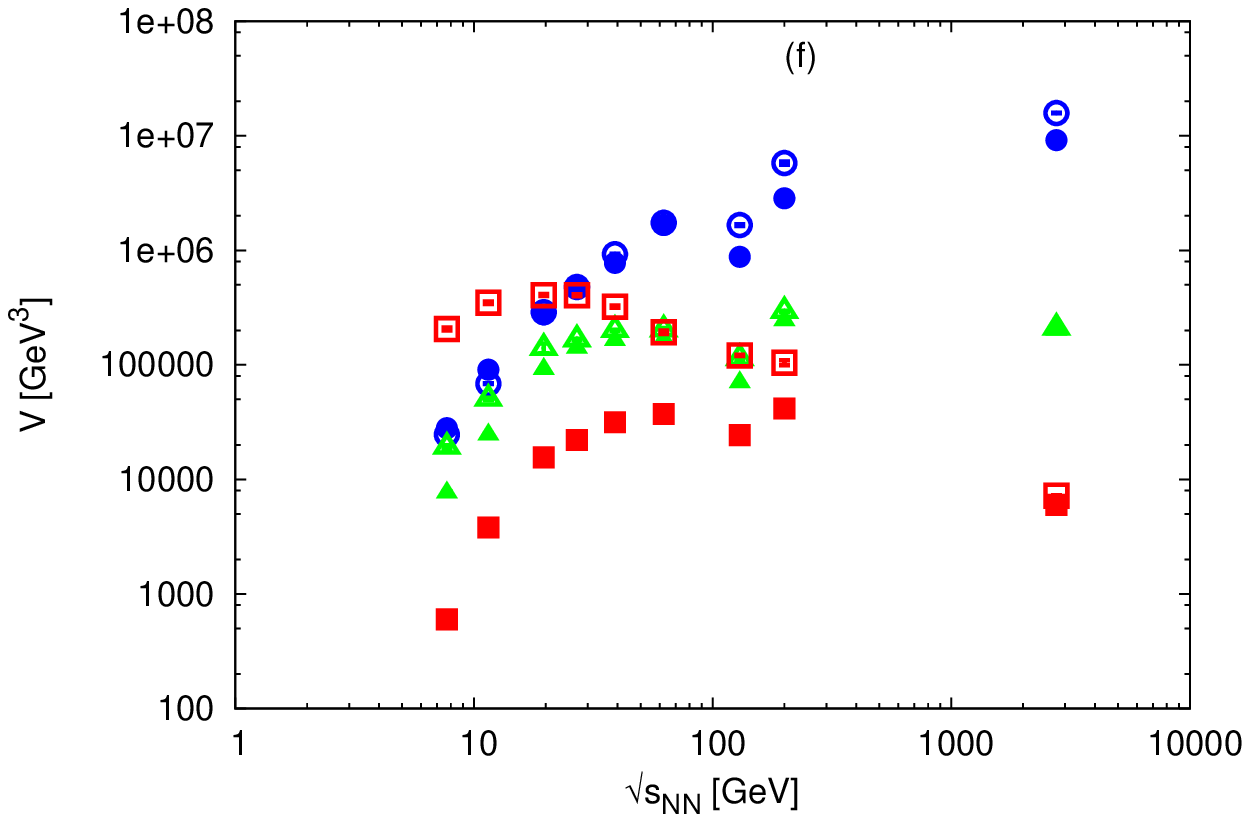} \\
\includegraphics[width=5cm,angle=-0]{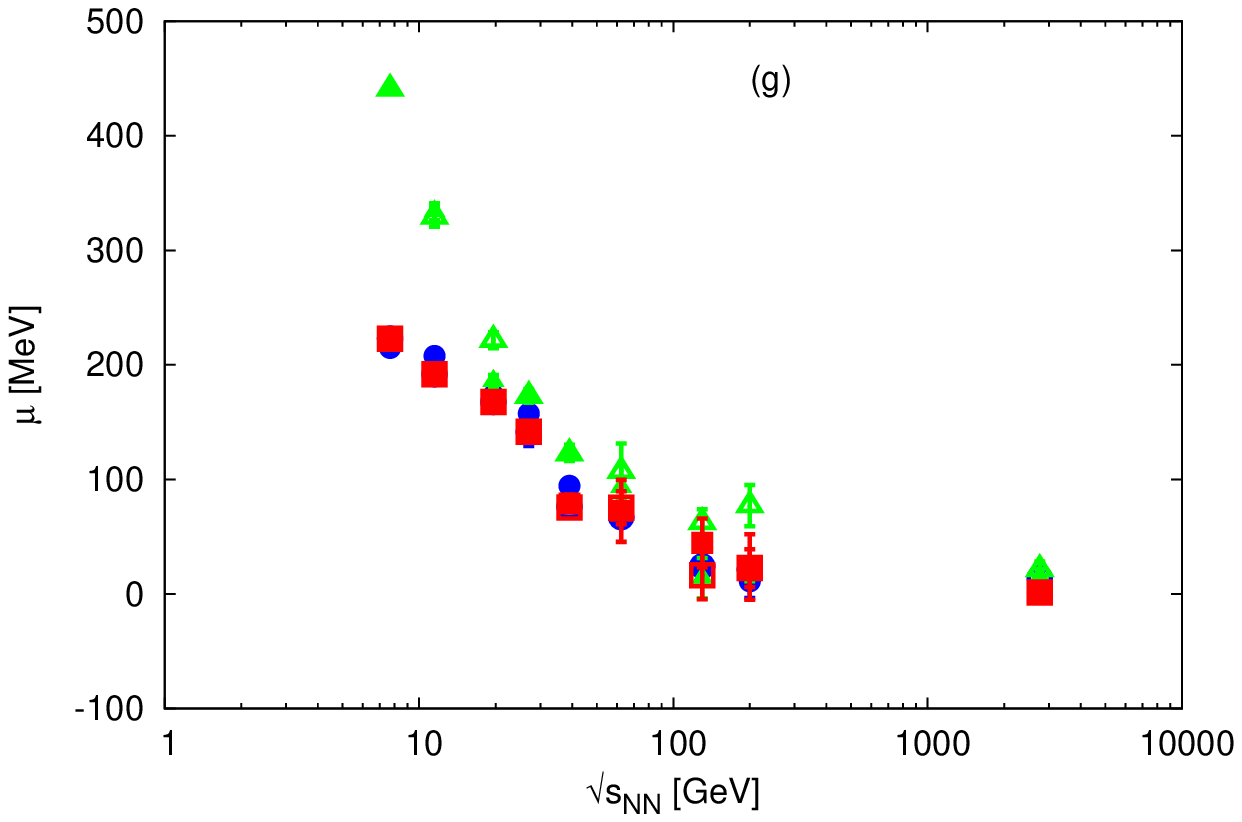}
\includegraphics[width=5cm,angle=-0]{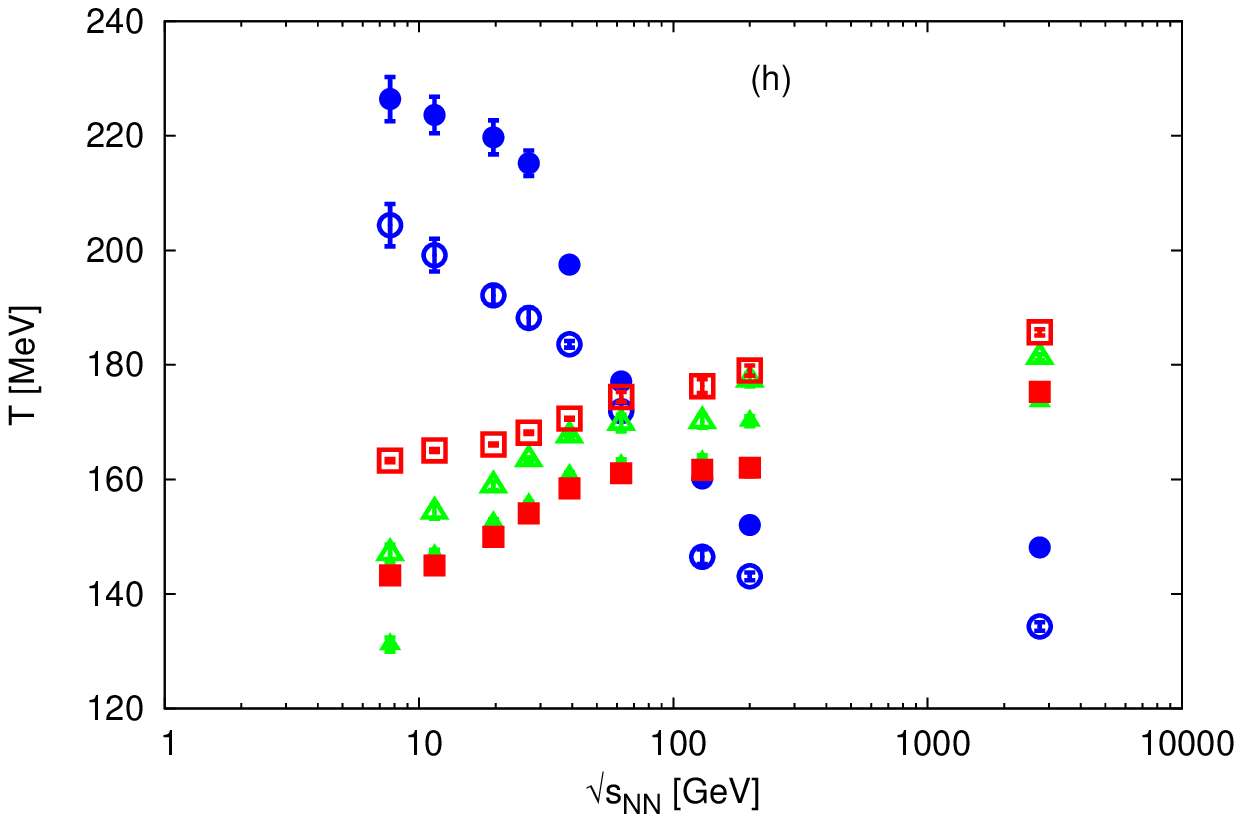}
\includegraphics[width=5cm,angle=-0]{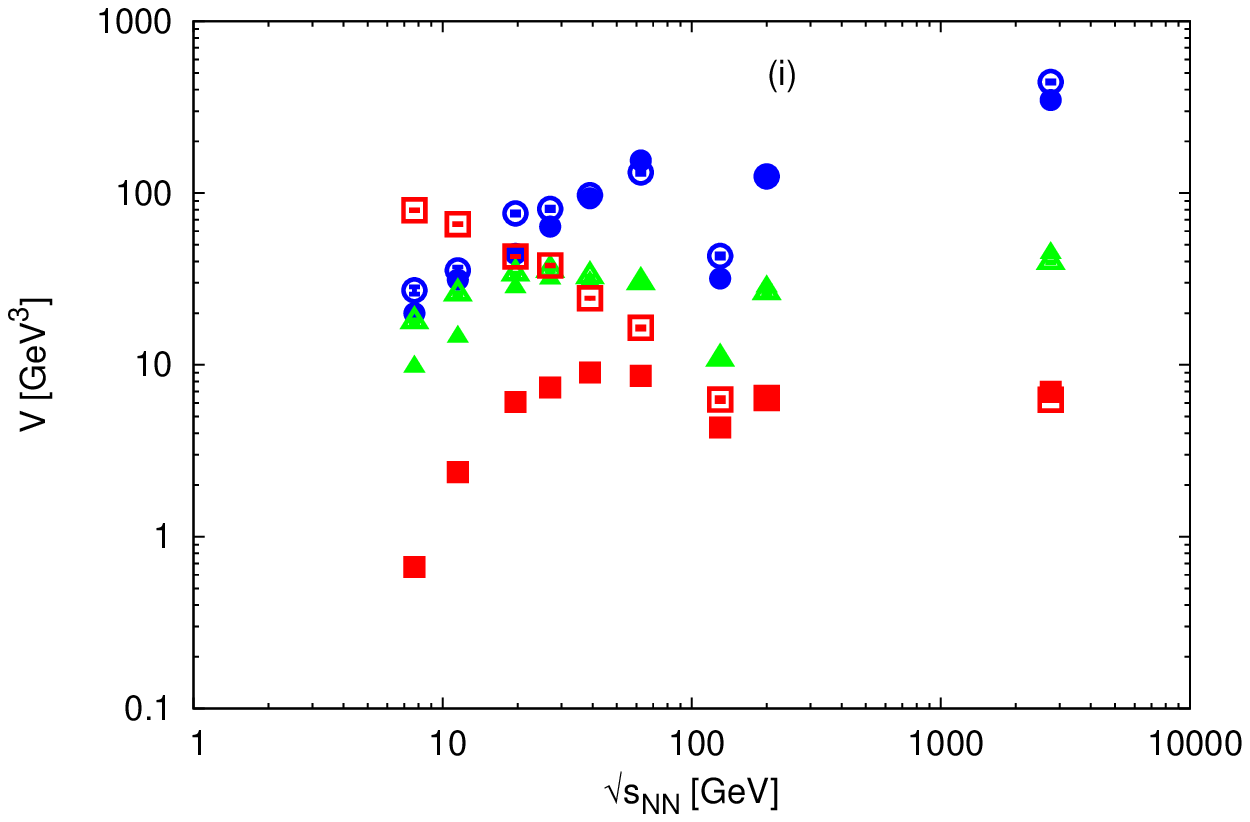}
\caption{(Color online) The various parameters obtained from the statistical fits within Boltzmann, Tsallis and generic axiomatic statistics for $p_{\mathtt{T}}$ measured in A$+$A collisions for various charged particles are depicted as functions of  energies, Appendices \ref{BoltzmannAA}, \ref{TsallisAA}, \ref{GenericAA}.
\label{fig:GenericAll}
}}
\end{figure}

\begin{figure}[htb]
\centering{
\includegraphics[width=8cm,angle=-0]{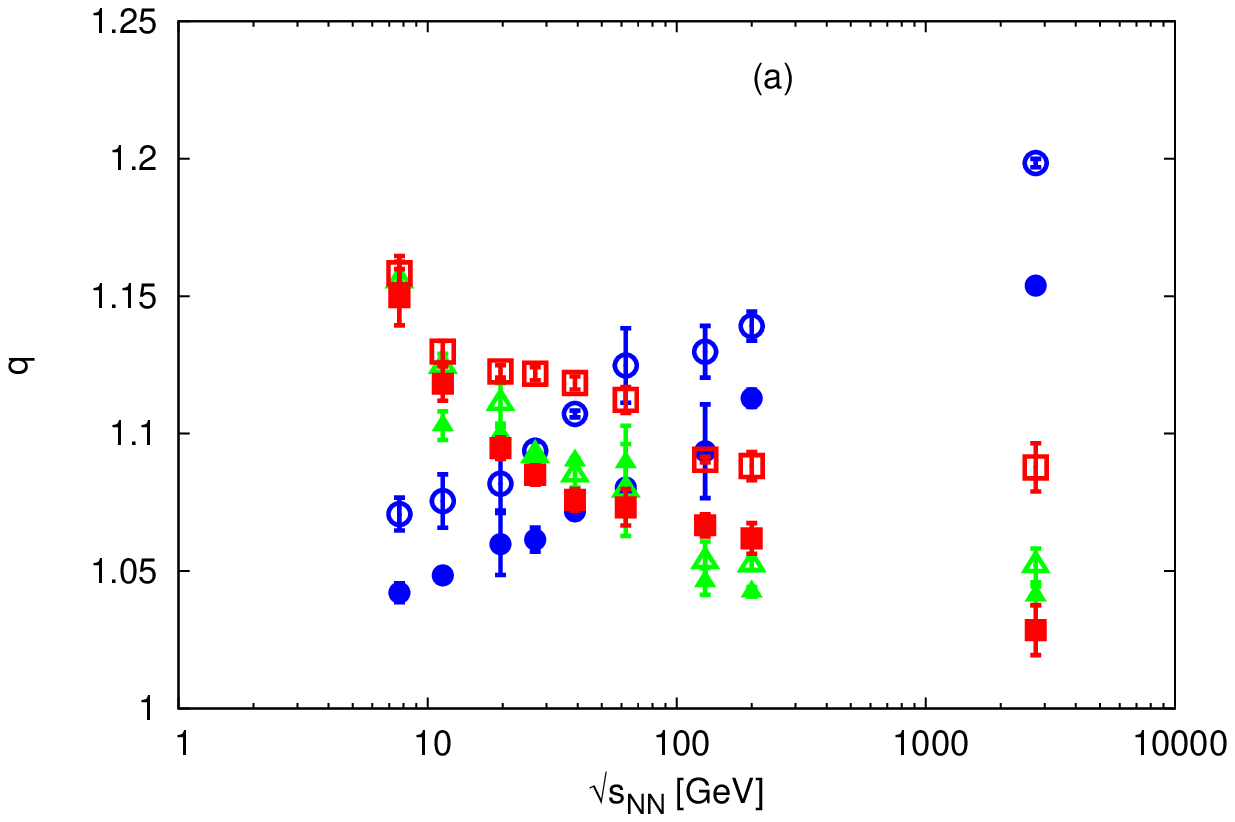}
\includegraphics[width=8cm,angle=-0]{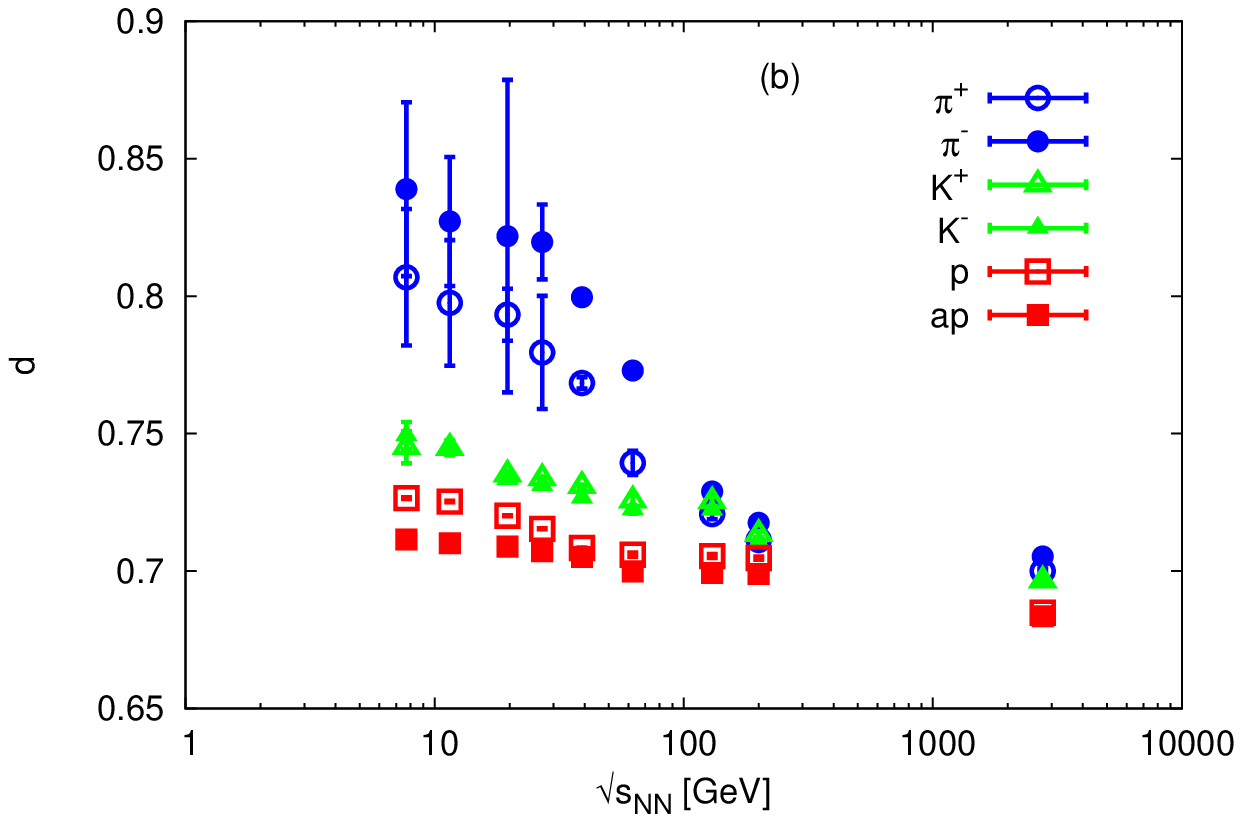}
\caption{(Color online) The nonextensive parameters $q$ and $d$ obtained from statistical fits of $p_{\mathtt{T}}$ measured in A$+$A collisions within Tsallis and generic axiomatic statistics, respectively, are depicted as functions of energy, Appendices \ref{TsallisAA}, \ref{GenericAA}.
\label{fig:Genericd}}
}
\end{figure}

\begin{figure}[htb]
\centering{
\includegraphics[width=5cm,angle=-0]{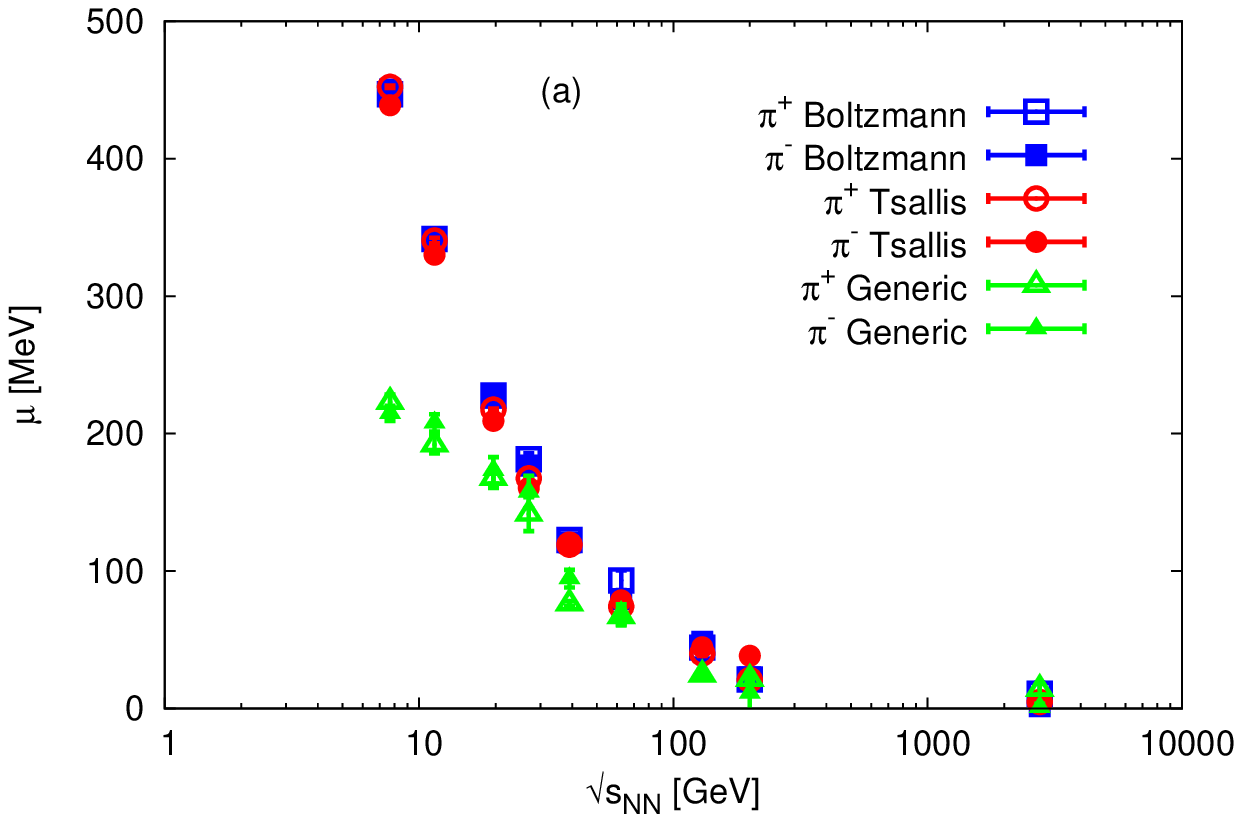}
\includegraphics[width=5cm,angle=-0]{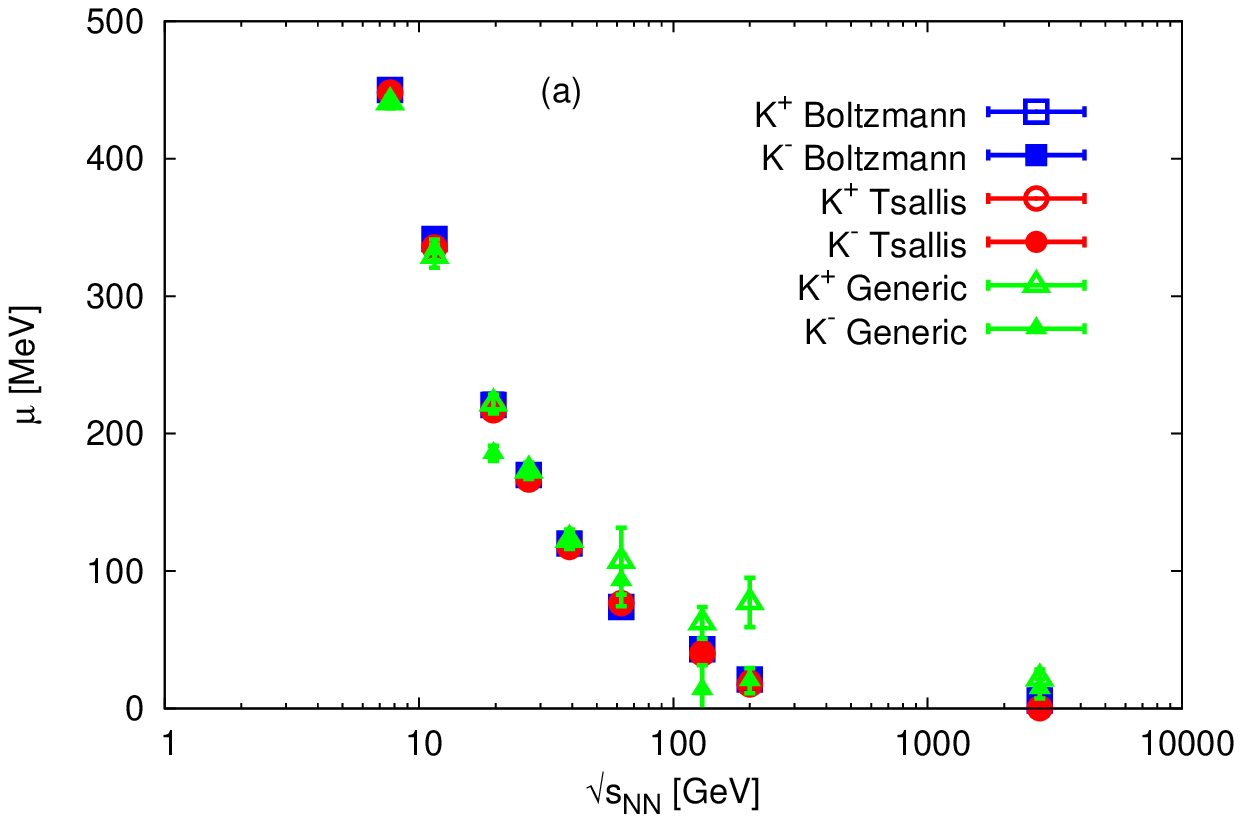}
\includegraphics[width=5cm,angle=-0]{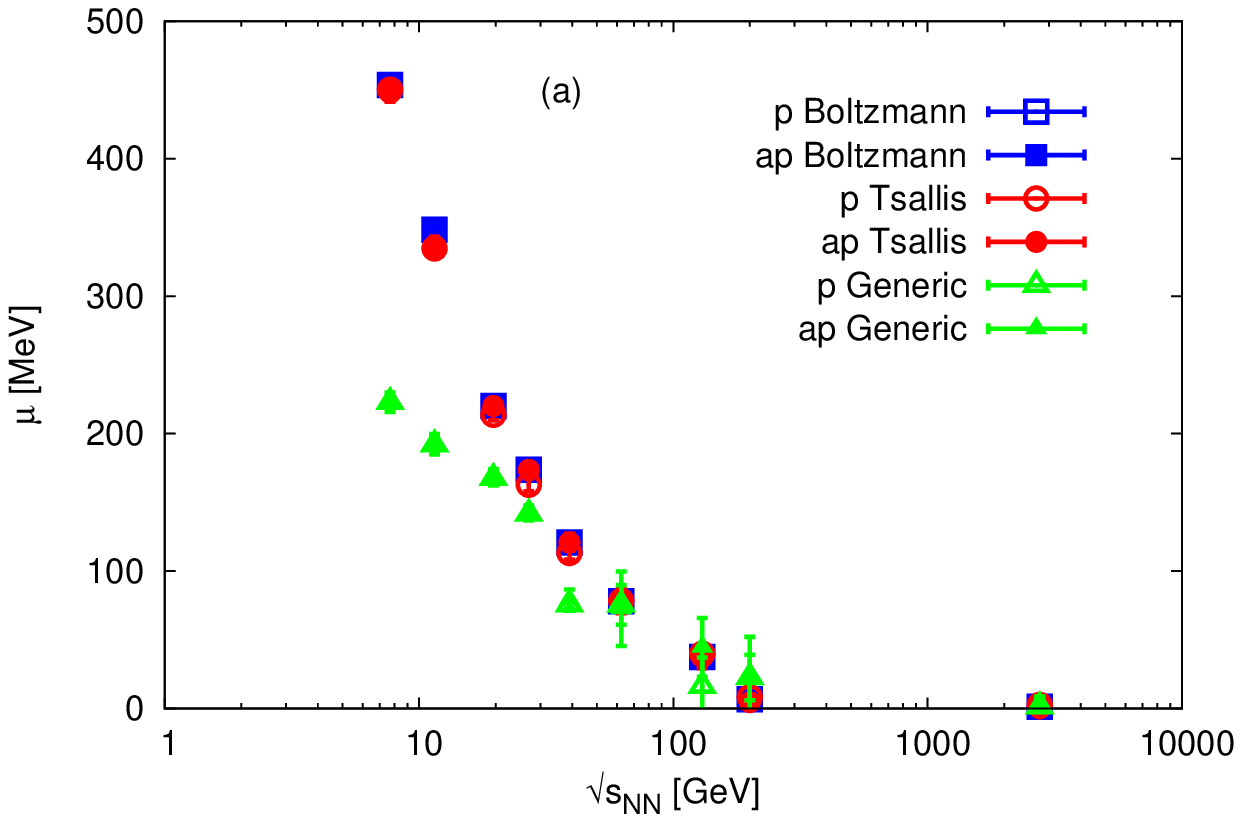} \\
\includegraphics[width=5cm,angle=-0]{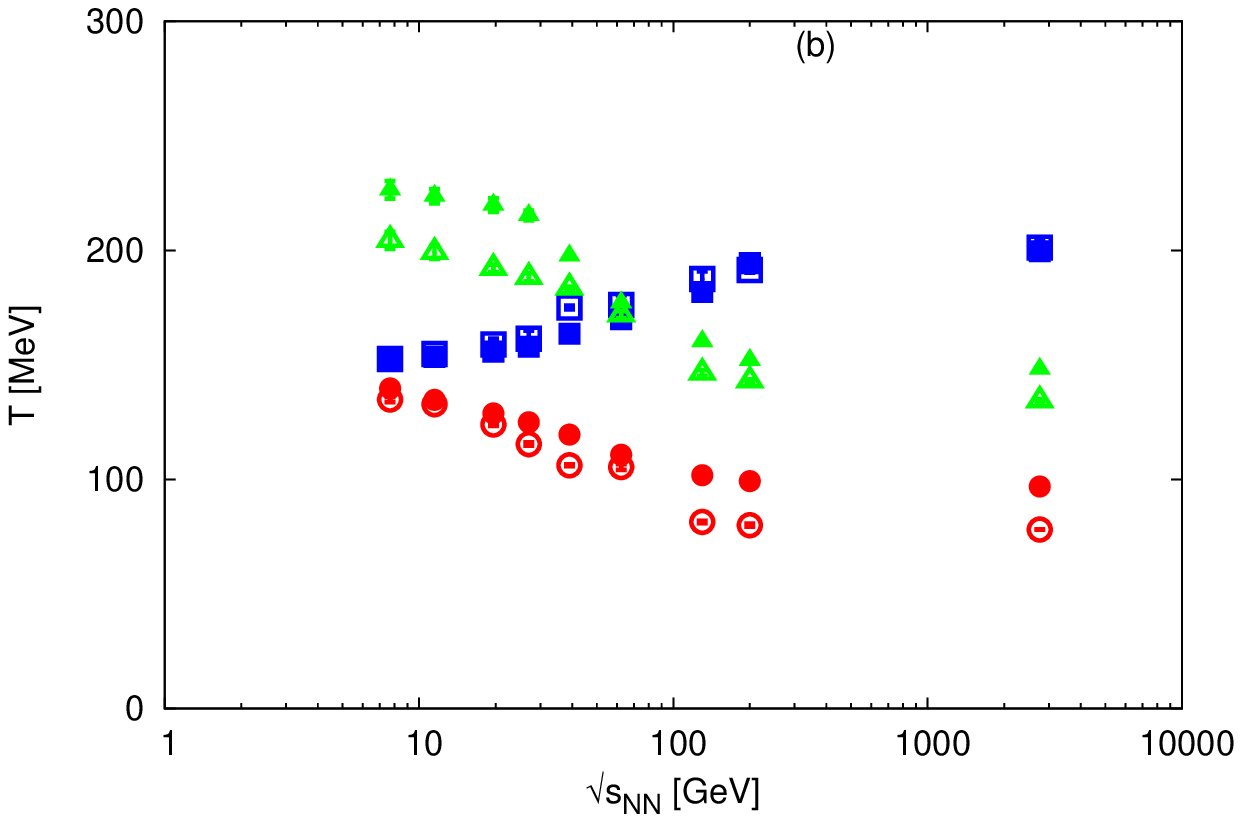}
\includegraphics[width=5cm,angle=-0]{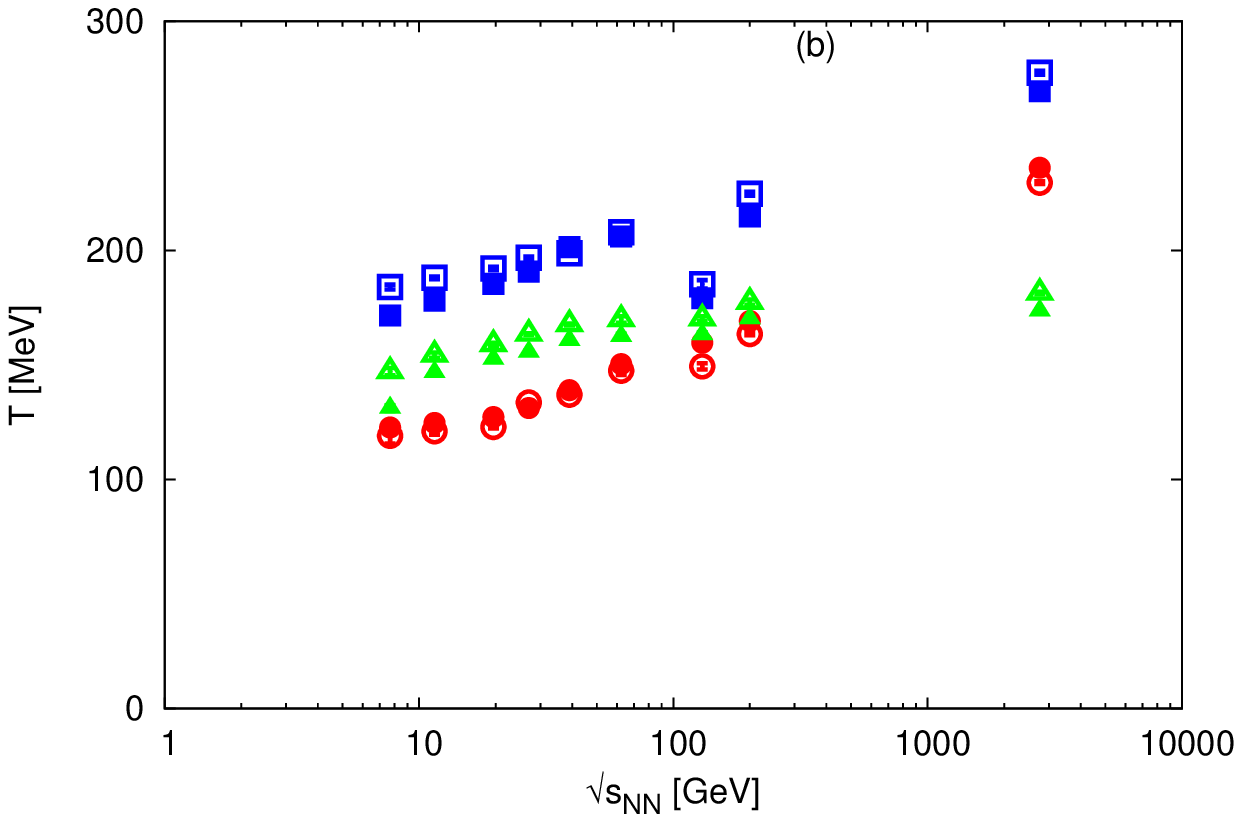}
\includegraphics[width=5cm,angle=-0]{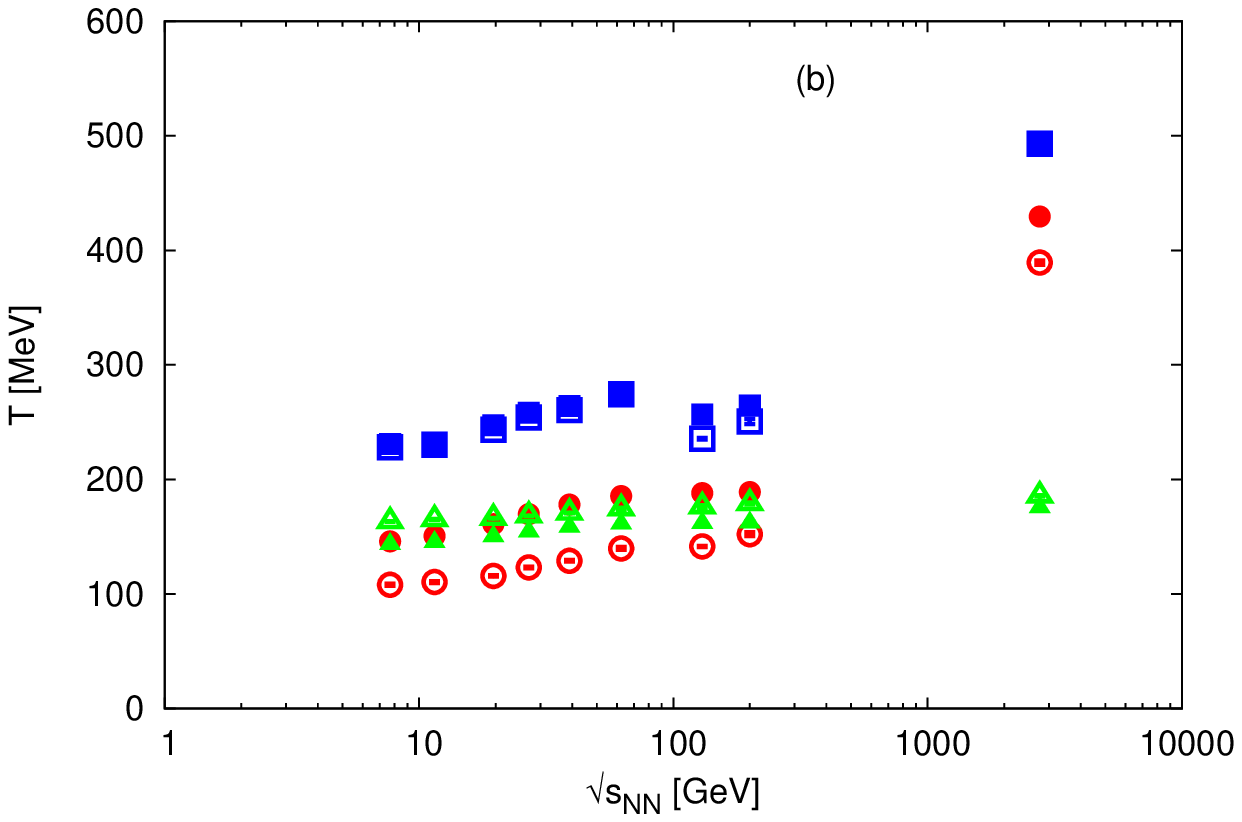} \\
\includegraphics[width=5cm,angle=-0]{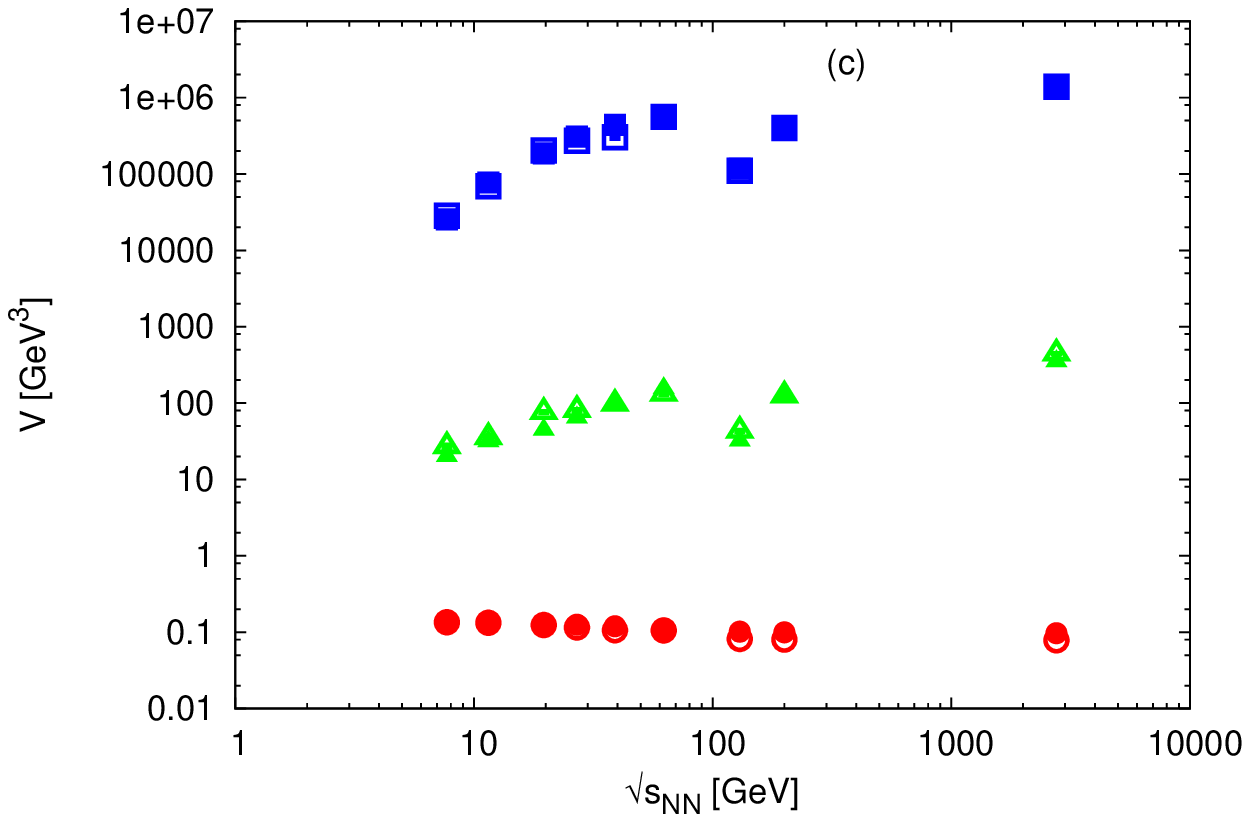}
\includegraphics[width=5cm,angle=-0]{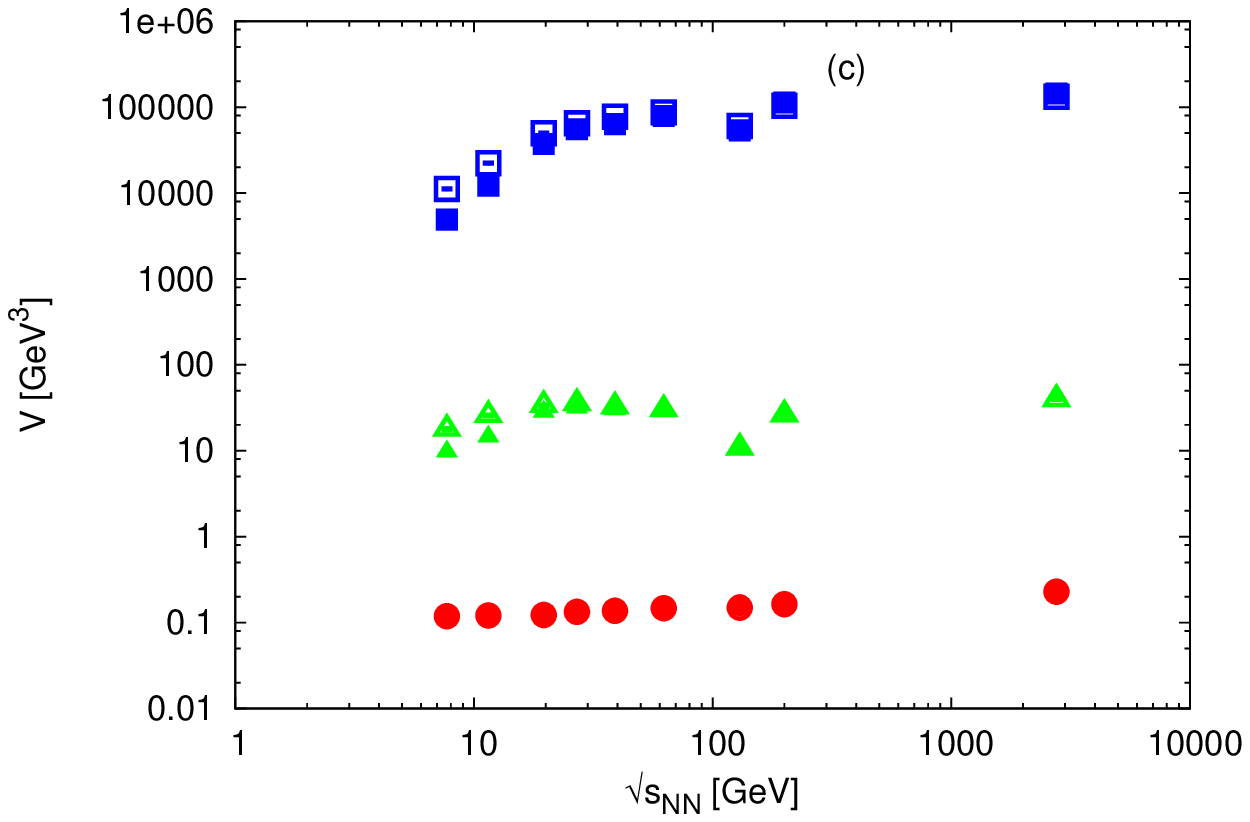}
\includegraphics[width=5cm,angle=-0]{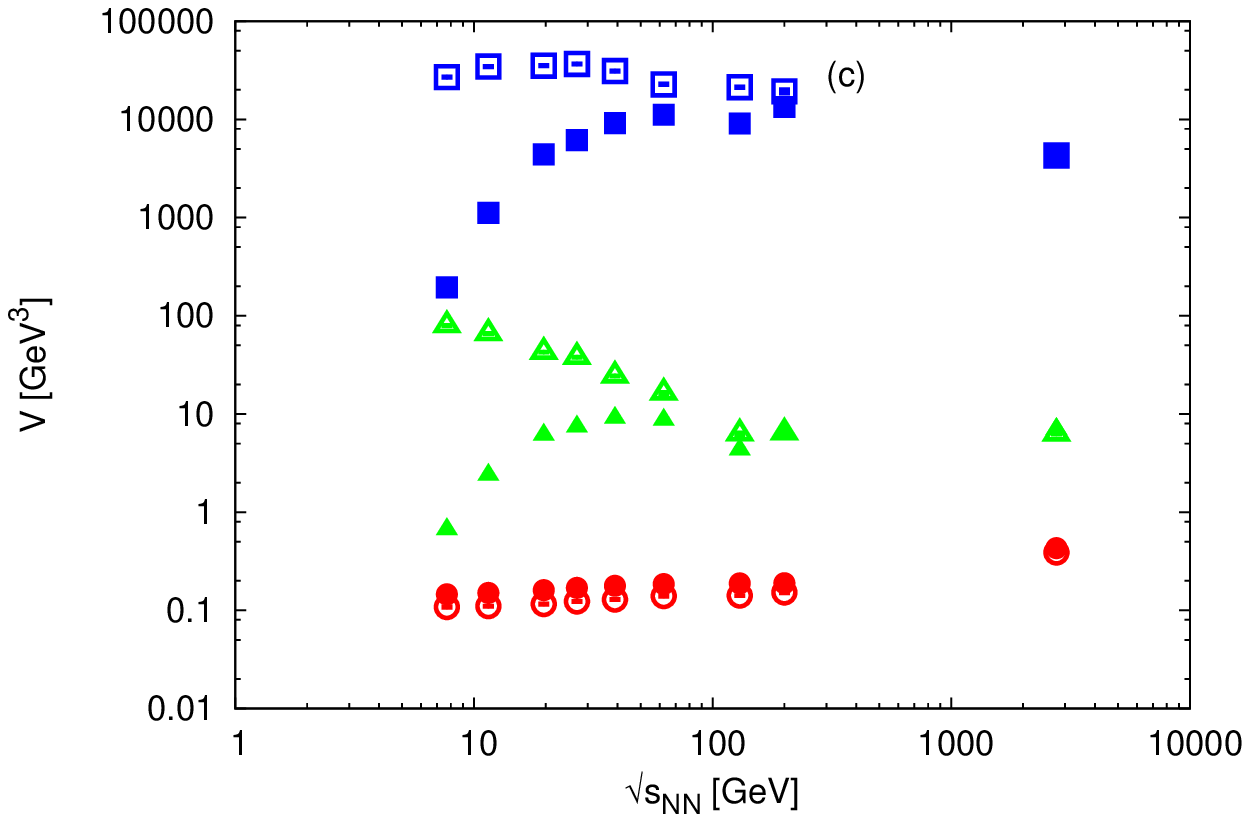} 
\caption{(Color online) A comparison between $\mu$, $T$, and $V$ obtained from the statistical fits within Boltzmann, Tsallis and generic axiomatic statistics for $p_{\mathtt{T}}$ spectra measured in A$+$A collisions for various charged particles in a wide of energies, compared with Fig. \ref{fig:GenericAll} and Appendices \ref{BoltzmannAA}, \ref{TsallisAA}, \ref{GenericAA}. Left panels for pions, middle panels for Kaons, and the right panels for protons and antiprotons.
\label{fig:AllStatisticsPerParticle}
}}
\end{figure} 

It was concluded that the $p_{\mathtt{T}}$ resolution of the CMS experiment doesn't affect the shape of the measured spectra \cite{Khachatryan:2010xs}. For a combination of all (most) charged particles detected in non-single-diffractive p$+$p collisions, it was found that 
\begin{itemize}
\item at $0.9~$TeV \cite{Khachatryan:2010xs}, $T=0.13\pm0.01~$GeV, $n=7.7\pm0.2$, $\left.d N_{\mathtt{ch}}/dy\right|_{|y|<0.5}=3.48\pm0.02\pm0.13$, and $\langle p_{\mathtt{T}}\rangle=0.46\pm0.01\pm0.01~$GeV, 
\item at $2.36~$TeV \cite{Khachatryan:2010xs}, $T=0.14\pm0.01~$GeV,  $n=6.7\pm0.2$, $\left.d N_{\mathtt{ch}}/dy\right|_{|y|<0.5}=4.47\pm0.04\pm0.16$, and $\langle p_{\mathtt{T}}\rangle=0.50\pm0.01\pm0.01~$GeV,
\item at $7~$TeV \cite{Khachatryan:2010us}, $T=0.145\pm0.005~$GeV,  $n=6.6\pm0.2$,  $\left.d N_{\mathtt{ch}}/dy\right|_{|y|<0.5}=5.78\pm0.01\pm0.23$, and  $\langle p_{\mathtt{T}}\rangle=0.545\pm0.005\pm0.015~$GeV.
\end{itemize}

\begin{figure}[htb]
\centering{
\includegraphics[width=5cm,angle=-0]{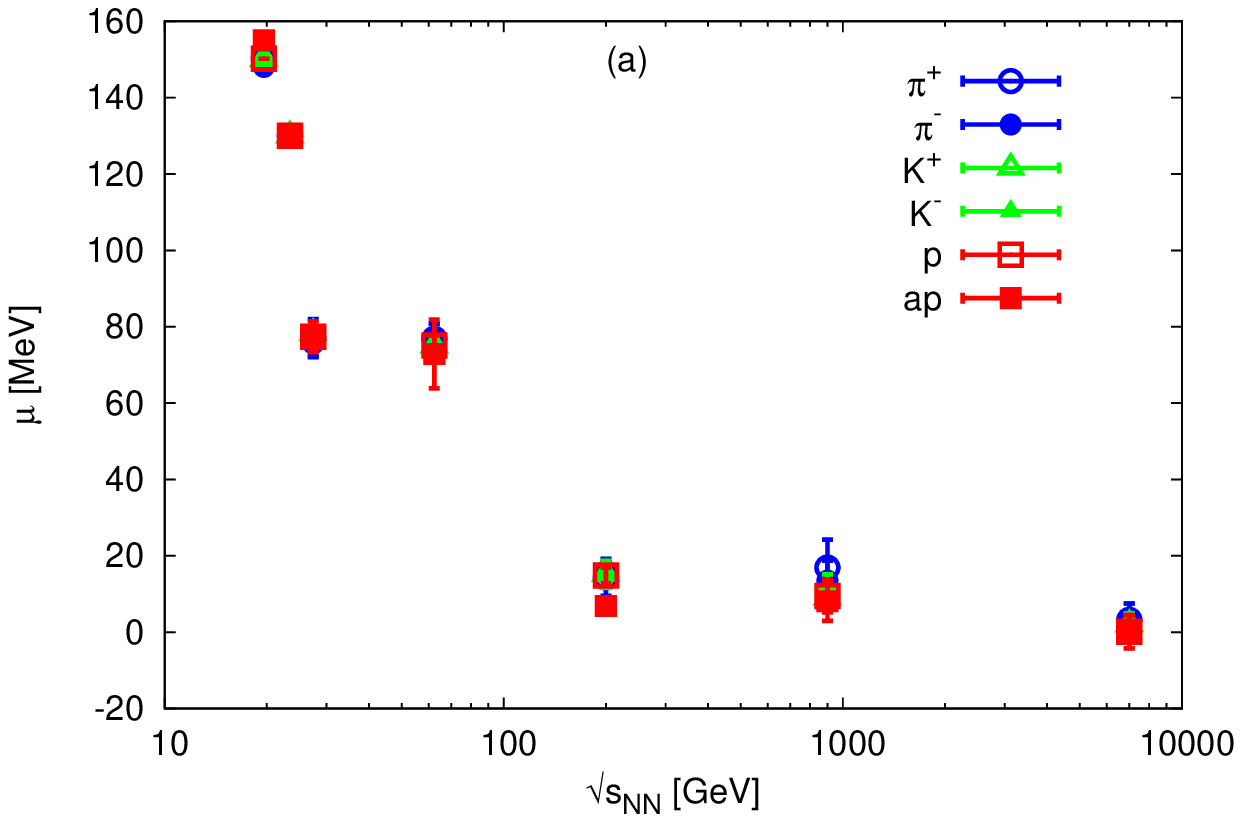}
\includegraphics[width=5cm,angle=-0]{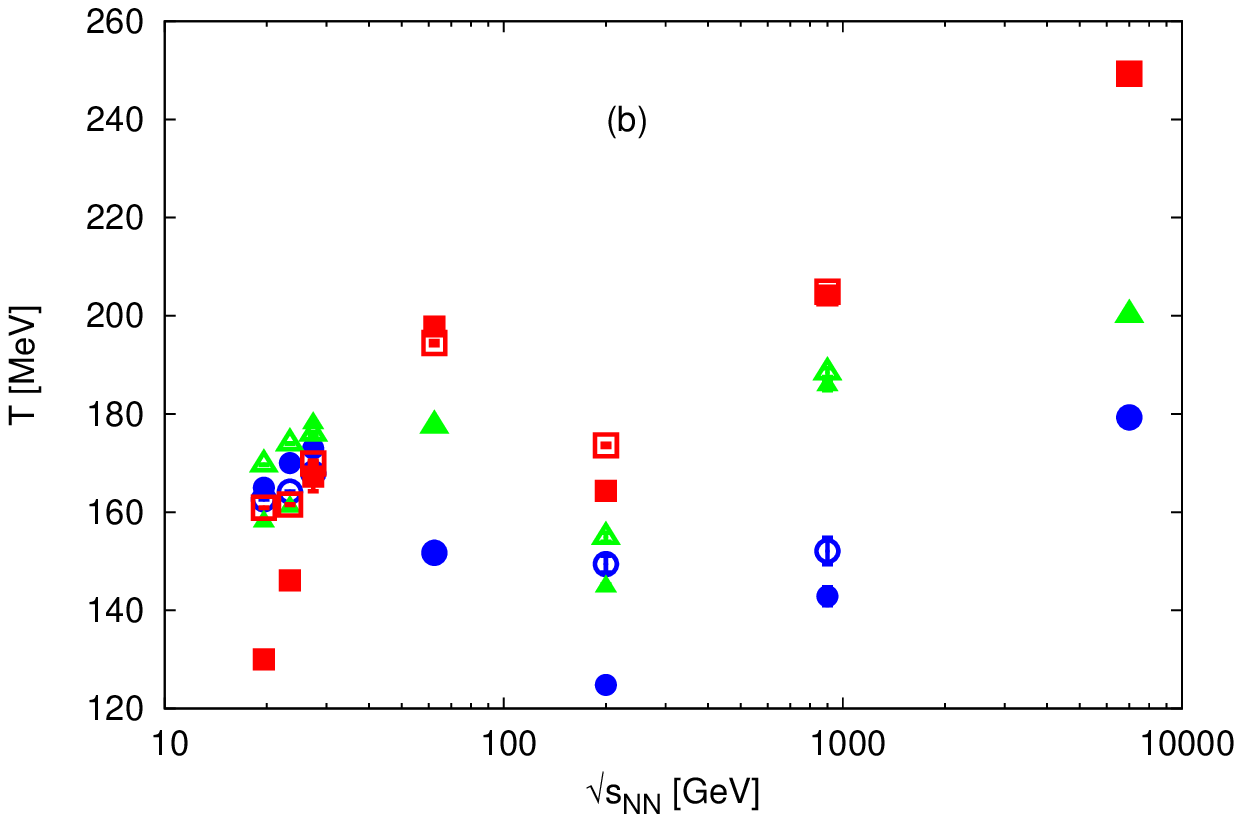}
\includegraphics[width=5cm,angle=-0]{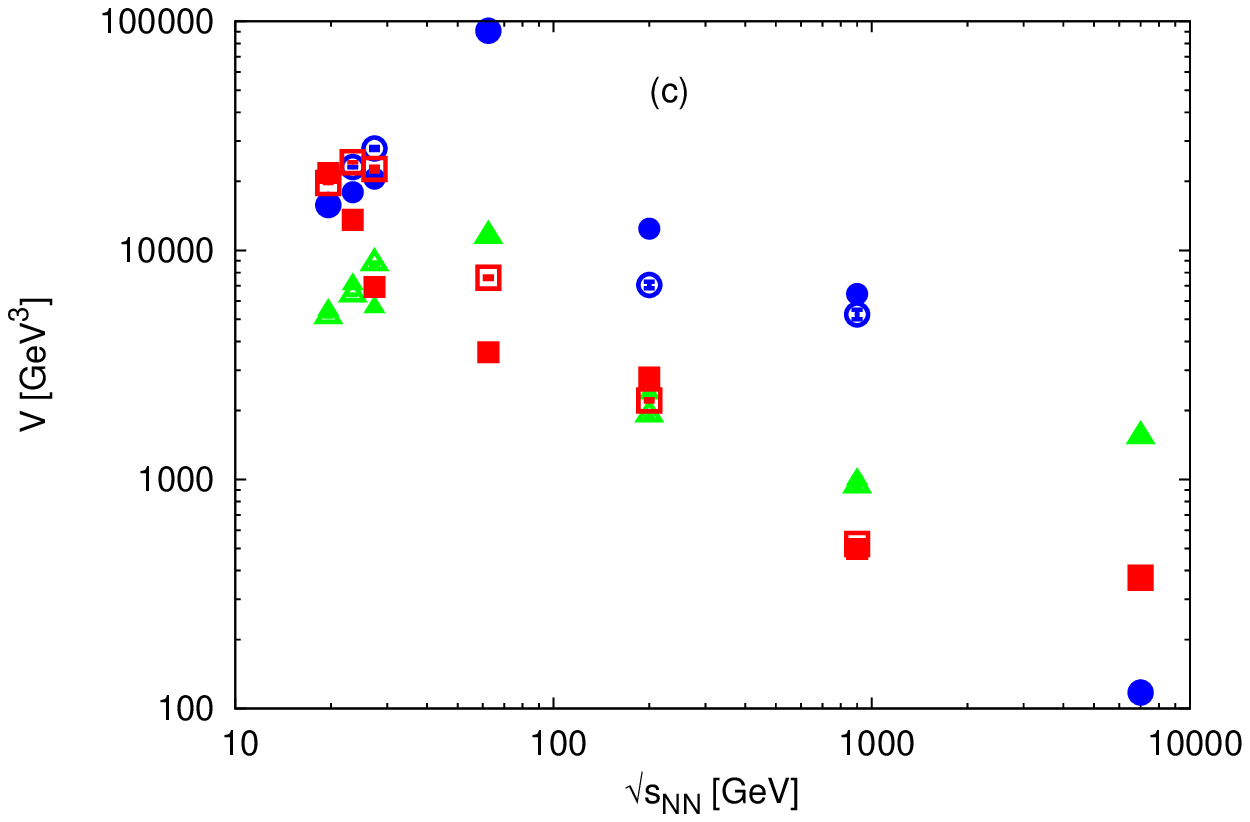}\\
\includegraphics[width=5cm,angle=-0]{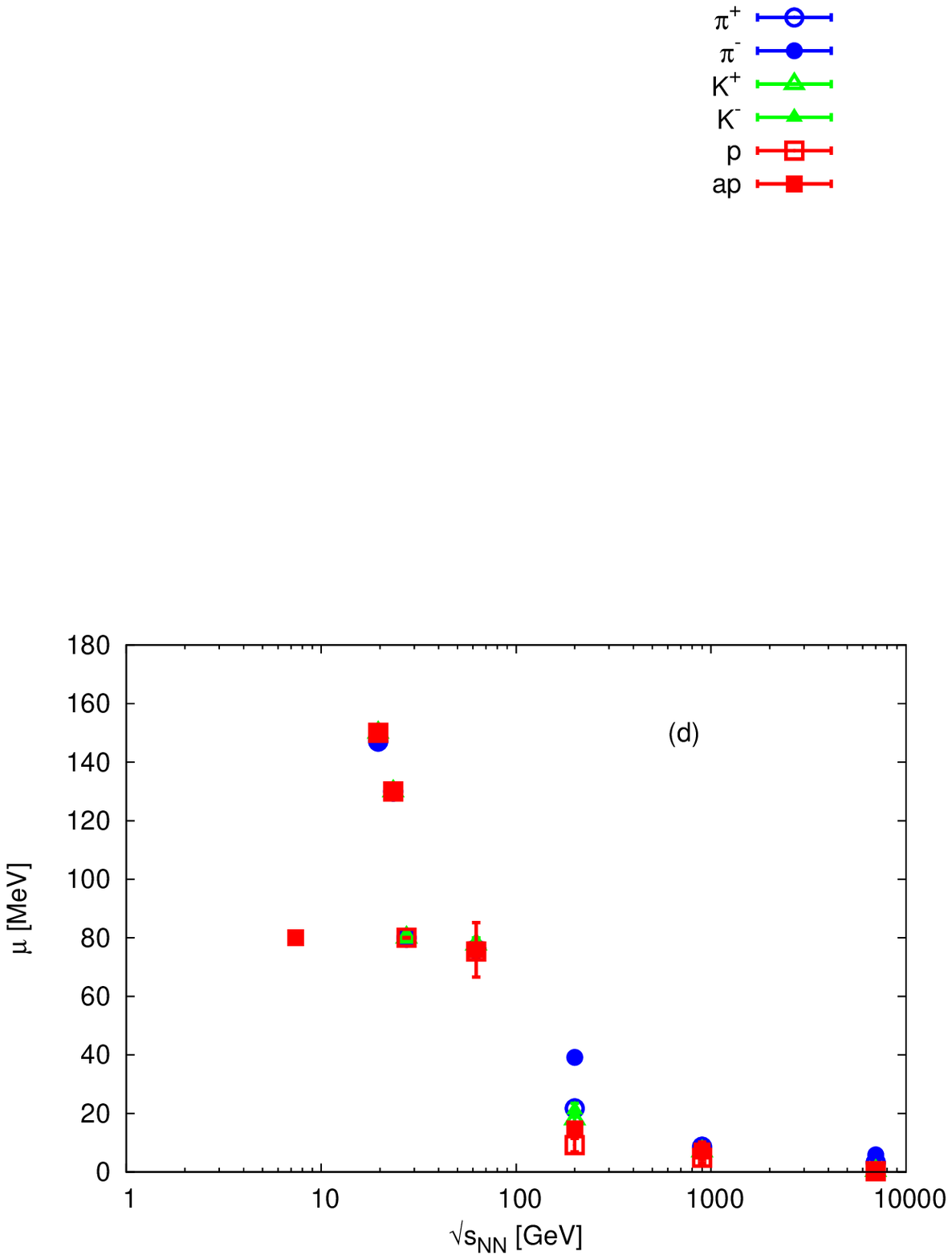}
\includegraphics[width=5cm,angle=-0]{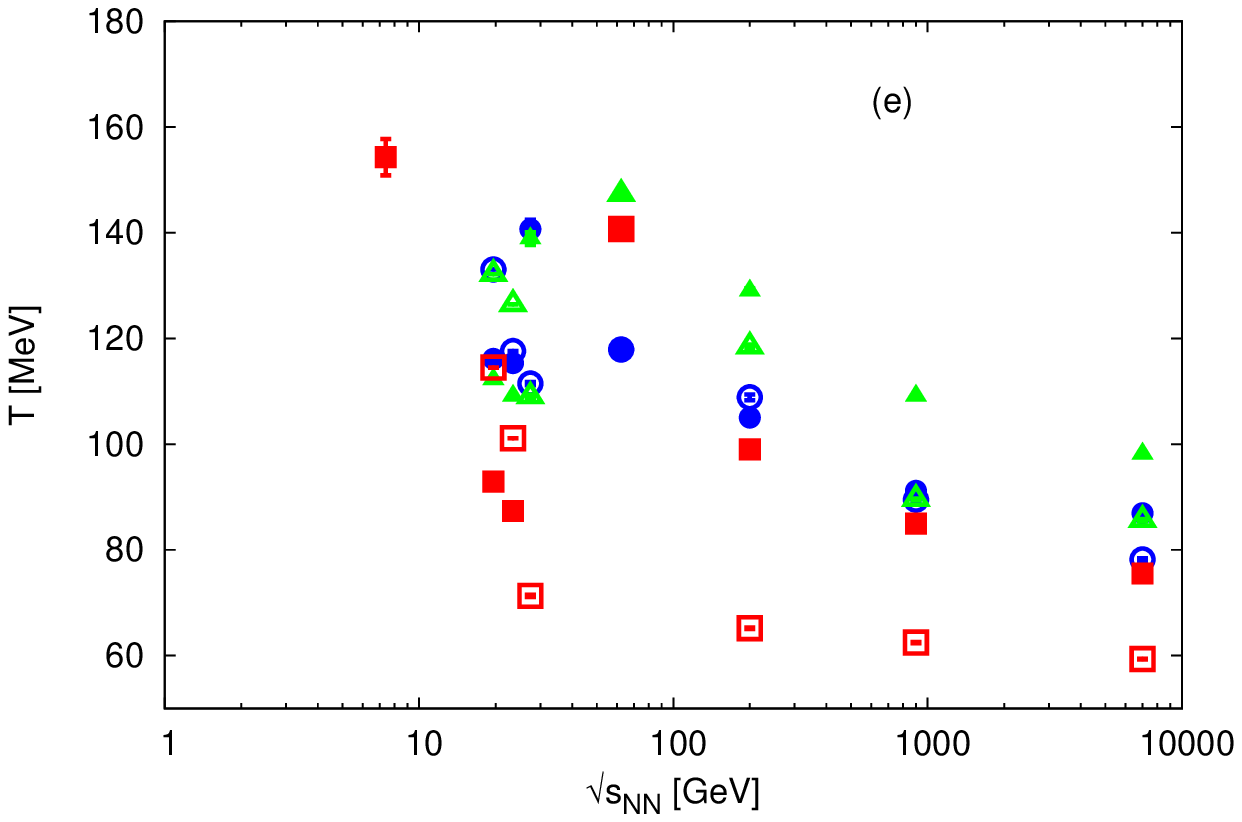}
\includegraphics[width=5cm,angle=-0]{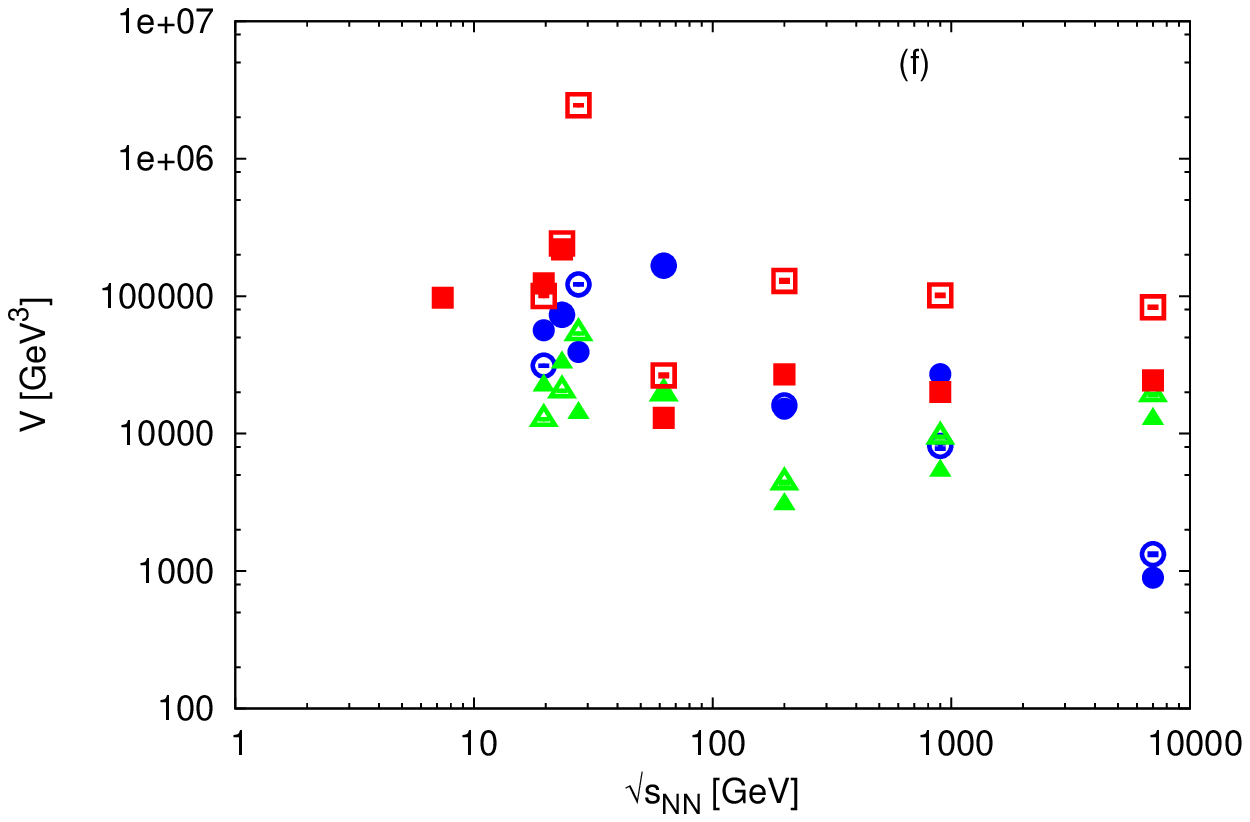} \\
\includegraphics[width=5cm,angle=-0]{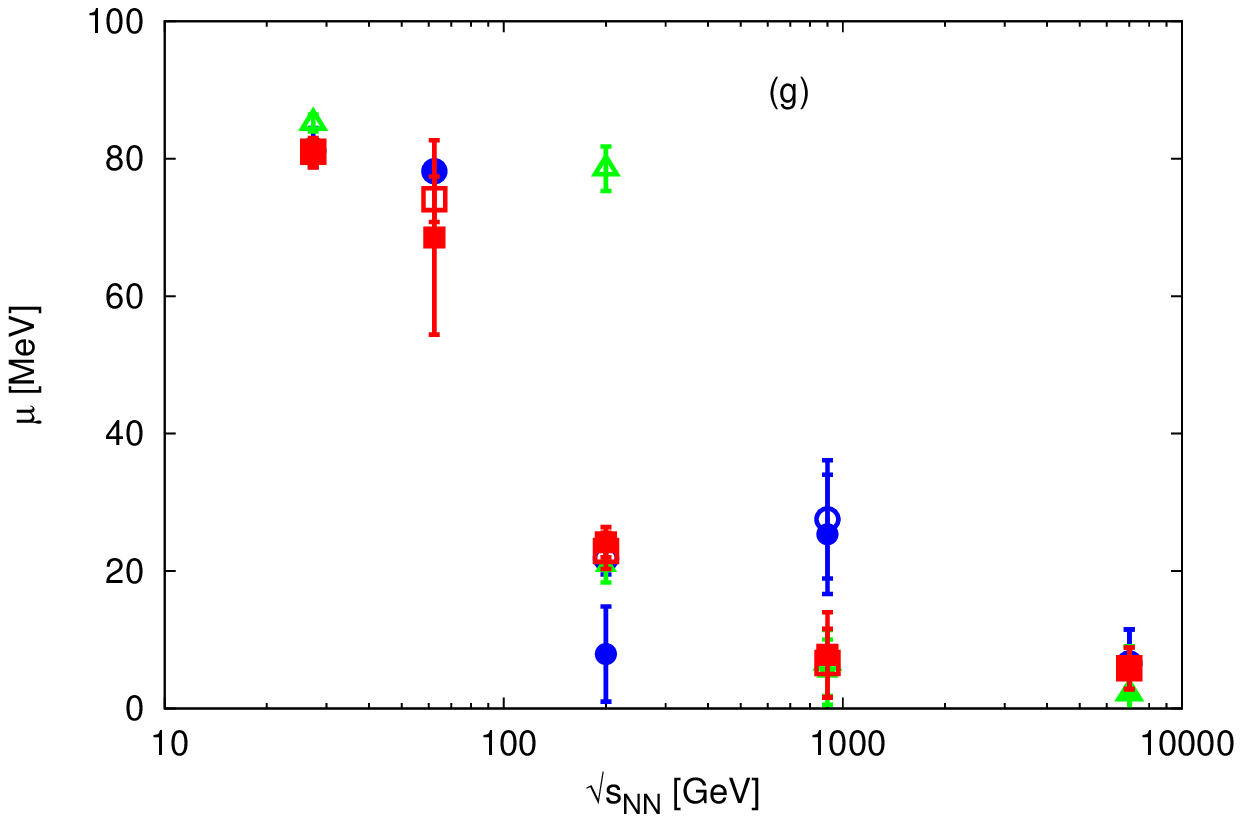}
\includegraphics[width=5cm,angle=-0]{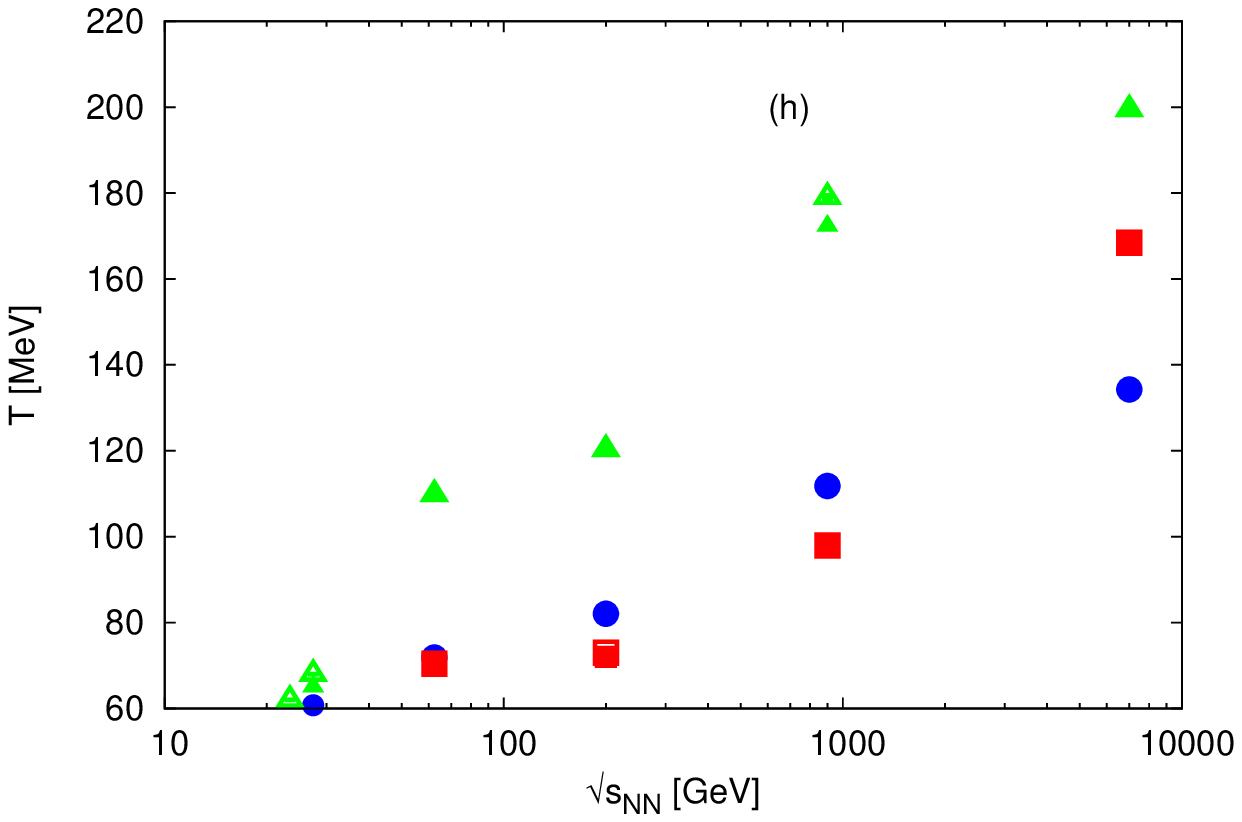}
\includegraphics[width=5cm,angle=-0]{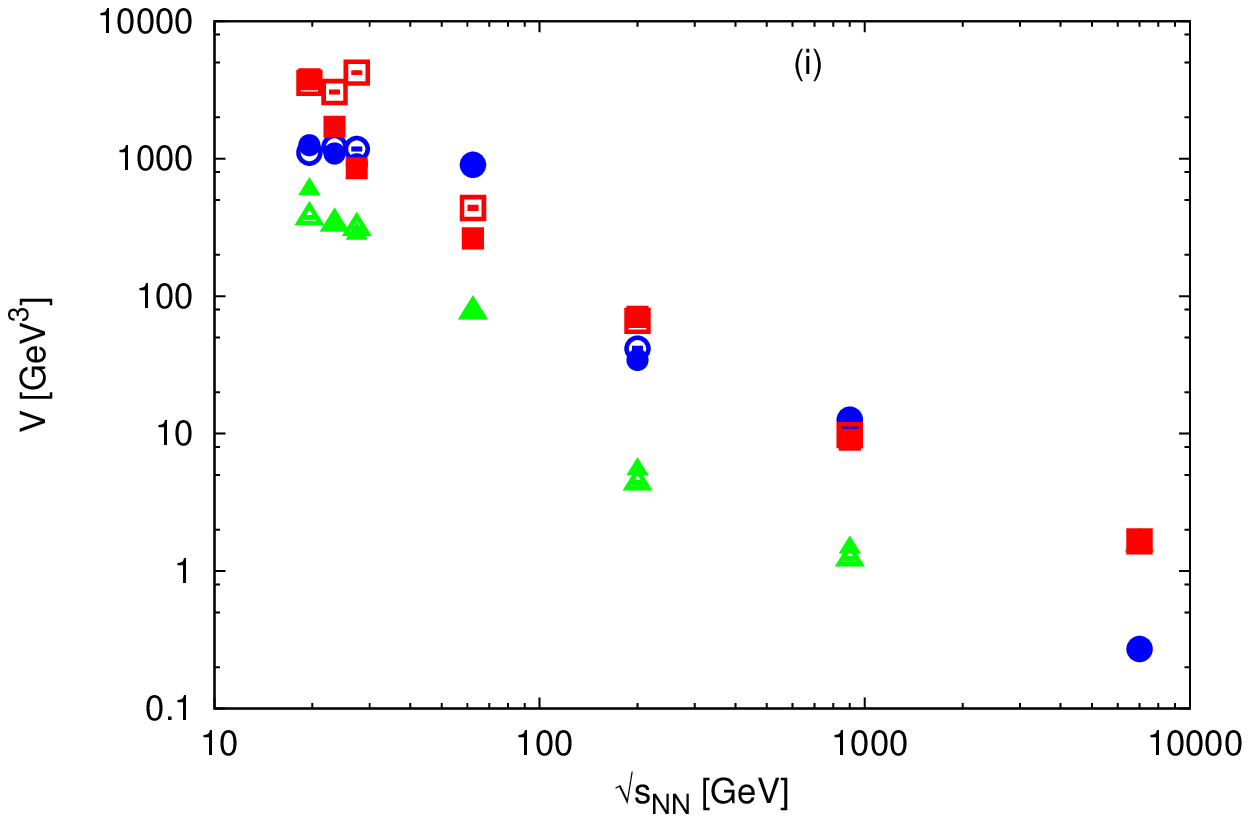}
\caption{(Color online) The various parameters obtained from statistical fits within Boltzmann, Tsallis and generic for $p_{\mathtt{T}}$ measured in p$+$p collisions for various charged particles in a wide range of energies, Appendices \ref{BoltzmannNN}, \ref{TsallisNN}, \ref{GenericNN}. 
\label{fig:GenericAllNN}
}}
\end{figure}

\begin{figure}[htb]
\centering{
\includegraphics[width=8cm,angle=-0]{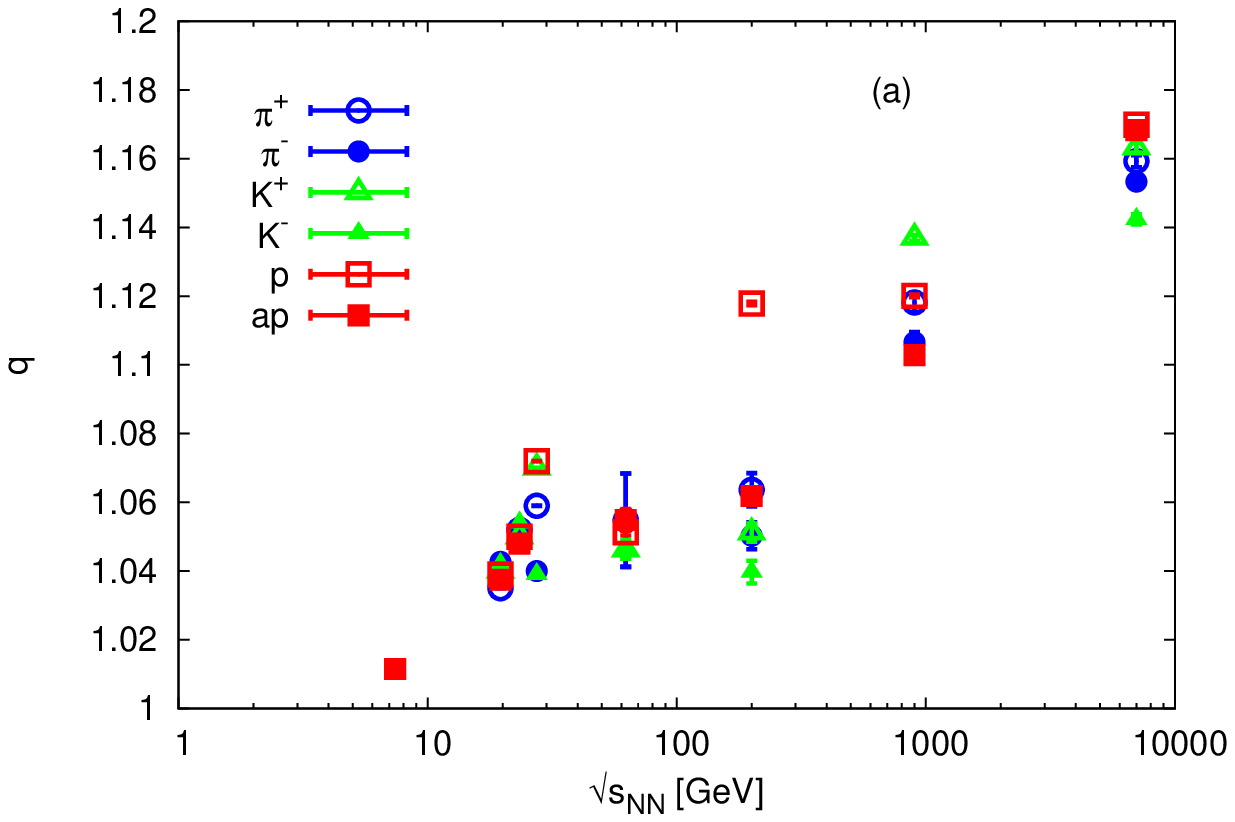}
\includegraphics[width=8cm,angle=-0]{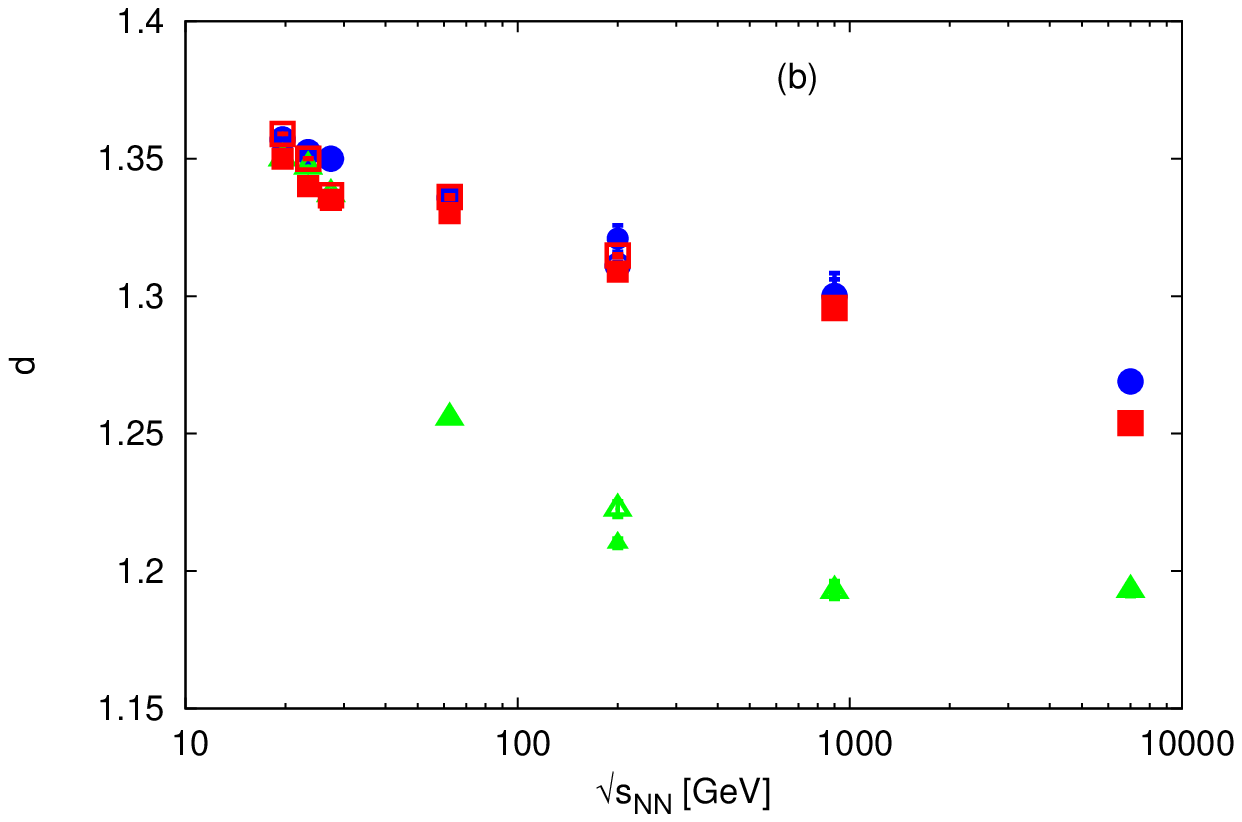}
\caption{(Color online) The nonextensive parameters $q$ and $d$ obtained from statistical fits within Tsallis and generic axiomatic statistics, respectively, for $p_{\mathtt{T}}$ measured in p$+$p collisions, Appendices \ref{TsallisNN}, \ref{GenericNN}.
\label{fig:GenericdNN}}
}
\end{figure} 

\begin{figure}
 \begin{center}
  \includegraphics[width=7cm]{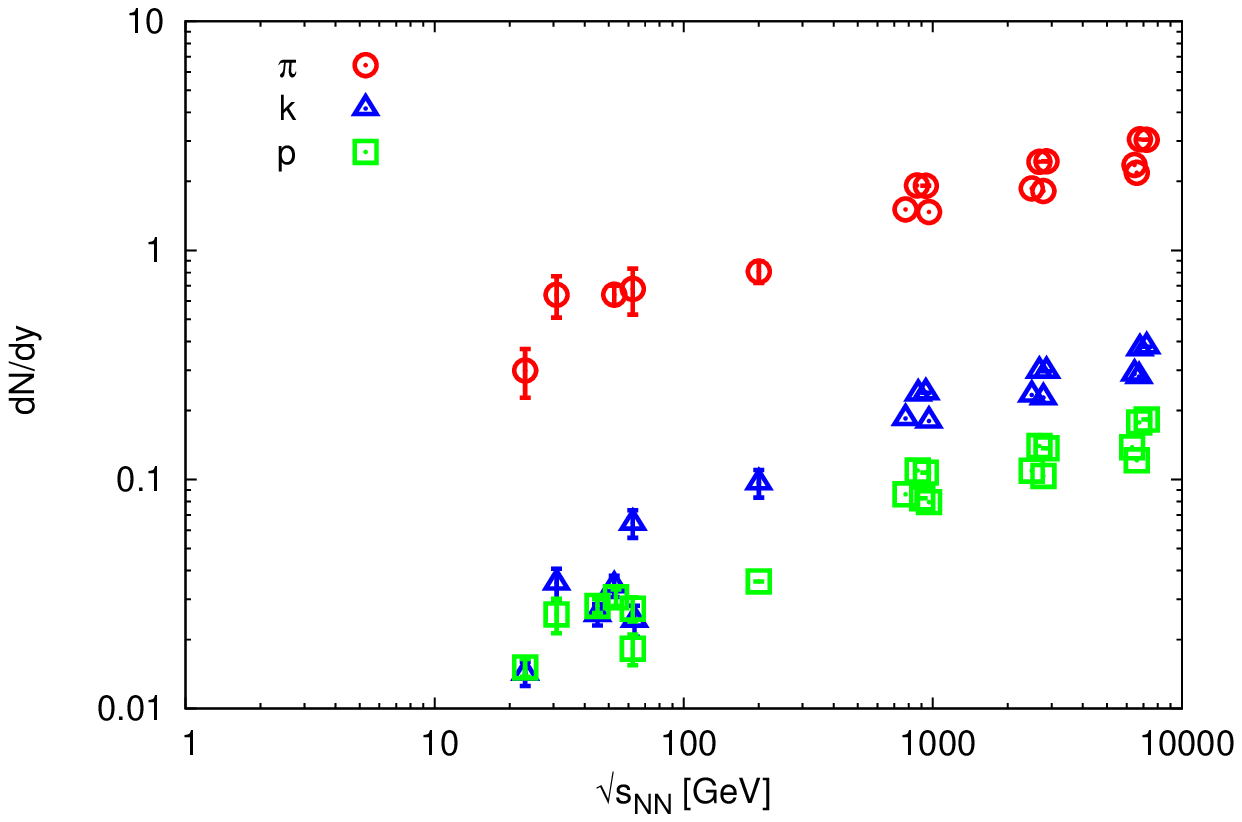}
  \includegraphics[width=7cm]{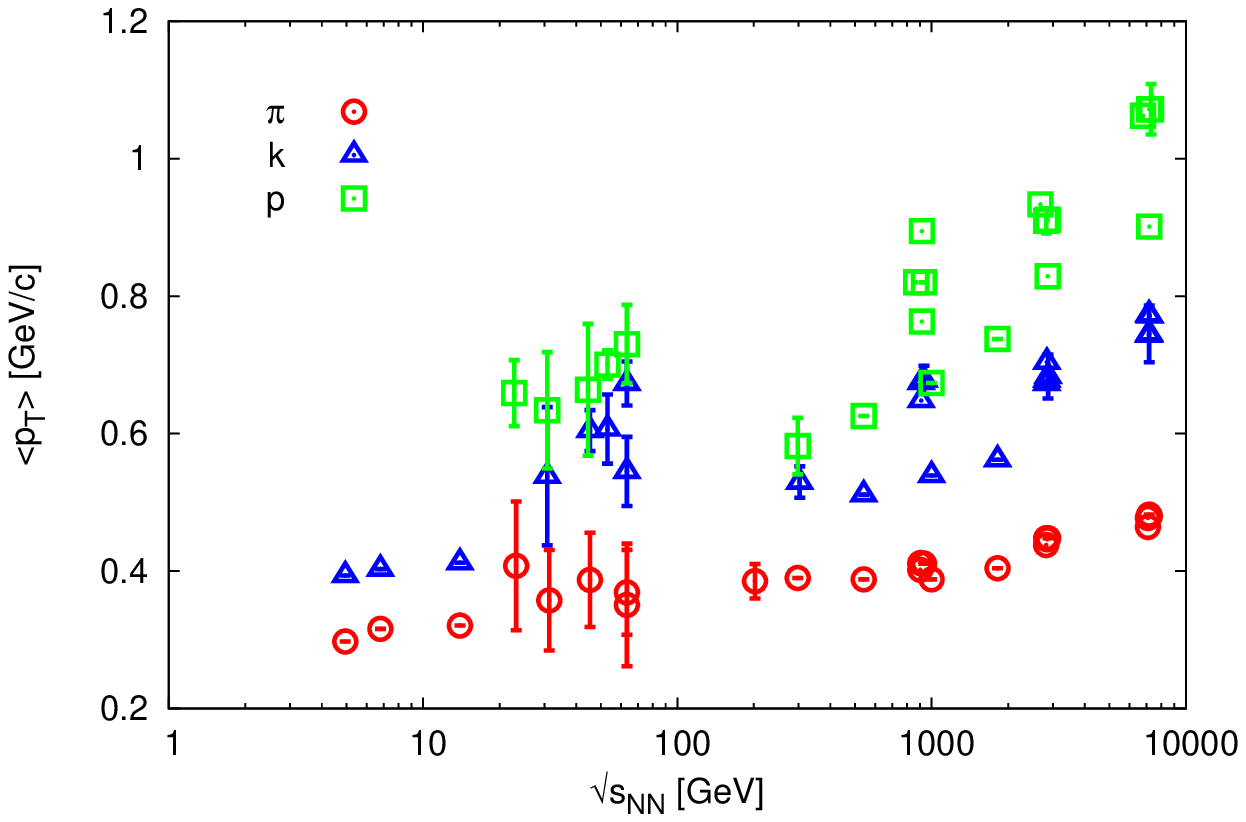} \hspace*{-2cm}
\caption{Left panel shows the central rapidity density $dN/dy$ as function of $\sqrt{s_{\mathtt{NN}}}$ as measured in p$+$p collisions. The right panel depicts  the average transverse momentum $\langle p_{\mathtt{T}}\rangle$ as function of $\sqrt{s_{\mathtt{NN}}}$ \cite{Khachatryan:2010us,Chatrchyan:2012qb,Adam:2015qaa}. Both graphs are taken from refs. \cite{Chatrchyan:2012qb,Adam:2015qaa}
\label{fig:CMSworld}}
\end{center} 
\end{figure}


\subsubsection{Universal trends of fit parameters obtained}

\begin{figure}[htb]
\centering{
\includegraphics[width=5cm,angle=-0]{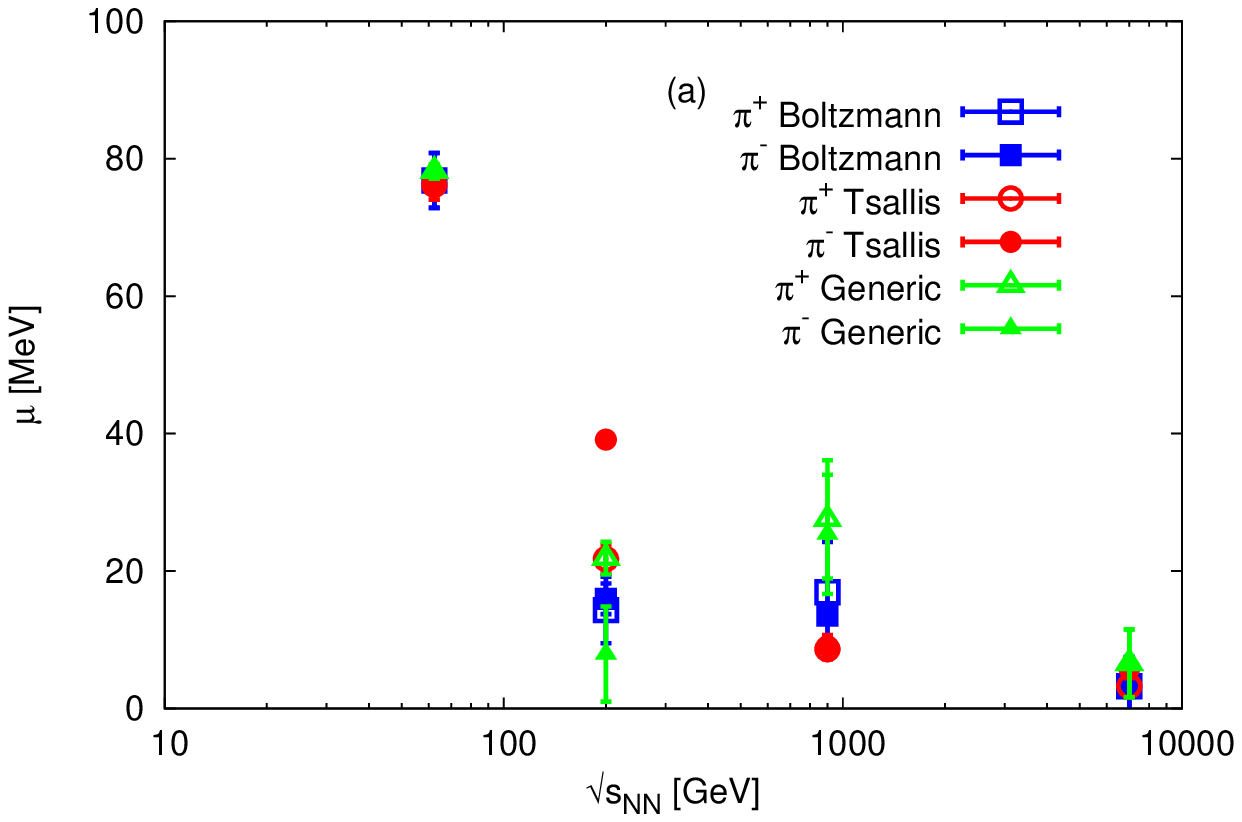}
\includegraphics[width=5cm,angle=-0]{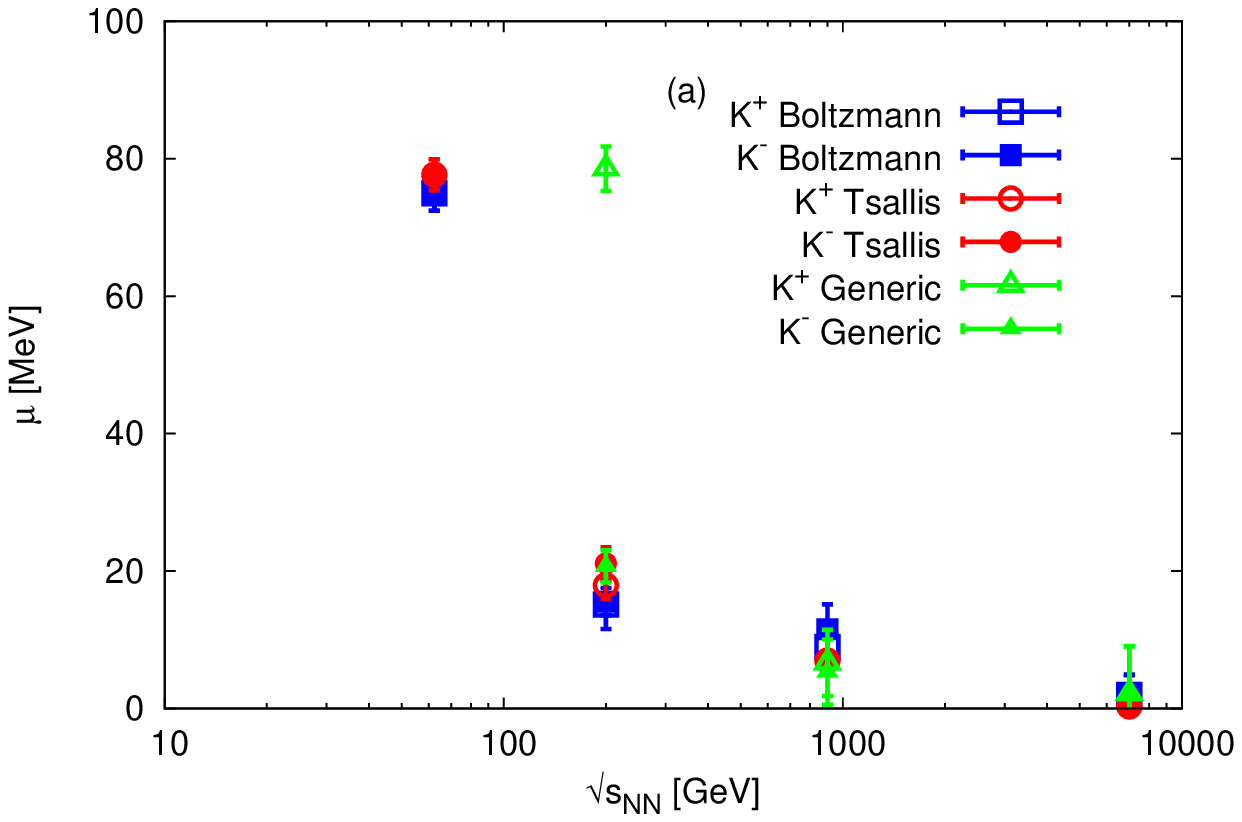}
\includegraphics[width=5cm,angle=-0]{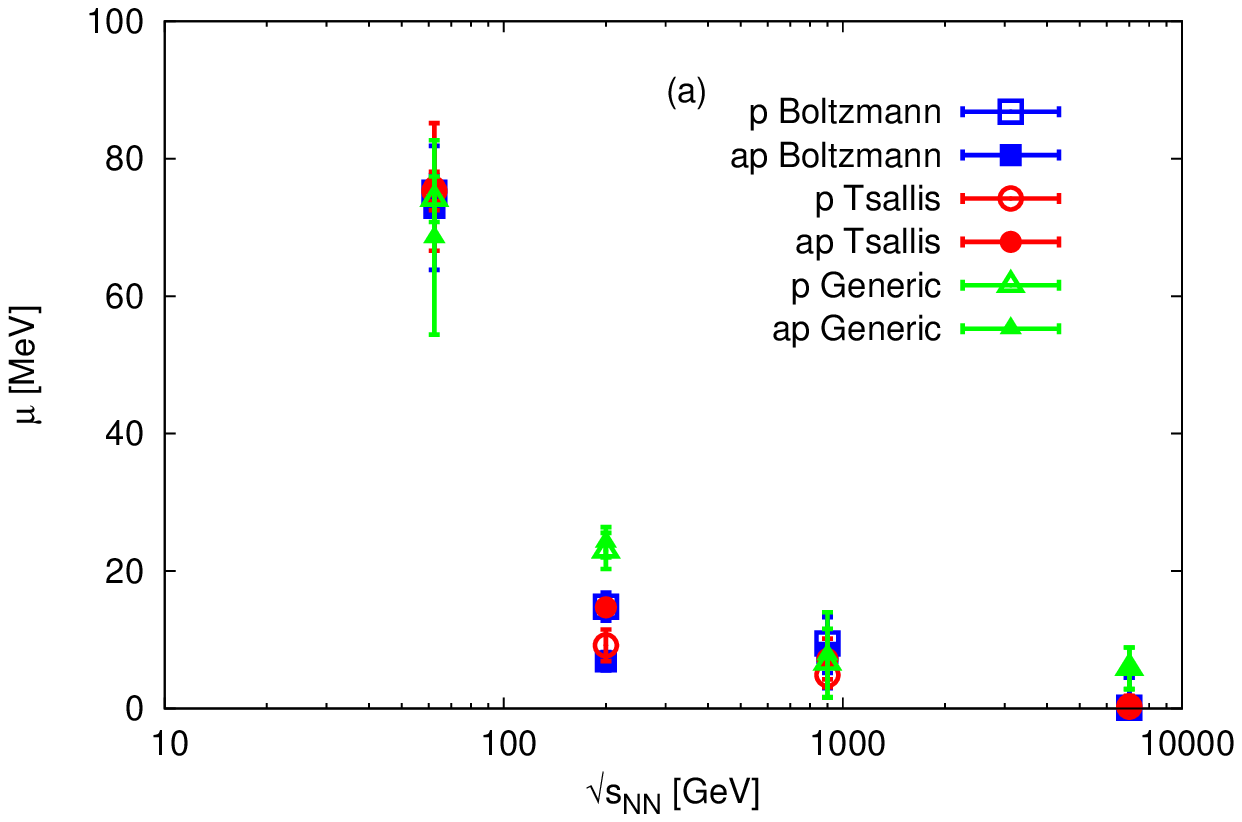} \\
\includegraphics[width=5cm,angle=-0]{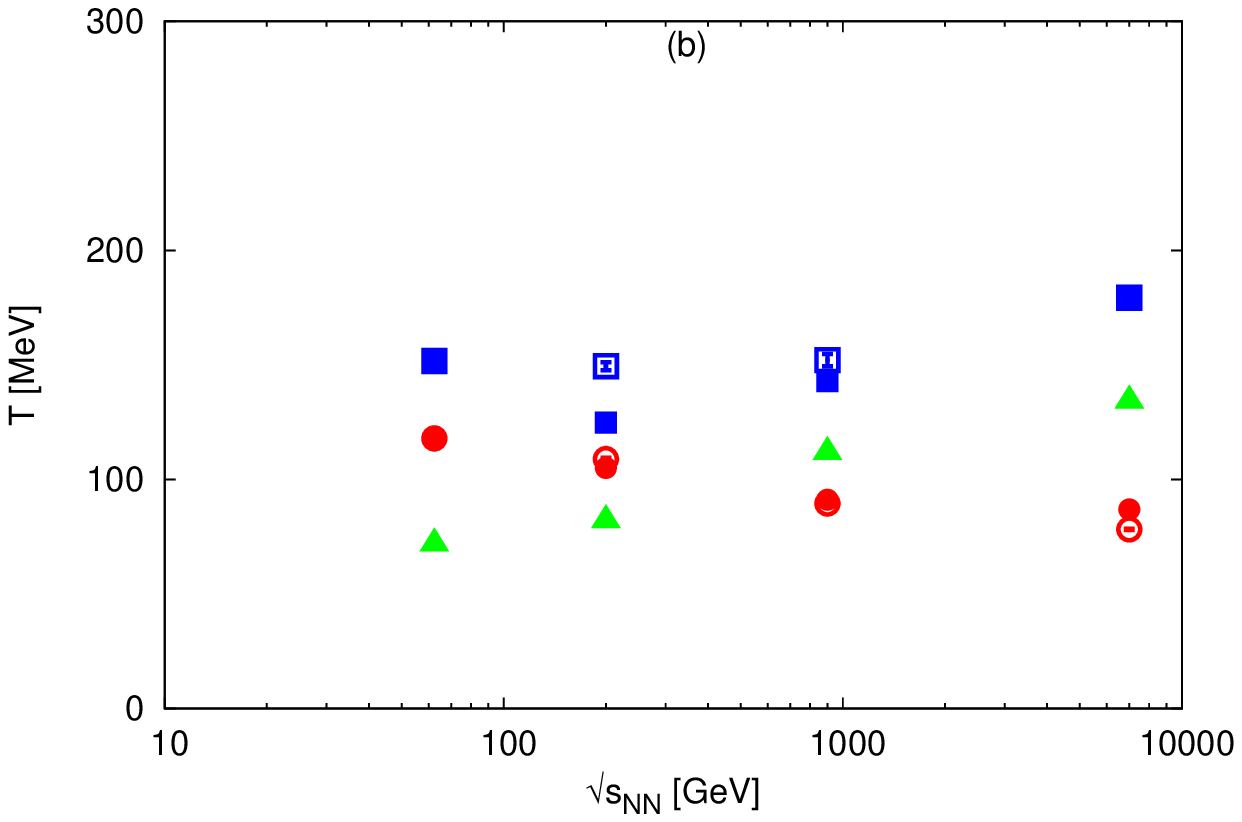}
\includegraphics[width=5cm,angle=-0]{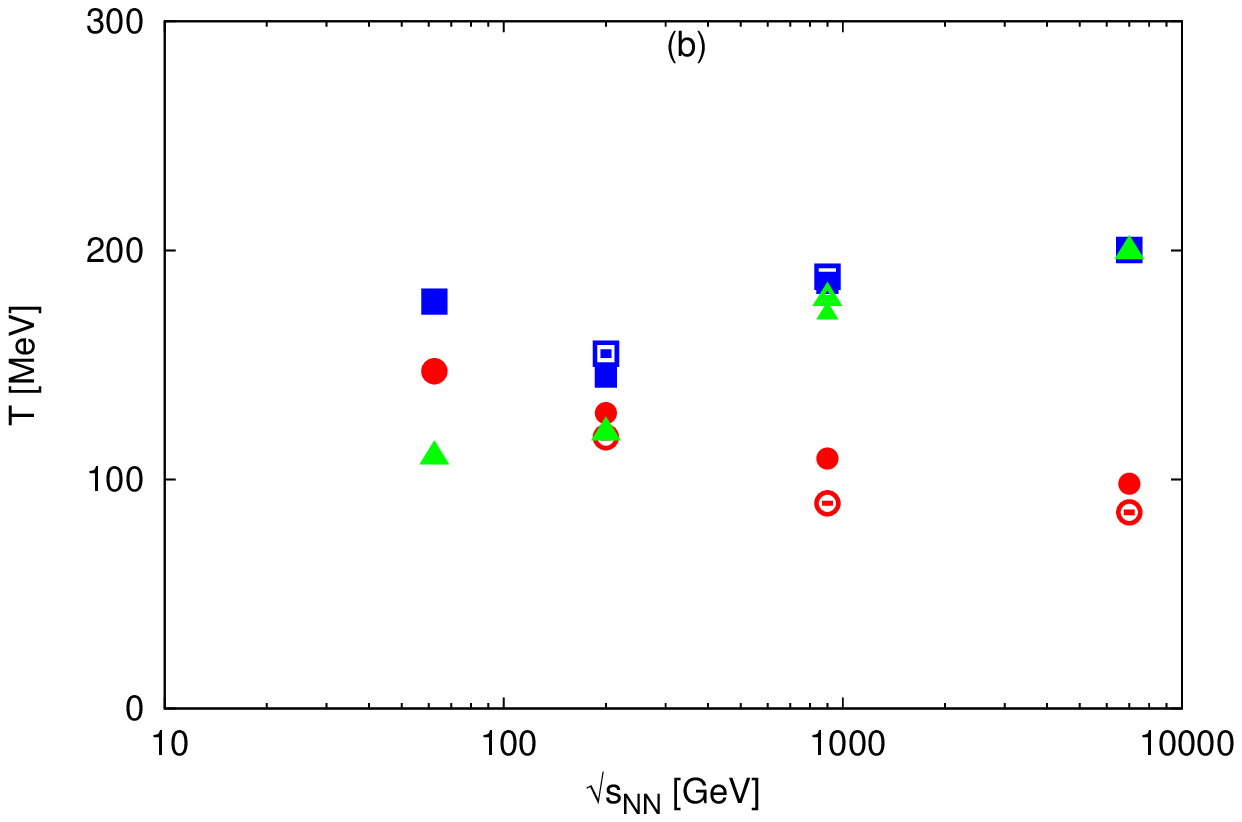}
\includegraphics[width=5cm,angle=-0]{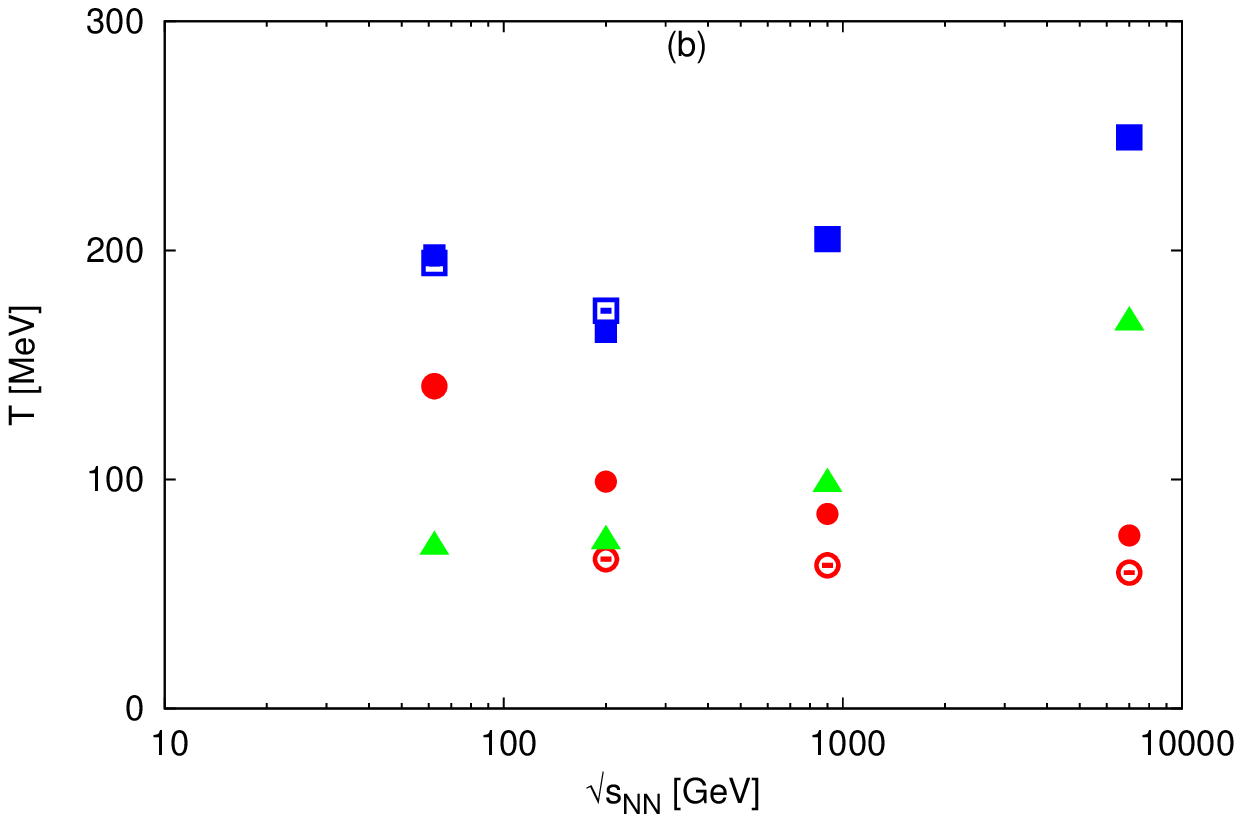} \\
\includegraphics[width=5cm,angle=-0]{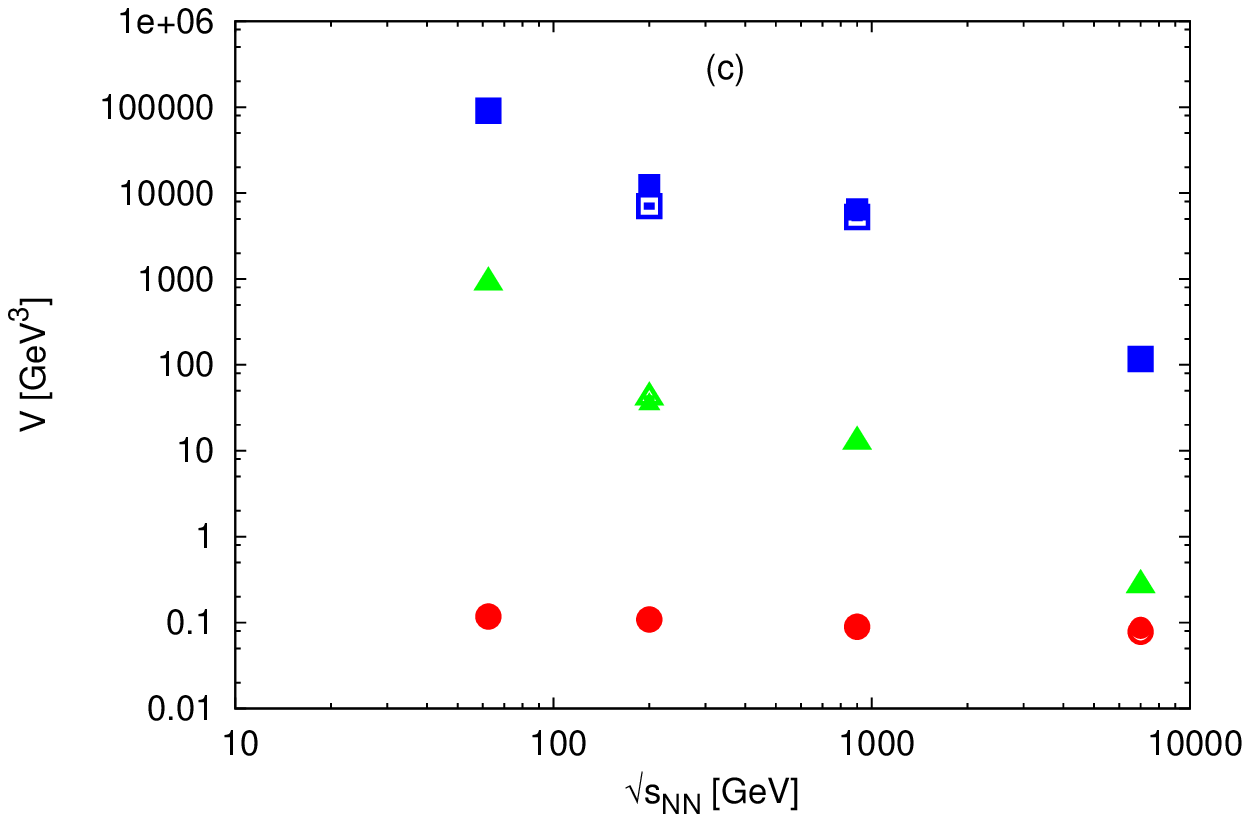}
\includegraphics[width=5cm,angle=-0]{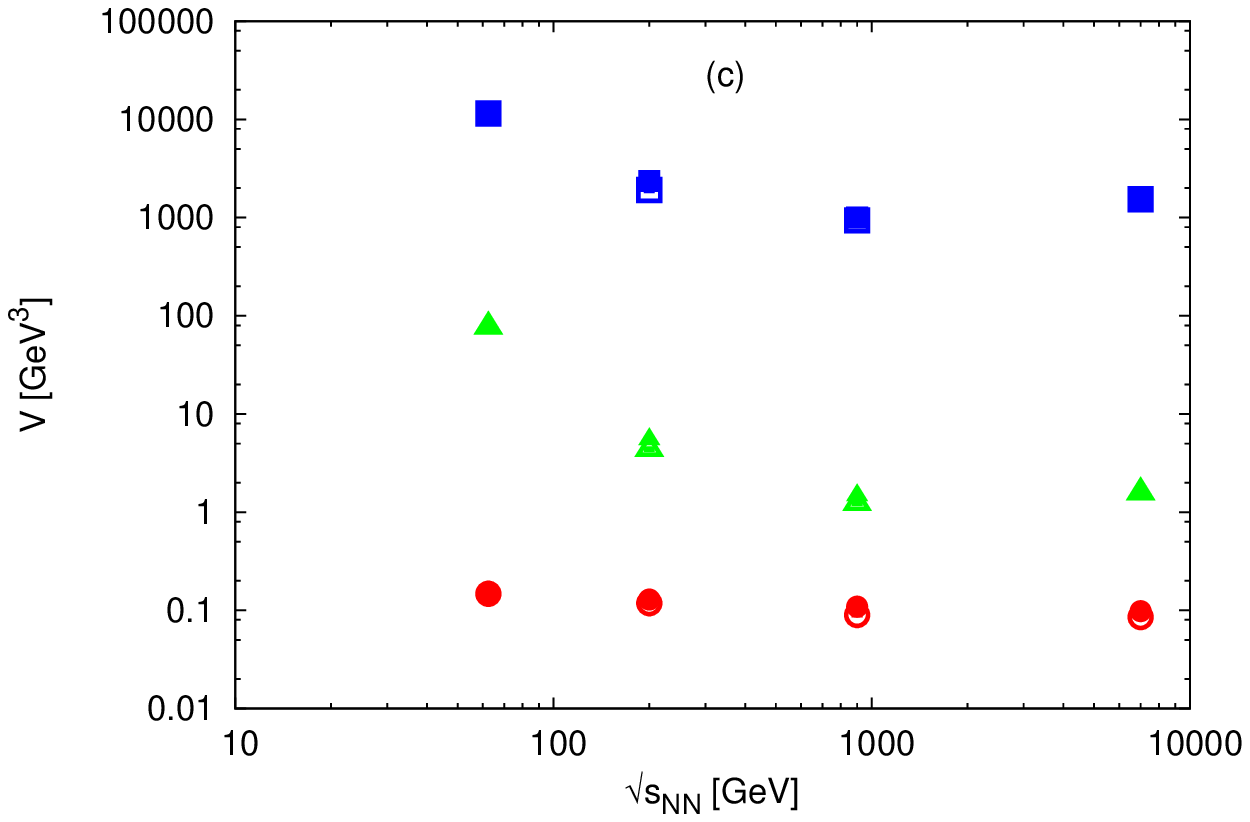}
\includegraphics[width=5cm,angle=-0]{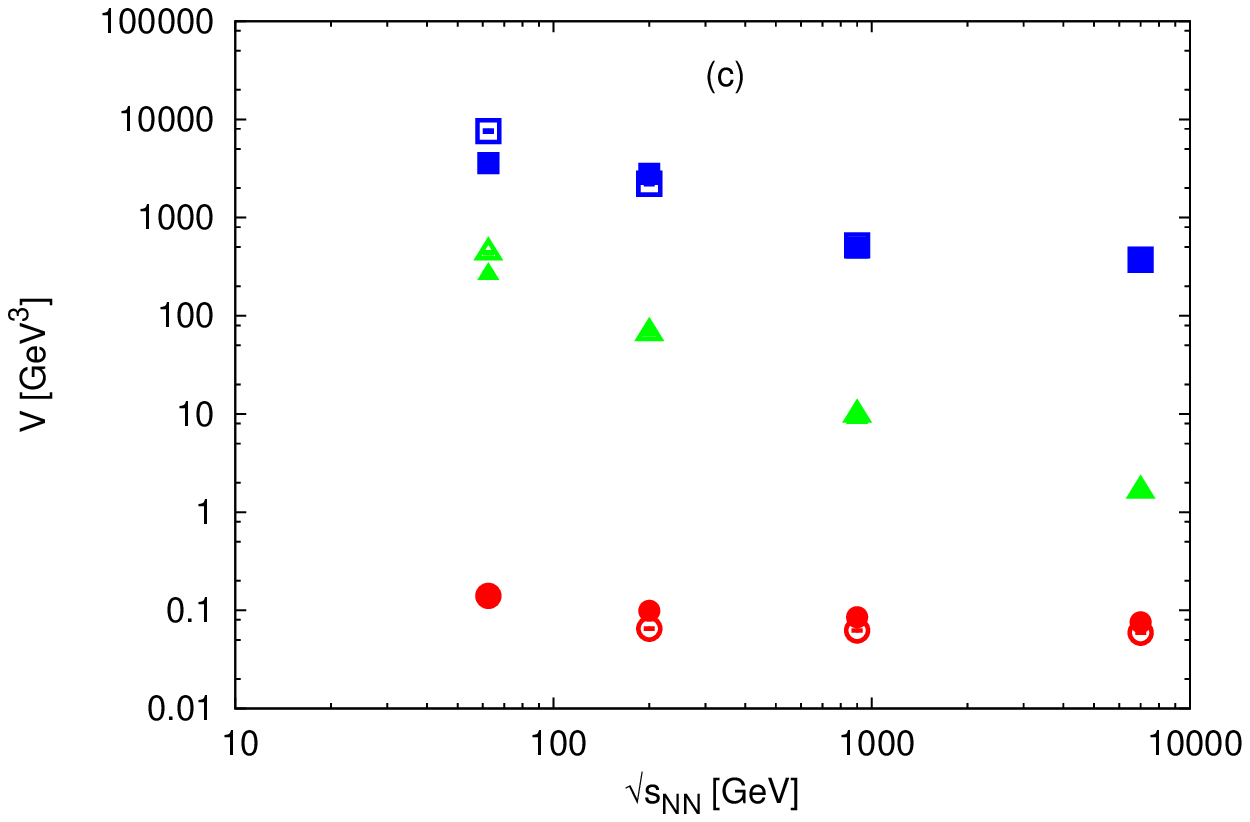} 
\caption{(Color online) A comparison between $\mu$, $T$, and $V$ obtained from the statistical fits within Boltzmann, Tsallis and generic axiomatic approaches for $p_{\mathtt{T}}$ spectra measured in p$+$p collisions for various charged particles in a wide range of energies, compare with Fig. \ref{fig:GenericAllNN}, Appendices \ref{BoltzmannNN}, \ref{TsallisNN}, \ref{GenericNN}. The results for pions, Kaons and protons are illustrated in left , middle and right panels, respectively. 
\label{fig:AllStatisticsPerParticleNN}
}}
\end{figure} 

At SPS energies ($6.27$ and $17.27~$GeV), RHIC energies ($62.4$ and $200$ GeV), and LHC energies ($0.9$, $2.76$, and $7$ TeV), the charged pion transverse momentum spectra, $p_{\mathtt{T}}$, in p$+$p collisions have been studied as functions of energy and multiplicity by means of Tsallis nonextensive approach \cite{Sett:2014csa}. The Tsallis parameters obtained have been parametrized as a function of the center-of-mass energy $\sqrt{s}$ as in the following expression:
\begin{eqnarray}
\label{par0}
f(\sqrt{s}) = \left[a + (\sqrt{s})^{-\alpha}\right]^{b}, \label{eq:discTsallis}
\end{eqnarray}
where $a=1.33 \pm 0.08$,  $\alpha =  0.22 \pm 0.06$ and $b=4.36 \pm 0.24$ for $n(\sqrt{s})$, 
$a=2.63 \pm 0.62$, $\alpha = 0.04 \pm 0.02$ and $b=3.76 \pm 0.49$ for $T(\sqrt{s})$ and 
$a=0.65 \pm 0.01$, $\alpha = 0.22 \pm 0.01$ and $b=-4.78 \pm 0.03$ for $dN(\sqrt{s})/dy$. 
Also, the charged pion spectra for different event multiplicities in p$+$p collisions at LHC energies was studied by using Tsallis distribution  \cite{Sett:2014csa}. Such as expression was being proposed due to its statistical fits to experimental results. We do the same procedure. The resulting expressions together with related graphs and tables for the parameters obtained are added to the end of this section.

Here, we analyze the transverse momentum spectra $p_{\mathtt{T}}$ for charged particles and anti-particles by using extensive and non-extensive statistics. Various parameters are obtained from the statistical fits in a wide range of center-of-mass energies by using three types of statistics: Boltzmann (extensive), Tsallis and generic statistics (non-extensive). It is concluded, sec. \ref{sec:res}, that Tsallis is more successful in describing p$+$p collisions than A$+$A collisions but BG has the reverse impact, i.e. better for A$+$A rather than for p$+$p. It was found that generic axiomatic approach is well applicable for both types of collisions. Boltzmann statistics can interpret the interaction between many particles so it can used excellently with a more crowded system, e.g. A$+$A collisions. But Tsallis is more accurate to explain the interaction between finite number of particles. This makes it good in explaining p$+$p collisions.

\subsubsection{Our expressions for fitting parameters}
\label{sec:outExprs}

For all particles produced the dependence of $\mu$ on the center-of-mass energies $\sqrt{s_{\mathtt{NN}}}$  as obtained from all types of statistical approaches for all types of collision can be expressed as 
\begin{equation}
\mu = a \sqrt{s_{\mathtt{NN}}}^b, \label{eq:musqrts}
\end{equation}
where $a$ and $b$ are given in Tables \ref{tab:fit_Boltzmann_AA}, \ref{tab:fit_Tsallis_AA}, \ref{tab:fit_Generic_AA}, \ref{tab:fit_Boltzmann_NN}, \ref{tab:fit_Tsallis_NN}, \ref{tab:fit_Generic_NN}, and taken from Figs. \ref{fig:GenericAll_fit}, \ref{fig:Genericd_NN_fit}}. For the seek of completeness, we refer to a widely used expression proposed in ref. \cite{Andronic:2005yp}, $\mu=a/(1+b \sqrt{s_{\mathtt{NN}}})$, where both parameters $a$ and $b$ differ from the ones in Eq. (\ref{eq:musqrts}). 

Other parameters have been deduced, as well. In the following, we review whether the size of the colliding system impacts the resulting quantities and how these vary with the energy and with the type of statistical approach applied.

\paragraph{A$+$A collision}

\begin{figure}[htb]
\centering{
\includegraphics[width=5cm,angle=-0]{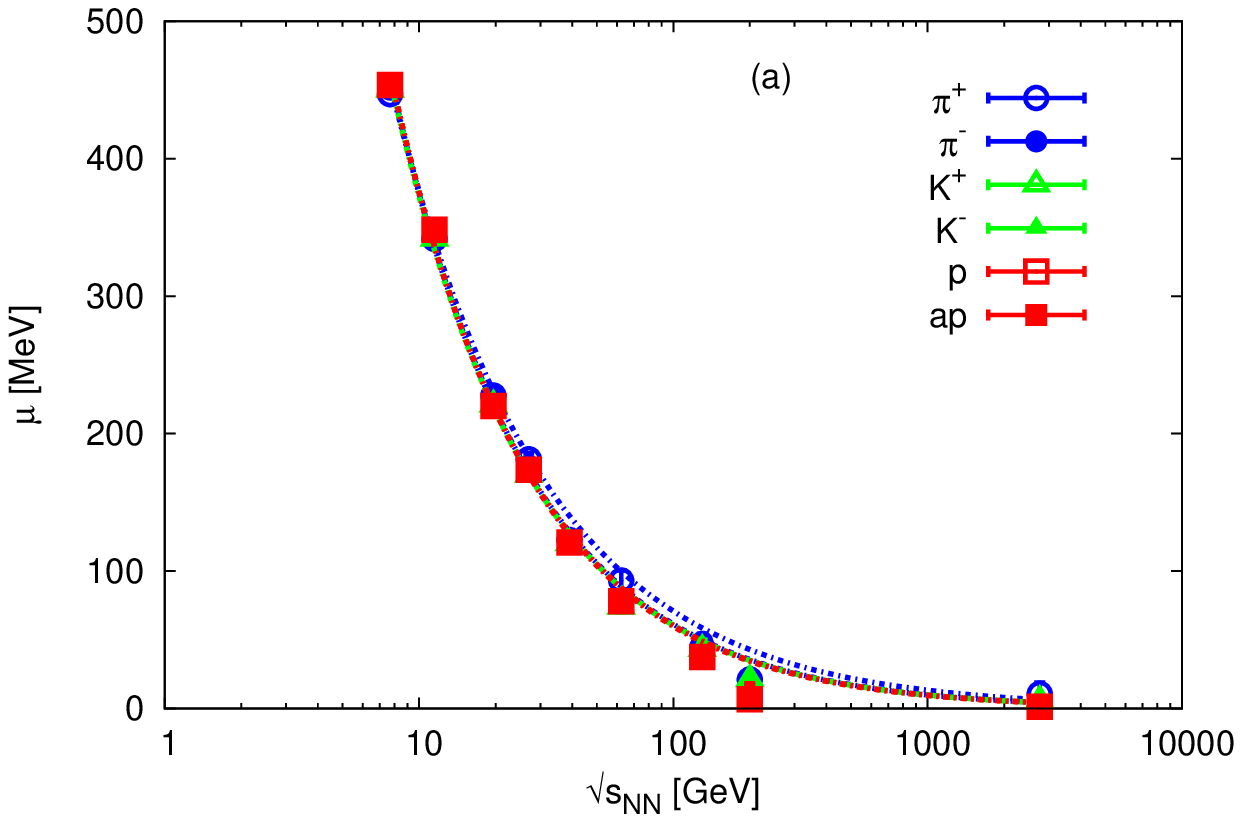}
\includegraphics[width=5cm,angle=-0]{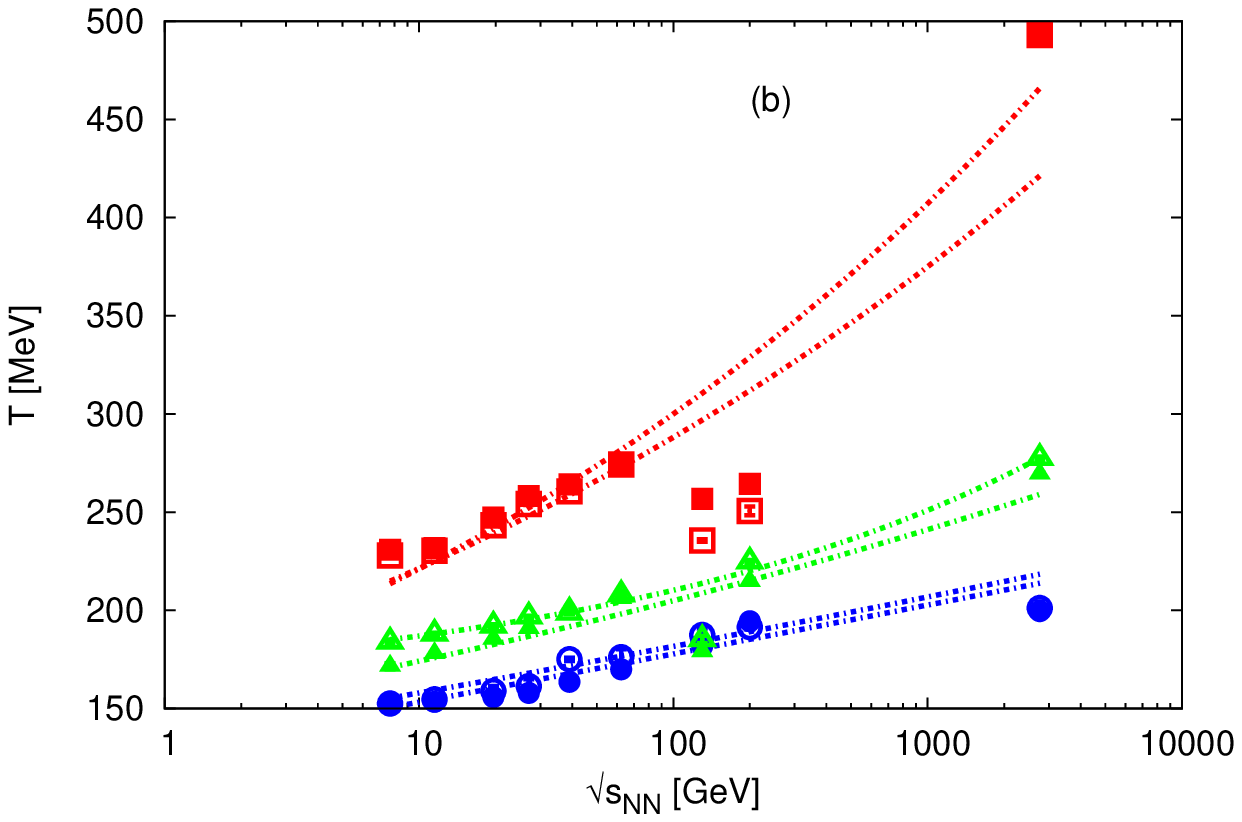}
\includegraphics[width=5cm,angle=-0]{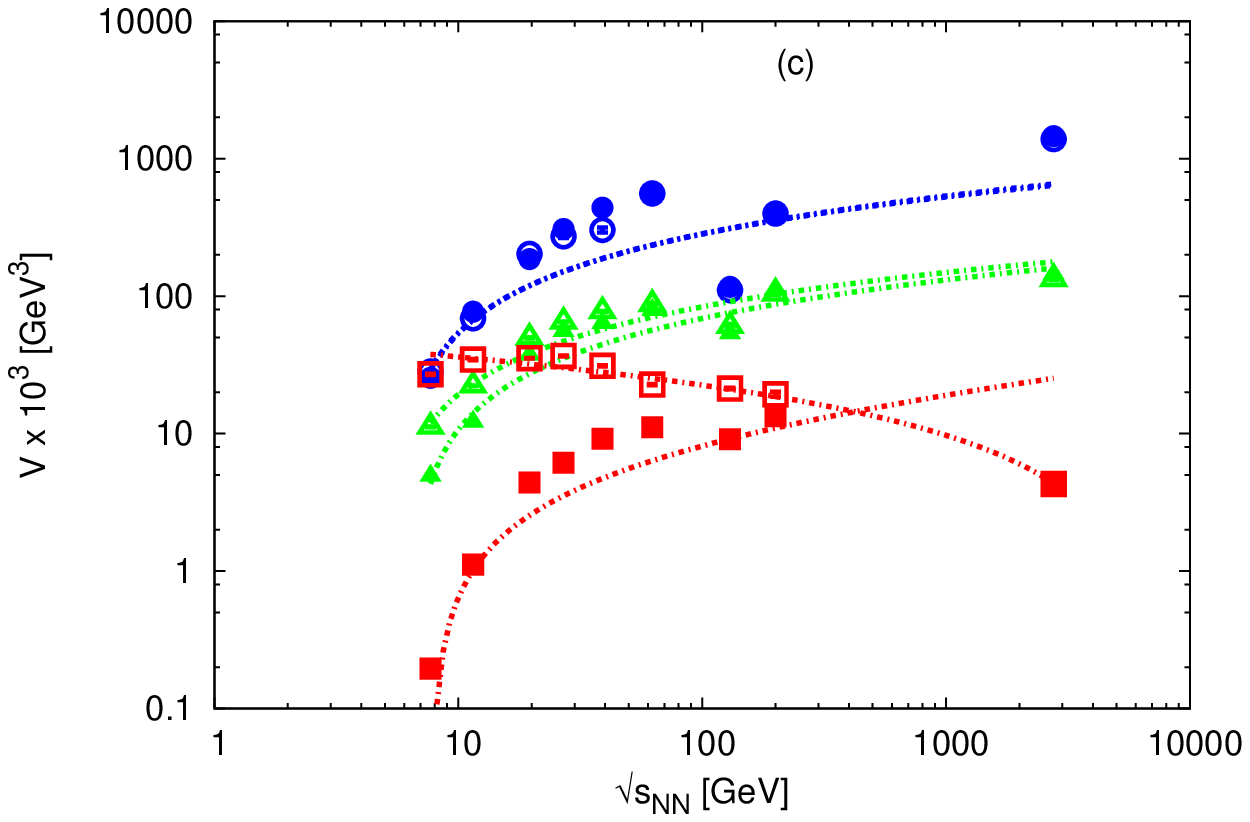}\\
\includegraphics[width=5cm,angle=-0]{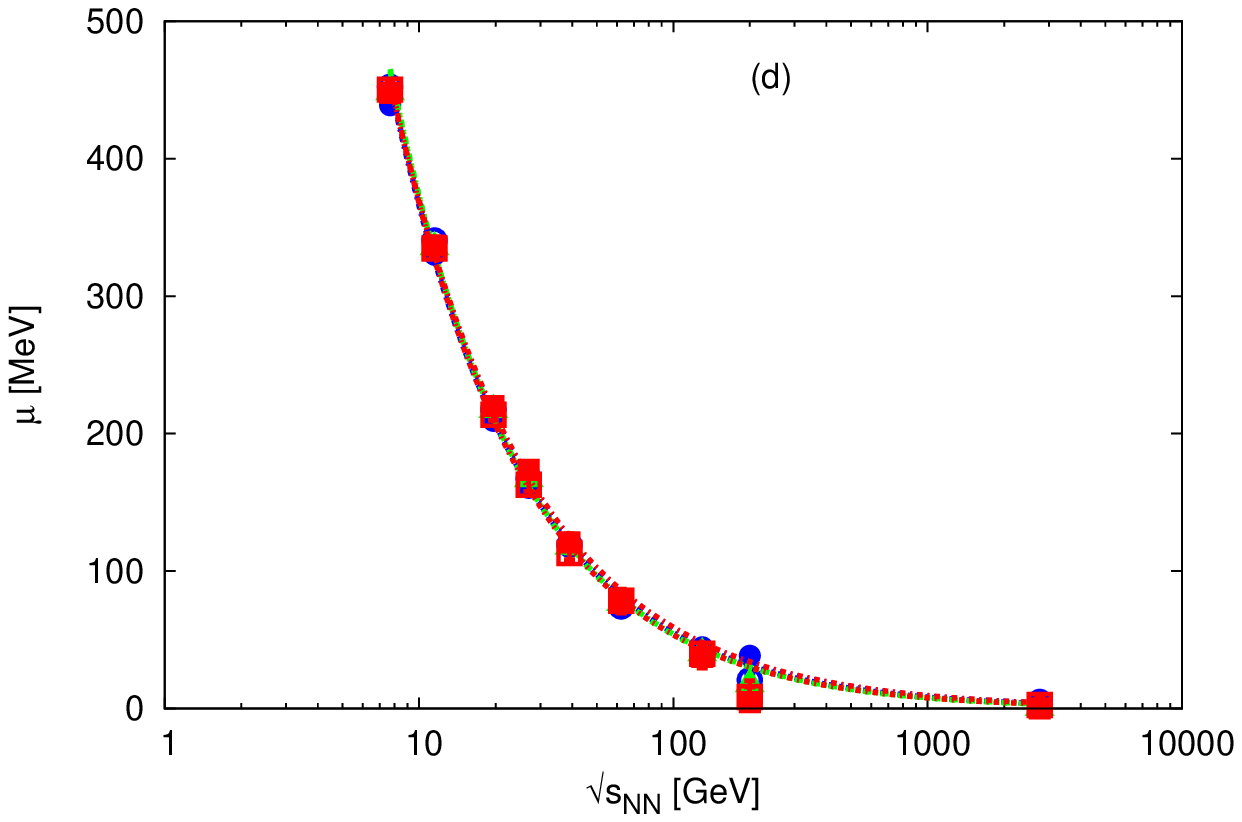}
\includegraphics[width=5cm,angle=-0]{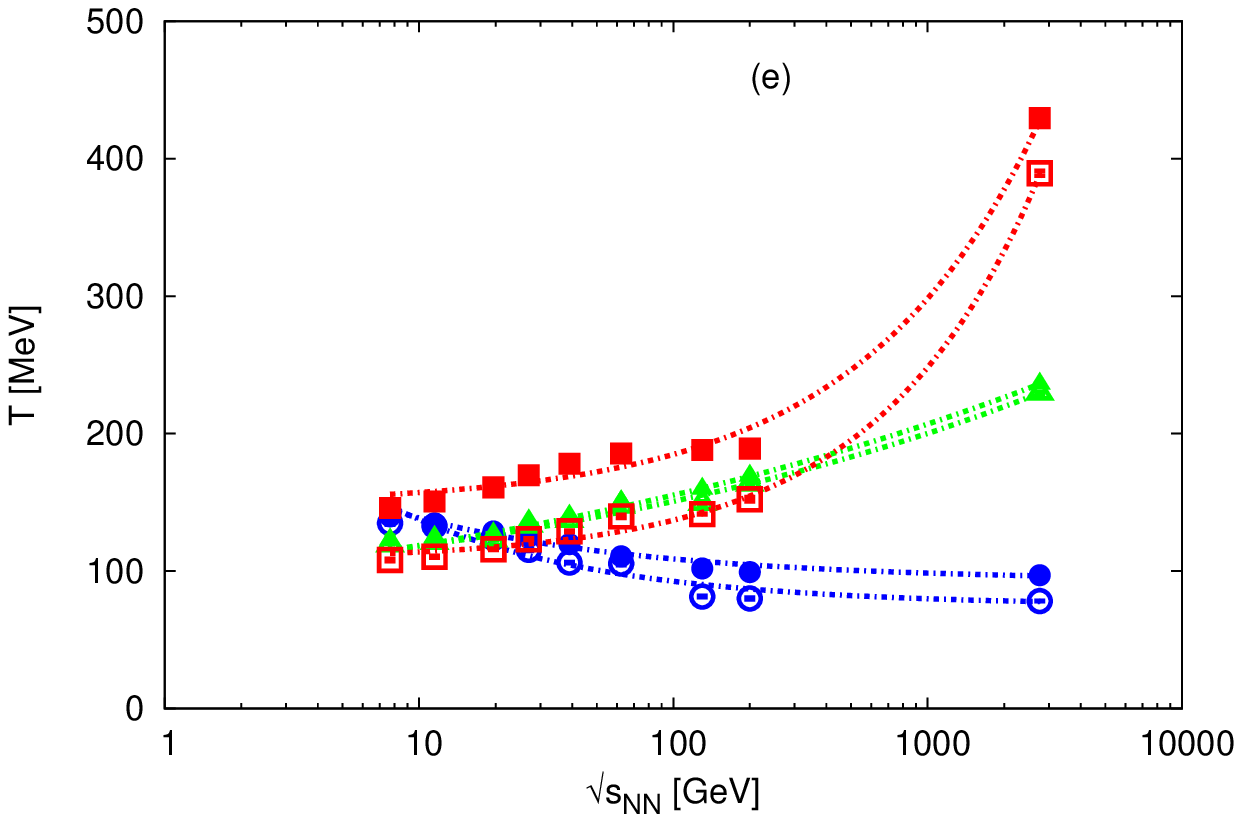}
\includegraphics[width=5cm,angle=-0]{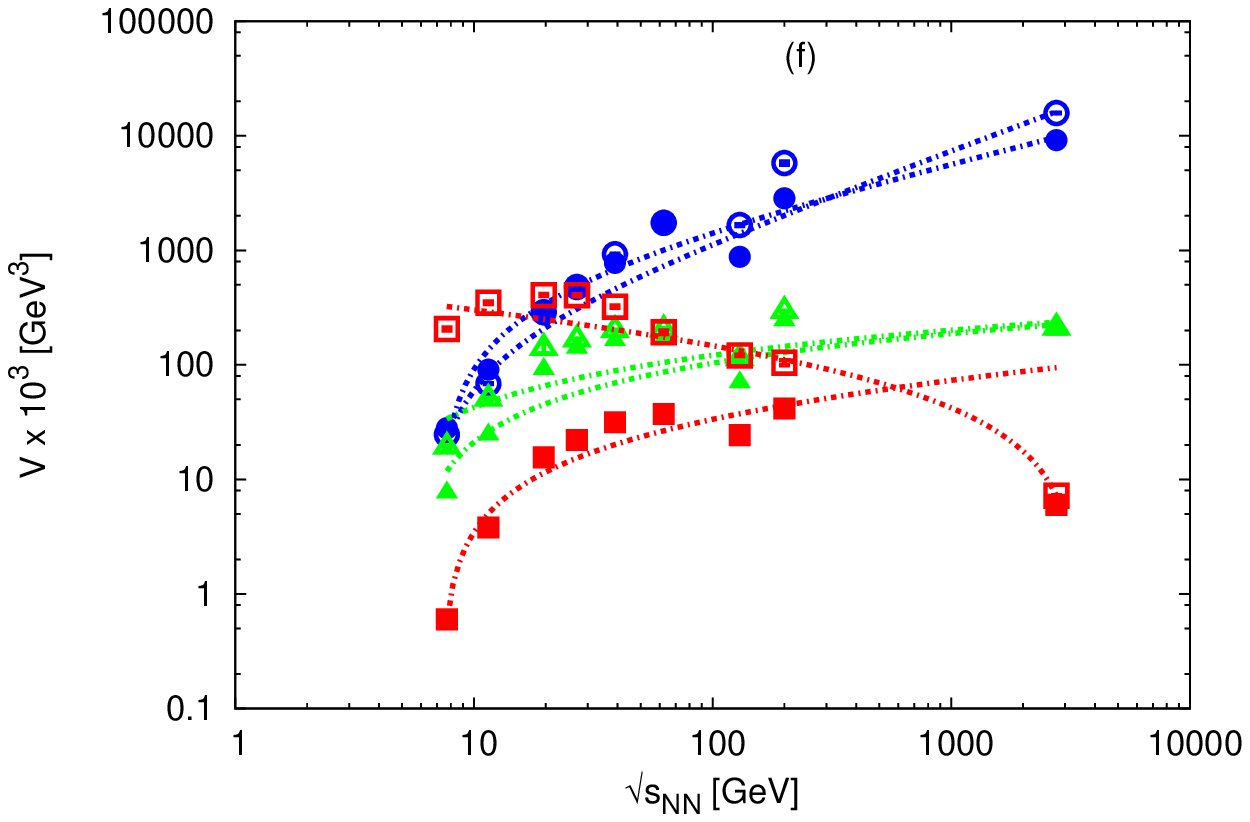} \\
\includegraphics[width=5cm,angle=-0]{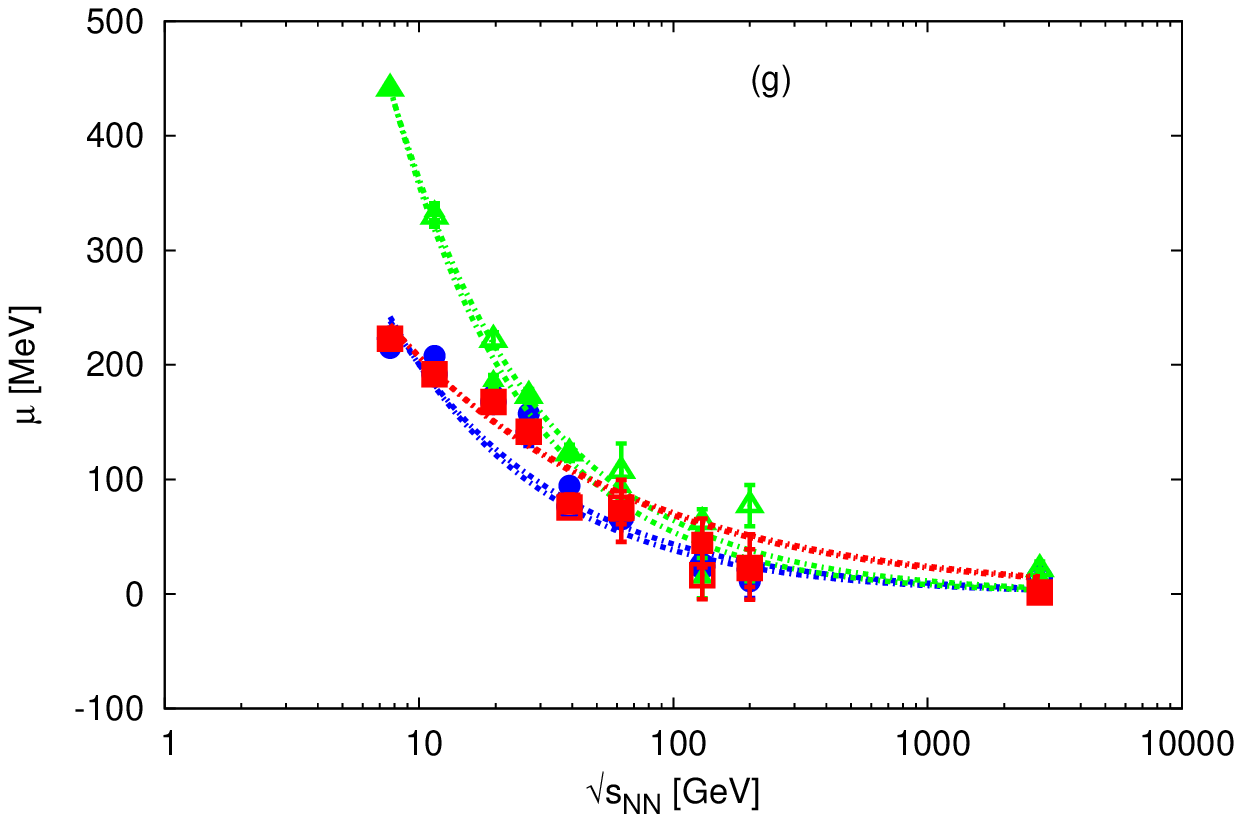}
\includegraphics[width=5cm,angle=-0]{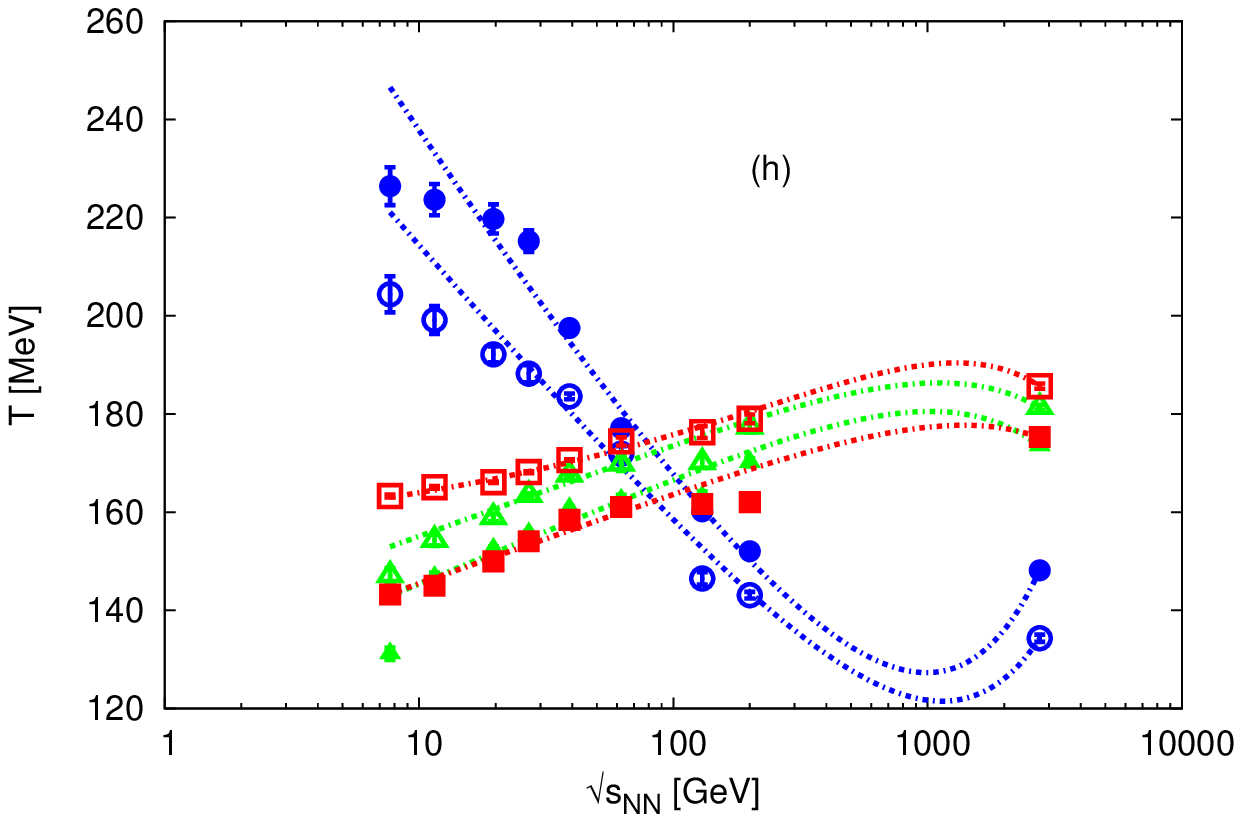}
\includegraphics[width=5cm,angle=-0]{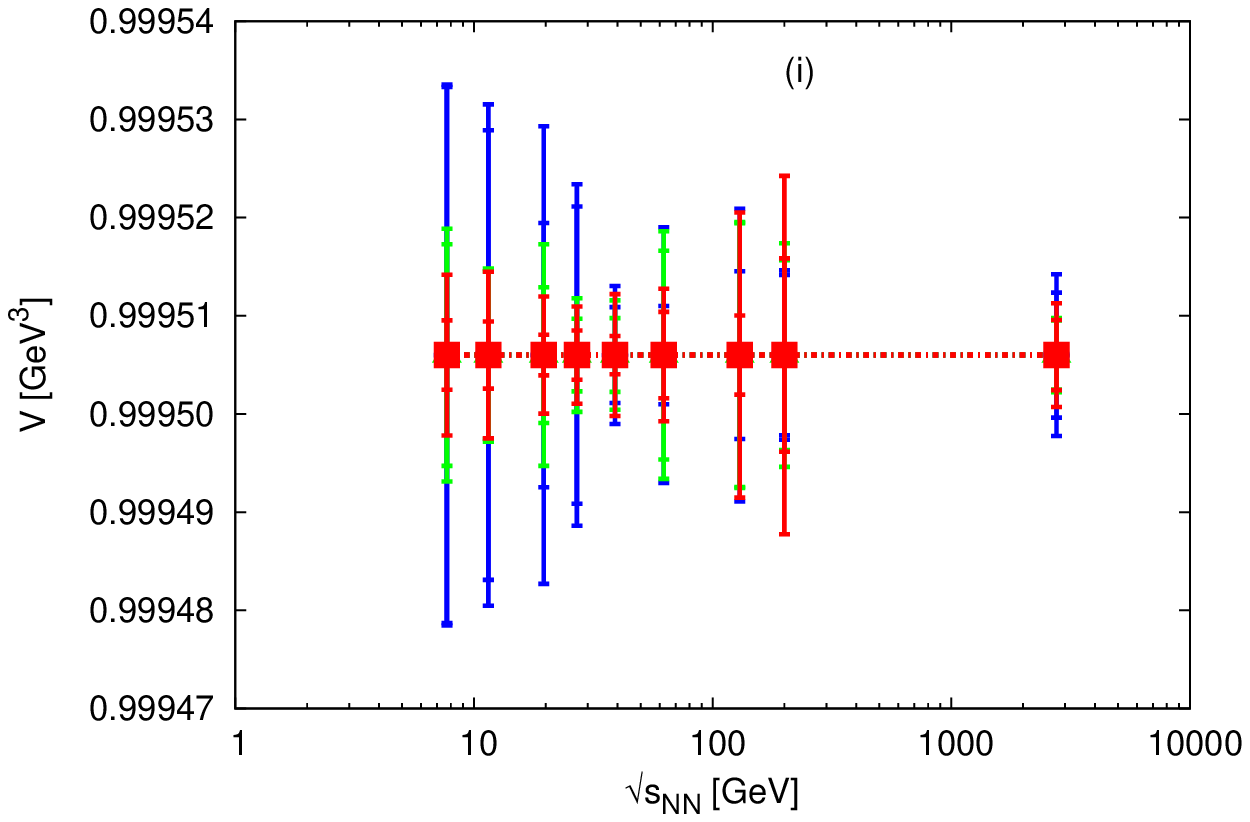}
\caption{(Color online) The various parameters obtained from the statistical fits within Boltzmann, Tsallis and generic axiomatic statistics for $p_{\mathtt{T}}$ measured in A$+$A collisions for various charged particles in a wide range of energies. The curves refer to the proposed expressions from the statistical fits. 
\label{fig:GenericAll_fit}
}}
\end{figure}

\begin{figure}[htb]
\centering{
\includegraphics[width=8cm,angle=-0]{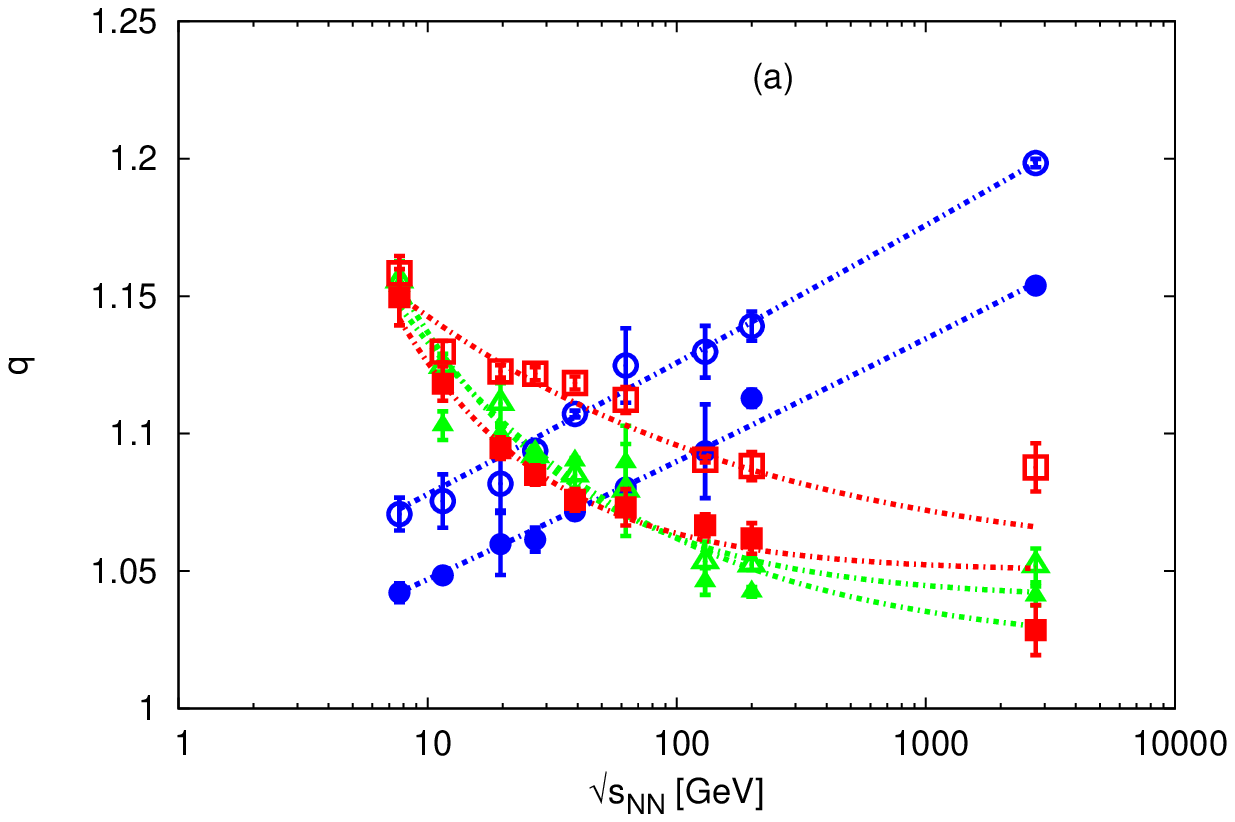}
\includegraphics[width=8cm,angle=-0]{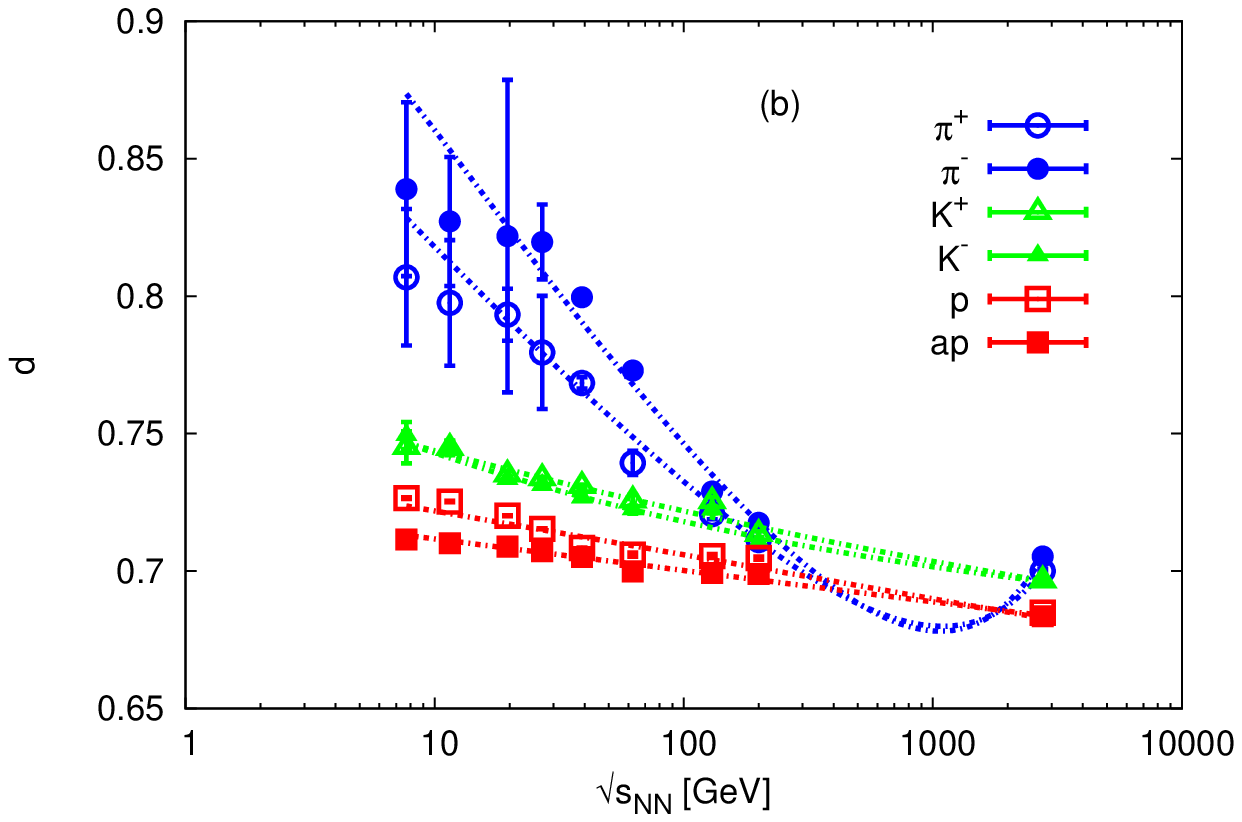}
\caption{(Color online) The nonextensive parameters $q$ and $d$ obtained from statistical fits within Tsallis and generic axiomatic statistics, respectively, for $p_{\mathtt{T}}$ measured in A$+$A collisions. The curves refer to the proposed expressions from the statistical fits. 
\label{fig:Genericd_fit}}
}
\end{figure}

Figures \ref{fig:GenericAll_fit}, \ref{fig:Genericd_fit} and Tables \ref{tab:fit_Boltzmann_AA}, \ref{tab:fit_Tsallis_AA}, and \ref{tab:fit_Generic_AA} illustrate and list out various fit parameters. In the following we summarize shortly the various dependences of these thermodynamical quantities on the centr-of-mass energy:
\begin{itemize}
\item Volume and temperature have a similar dependence on $\sqrt{s_{\mathtt{NN}}}$. This was obtained by using Boltzmann and Tsallis statistics for all particles 
\begin{equation}
V \; \mathtt{and} \; T = a \sqrt{s_{\mathtt{NN}}}^b  + c
\end{equation}
where the values of $a$, $b$, and $c$, are taken from Fig. \ref{fig:GenericAll_fit}, see Tab. \ref{tab:fit_Boltzmann_AA}, \ref{tab:fit_Tsallis_AA}.

\item By using the generic axiomatic statistical approach, we could estimate the dependence of temperature on $\sqrt{s_{\mathtt{NN}}}$ for all particles 
\begin{equation}
T = a \sqrt{s_{\mathtt{NN}}}^b  + c \sqrt{s_{\mathtt{NN}}} + f
\end{equation}
where the values of $a$, $b$, $c$, and $f$ are shown in Fig. \ref{fig:GenericAll_fit}, see Tab. \ref{tab:fit_Generic_AA}.

\item The dependence of volume on $\sqrt{s_{\mathtt{NN}}}$ for all particles
\begin{equation}
V = (a + \sqrt{s_{\mathtt{NN}}}^b )^ c
\end{equation}
where the values of $a$, $b$, and $c$ are illustrated in Fig. \ref{fig:GenericAll_fit}, see Tab. \ref{tab:fit_Generic_AA}.

\item The nonextensivity parameter $q$ obtained by means of statistical fits by using Tsallis approach for all particles except pions reads 
\begin{equation}
q = (a + \sqrt{s_{\mathtt{NN}}}^b)^c
\end{equation}
where the values of $a$, $b$, and $c$, are taken from  Fig. \ref{fig:Genericd_fit}, see Tab. \ref{tab:fit_Tsallis_AA}.

\item For pions, this reads 
\begin{equation}
q = a \sqrt{s_{\mathtt{NN}}}^b
\end{equation}
where the values of $a$ and $b$, are as depicted in left panel of Fig. \ref{fig:Genericd_fit}, see Tab. \ref{tab:fit_Tsallis_AA}.

\item the dependence of the equivalent class $d$ for all particles except pions on $\sqrt{s_{\mathtt{NN}}}$  can be given as 
\begin{equation}
d = a \sqrt{s_{\mathtt{NN}}}^b  + c
\end{equation}
where the values of $a$, $b$, and $c$, are shown in Fig. \ref{fig:Genericd_fit}, see Tab. \ref{tab:fit_Generic_AA}. 

\item For pions, this reads
\begin{equation}
d = (a + \sqrt{s_{\mathtt{NN}}}^b)^c + f \sqrt{s_{\mathtt{NN}}}
\end{equation}
where the values of $a$, $b$, $c$, and $f$, are shown in Fig. \ref{fig:Genericd_fit}, see Tab. \ref{tab:fit_Generic_AA}.
\end{itemize}

\begin{table}[h!]
\centering
  \begin{tabular}{|c|c|c|c|c|c|c|c|c|}   \hline
    &  & $\pi^+$ & $\pi^-$ & $K^+$ & $K^-$ & $p$ & $\bar{p}$ \\ \hline
    \multirow{3}{*}{$T$} & $a$ & $366.627\pm1.463$ &  $1625.14\pm1.914$ &  $17.6137\pm0.482$ &  $147.901\pm3.003$ &  $169.664\pm5.406$ &  $162.38\pm4.27$ \\
     \cline{2-8} & $b$ & $0.0257\pm0.001$ &  $0.0064\pm0.0003$ &  $0.2437\pm0.0971$ &  $0.0707\pm0.0042$ &  $0.1146\pm0.0081$ &  $0.1329\pm0.0056$ \\  
      \cline{2-8} & $c$ & $-231.031\pm1.615$ &  $-1496.04\pm1.958$ &  $156.121\pm1.469$ &  $0.0802\pm4.078$ &  $0.4695\pm8.082$ &  $0.6359\pm7.204$ \\ \hline
   \multirow{2}{*}{$\mu$} & $a$  & $2285.34\pm167.5$ &  $1999.9\pm101.3$ &  $2290.63\pm116.7$ &  $2337.01\pm154.2$ &  $2329.17\pm121.8$ &  $2358.88\pm140$ \\
        \cline{2-8} & $b$ & $-0.7854\pm0.0304$ &  $\-0.7254\pm0.0216$ &  $-0.7889\pm0.021$ &  $-0.7939\pm0.0268$ &  $-0.7923\pm0.0197$ &  $-0.7982\pm0.0227$ \\   \hline
     \multirow{3}{*}{$V$} & $a$ & $2237.37\pm4.01$ &  $3363.36\pm4.979$ &  $6624.89\pm1.773$ &  $780.821\pm1.024$ &  $240.72\pm0.60217$ &  $11.2916\pm0.8055$ \\ 
     \cline{2-8} & $b$ & $0.039\pm0.0008$ &  $0.027\pm0.0007$ &  $0.0042\pm0.0001$ &  $0.0293\pm0.0006$ &  $-0.027\pm0.0003$ &  $0.1629\pm0.0225$ \\   
    \cline{2-8} & $c$ & $-2393.41\pm4.368$ &  $-3525.19\pm5.276$ &  $-6669.5\pm1.79$ &  $-824.577\pm1.093$ &  $-190.086\pm0.4921$ &  $-15.7913\pm0.8193$ \\  \hline
    \end{tabular}
      \caption{Fit parameters obtained from Boltzmann statistics to results from A$+$A collisions. 
     \label{tab:fit_Boltzmann_AA} }
\end{table}

\begin{table}[h!]
\centering
  \begin{tabular}{|c|c|c|c|c|c|c|c|c|}   \hline
    &  & $\pi^+$ & $\pi^-$ & $K^+$ & $K^-$ & $p$ & $\bar{p}$ \\ \hline
    \multirow{3}{*}{$q$} & $a$  & $1.0321\pm0.0018$ &  $1.0058\pm0.0027$ &  $1.0849\pm0.0087$ &  $1.0575\pm0.0494$ &  $1.2096\pm0.0078$ &  $1.1079\pm0.0042$ \\ 
   \cline{2-8} & $b$ & $0.0189\pm0.0003$ &  $0.0174\pm0.0005$ &  $0.65134\pm0.07781$ &  $0.4631\pm0.1519$ &  $-0.337\pm0.0079$ &  $-0.7548\pm0.017$ \\    
     \cline{2-8} & $c$ & $ $ &  $ $ &  $0.4776\pm0.0682$ &  $0.3704\pm0.0894$ &  $0.2601\pm0.0036$ &  $0.4746\pm0.0099$ \\  \hline
    \multirow{3}{*}{$T$} & $a$ & $221.649\pm14.25$ &  $131.978\pm3.972$ &  $41.1364\pm0.282$ &  $62.3074\pm0.2107$ &  $1.268\pm0.0399$ &  $2.1977\pm0.1054$ \\
     \cline{2-8} & $b$ & $-0.5487\pm0.0195$ &  $-0.4618\pm0.009$ &  $0.1819\pm0.0011$ &  $0.1507\pm0.0006$ &  $0.6814\pm0.0046$ &  $0.6115\pm0.0065$ \\  
      \cline{2-8} & $c$ & $74.8774\pm1.143$ &  $93.1116\pm0.6771$ &  $55.6622\pm0.7537$ &  $30.5038\pm0.4605$ &  $107.78\pm1.26$ &  $148.16\pm2.544$ \\ \hline
   \multirow{2}{*}{$\mu$} & $a$  & $2335.51\pm67.11$ &  $2423.89\pm70.4$ &  $2558.91\pm188.6$ &  $2597.82\pm189.6$ &  $2525.43\pm109.1$ &  $2309.37\pm153.5$ \\
        \cline{2-8} & $b$ & $-0.8018\pm0.0117$ &  $-0.8215\pm0.0096$ &  $-0.8372\pm0.0268$ &  $-0.8427\pm0.0273$ &  $-0.8368\pm0.0167$ &  $-0.7958\pm0.026$ \\  \hline
    \multirow{3}{*}{$V$} & $a$ & $36.5323\pm1.103$ &  $188.569\pm4.098$ &  $4873.72\pm13.48$ &  $4869\pm7.284$ &  $710.354\pm4.266$ &  $68.9332\pm238.9$ \\
     \cline{2-8} & $b$ & $0.771\pm0.0113$ &  $0.504\pm0.0084$ &  $0.0068\pm0.0007$ &  $0.0072\pm0.0005$ &  $-0.1682\pm0.0008$ &  $0.1231\pm0.3134$ \\   
    \cline{2-8} & $c$ & $-155.883\pm6.116$ &  $-510.2\pm12.79$ &  $-4908.35\pm13.78$ &  $-4928.96\pm7.416$ &  $-180.231\pm1.13$ &  $-88.0645\pm250.6$ \\  \hline
    \end{tabular}
      \caption{The same as in Tab. \ref{tab:fit_Boltzmann_AA} but here for the Tsallis statistical approach. 
     \label{tab:fit_Tsallis_AA} }
\end{table}

\begin{table}[h!]
\centering
\tiny{
  \begin{tabular}{|c|c|c|c|c|c|c|c|c|}   \hline
    &  & $\pi^+$ & $\pi^-$ & $K^+$ & $K^-$ & $p$ & $\bar{p}$ \\ \hline
    \multirow{4}{*}{$d$} & $a$  & $0.0985\pm0.0014$ &  $18.8923\pm455.8$ &  $0.1595\pm0.0007$ &  $0.1111\pm0.0015$ &  $\-3.0078\pm0.0011$ &  $1.7382\pm0.0003$ \\ 
   \cline{2-8} & $b$ & $0.0567\pm0.0002$ &  $2.3948\pm0.0126$ &  $-0.0781\pm0.0008$ &  $-0.1814\pm0.0026$ &  $0.0023\pm0.00009$ &  $-0.0029\pm0.00003$ \\    
     \cline{2-8} & $c$ & $-0.9432\pm0.0022$ &  $-0.027\pm0.0001$ &  $0.6106\pm0.0004$ &  $0.6698\pm0.0006$ &  $3.7456\pm0.0011$ &  $-1.015\pm0.0003$ \\ 
     \cline{2-8} & $f$ & $(2.972\pm0.0357)\times10^{-5}$ &  $(3.841\pm0.0858)\times10^{-5}$ &   &  &   &   \\ \hline
    \multirow{4}{*}{$T$} & $a$ & $1452.9\pm1.315$ &  $733.242\pm1.448$ &  $507.126\pm0.5621$ &  $572.472\pm0.9146$ &  $6.4303\pm0.059$ &  $644.535\pm0.539$ \\
     \cline{2-8} & $b$ & $-0.0184\pm0.0001$ &  $-0.0512\pm0.0004$ &  $0.0156\pm0.0002$ &  $0.0158\pm0.0003$ &  $0.3024\pm0.0018$ &  $0.0121\pm0.0002$ \\  
      \cline{2-8} & $c$ & $0.0206\pm0.001$ &  $0.0267\pm0.001$ &  $-0.008\pm0.0005$ &  $-0.0099\pm0.0007$ &  $-0.0131\pm0.0004$ &  $-0.0061\pm0.0008$ \\
      \cline{2-8} & $f$ & $-1178.47\pm1.202$ &  $-414.133\pm1.136$ &  $-370.462\pm0.6012$ &  $-448.237\pm0.9873$ &  $151.206\pm0.1855$ &  $-517.188\pm0.5623$ \\  \hline
   \multirow{2}{*}{$\mu$} & $a$  & $1051.61\pm171$ &  $920.591\pm211$ &  $2038.82\pm102.7$ &  $2359.56\pm210$ &  $625.662\pm139.4$ &  $605.863\pm111.5$ \\
        \cline{2-8} & $b$ & $-0.72\pm0.0512$ &  $-0.6621\pm0.0872$ &  $-0.7502\pm0.0212$ &  $-0.8202\pm0.0393$ &  $-0.4797\pm0.0817$ &  $-0.466\pm0.0674$ \\   \hline
      \multirow{3}{*}{$V$} & $a$ & $1.0944\pm0.0978$ &  $1.0843\pm0.0427$ &  $0.3676\pm2.532\times10^{-9}$ &  $0.3676\pm1.785\times10^{-9}$ &  $3.6721\pm3.624\times10^{-8}$ &  $3.5195\pm3.507\times10^{-8}$ \\
      \cline{2-8} & $b$ & $(-3.806\pm0.474)\times10^{-7}$ &  $(-1.45\pm0.169)\times10^{-7}$ &  $-7.337\pm1.017$ &  $-7.061\pm0.7599$ &  $-7.2576\pm22.97$ &  $-7.1715\pm18.94$ \\   
     \cline{2-8} & $c$ & $(-6.684\pm0.422)\times10^{-4}$ &  $(-6.728\pm0.188)\times10^{-4}$ &  $0.0005\pm2.186\times10^{-11}$ &  $0.0005\pm5.097\times10^{-11}$ &  $-0.0004\pm2.881\times10^{-12}$ &  $-0.0004\pm3.108\times10^{-12}$ \\  \hline
\end{tabular}
\caption{The same as in Tab. \ref{tab:fit_Boltzmann_AA} but here for the generic axiomatic statistical approach. 
\label{tab:fit_Generic_AA} }
}
\end{table}

So far we can conclude that the fit parameters obtained, even with this system size, depend on the energy but as well as on the type of statistical approach applied, especially when moving from Boltzmann to Tsallis approach.

\paragraph{p$+$p collisions}

\begin{figure}[htb]
\centering{
\includegraphics[width=5cm,angle=-0]{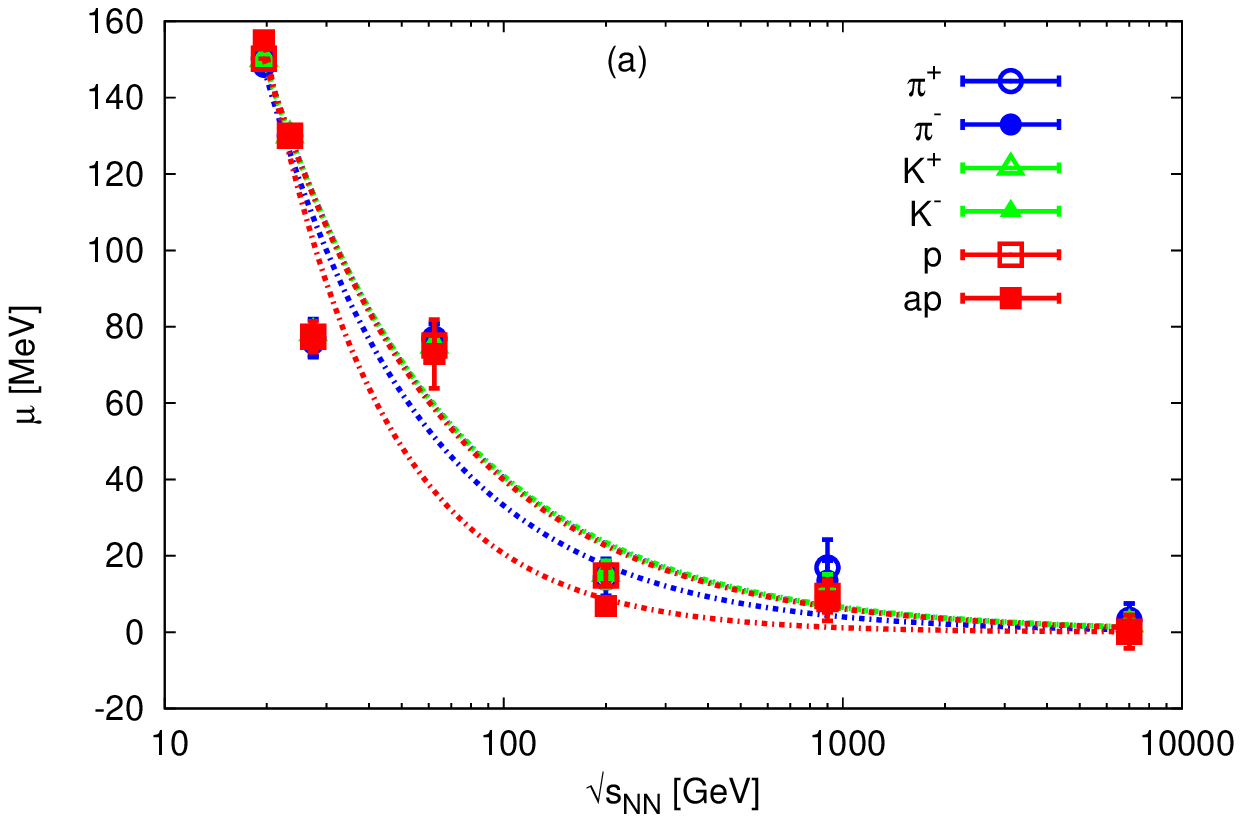}
\includegraphics[width=5cm,angle=-0]{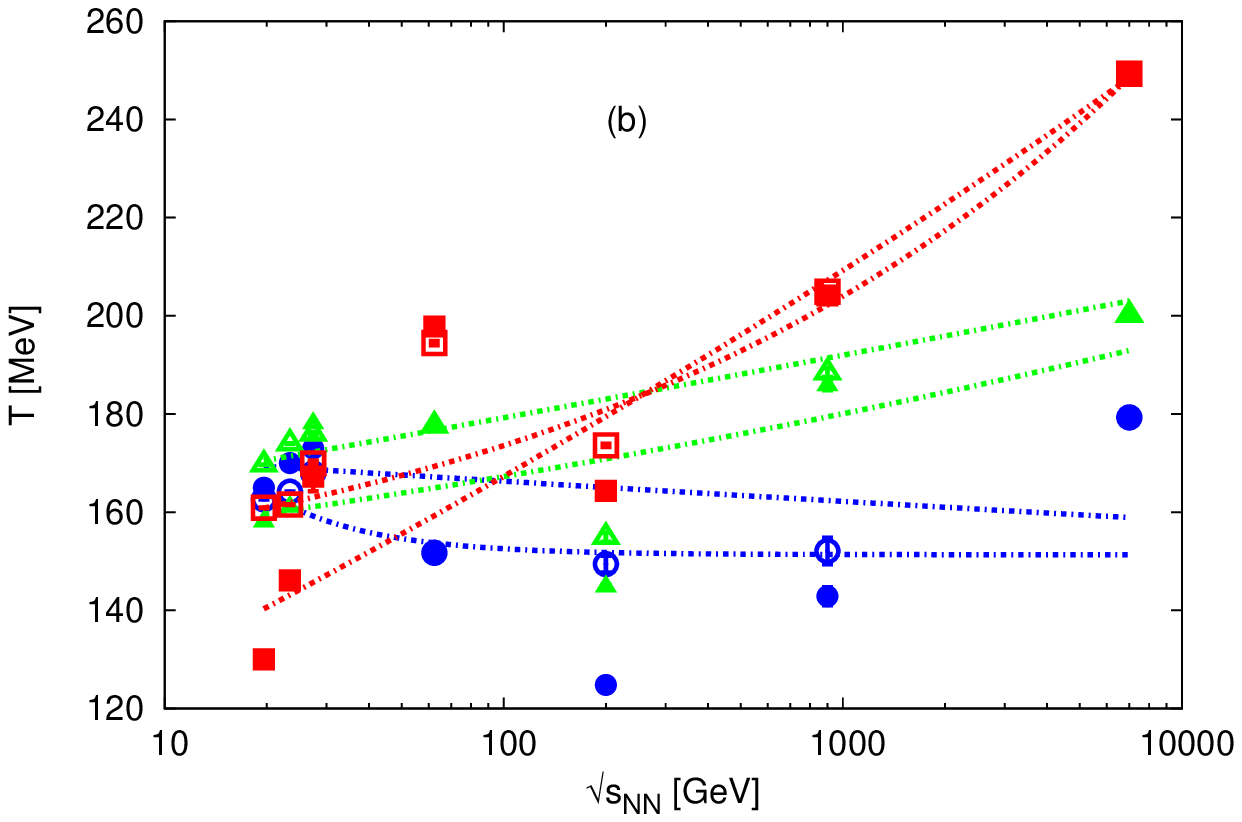}
\includegraphics[width=5cm,angle=-0]{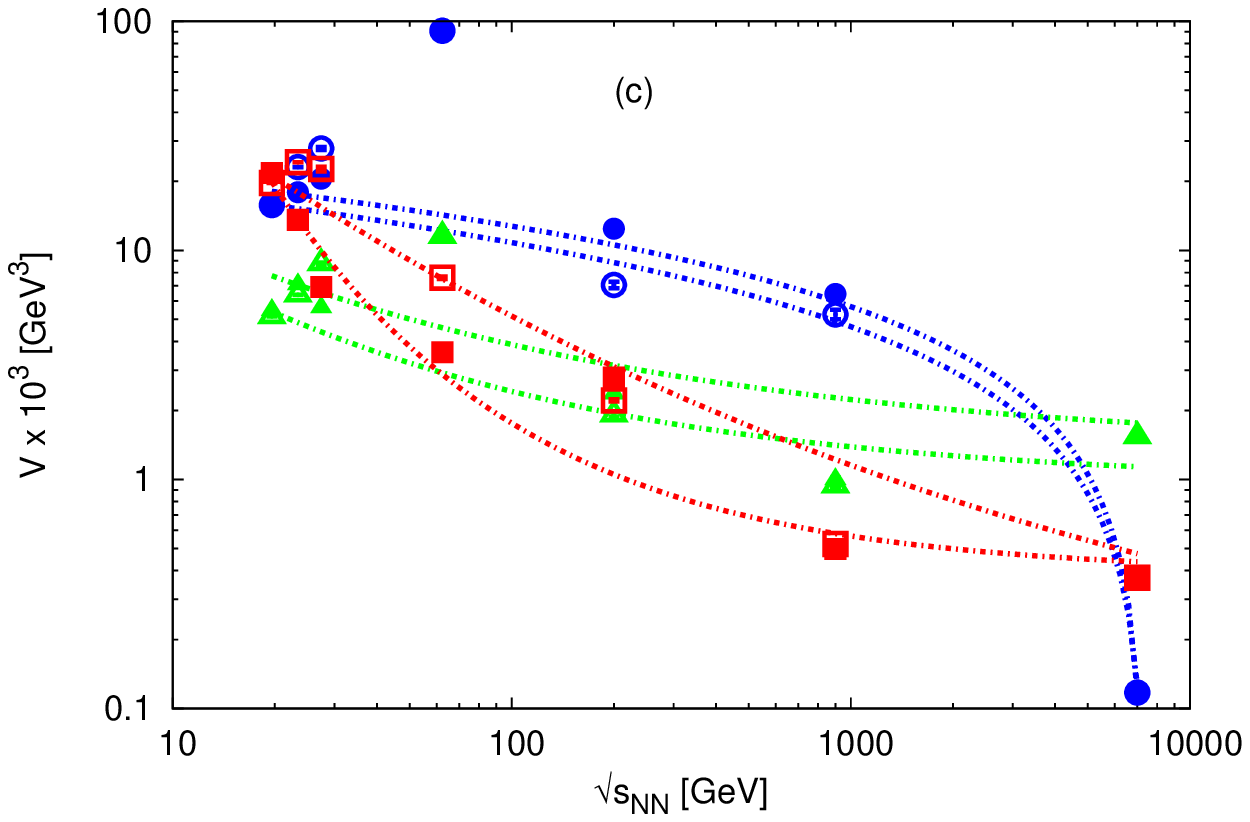}\\
\includegraphics[width=5cm,angle=-0]{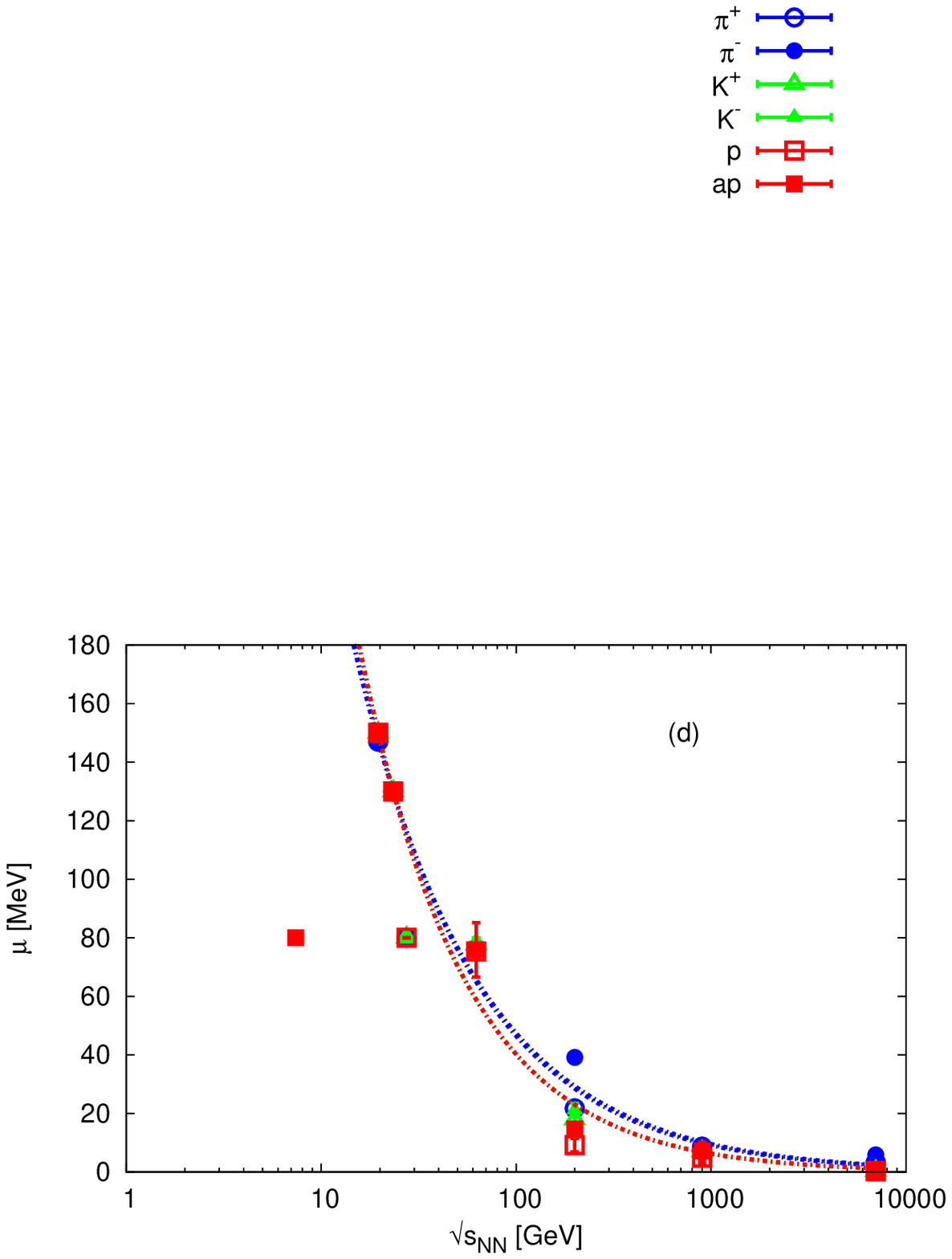}
\includegraphics[width=5cm,angle=-0]{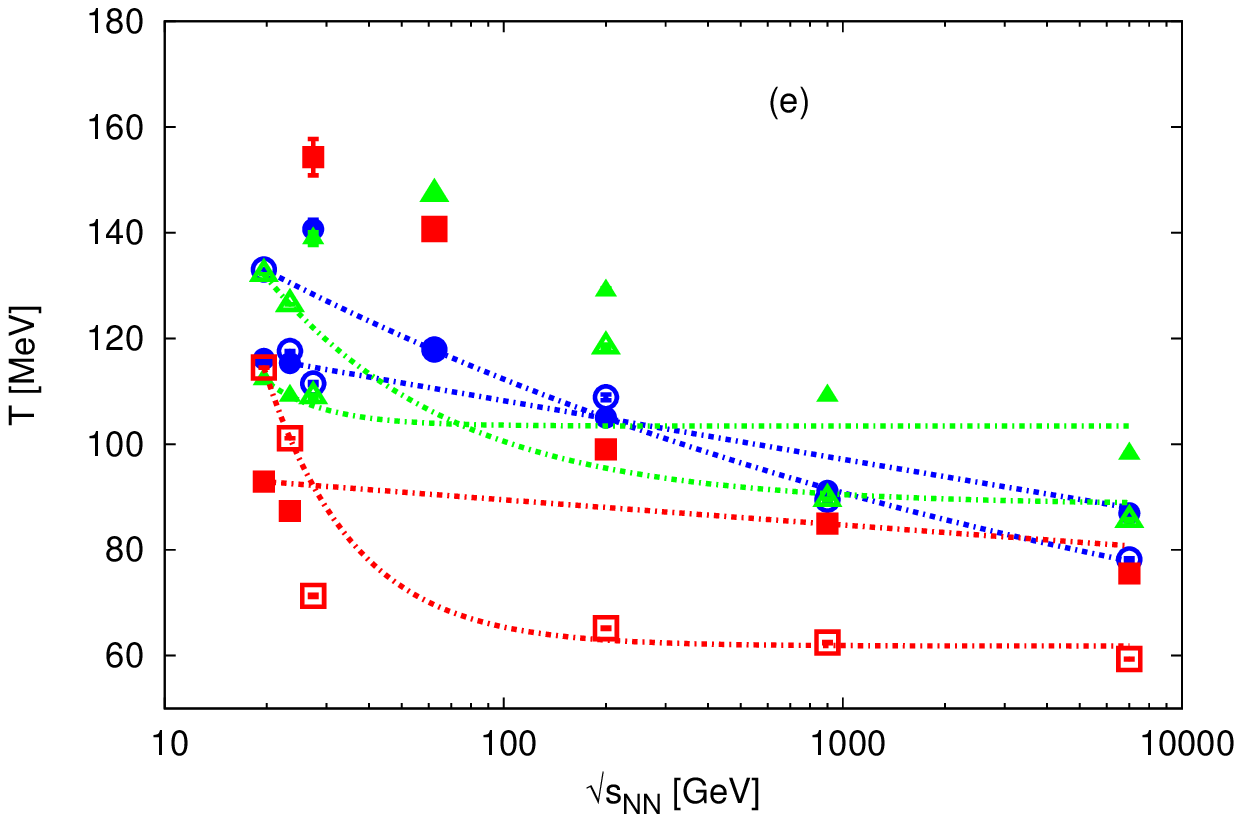}
\includegraphics[width=5cm,angle=-0]{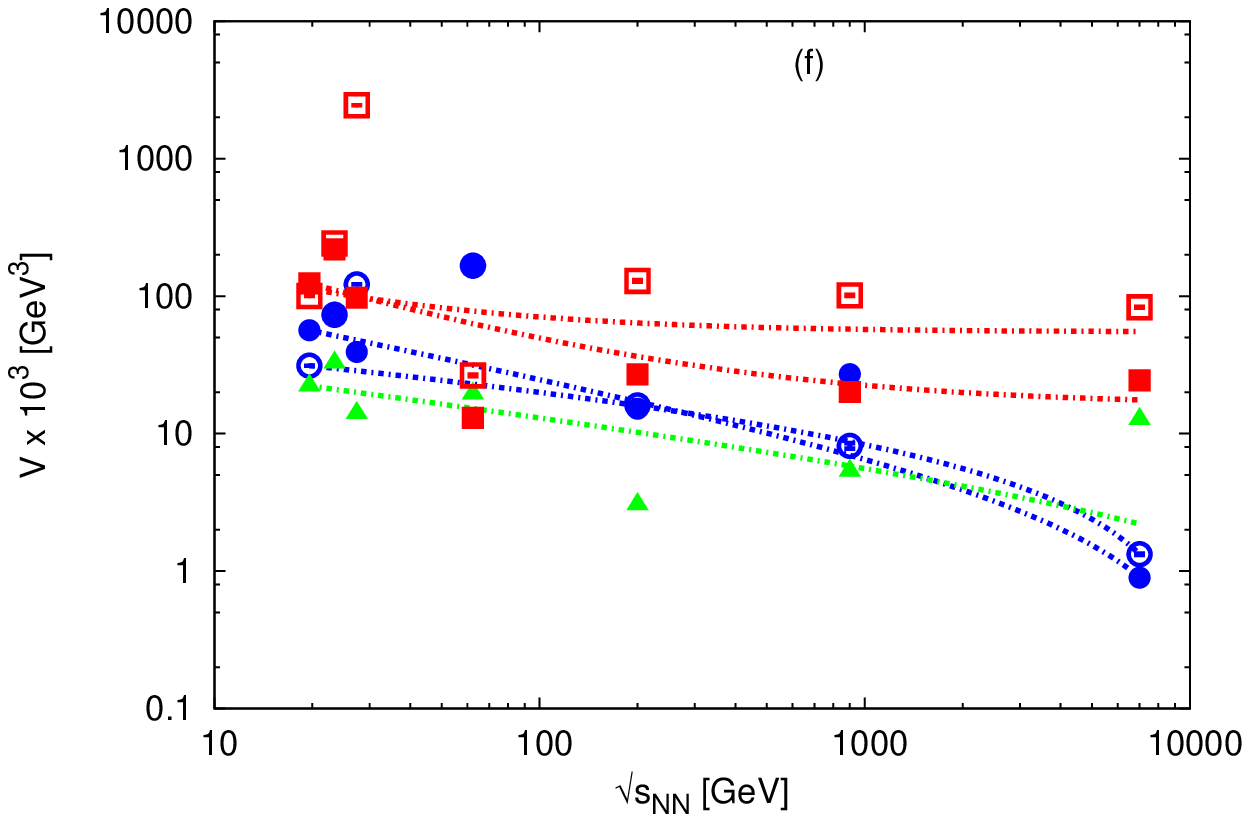} \\
\includegraphics[width=5cm,angle=-0]{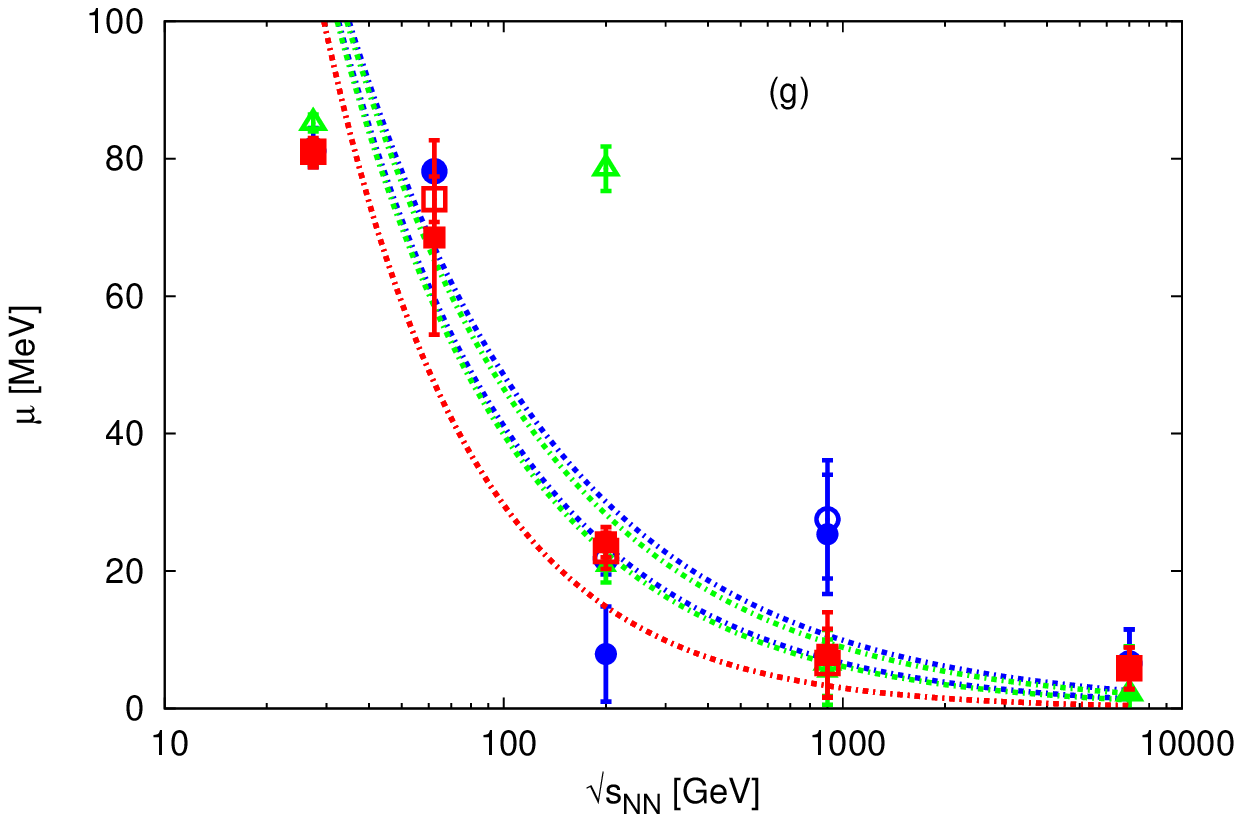}
\includegraphics[width=5cm,angle=-0]{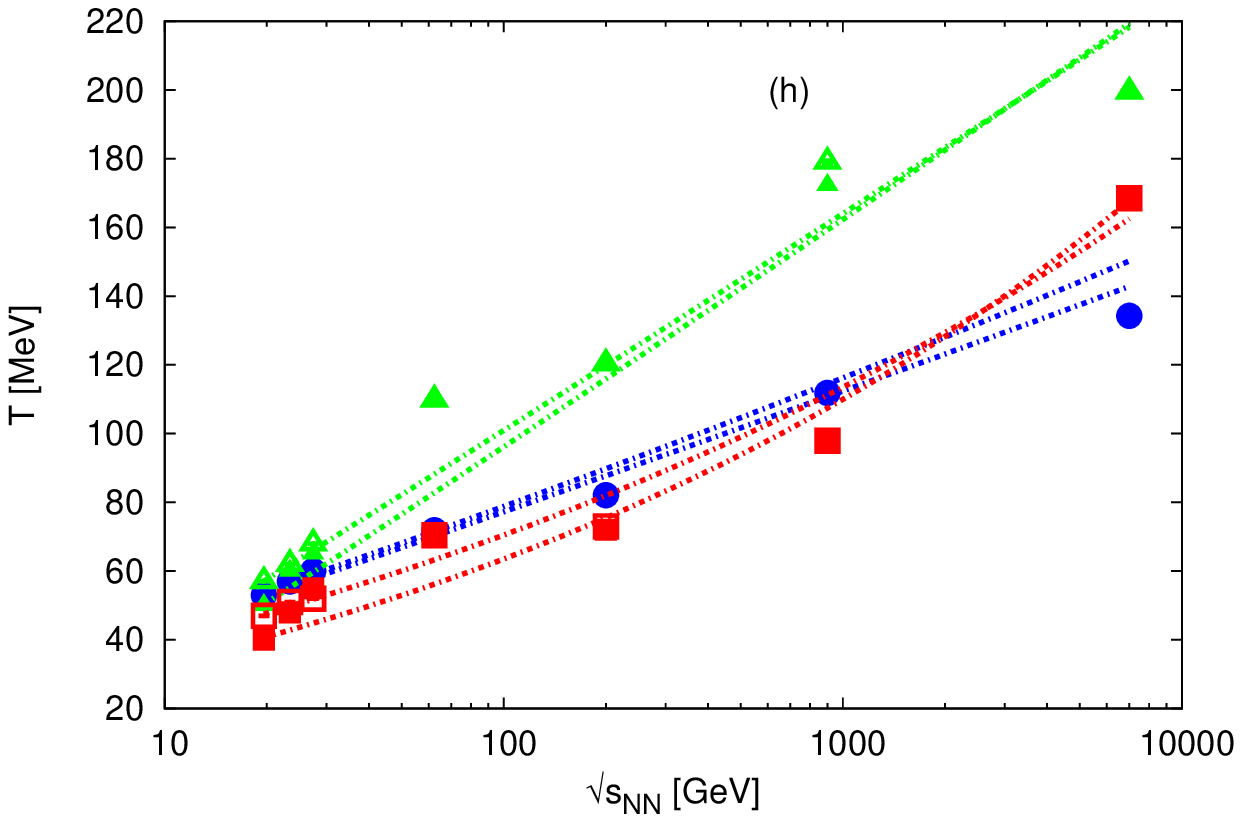}
\includegraphics[width=5cm,angle=-0]{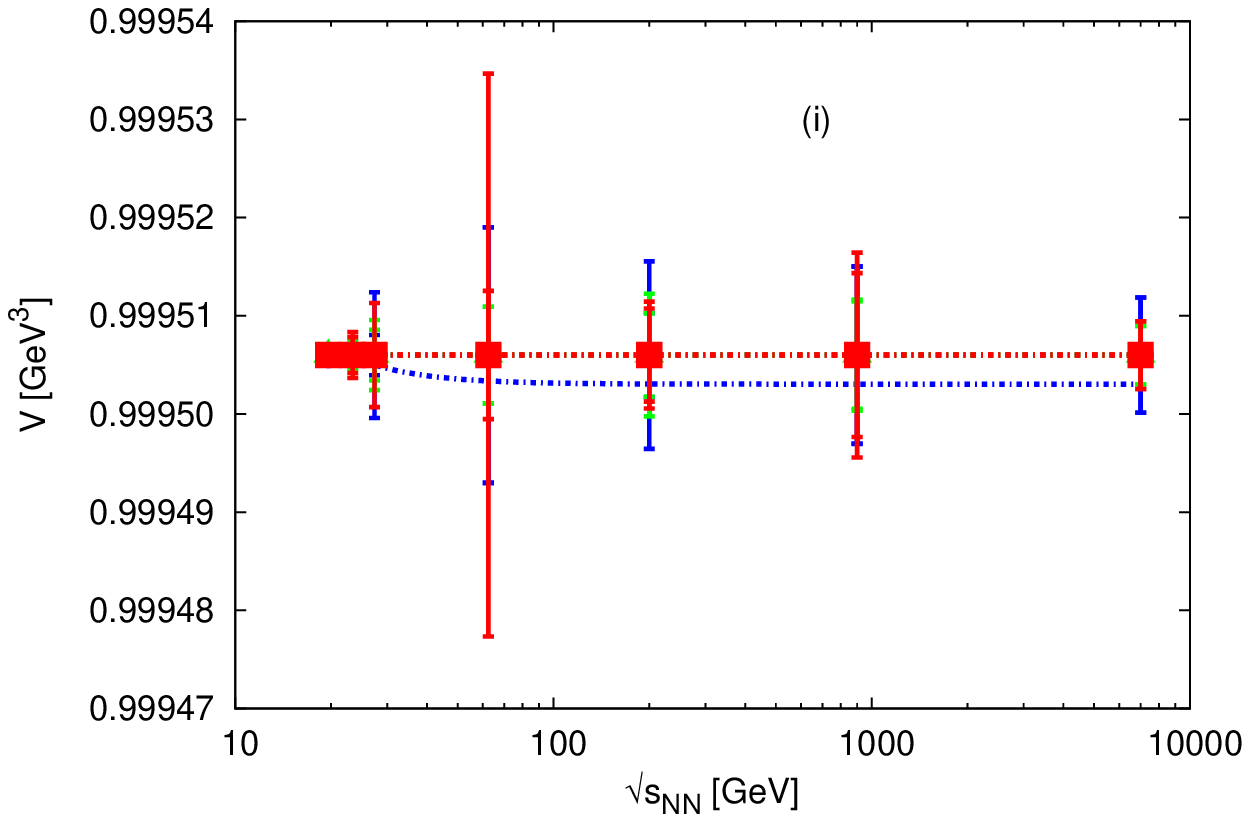}
\caption{(Color online) The various parameters obtained from the statistical fits within Boltzmann, Tsallis and generic axiomatic statistics for $p_{\mathtt{T}}$ measured in p$+$p collisions for various charged particles in a wide range of energies. The curves refer to the proposed expressions for statistical fits.
\label{fig:GenericAll_NN_fit}
}}
\end{figure}

\begin{figure}[htb]
\centering{
\includegraphics[width=8cm,angle=-0]{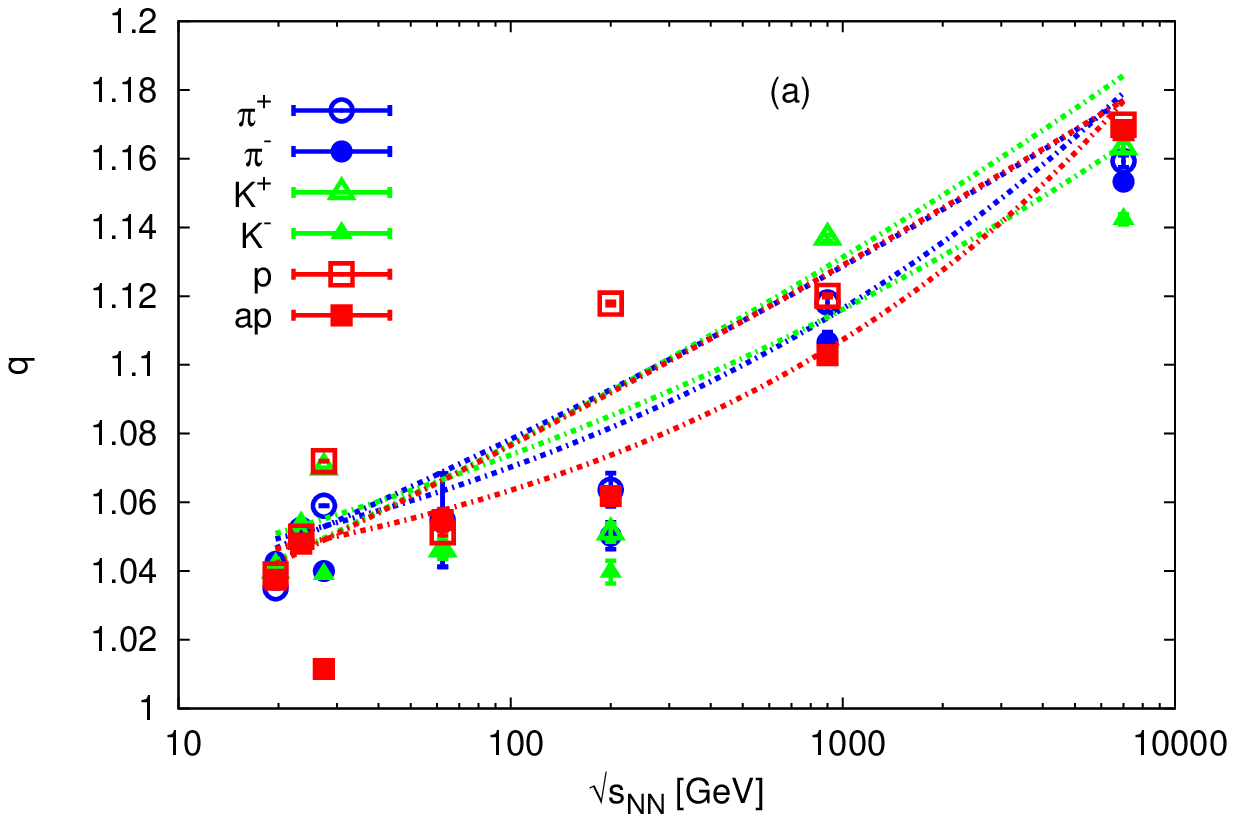}
\includegraphics[width=8cm,angle=-0]{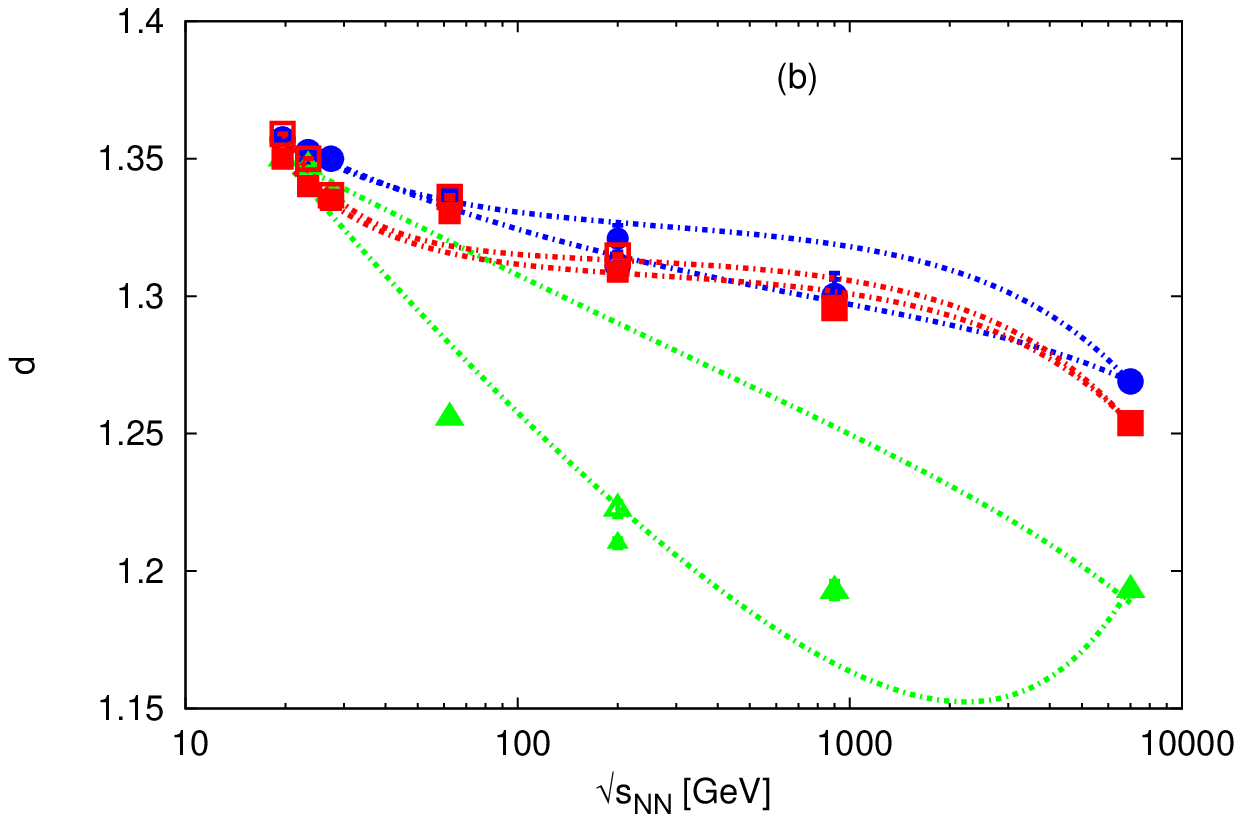}
\caption{(Color online) The nonextensive parameters $q$ and $d$ obtained from statistical fits within Tsallis and generic axiomatic statistics, respectively, for $p_{\mathtt{T}}$ measured in p$+$p collisions. The curves refer to the proposed expressions for statistical fits.
\label{fig:Genericd_NN_fit}}
}
\end{figure}

Details of this type of collisions are depicts in Figs. \ref{fig:GenericAll_NN_fit}, \ref{fig:Genericd_NN_fit} and listed out in Tables \ref{tab:fit_Boltzmann_NN}, \ref{tab:fit_Tsallis_NN}, and \ref{tab:fit_Generic_NN}. The thermodynamic quantities obtained can be summarized as follows.

\begin{itemize}
\item By using Boltzmann, the temperature obtained for all particles except $\pi^+$ can be related to $\sqrt{s_{\mathtt{NN}}}$ as 
\begin{equation}
T = a \sqrt{s_{\mathtt{NN}}}^b  + c
\end{equation}
where the values of $a$, $b$, and $c$ are shown in Fig. \ref{fig:GenericAll_NN_fit}, see Tab. \ref{tab:fit_Boltzmann_NN}.

\item For $\pi^+$ and by using Boltzmann, the temperature can be related to $\sqrt{s_{\mathtt{NN}}}$ as
\begin{equation}
T = (a + \sqrt{s_{\mathtt{NN}}}^b )^ c
\end{equation}
where the values of $a$, $b$, and $c$, are depicted in Fig. \ref{fig:GenericAll_NN_fit}, see Tab. \ref{tab:fit_Boltzmann_NN}.

\item The volume obtained by using Boltzmann for all particles except pions can be expressed as
\begin{equation}
V = (a + \sqrt{s_{\mathtt{NN}}}^b )^ c
\end{equation}
where the values of $a$, $b$, and $c$ are shown in Fig. \ref{fig:GenericAll_NN_fit}, see Tab. \ref{tab:fit_Boltzmann_NN}.

\item For $\pi^+$ and by using Boltzmann, the volume can be related to $\sqrt{s_{\mathtt{NN}}}$,
\begin{equation}
V = a \sqrt{s_{\mathtt{NN}}}^b  + c
\end{equation}
where the values of $a$, $b$, and $c$ are given in Fig. \ref{fig:GenericAll_NN_fit}, see Tab. \ref{tab:fit_Boltzmann_NN}.

\item By using Tsallis for all particles, the temperature and volume have a similar dependence $\sqrt{s_{\mathtt{NN}}}$ 
\begin{equation}
T \; \mathtt{and} \; V = a \sqrt{s_{\mathtt{NN}}}^b  + c
\end{equation}
where the values of $a$, $b$, and $c$ are shown in Fig. \ref{fig:GenericAll_NN_fit}, see Tab. \ref{tab:fit_Tsallis_NN}.

\item The nonextensivity parameter $q$ can be given in dependence on $\sqrt{s_{\mathtt{NN}}}$ as 
\begin{equation}
q = a \sqrt{s_{\mathtt{NN}}}^b + c
\end{equation}
where the values of $a$, $b$, and $c$, are presented in Fig. \ref{fig:Genericd_NN_fit}, see Tab. \ref{tab:fit_Tsallis_NN}.
 
\item By using the generic axiomatic approach, the temperature obtained for all particles is to be expressed as 
\begin{equation}
T = a \sqrt{s_{\mathtt{NN}}}^b  + c
\end{equation}
where the values of $a$, $b$, and $c$ are depicted in Fig. \ref{fig:GenericAll_NN_fit}, see Tab. \ref{tab:fit_Generic_NN}.

\item Also by using the generic axiomatic approach, the volume varies with $sqrt{s_{\mathtt{NN}}}$ as follows 
\begin{equation}
V = (a + \sqrt{s_{\mathtt{NN}}}^b )^ c
\end{equation}
where the values of $a$, $b$, and $c$ are shown in Fig. \ref{fig:GenericAll_NN_fit}, see Tab. \ref{tab:fit_Generic_NN}.

\item The equivalent class $d$ for all particles reads 
\begin{equation}
d = a \sqrt{s_{\mathtt{NN}}}^b  + c \sqrt{s_{\mathtt{NN}}} + f
\end{equation}
where the values of $a$, $b$, $c$, and $f$ are shown in Fig. \ref{fig:Genericd_NN_fit}, see Tab. \ref{tab:fit_Generic_NN}. 

\end{itemize}

\begin{table}[h!]
\centering
  \begin{tabular}{|c|c|c|c|c|c|c|c|c|}   \hline
    &  & $\pi^+$ & $\pi^-$ & $K^+$ & $K^-$ & $p$ & $\bar{p}$ \\ \hline
     \multirow{3}{*}{$T$} & $a$ & $1.6877\pm0.001$ &  $88.2306\pm2.545$ &  $403.785\pm0.6955$ &  $40.2961\pm0.9884$ &  $10.9575\pm0.3134$ &  $290.415\pm0.7191$ \\
     \cline{2-8} & $b$ & $-1.4247\pm0.0252$ &  $-0.0231\pm0.0089$ &  $0.0127\pm0.0006$ &  $0.0845\pm0.0077$ &  $0.2623\pm0.0079$ &  $0.0476\pm0.0021$ \\  
      \cline{2-8} & $c$ & $9.5911\pm0.0107$ &  $86.96\pm2.364$ &  $-248.923\pm0.723$ &  $107.799\pm1.289$ &  $136.919\pm0.7191$ &  $-194.302\pm2.288$ \\ \hline
   \multirow{2}{*}{$\mu$} & $a$  & $1653.35\pm107.1$ &  $2264.02\pm1346$ &  $1660.09\pm15.36$ &  $1604.13\pm648.1$ &  $1693.96\pm7.738$ &  $6210.22\pm2995$ \\
      \cline{2-8} & $b$ & $-0.80615\pm0.02179$ &  $-0.91734\pm0.1988$ &  $-0.8079\pm0.00308$ &  $-0.79703\pm0.1283$ &  $-0.81431\pm0.00152$ &  $-1.24029\pm0.1623$ \\   
 \hline
     \multirow{3}{*}{$V$} & $a$ & $58.7295\pm0.0383$ &  $110.746\pm0.1092$ &  $0.9951\pm0.0094$ &  $1.0541\pm0.0005$ &  $0.7857\pm0.0075$ &  $0.9643\pm0.00003$ \\
    \cline{2-8} & $b$ & $-0.0669\pm0.0002$ &  $-0.0336\pm0.0001$ &  $-0.4227\pm0.0112$ &  $-0.3495\pm0.0005$ &  $-0.2145\pm0.0048$ &  $-0.6189\pm0.00006$ \\   
    \cline{2-8} & $c$ & $-32.3597\pm0.0314$ &  $-82.102\pm0.0812$ &  $6.8612\pm0.2081$ &  $5.9862\pm0.0068$ &  $11.1838\pm0.2391$ &  $25.81\pm0.0065$ \\  \hline
    \end{tabular}
      \caption{Fit parameters obtained from Boltzmann statistics to results from p$+$p collisions. 
     \label{tab:fit_Boltzmann_NN} }
\end{table}

\begin{table}[h!]
\centering
  \begin{tabular}{|c|c|c|c|c|c|c|c|c|}   \hline
    &  & $\pi^+$ & $\pi^-$ & $K^+$ & $K^-$ & $p$ & $\bar{p}$ \\ \hline
    \multirow{3}{*}{$q$} & $a$  & $0.2638\pm0.0017$ &  $0.0241\pm0.0004$ &  $0.2506\pm0.0011$ &  $0.0613\pm0.0007$ &  $0.499\pm0.0013$ &  $0.0112\pm0.0001$ \\ 
   \cline{2-8} & $b$ & $0.059\pm0.002$ &  $0.225\pm0.0058$ &  $0.0649\pm0.0014$ &  $0.1364\pm0.0038$ &  $0.0369\pm0.0009$ &  $0.2985\pm0.0039$ \\    
     \cline{2-8} & $c$ & $0.7322\pm0.002$ &  $1.0022\pm0.0009$ &  $0.7389\pm0.0013$ &  $0.959\pm0.0011$ &  $0.4853\pm0015$ &  $1.0192\pm0.0004$ \\  \hline
    \multirow{3}{*}{$T$} & $a$ & $145.732\pm9.79$ &  $544.518\pm0.0231$ &  $460.56\pm0.2756$ &  $13661.9\pm215.2$ &  $7144.58\pm23.78$ &  $195.214\pm1756$ \\
     \cline{2-8} & $b$ & $-0.1596\pm0.0618$ &  $-0.0093\pm0.00001$ &  $-0.792\pm0.0002$ &  $-2.4715\pm0.0051$ &  $-1.6503\pm0.0011$ &  $-0.0113\pm1.087$ \\  
      \cline{2-8} & $c$ & $42.4196\pm22.44$ &  $-413.4\pm0.0224$ &  $88.5391\pm0.025$ &  $103.443\pm0.0931$ &  $61.8312\pm0.1612$ &  $-95.8192\pm1758$ \\ \hline
   \multirow{2}{*}{$\mu$} & $a$  & $1157.72\pm14.3$ &  $1212.68\pm1.873$ &  $1658.21\pm3.972$ &  $1658.21\pm4.229$ &  $1658.69\pm23.16$ &  $1658.17\pm4.187$ \\
     \cline{2-8} & $b$ & $-0.6936\pm0.004$ &  $-0.7083\pm0.0005$ &  $-0.8075\pm0.0008$ &  $-0.8075\pm0.0008$ &  $-0.8076\pm0.0046$ &  $-0.8075\pm0.0008$ \\ \hline
    \multirow{3}{*}{$V$} & $a$ & $78.8988\pm22.24$ &  $240.609\pm109.8$ &  $22.0623\pm3.984$ &  $58.2233\pm48.84$ &  $628.046\pm187.4$ &  $904.151\pm855.8$ \\
    \cline{2-8} & $b$ & $-0.1607\pm0.3684$ &  $-0.4709\pm0.1771$ &  $-0.3854\pm0.0596$ &  $-0.2917\pm0.776$ &  $-0.7999\pm0.0989$ &  $-0.7149\pm0.348$ \\   
   \cline{2-8} & $c$ & $-17.6903\pm67.37$ &  $-2.8175\pm4.21$ &  $7.0566\pm1.241$ &  $-2.1949\pm37.23$ &  $54.7858\pm16.97$ &  $15.9831\pm12.73$ \\  \hline
    \end{tabular}
      \caption{The same as in Tab. \ref{tab:fit_Boltzmann_NN} but here by using the Tsallis statistical approach.
     \label{tab:fit_Tsallis_NN} }
\end{table}

\begin{table}[h!]
\centering
{\tiny
  \begin{tabular}{|c|c|c|c|c|c|c|c|c|}   \hline
    &  & $\pi^+$ & $\pi^-$ & $K^+$ & $K^-$ & $p$ & $\bar{p}$ \\ \hline
    \multirow{4}{*}{$d$} & $a$  & $0.205\pm0.0003$ &  $0.7465\pm0.0027$ &  $0.6784\pm0.0009$ &  $0.8667\pm0.0007$ &  $17.0854\pm0.3639$ &  $3.3013\pm0.0154$ \\ 
   \cline{2-8} & $b$ & $-0.316\pm0.0004$ &  $-1.0578\pm0.0012$ &  $-0.0459\pm0.0004$ &  $-0.0952\pm0.0002$ &  $-1.9958\pm0.0069$ &  $-1.4706\pm0.0016$ \\    
     \cline{2-8} & $c \times10^{-6}$ & $-2.9157\pm0.925$ &  $-8.1537\pm0.5914$ &  $-3.197\pm7.312$ &  $17.7573\pm3.1$ &  $-8.721\pm1.839$ &  $-7.8582\pm0.3498$ \\  
     \cline{2-8} & $f$ & $1.2768\pm0.0001$ &  $1.3257\pm0.0001$ &  $0.7589\pm0.0008$ &  $0.6968\pm0.0004$ &  $1.3144\pm0.0007$ &  $1.3086\pm0.0002$ \\ \hline
    \multirow{3}{*}{$T$} & $a$ & $383.278\pm0.2362$ &  $1203.82\pm0.4721$ &  $2979.59\pm0.2551$ &  $2764.62\pm0.542$ &  $57.9143\pm0.2225$ &  $38.3409\pm0.4858$ \\
     \cline{2-8} & $b$ & $0.0347\pm0.0002$ &  $0.0118\pm0.0001$ &  $0.0087\pm0.00003$ &  $0.0098\pm0.00007$ &  $0.1418\pm0.0012$ &  $0.1826\pm0.004$ \\  
      \cline{2-8} & $c$ & $-370.846\pm0.2633$ &  $-1194.03\pm0.4892$ &  $-3000.95\pm0.262$ &  $-2796.31\pm0.558$ &  $-40.8007\pm0.3511$ &  $-25.3621\pm0.8487$ \\ \hline
   \multirow{2}{*}{$\mu$} & $a$  & $1576.51\pm774.6$ &  $1155.98\pm294.4$ &  $1268.41\pm495.8$ &  $1685.21\pm1106$ &  $2911.71\pm1637$ &  $2911.71\pm1637$ \\
        \cline{2-8} & $b$ & $-0.792\pm0.1579$ &  $-0.6887\pm0.0836$ &  $-0.7179\pm0.1288$ &  $-0.8132\pm0.2202$ &  $0.9965\pm0.188$ &  $-0.9965\pm0.188$ \\ \hline
      \multirow{3}{*}{$V$} & $a$ & $0.7905\pm0.00009$ &  $3.9914\pm7.746\times10^{-8}$ &  $0.3676\pm5.135\times10^{-9}$ &  $1.0554\pm1.959\times10^{-11}$ &  $3.4952\pm2.91\times10^{-8}$ &  $3.6782\pm1.756\times10^{-8}$ \\
      \cline{2-8} & $b$ & $-2.1787\pm0.0248$ &  $-4.8375\pm0.055$ &  $-5.3647\pm0.0594$ &  $(7.7117\pm0.7752)\times10^{-8}$ &  $-4.1548\pm0.096$ &  $-5.3507\pm14.57$ \\   
     \cline{2-8} & $c$ & $0.0021\pm9.773\times10^{-7}$ &  $-0.0004\pm4.379\times10^{-12}$ &  $0.0005\pm2.021\times10^{-11}$ &  $-0.0007\pm1.117\times10^{-11}$ &  $-0.0004\pm5.505\times10^{-11}$ &  $-0.0004\pm1.39\times10^{-12}$ \\  \hline
    \end{tabular}
      \caption{The same as in Tab. \ref{tab:fit_Boltzmann_NN} but here by using the generic axiomatic statistical approach. 
     \label{tab:fit_Generic_NN} }
     }
\end{table}

Now we can draw the conclusion that the fit parameters obtained depend on the size of the colliding system, the center-of-mass energy, and the type of the statistical approach applied, especially when moving from extensive to Tsallis-type approach.

\subsection{Empirical parameterization}

The transverse momentum spectra of the charged particles produced in A$+$A collisions in the STAR experiment are well described in refs. \cite{Adamczyk:2017iwn,Abelev:2008ab} by using different parametrizations at low $p_T$ ranges. These parametrizations are known as Bose-Einstein, $m_T$-exponential, double-exponential, Boltzmann, $p_T$-exponential, $p_T$-Gaussian, and $p_T^3$-exponential. Each of these expressions suggests different power-scales which are compared with Tsallis statistics at various ranges of $p_T$ at a wide range of energies \cite{Abelev:2008ab,Bialas:2015pla}. It was noticed that the temperatures $T$ of the $p_T$ spectra of pions are smaller than that of the Kaons, which in turn are smaller than that of the protons and there is a linear dependence of $T$ on the energies except for protons and anti-protons at $19.6$ GeV or $27$ GeV where a rapid increase or decrease is found. Also, it was concluded that the temperature of charged particles are slightly greater than the ones for the anti-charged particles.

The high transverse momentum $p_T$ spectra of hadrons produced in A$+$A collisions at top RHIC energies \cite{dEnterria2005} are found strongly suppressed by a factor of $4-5$ \cite{Adcox:2001jp,Adler:2002xw,Adler:2003qi,Adams:2003kv,Back:2003qr,Arsene:2003yk} comparing to the results for the same hadrons in p$+$p collisions at the same energies \cite{Adams:2003kv,Adler:2003pb,Aggarwal:2001gn,ReffInt15b,Aggarwal:1998vh}. Different parametrization have been used to describe the $p_T$ spectra at LHC energies \cite{Abelev:2013ala,Chatrchyan:2012qb,Aamodt:2011zj,Khachatryan:2010xs,Khachatryan:2010us,Adams:2004ep}.

Tsallis statistics is more effective within the low $p_T$ range rather than in the high $p_T$. Also we conclude that Tsallis statistics is remarkably successful in describing p$+$p collisions rather than A$+$A collisions. The results obtained from these parameterizations are given in Tabs. \ref{tab:1} - \ref{tab:2}.

\subsection{Other models for $p_{\mathtt{T}}$ spectra distributions}

Different parametrizations as Tsallis-Pareto, Tsallis-Levy \cite{Biro:2008hz}, and multicomponent Erlang $p_T$ spectra distribution have been proposed to analyze the transverse momentum spectra in various collisions at different centralities \cite{Wei:2015yut,Wei:2016ihj} in order to determine the effective temperature of the interacting system. Also, other thermodynamic quantities obtained have been summarized in Tab. \ref{Enlarge}, \cite{Wei:2016ihj}

\begin{table*}
\centering
\begin{tabular}{|c|c|c|c|c|c|}
\hline  $\sqrt{s_{\mathtt{NN}}}$ (TeV) & Type & $T$ (GeV)& $q$& $N_{\mathtt{T}0}$ & $\chi^2$/dof \\
\hline
      & $\pi^{+}$ & $0.088\pm0.002$ & $1.095\pm0.010$ & $1.048\pm0.113$ & 0.018 \\
$0.2$ & $K^{+}$ & $0.105\pm0.003$ & $1.074\pm0.010$ & $0.081\pm0.012$   & 0.037 \\
      & $p$ & $0.135\pm0.003$ & $1.045\pm0.008$ & $0.051\pm0.006$ & 0.085 \\ \hline
      & $\pi^+$  & $0.092\pm0.003$ & $1.127\pm0.007$ & $1.852\pm0.308$   & 0.077 \\
$0.9$ & $K^+$ & $0.122\pm0.004$ & $1.107\pm0.007$ & $0.229\pm0.029$ & 0.017 \\
      & $p$ & $0.159\pm0.004$ & $1.069\pm0.007$ & $0.103\pm0.015$   & 0.084 \\ \hline
       & $\pi^+$ & $0.092\pm0.003$ & $1.140\pm0.007$ & $2.305\pm0.360$  & 0.131 \\
$2.76$ & $K^+$  & $0.123\pm0.003$ & $1.121\pm0.008$ & $0.302\pm0.040$  & 0.014 \\
       & $p$ & $0.159\pm0.003$ & $1.082\pm0.007$ & $0.134\pm0.021$ & 0.081 \\ \hline
      & $\pi^+$  & $0.090\pm0.003$ & $1.150\pm0.005$ & $2.920\pm0.319$  & 0.143 \\
$7.0$ & $K^+$  & $0.127\pm0.003$ & $1.124\pm0.005$ & $0.376\pm0.036$  & 0.022 \\
      & $p$  & $0.163\pm0.003$ & $1.098\pm0.005$ & $0.168\pm0.016$          & 0.042 \\
\hline
\end{tabular}%
\caption{Estimations for $T$, $q$, and $N_{\mathtt{T}0}$ as obtained from Tsallis distribution at various energies \cite{Wei:2016ihj}. \label{Enlarge}}
\end{table*}

The values of the free parameters $T$ and $q$, the normalization constant $N_{\mathtt{T0}}$, and $\chi^2$/dof corresponding to Tsallis statistics are given in Tab. \ref{Enlarge}. It is noted that, the effective temperature $T$ increases while the parameter $q$ decreases with increase of the rest mass, which refers to the non-simultaneous productions of different types of particles, while $T$ and $q$ increase with increase of $\sqrt{s_{\mathtt{NN}}}$. The normalization constant $N_{\mathtt{T0}}$ decreases with the increase of rest mass, and increase with increase of $\sqrt{s_{\mathtt{NN}}}$.

Also, by using two types of Tsallis-type approaches, e.g. with and without thermodynamic description, the energy dependences of the transverse momentum spectra for charged particles in p$+$p collisions were analyzed \cite{Zheng:2015mhz}. The obtained temperatures from the thermodynamic description is noted to be smaller than the ones without such description.

The discrepancies of the temperatures using the Tsallis distribution with thermodynamical description is related to the introduction of the extra term $m_T$ to the distribution. The effect of the choice of $E_T$ in the Tsallis distribution beats the effect of the extra term $m_T$ for the heavier particles.

There are various types of Tsallis distribution proposed. These can be summarized as follow. \begin{itemize}
\item Type-A Tsallis distribution introduced in ref. \cite{Zheng:2015mhz} 
\begin{equation}
\left(E\frac{d^3N}{dp^3}\right)_{|\eta|<a} = A\left(1+\frac{E_T}{nT_5}\right)^{-n}, \label{tsallisus}
\end{equation}
where $E_T=m_T - m$. $A$, $n$ and $T$ are the free fit parameters. The transverse mass reads $m_T = \sqrt{p^2_{\mathtt{T}} + m^2}$, where $m$ is the rest mass of the particle. $n$ is a fitting parameter known as the nonextensive power (can be related to Tsallis parameter $q$), $T_5$ is the temperature, and $A$ is an additional fit parameter \cite{Zheng:2015tua}.

\item If self-consistent thermodynamical description was taken into consideration, then the Tsallis distribution at mid-rapidity becomes
\begin{equation}
E\frac{d^3N}{dp^3} = gV\frac{m_T}{(2\pi)^3} \left[1+(q-1)\frac{m_T}{T_1}\right]^{-q/(q-1)}. \label{tsallisBR}
\end{equation}
where $g$ is defined as the degeneracy of the particle and $V$ is the volume. This equation, known as Type-B Tsallis distribution, is very similar to Eq. (\ref{tsallisus}), but $m_T$ was replaced by $E_T$ and also there was an extra term $m_T$ in front of the bracket \cite{Zheng:2015mhz}. 
\end{itemize}

Between these two types of Tsallis distribution, namely A and B, there are three stages or transitions. 
\begin{itemize}
\item The first one is known as a Tsallis-like distribution which was obtained in the scope of nonextensive statistics for the particle yield at mid-rapidity \cite{Alberico:2009gj}
\begin{equation}
E\frac{d^3N}{dp^3}=A m_T \left[1+(q-1)\frac{m_T}{T_2}\right]^{-\frac{1}{q-1}}, \label{transeq1}
\end{equation}
where $A$, $q$ and $T_2$ are fit parameters. $q$ is a fit parameter which gives the nonextensive power, $T_2$ is the temperature, and $A$ is another fit parameter \cite{Zheng:2015tua}. Comparing Eq. (\ref{transeq1}) and Eq. (\ref{tsallisBR}), the only difference is the power of the distribution function, i.e. $q$ for Eq. (\ref{tsallisBR}) and $1$ for Eq. (\ref{transeq1}). 

\item The second stage is expressed as \cite{Zheng:2015mhz}
\begin{equation}
E\frac{d^3N}{dp^3}=A \left[1+(q-1)\frac{m_T}{T_3}\right]^{-\frac{q}{q-1}}, \label{transeq2}
\end{equation}
where the term $m_T$ outside of the bracket in Eq. (\ref{tsallisBR}) was neglected and the constants are absorbed into the new parameter $A$ \cite{Zheng:2015mhz}. 
\item The third stage reads \cite{Zheng:2015mhz}
\begin{equation}
E\frac{d^3N}{dp^3}=A \left[1+(q-1)\frac{m_T}{T_4}\right]^{-\frac{1}{q-1}}. \label{transeq3}
\end{equation}
where $m_T = \sqrt{p^2_{\mathtt{T}} + m^2}$, $m$ is the rest mass of the particle, $q$ is a fit parameter giving the nonextensive power, $T_4$ is the temperature, and $A$ is a free fit parameter \cite{Zheng:2015tua}.
\end{itemize}

\begin{table}[hbt]
\centering
\resizebox{0.9\textwidth}{!}{%
\begin{tabular}{|c|c|c|c|c|c|c|c|c|c|c|c|c|}
\hline
Experiment     & $\sqrt{s}$(GeV) & particle & $T_1$    & $\chi_1^2$/ndf     & $T_2$ & $\chi_2^2$/ndf & $T_3$ & $\chi_3^2$/ndf &$T_4$ & $\chi_4^2$/ndf & $T_5$  & $\chi_5^2$/ndf   \\
\hline
PHENIX \cite{Adare:2011vy} &    62.4        & $\pi^+$   &  0.0927   &  6.857/23     & 0.085 & 6.866/23 & 0.133     & 4.767/23  & 0.123 & 4.784/23 & 0.132 & 4.779/23 \\
  &                   & $\pi^- $   &   0.0898  &  8.049/23    & 0.0824 & 8.045/23 & 0.128     &  5.173/23 & 0.118 & 5.198/23 & 0.130 & 5.194/23 \\     
  &                  & $K^+$    &   0.0856   &   4.837/13   & 0.0775 & 4.822/13 & 0.122    &  5.141/13  & 0.105 & 5.349/13 & 0.160 & 5.121/13\\
  &                   & $K^-$     &   0.0936   &   2.002/13   & 0.0851& 2.006/13 & 0.130   &  2.199/13   & 0.119 & 2.203/13  & 0.163 & 2.186/13\\
  &                   & $p$        &    0.106    &   7.017/ 24  & 0.101 & 7.075/24 &  0.133  & 6.934/24  & 0.125 & 6.945/24 & 0.179 & 6.966/24 \\
  &                   & $\bar p$ &    0.0635  &   6.605/22  & 0.0588 & 6.563/22 &  0.0831 & 6.037/22     & 0.0817 & 5.079/22 & 0.148 & 7.178/22 \\
\hline
PHENIX \cite{Adare:2011vy} &    200     &  $\pi^+$   &  0.0741  &  5.278/24   &  0.0657 & 5.275/24  &  0.111  &  4.491/24  &  0.0981 & 4.515/24 & 0.114  & 4.485/24 \\
 &               & $\pi^-$     &  0.0811  &  4.710/24    & 0.0725 & 4.703/24 &  0.121 &  3.350/24 & 0.108 & 3.372/24 & 0.123 & 3.354/24 \\
 &               & $K^+$     &  0.0473  &  1.561/13    &  0.0418 & 1.591/13 & 0.0729 & 1.634/13  & 0.0601 & 1.602/13 & 0.138  & 1.587/13\\
 &              & $K^-$      &0.0621   &    3.013/13    &  0.0542 &  3.010/13 & 0.0913   &   3.004/13  &  0.0781 & 2.999/13 & 0.147  & 2.999/13 \\
 &               & $p$         & 0.0311   &   23.832/31  &  0.0279& 23.659/31 & 0.0404  & 24.004/31 & 0.0350 & 24.272/31 & 0.145 &  24.581/31\\
 &                 & $\bar p$  & 0.0473 &    12.902/31  &  0.0426 & 12.970/31 & 0.0609 & 13.240/31 & 0.0547 & 13.153/31 & 0.154  & 13.535/31\\
\hline
STAR \cite{Adams:2006nd} &    200     &   $\pi^+$   &  0.0895 & 6.545/20 &  0.0809 & 6.539/20 &    0.126 &  5.032/20  & 0.113 & 5.008/20 & 0.128 & 5.009/20\\
 &                & $\pi^-$     &  0.0900  &     6.855/20  & 0.0814 & 6.854/20 &  0.127  &  4.700/20 & 0.114 & 4.718/20 & 0.128 & 4.705/20 \\
 &                 & $p$         & 0.0804   &     10.683/17  & 0.0735 & 10.653/17 & 0.104  & 10.375/17 & 0.0950 & 10.396& 0.180  & 10.359/17\\
 &                 & $\bar p$  & 0.0765 &   10.380/17    & 0.0695 & 10.318/17 & 0.0995 & 10.079/17 & 0.0901 & 10.076/17 & 0.177 & 9.991/17\\
 \hline
ALICE \cite{Aamodt:2011zj}    &   900  & $\pi^+$ & 0.0716 &  24.640/30  & 0.0627&  25.530/30 &  0.123 &  13.528/30 & 0.107 & 13.749/30 & 0.125 & 13.460/30           \\
 &          & $\pi^-$   & 0.0727 &   17.138/30 & 0.0636 & 17.602/30 &  0.125  & 12.394/30 & 0.109 & 12.645/30 & 0.126 & 12.483/30        \\
 &          & $K^+$    & 0.0568 & 12.790/24 & 0.0488 & 12.807/24  & 0.0904 & 13.034/24 & 0.0749 & 13.069/24 & 0.159 & 12.980/24\\
 &           & $K^-$    &   0.0624 & 6.457/24 &   0.0538   & 6.552/24 & 0.0968 & 6.641/24 & 0.0820 & 6.636/24  & 0.161 & 6.609/24\\
 &          & $p$    &0.0397 & 13.879/21 & 0.0358 & 13.908/21 & 0.0522 &13.816/21 & 0.0460 & 13.849/21 & 0.175 & 13.974/21\\
 &          & $\bar p$& 0.0649 & 13.586/21 &0.0568 &13.674/21 & 0.119 & 14.860/21 & 0.0769 &13.544/21 & 0.188 &13.675/21  \\
\hline
  \end{tabular}
}
\caption{The fit parameters $T$ and the corresponding $\chi^2$/ndf in $p_{\mathtt{T}}$ distributions,  Eqs. (\ref{tsallisBR}), (\ref{transeq1}), (\ref{transeq2}), (\ref{transeq3}), (\ref{tsallisus}), measured in p$+$p collisions \cite{Zheng:2015mhz}. \label{fitpara}}
\end{table} 

From Table. \ref{fitpara}, we notice that all distributions have almost the same fitting goodness to the particle spectra. The temperatures of Type-A and Type-B Tsallis distribution are known as $T_5$ and $T_1$, respectively. While, $T_2$, $T_3$ and $T_4$ are deduced from the stages or transitions between both types of distribution. The Type-A Tsallis distribution gives higher temperature than the Type-B Tsallis distribution. For all the particles,  $T_1$ and $T_2$ from distributions with extra $m_T$ term are lower than temperatures $T_3$, $T_4$ and $T_5$ from the distributions without it. Also, it was found that $T_1$ is larger than $T_2$, the parameter $q$ in Eq. (\ref{tsallisBR}) leads to larger $T$. Similarly, $T_4$ is smaller than $T_5$, the $m_T$ in Eq. (\ref{transeq3}) leads to smaller the temperature. Finally, the temperature $T_3$ and $T_5$ for pions are similar which result from cancelling the effects of $q$ and $m_T$ on each other. But for Kaons and protons, the effect of $m_T$ in Eq. (\ref{transeq2}) overcome on the effect of $q$ so $T_3$ was smaller than $T_5$.

In comparing these temperatures with our results, section \ref{sec:outExprs}, we find that the most temperatures which agree well with most of our results by using Boltzmann statistics are $T_5$ which refer to Type-A Tsallis distribution. 
\begin{itemize}
\item For pions: \\
At $\sqrt{s_{\mathtt{NN}}}=62.4~$GeV, $T_5$ agree well with our results by using Boltzmann statistics while $T_2$ agrees well with our results by using generic axiomatic statistics. At $\sqrt{s_{\mathtt{NN}}}=200~$GeV, also $T_2$ agrees well with our results by using generic axiomatic statistics but our results by using Boltzmann statistics are greater than results obtained by using different types of Tsallis distributions. At $\sqrt{s_{\mathtt{NN}}}=900~$GeV, $T_4$ agrees well with our results from generic axiomatic statistics but our results deduced from Boltzmann statistics are greater than results taken from different types of Tsallis distributions.

\item For Kaons: \\
At $\sqrt{s_{\mathtt{NN}}}=62.4~$GeV, $T_5$ agree well with our results by using Boltzmann statistics while $T_4$ agrees well with our results from generic axiomatic  statistics. At $\sqrt{s_{\mathtt{NN}}}=200~$ and $900~$GeV, $T_5$ for Kaons agrees well with our results deduced from both Boltzmann and generic axiomatic statistics.

\item For protons and antiprotons: \\
At $\sqrt{s_{\mathtt{NN}}}=62.4~$GeV, the results of $T_1$ for antiprotons agrees well with our results by using generic axiomatic statistics but our results by using Boltzmann statistics are greater than the results taken from different types of Tsallis distributions. Our results for protons by using generic axiomatic  statistics are lower than the results taken from any type of the Tsallis distributions. At $\sqrt{s_{\mathtt{NN}}}=200~$GeV, $T_5$ agrees well with our results by using Boltzmann statistics but our results using generic axiomatic  statistics are less than results from the different types of Tsallis distributions. At $\sqrt{s_{\mathtt{NN}}}=900~$GeV, $T_5$ agrees well with our results by using Boltzmann statistics, while $T_4$ agrees well with our results determined from generic axiomatic statistics.  
\end{itemize}

In comparing our fit parameters with the ones shown in Table \ref{Enlarge} taken from ref. \cite{Wei:2016ihj}, we find that our results for the nonextensive parameter $q$ has the same trend as the later for all particles except for pions. Our results for all particles are greater than the results listed out in Tab. \ref{Enlarge}, especially for protons at all center-of-mass energies.

\section{Conclusions}

The transverse momentum spectra $p_{\mathtt{T}}$ for the charged particles and anti-particles at a wide range of center-of-mass energies from various high-energy collisions are studied. Extensive and non-extensive statistics were used to analyze the $p_{\mathtt{T}}$ spectra for charged particles. $p_{\mathtt{T}}$ spectra for pions, Kaons, and protons and their anti-particles are well reproduced by using Maxwell-Boltzmann, Tsallis, and generic axiomatic statistics. Different fit parameters are estimated for each particle by using the three types of statistics for A$+$A and p$+$p collisions. Concretely, for p$+$p collisions, there is a general trend for the temperature $T$, namely $T$ increases with the increase in the energies as deduced by using all types of statistics (extensive and nonextensive) for all particles except pions, where the temperature is found decreasing with the increase in the energies, especially  by using the nonextensive statistics. We also noticed that the values of $T$ for anti-particles are slightly greater than the ones from the particles. For the three types of statistics, the fit parameter $\mu$ decreases with the increase in the energies for all particles and anti-particles. By using Boltzmann and generic axiomatic statistics, we noticed that the volume of the system decreases with the increase in energies for all particles and their anti-particles. For protons and Kaons and their anti-particles, the values of $V$ obtained by using Tsallis statistics are found nearly independent on energies, while for pions decrease slightly. The nonextensive parameter $q$ from Tsallis statistics increases with the increase in the energies for all particles and their anti-particles except at energies $62.4$ and $200$ GeV. Also, the equivalent class, similar to a nonextensive parameter, $d$ which is obtained by using generic axiomatic statistics, decreases with the increase in the energies for all types of particles.

But for A$+$A collisions, it was noticed that by using Boltzmann statistics there is a general behavior that the temperature increases with the increase in energies for all particles. At $200~$GeV, the temperature becomes smaller than the ones at lower energies. Also, we conclude that the temperature obtained from anti-particles are slightly greater than the ones from the particles. But for nonextensive statistics, the temperature is found increasing with the increase in energies for all particles except for pions. Another exception for pions could be highlighted that there is a reverse proportionality between the resulting temperature and the energies. The temperature deduced from the $p_{\mathtt{T}}$ spectra of anti-pions, anti-Kaons, and anti-protons are slightly greater than the ones from the $p_{\mathtt{T}}$ spectra of their particles.

Using various parametrization expressions, it was noticed that the Tsallis statistics is more effective within the low $p_T$ range rather than in the high $p_T$. Also we conclude that the Tsallis statistics is successful in describing p$+$p collisions rather than A$+$A collisions, while Boltzmann statistics describes well the latter more than the earlier. We conclude that generic axiomatic statistics is well applicable in describing both types of collisions at all energies for charged particles and their anti-particles.  

The values obtained for the equivalent classes $(c,d)$ that $c\rightarrow 1$, while $0<d<1$ warrens the conclusion that the resulting Lambert-$W$ exponentials characterize entropic equivalence classes. This means that the fractional power-law and the entropy of the system of interest are characterized by delayed relaxation. 

Analytical expressions for the dependence of the various fit parameters otained by using different types of statistical approaches are proposed for each particle. We compare these with the ones deduced from other models. It is found that our nonextensive fit parameter $q$ for all particles except for pions agree well with the one deduced in ref. \cite{Wei:2016ihj}. Also, in comparing our results for the fit parameter $T$ with the ones deduced in ref. \cite{Zheng:2015mhz}, we found that our temperature obtained by using Boltzmann statistics is in a good agreement with Type-A Tsallis distribution $T_5$ for pions and Kaons at $\sqrt{s_{\mathtt{NN}}}=62.4~$GeV and also for Kaons, protons, and anti-protons at $\sqrt{s_{\mathtt{NN}}}=200~$ and $900~$GeV. While our results for pions and anti-protons obtained from Boltzmann statistics are greater than the temperatures deduced from all types of Tsallis distribution at $\sqrt{s_{\mathtt{NN}}}=200,~900~$GeV. For pions by using generic axiomatic statistics, it is found that our results are in a good agreement with $T_2$ at $\sqrt{s_{\mathtt{NN}}}=62.4,~200~$GeV. But our results for all studied particles agree well with $T_4$ at $\sqrt{s_{\mathtt{NN}}}=900~$GeV and $\sqrt{s_{\mathtt{NN}}}=62.4~$GeV, respectively. The temperature we estimated for Kaons agree well with $T_5$ at $\sqrt{s_{\mathtt{NN}}}=200~$ and $900~$GeV. For anti-protons and by using generic axiomatic statistics, our fit parameter agrees well with $T_1$ at $\sqrt{s_{\mathtt{NN}}}=62.4~$GeV. Our results for protons and anti-protons by using generic axiomatic statistics are lower than results taken from any type of the Tsallis distributions at $\sqrt{s_{\mathtt{NN}}}=200~$GeV.

Last but not least, the dependence of the various fit parameters on the types of the statistical approaches implemented, especially between Boltzmann and Tsallis approaches, arises when {\it ad hoc} a specific degree of extensivity or nonextensivity is applied. This can be seen when comparing the obtained results by the ones deduced from the empirical parameterizations, the other models, and the proposed generic (non)extensive approach, where the system, in our case, transverse momentum spectra, for instance, determines {\it almost} alone the degree of its extensivity or nonextensivity.

\bibliographystyle{aip}
\bibliography{Tsallis_pT8}

\begin{thebibliography}{100}

\bibitem{Tawfik:2014eba}
A.~N. Tawfik,
\newblock Int. J. Mod. Phys. {\bf A29}, 1430021 (2014).

\bibitem{Tawfik:2013tza}
A.~N. Tawfik,
\newblock Z. Naturforsch. {\bf A69}, 106 (2014).

\bibitem{Fermi1950}
E.~Fermi,
\newblock Prog. Theor. Phys. {\bf 5}, 570 (1950).

\bibitem{Fermi1950book}
E.~Fermi,
\newblock {\em Elementary Particles},
\newblock Yale University Press, New Haven, 1951.

\bibitem{Magalinskii1957}
V.~B. Magalinskii and I.~P. Terletskii,
\newblock Zh. Eksp. Teor. Fiz. {\bf 32}, 584 (1957).

\bibitem{Hagedorn1963}
G.~Fast and R.~Hagedorn,
\newblock Nuono Cimento {\bf 27}, 208 (1963).

\bibitem{Hagedorn1963b}
G.~Fast, R.~Hagedorn, and L.~W. Jones,
\newblock Nuovo Cimento {\bf 27}, 856.

\bibitem{Tsallis:1987eu}
C.~Tsallis,
\newblock J. Statist. Phys. {\bf 52}, 479 (1988).

\bibitem{Bediaga:1999hv}
I.~Bediaga, E.~M.~F. Curado, and J.~M. de~Miranda,
\newblock Physica {\bf A286}, 156 (2000).

\bibitem{Parvan:2016rln}
A.~S. Parvan, O.~V. Teryaev, and J.~Cleymans,
\newblock Eur. Phys. J. {\bf A53}, 102 (2017).

\bibitem{Beck:2000nz}
C.~Beck,
\newblock Physica {\bf A286}, 164 (2000).

\bibitem{Wilk:1999dr}
G.~Wilk and Z.~Wlodarczyk,
\newblock Phys. Rev. Lett. {\bf 84}, 2770 (2000).

\bibitem{Walton:1999dy}
D.~B. Walton and J.~Rafelski,
\newblock Phys. Rev. Lett. {\bf 84}, 31 (2000).

\bibitem{Alberico:2000nc}
W.~M. Alberico, A.~Lavagno, and P.~Quarati,
\newblock Nucl. Phys. {\bf A680}, 94 (2000).

\bibitem{Zimanyi:2005nn}
J.~Zimanyi, P.~Levai, and T.~S. Biro,
\newblock J. Phys. {\bf G31}, 711 (2005).

\bibitem{Trainor:2007zj}
T.~A. Trainor,
\newblock Int. J. Mod. Phys. {\bf E17}, 1499 (2008).

\bibitem{Wilk:2008ue}
G.~Wilk and Z.~Wlodarczyk,
\newblock Eur. Phys. J. {\bf A40}, 299 (2009).

\bibitem{Biro:2008er}
T.~S. Biro and K.~Urmossy,
\newblock J. Phys. {\bf G36}, 064044 (2009).

\bibitem{at14}
Q.~A. Wong,
\newblock Entropy {\bf 5}, 220 (2003).

\bibitem{Tripathy:2016hlg}
S.~Tripathy et~al.,
\newblock Eur. Phys. J. {\bf A52}, 289 (2016).

\bibitem{Khuntia:2016ikm}
A.~Khuntia, P.~Sahoo, P.~Garg, R.~Sahoo, and J.~Cleymans,
\newblock Eur. Phys. J. {\bf A52}, 292 (2016).

\bibitem{Bhattacharyya:2015hya}
T.~Bhattacharyya, J.~Cleymans, A.~Khuntia, P.~Pareek, and R.~Sahoo,
\newblock Eur. Phys. J. {\bf A52}, 30 (2016).

\bibitem{Deppman:2012qt}
A.~Deppman,
\newblock J. Phys. {\bf G41}, 055108 (2014).

\bibitem{Alberico:2005vu}
W.~M. Alberico, P.~Czerski, A.~Lavagno, M.~Nardi, and V.~Soma,
\newblock Physica {\bf A387}, 467 (2008).

\bibitem{Tawfik:2016pwz}
A.~Nasser~Tawfik,
\newblock Eur. Phys. J. {\bf A52}, 253 (2016).

\bibitem{Thurner:2010if}
S.~Thurner and R.~Hanel,
\newblock Eur. Phys. J. {\bf B84}, 707 (2011).

\bibitem{Tawfik:2017bul}
A.~N. Tawfik, H.~Yassin, and E.~R. Abo~Elyazeed,
\newblock Chin. Phys. {\bf C41}, 053107 (2017).

\bibitem{Weisskopf1937}
V.~F. Weisskopf,
\newblock Phys. Rev. {\bf 52}, 295 (1937).

\bibitem{Cheng:2001dz}
S.~Cheng et~al.,
\newblock Phys. Rev. {\bf C65}, 024901 (2002).

\bibitem{Tounsi:2001ck}
A.~Tounsi and K.~Redlich,
\newblock (2001).

\bibitem{Kassner:2016obj}
K.~Kassner,
\newblock Eur. J. Phys. {\bf 38}, 015605 (2017).

\bibitem{Zheng:2015mhz}
H.~Zheng and L.~Zhu,
\newblock Adv. High Energy Phys. {\bf 2016}, 9632126 (2016).

\bibitem{Gao:2015qsq}
Y.-Q. Gao and F.-H. Liu,
\newblock Indian J. Phys. {\bf 90}, 319 (2016).

\bibitem{Zheng:2015gaa}
H.~Zheng and L.~Zhu,
\newblock Adv. High Energy Phys. {\bf 2015}, 180491 (2015).

\bibitem{Zheng:2015tua}
H.~Zheng, L.~Zhu, and A.~Bonasera,
\newblock Phys. Rev. {\bf D92}, 074009 (2015).

\bibitem{Wilk:2015pva}
G.~Wilk and Z.~Wlodarczyk,
\newblock (2015),
\newblock [Entropy17,384(2015)].

\bibitem{Marques:2015mwa}
L.~Marques, J.~Cleymans, and A.~Deppman,
\newblock Phys. Rev. {\bf D91}, 054025 (2015).

\bibitem{Urmossy:2015kva}
K.~Urmossy, G.~G. Barnaföldi, S.~Harangozó, T.~S. Biró, and Z.~Xu,
\newblock J. Phys. Conf. Ser. {\bf 805}, 012010 (2017).

\bibitem{Cleymans:2013rfq}
J.~Cleymans et~al.,
\newblock Phys. Lett. {\bf B723}, 351 (2013).

\bibitem{Rybczynski:2014cha}
M.~Rybczynski and Z.~Wlodarczyk,
\newblock Eur. Phys. J. {\bf C74}, 2785 (2014).

\bibitem{Deppman:2012us}
A.~Deppman,
\newblock Physica {\bf A391}, 6380 (2012).

\bibitem{Kataja:1990tp}
M.~Kataja and P.~V. Ruuskanen,
\newblock Phys. Lett. {\bf B243}, 181 (1990).

\bibitem{Turbide:2003si}
S.~Turbide, R.~Rapp, and C.~Gale,
\newblock Phys. Rev. {\bf C69}, 014903 (2004).

\bibitem{Banerjee:2010}
A.~Banerjee and V.~M. Yakovenko,
\newblock New J. Phys. {\bf 12}, 075032 (2010).

\bibitem{Khandai:2013gva}
P.~K. Khandai, P.~Sett, P.~Shukla, and V.~Singh,
\newblock Int. J. Mod. Phys. {\bf A28}, 1350066 (2013).

\bibitem{Saraswat:2017kpg}
K.~Saraswat, P.~Shukla, and V.~Singh,
\newblock (2017).

\bibitem{Parvan:2016mbv}
A.~S. Parvan,
\newblock Eur. Phys. J. {\bf A52}, 355 (2016).

\bibitem{Buyukkilic1995}
F.~Buyukkilic, D.~Demirhan, , and A.~Gulec,
\newblock Phys. Lett. A {\bf 197}, 209 (1995).

\bibitem{Stachel:2013zma}
J.~Stachel, A.~Andronic, P.~Braun-Munzinger, and K.~Redlich,
\newblock J. Phys. Conf. Ser. {\bf 509}, 012019 (2014).

\bibitem{imcm}
S.~Ban-Hao and C.-Y. Wong,
\newblock Phys. Rev. D {\bf 32}, 1706 (1985).

\bibitem{Danielewicz:1984ww}
P.~Danielewicz and M.~Gyulassy,
\newblock Phys. Rev. {\bf D31}, 53 (1985).

\bibitem{Siemens:1978pb}
P.~J. Siemens and J.~O. Rasmussen,
\newblock Phys. Rev. Lett. {\bf 42}, 880 (1979).

\bibitem{Westfall:1976fu}
G.~D. Westfall et~al.,
\newblock Phys. Rev. Lett. {\bf 37}, 1202 (1976).

\bibitem{Schnedermann:1993ws}
E.~Schnedermann, J.~Sollfrank, and U.~W. Heinz,
\newblock Phys. Rev. {\bf C48}, 2462 (1993).

\bibitem{Anderlik:1998cb}
C.~Anderlik et~al.,
\newblock Phys. Rev. {\bf C59}, 388 (1999).

\bibitem{Lokhtin:1996ht}
I.~P. Lokhtin and A.~M. Snigirev,
\newblock Phys. Lett. {\bf B378}, 247 (1996).

\bibitem{biroBook}
T.~D. Biro,
\newblock {\em Is there a temperature: Conceptual Challenges at High Energy,
  Acceleration and Complexity},
\newblock Springer-Verlag, New York, 2011.

\bibitem{Wei:2016ihj}
H.-R. Wei, F.-H. Liu, and R.~A. Lacey,
\newblock Eur. Phys. J. {\bf A52}, 102 (2016).

\bibitem{Lao:2015zgd}
H.-L. Lao, H.-R. Wei, F.-H. Liu, and R.~A. Lacey,
\newblock Eur. Phys. J. {\bf A52}, 203 (2016).

\bibitem{Huovinen:2006jp}
P.~Huovinen and P.~V. Ruuskanen,
\newblock Ann. Rev. Nucl. Part. Sci. {\bf 56}, 163 (2006).

\bibitem{Adamczyk:2017iwn}
L.~Adamczyk et~al.,
\newblock Phys. Rev. {\bf C96}, 044904 (2017).

\bibitem{Abelev:2008ab}
B.~I. Abelev et~al.,
\newblock Phys. Rev. {\bf C79}, 034909 (2009).

\bibitem{Bialas:2015pla}
A.~Bialas,
\newblock Phys. Lett. {\bf B747}, 190 (2015).

\bibitem{Gyulassy:2003mc}
M.~Gyulassy, I.~Vitev, X.-N. Wang, and B.-W. Zhang,
\newblock page 123 (2003).

\bibitem{dEnterria2005}
D.~d’Enterria,
\newblock J. Phys. G: Nucl. Part. Phys. {\bf 31}, S491 (2005).

\bibitem{Adcox:2001jp}
K.~Adcox et~al.,
\newblock Phys. Rev. Lett. {\bf 88}, 022301 (2002).

\bibitem{Adler:2002xw}
C.~Adler et~al.,
\newblock Phys. Rev. Lett. {\bf 89}, 202301 (2002).

\bibitem{Adler:2003qi}
S.~S. Adler et~al.,
\newblock Phys. Rev. Lett. {\bf 91}, 072301 (2003).

\bibitem{Adams:2003kv}
J.~Adams et~al.,
\newblock Phys. Rev. Lett. {\bf 91}, 172302 (2003).

\bibitem{Back:2003qr}
B.~B. Back et~al.,
\newblock Phys. Lett. {\bf B578}, 297 (2004).

\bibitem{Arsene:2003yk}
I.~Arsene et~al.,
\newblock Phys. Rev. Lett. {\bf 91}, 072305 (2003).

\bibitem{Adler:2003pb}
S.~S. Adler et~al.,
\newblock Phys. Rev. Lett. {\bf 91}, 241803 (2003).

\bibitem{Aggarwal:2001gn}
M.~M. Aggarwal et~al.,
\newblock Eur. Phys. J. {\bf C23}, 225 (2002).

\bibitem{ReffInt15b}
M.~M. Aggarwal et~al.,
\newblock Phys. Rev. Lett. {\bf 84}, 578 (2000).

\bibitem{Aggarwal:1998vh}
M.~M. Aggarwal et~al.,
\newblock Phys. Rev. Lett. {\bf 81}, 4087 (1998),
\newblock [Erratum: Phys. Rev. Lett.84,578(2000)].

\bibitem{Wang:1998hs}
X.-N. Wang,
\newblock Phys. Rev. Lett. {\bf 81}, 2655 (1998).

\bibitem{Wang:1998ww}
X.-N. Wang,
\newblock Phys. Rev. {\bf C61}, 064910 (2000).

\bibitem{Wang:2001cy}
E.~Wang and X.-N. Wang,
\newblock Phys. Rev. {\bf C64}, 034901 (2001).

\bibitem{Cronin:1974zm}
J.~W. Cronin et~al.,
\newblock Phys. Rev. {\bf D11}, 3105 (1975).

\bibitem{Kopeliovich:2002yh}
B.~Z. Kopeliovich, J.~Nemchik, A.~Schafer, and A.~V. Tarasov,
\newblock Phys. Rev. Lett. {\bf 88}, 232303 (2002).

\bibitem{Gribov:1967hh}
V.~N. Gribov, B.~L. Ioffe, and I.~{\relax Ya}. Pomeranchuk,
\newblock Sov. J. Nucl. Phys. {\bf 6}, 427 (1968),
\newblock [Phys. Lett.24B,554(1967)].

\bibitem{Geiger:1995ak}
K.~Geiger,
\newblock Phys. Rev. {\bf D54}, 949 (1996).

\bibitem{Ayala:1995kg}
A.~Ayala, J.~Jalilian-Marian, L.~D. McLerran, and R.~Venugopalan,
\newblock Phys. Rev. {\bf D52}, 2935 (1995).

\bibitem{Liang:2004ph}
Z.-T. Liang and X.-N. Wang,
\newblock Phys. Rev. Lett. {\bf 94}, 102301 (2005),
\newblock [Erratum: Phys. Rev. Lett.96,039901(2006)].

\bibitem{Diehl:2017wew}
M.~Diehl and J.~R. Gaunt,
\newblock Adv. Ser. Direct. High Energy Phys. {\bf 29}, 7 (2018).

\bibitem{Kasemets:2017vyh}
T.~Kasemets and S.~Scopetta,
\newblock Adv. Ser. Direct. High Energy Phys. {\bf 29}, 49 (2018).

\bibitem{Antreasyan:1978cw}
D.~Antreasyan et~al.,
\newblock Phys. Rev. {\bf D19}, 764 (1979).

\bibitem{Straub:1992xd}
P.~B. Straub et~al.,
\newblock Phys. Rev. Lett. {\bf 68}, 452 (1992).

\bibitem{Angelis:1985fk}
A.~L.~S. Angelis et~al.,
\newblock Phys. Lett. {\bf B185}, 213 (1987).

\bibitem{Abelev:2013ala}
B.~B. Abelev et~al.,
\newblock Eur. Phys. J. {\bf C73}, 2662 (2013).

\bibitem{Sassot:2010bh}
R.~Sassot, P.~Zurita, and M.~Stratmann,
\newblock Phys. Rev. {\bf D82}, 074011 (2010).

\bibitem{Chatrchyan:2012qb}
S.~Chatrchyan et~al.,
\newblock Eur. Phys. J. {\bf C72}, 2164 (2012).

\bibitem{Aamodt:2011zj}
K.~Aamodt et~al.,
\newblock Eur. Phys. J. {\bf C71}, 1655 (2011).

\bibitem{Khachatryan:2010xs}
V.~Khachatryan et~al.,
\newblock JHEP {\bf 02}, 041 (2010).

\bibitem{Khachatryan:2010us}
V.~Khachatryan et~al.,
\newblock Phys. Rev. Lett. {\bf 105}, 022002 (2010).

\bibitem{Adams:2004ep}
J.~Adams et~al.,
\newblock Phys. Rev. {\bf C71}, 064902 (2005).

\bibitem{Biro:2008hz}
T.~S. Biro, G.~Purcsel, and K.~Urmossy,
\newblock Eur. Phys. J. {\bf A40}, 325 (2009).

\bibitem{Liu:2008am}
F.-H. Liu,
\newblock Nucl. Phys. {\bf A810}, 159 (2008).

\bibitem{Liu:2008ar}
F.-H. Liu and J.-S. Li,
\newblock Phys. Rev. {\bf C78}, 044602 (2008).

\bibitem{Liu:2014nra}
F.-H. Liu, Y.-Q. Gao, T.~Tian, and B.-C. Li,
\newblock Eur. Phys. J. {\bf A50}, 94 (2014).

\bibitem{YA-QIN:2012}
Y.-Q. Gaoa, C.-X. Tiana, F.-H. Liua, M.~A. Rahimb, and S.~Fakhraddin,
\newblock PRAMANA J. phys. {\bf 79}, 1407 (2012).

\bibitem{Wei:2015yut}
H.-R. Wei and F.-H. Liu,
\newblock Adv. High Energy Phys. {\bf 2015}, 263135 (2015).

\bibitem{Wei:2015oha}
H.-R. Wei, F.-H. Liu, and R.~A. Lacey,
\newblock J. Phys. {\bf G43}, 125102 (2016).

\bibitem{Adler:2003cb}
S.~S. Adler et~al.,
\newblock Phys. Rev. {\bf C69}, 034909 (2004).

\bibitem{Takeuchi:2015ana}
S.~Takeuchi, K.~Murase, T.~Hirano, P.~Huovinen, and Y.~Nara,
\newblock Phys. Rev. {\bf C92}, 044907 (2015).

\bibitem{Heiselberg:1998es}
H.~Heiselberg and A.-M. Levy,
\newblock Phys. Rev. {\bf C59}, 2716 (1999).

\bibitem{Heinz:2004qz}
U.~W. Heinz,
\newblock {Concepts of heavy ion physics},
\newblock in {\em {2002 European School of high-energy physics, Pylos, Greece,
  25 Aug-7 Sep 2002: Proceedings}}, pages 165--238, 2004.

\bibitem{Russo:2015xtz}
R.~Russo,
\newblock {\em {Measurement of D$^{+}$ meson production in p-Pb collisions with
  the ALICE detector}},
\newblock PhD thesis, Turin U., 2015.

\bibitem{Adam:2015qaa}
J.~Adam et~al.,
\newblock Eur. Phys. J. {\bf C75}, 226 (2015).

\bibitem{Andrei:2014vaa}
C.~Andrei,
\newblock Nucl. Phys. {\bf A931}, 888 (2014).

\bibitem{Suzuki:2005zb}
N.~Suzuki and M.~Biyajima,
\newblock Int. J. Mod. Phys. {\bf E16}, 133 (2007).

\bibitem{Liu:2014nsa}
F.-H. Liu, Y.-Q. Gao, and B.-C. Li,
\newblock Eur. Phys. J. {\bf A50}, 123 (2014).

\bibitem{Yassin:2018svv}
H.~Yassin and E.~R.~A. Elyazeed,
\newblock A. Phys. Pol. {\bf B50}, 37 (2018).

\bibitem{AP1999}
C.~Anteneodo and A.~R. Plastino,
\newblock J. Phys. A: Math. Gen. {\bf 32}, 1089 (1999).

\bibitem{Tawfik:2018ahq}
A.~N. Tawfik, H.~Yassin, and E.~R. Abo~Elyazeed,
\newblock Indian J. Phys. {\bf 92}, 1325 (2018).

\bibitem{Banner:1983jq}
M.~Banner et~al.,
\newblock Phys. Lett. {\bf 122B}, 322 (1983).

\bibitem{Alexopoulos:1993wt}
T.~Alexopoulos et~al.,
\newblock Phys. Rev. {\bf D48}, 984 (1993).

\bibitem{Adare:2011vy}
A.~Adare et~al.,
\newblock Phys. Rev. {\bf C83}, 064903 (2011).

\bibitem{Abelev:2006cs}
B.~I. Abelev et~al.,
\newblock Phys. Rev. {\bf C75}, 064901 (2007).

\bibitem{Sett:2014csa}
P.~Sett and P.~Shukla,
\newblock Adv. High Energy Phys. {\bf 2014}, 896037 (2014).

\bibitem{Andronic:2005yp}
A.~Andronic, P.~Braun-Munzinger, and J.~Stachel,
\newblock Nucl. Phys. {\bf A772}, 167 (2006).

\bibitem{Alberico:2009gj}
W.~M. Alberico and A.~Lavagno,
\newblock Eur. Phys. J. {\bf A40}, 313 (2009).

\bibitem{Adams:2006nd}
J.~Adams et~al.,
\newblock Phys. Lett. {\bf B637}, 161 (2006).

\end{thebibliography}

\newpage

\appendix

\section{Statistical-thermal approaches}
\subsection{A$+$A collisions}

\subsubsection{Maxwell-Boltzmann statistical fits} 
\label{BoltzmannAA}

\begin{figure}[htb]
\centering{
\includegraphics[width=5cm,angle=-0]{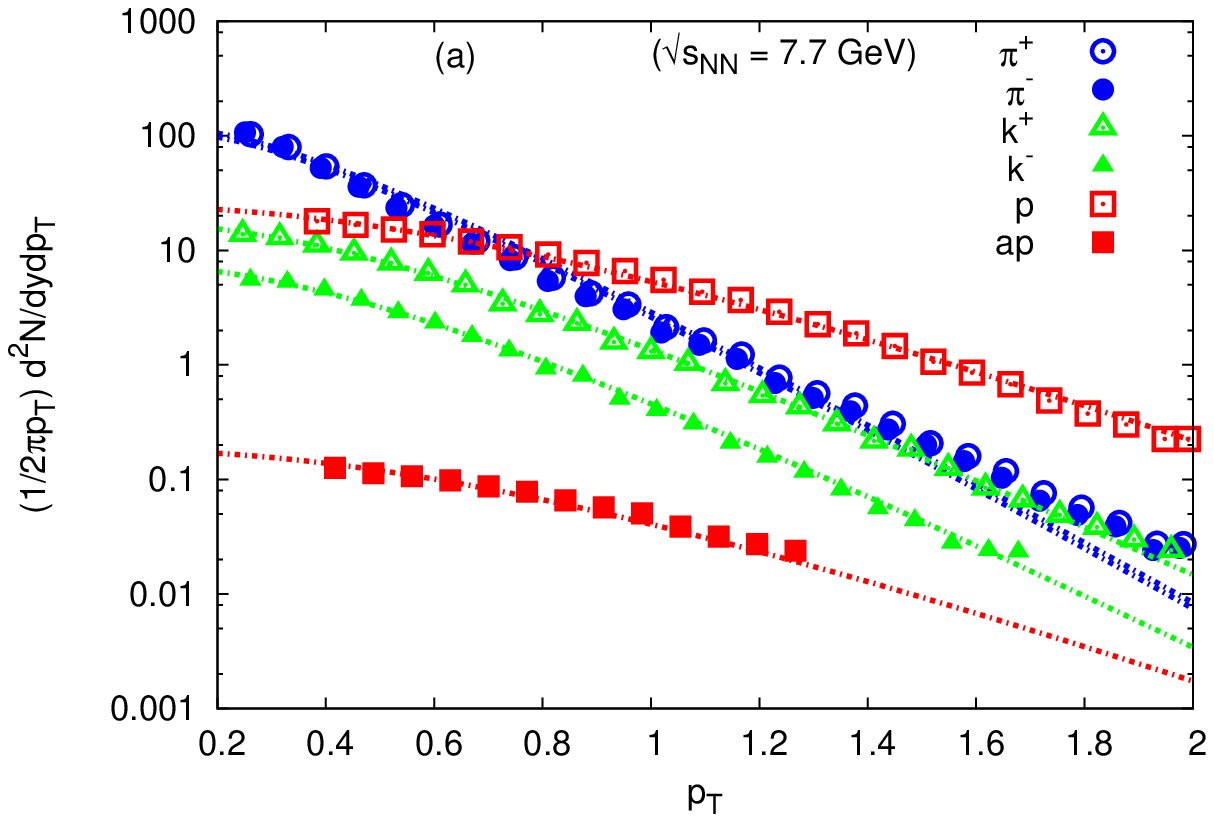}
\includegraphics[width=5cm,angle=-0]{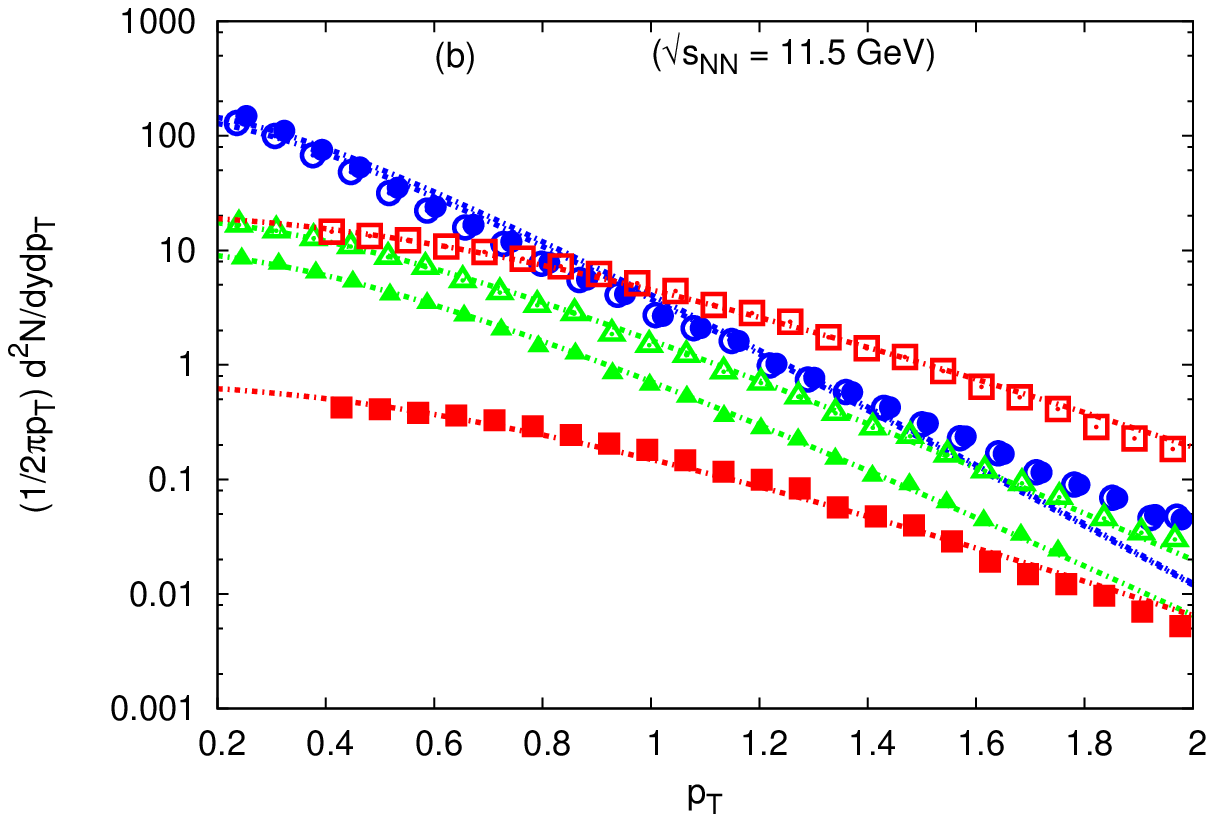}
\includegraphics[width=5cm,angle=-0]{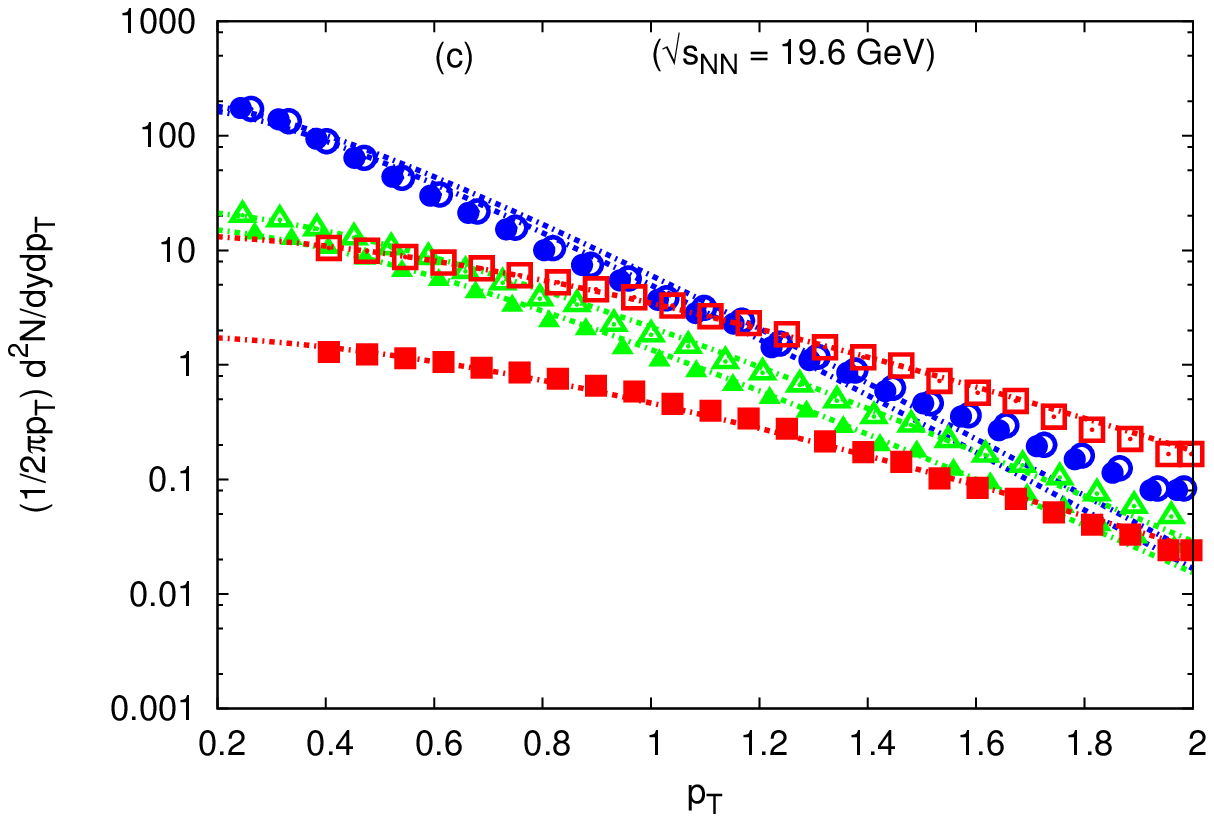}
\includegraphics[width=5cm,angle=-0]{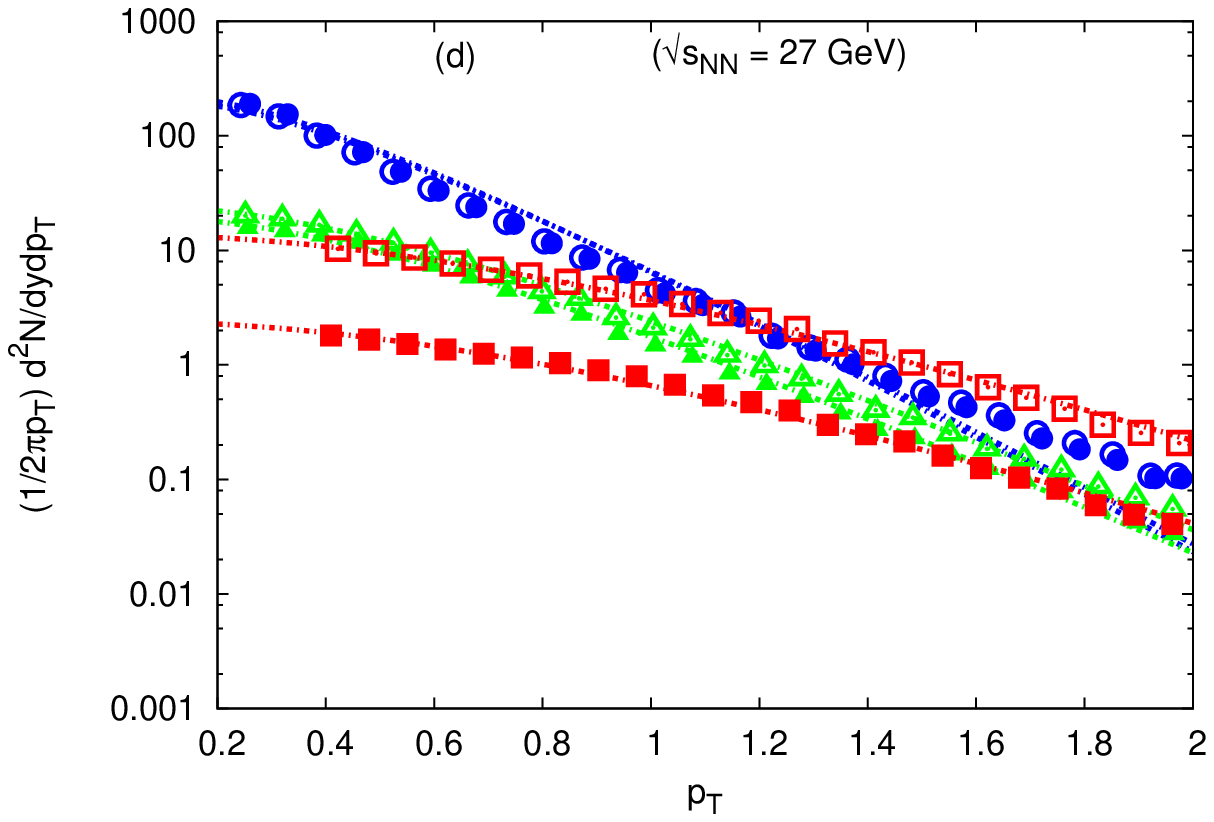}
\includegraphics[width=5cm,angle=-0]{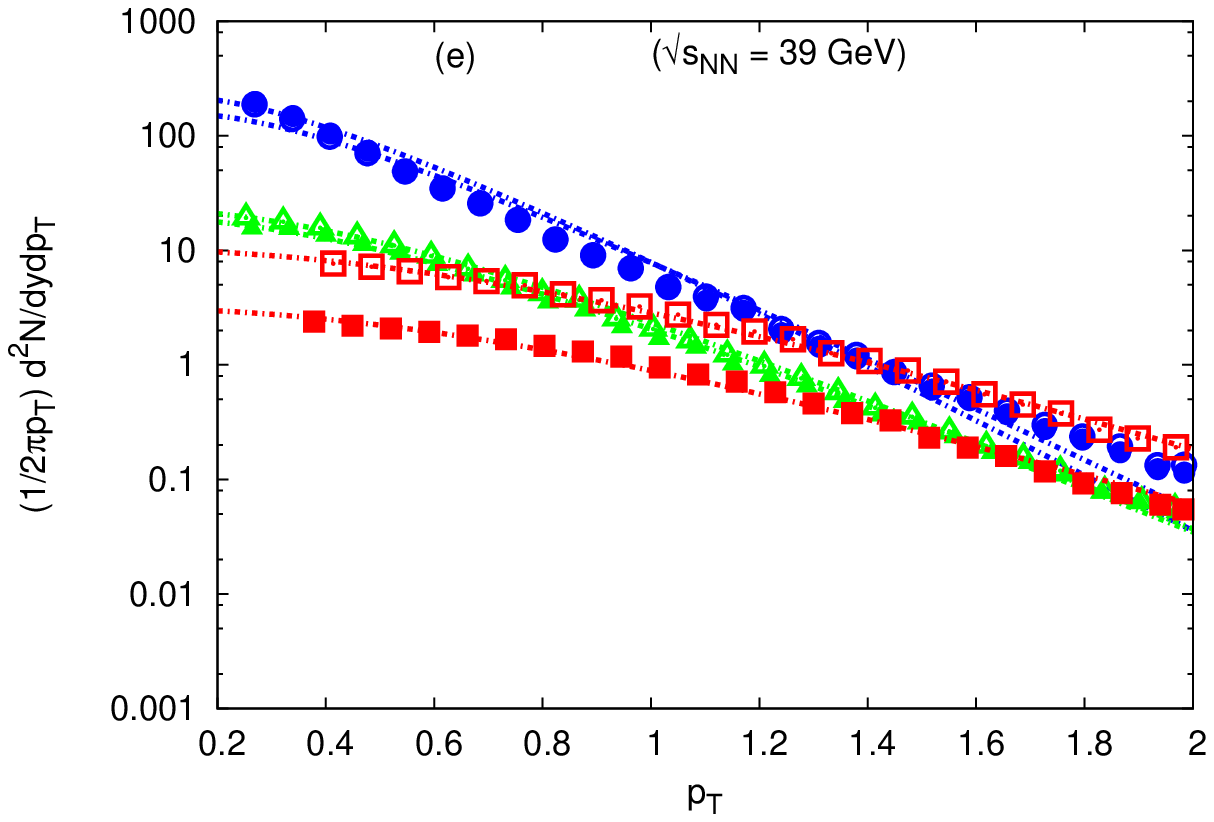}
\includegraphics[width=5cm,angle=-0]{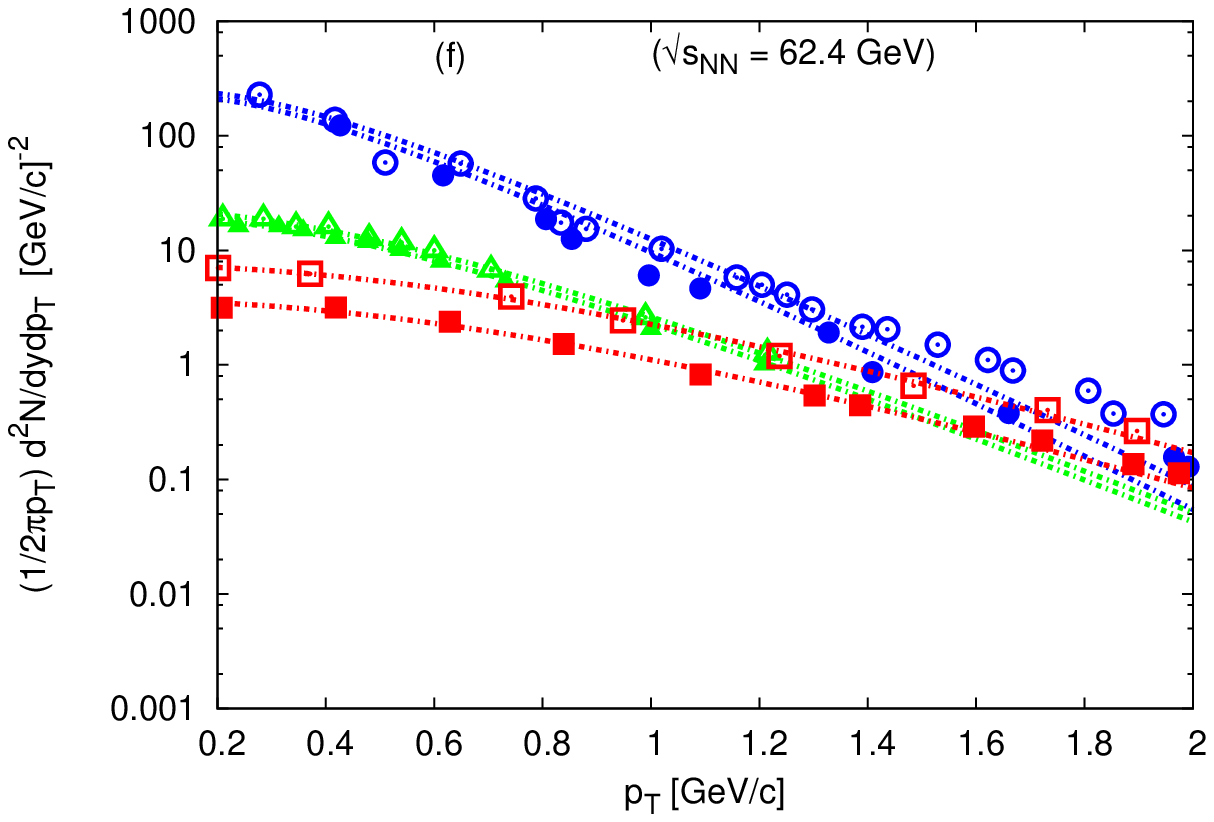}
\includegraphics[width=5cm,angle=-0]{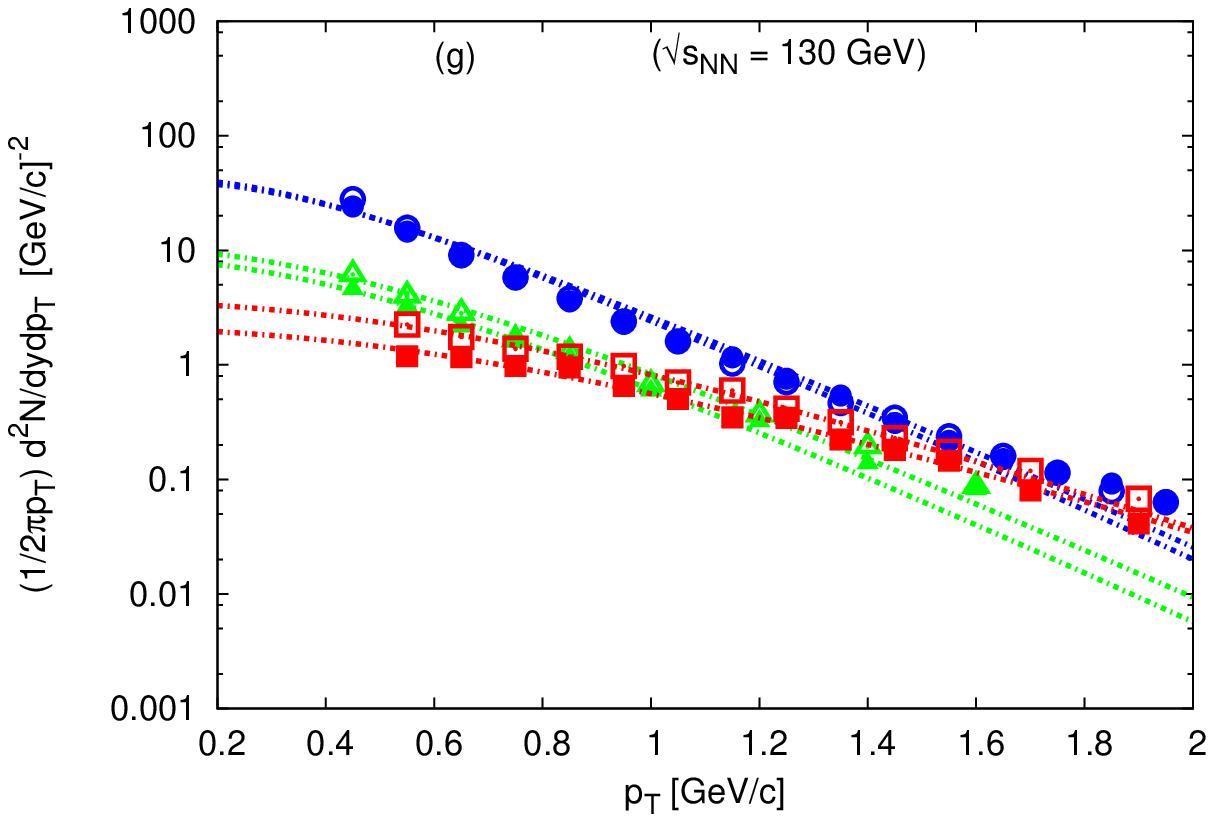}
\includegraphics[width=5cm,angle=-0]{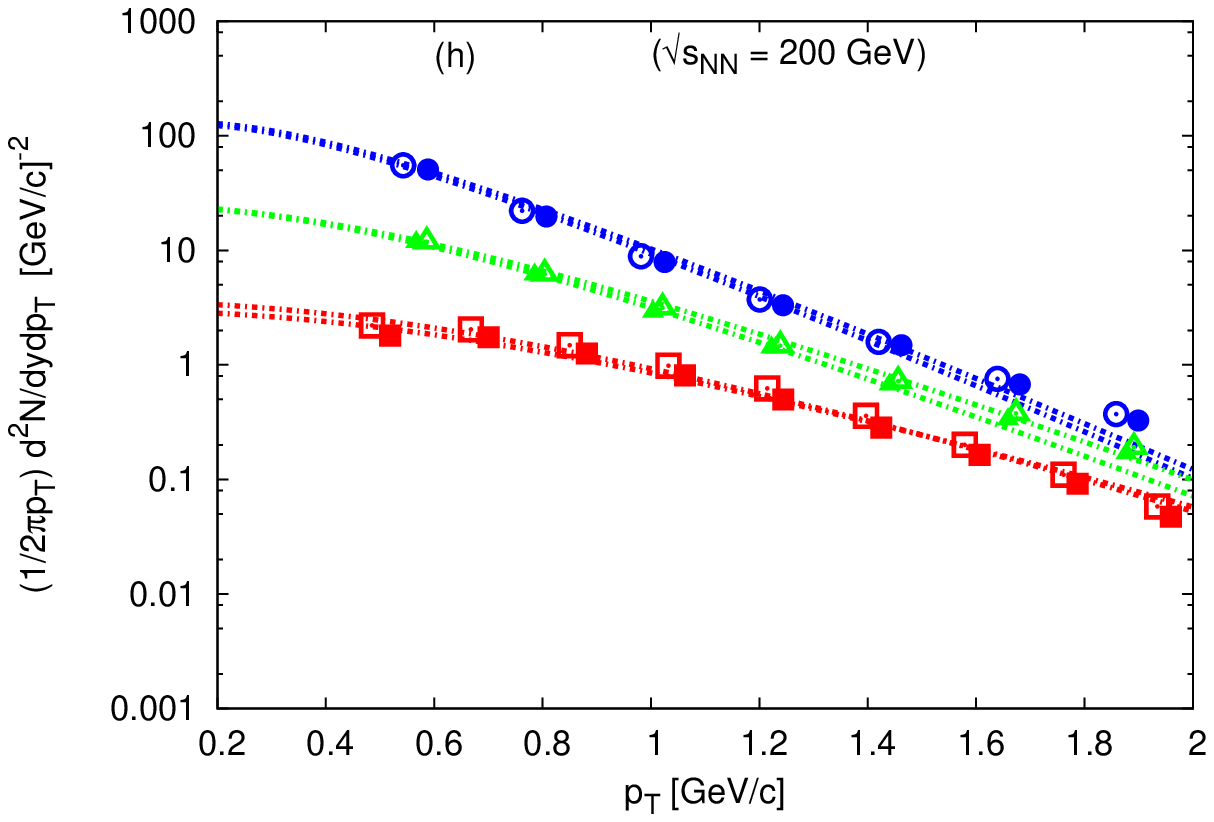}
\includegraphics[width=5cm,angle=-0]{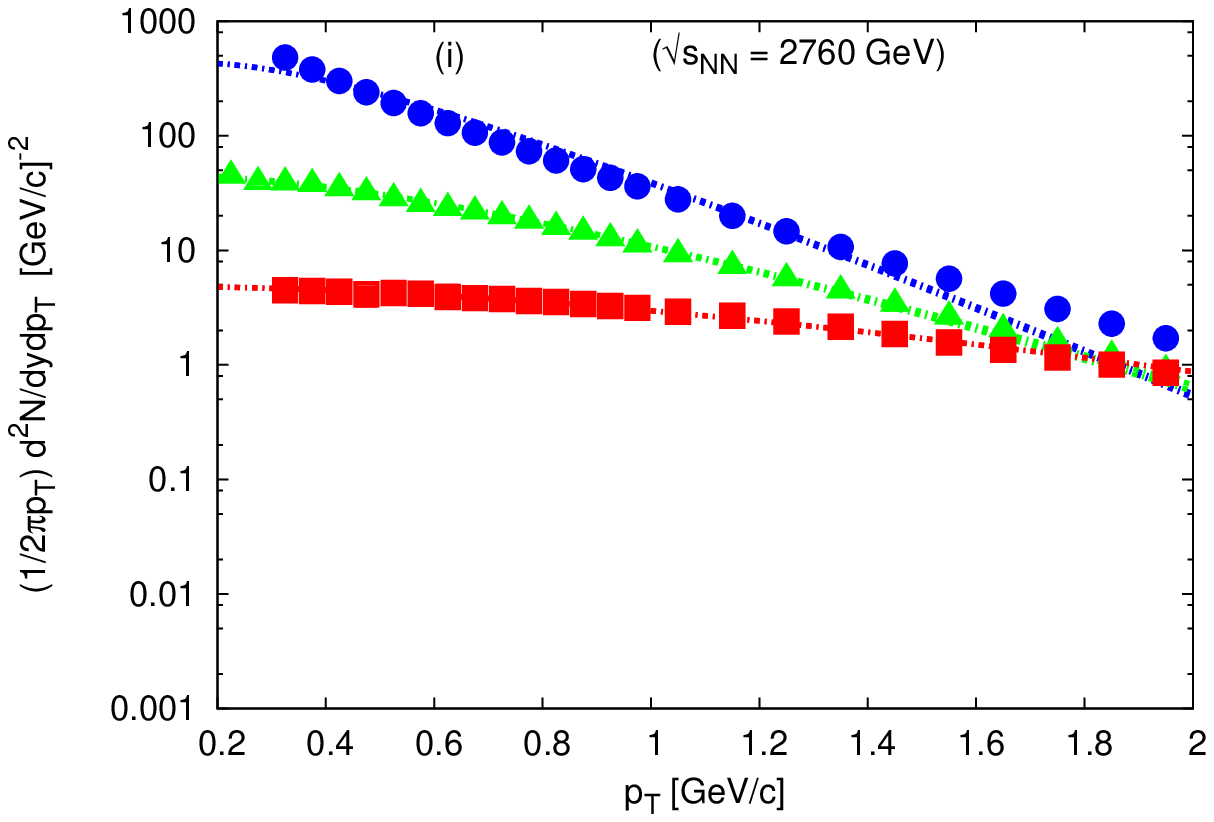}
\caption{(Color online) The transverse momentum spectra $p_{\mathtt{T}}$ for charged particles measured from A$+$A collisions in a wide range of energies are analyzed by using Maxwell-Boltzmann statistics. Panels (a)-(i) refer to the change in the center-of-mass energies from $7.7$ to $11.5$ to $19.6$ to $27$ to $39$ to $62.4$ to $130$ to $200$ and to $2760~$GeV, respectively. The fit parameters are shown in Fig. \ref{fig:AllStatisticsPerParticle}.
\label{fig:3A}
}}
\end{figure}

Figure \ref{fig:3A} presents an analysis for the transverse momentum spectra $p_{\mathtt{T}}$ for charged particles and their anti-particles at energies ranging from $7.7$ to $2760.0~$GeV by using Maxwell-Boltzmann statistics. The experimental results measured in A$+$A collisions are represented by symbols. Open symbols refer to the charged particles while the closed symbols refer to the anti-particles. The fitted spectra by using Maxwell-Boltzmann statistics are depicted by curves. It is found that Maxwell-Boltzmann statistics fits well the transverse momentum distributions for all particles. The fit parameters are compared to those obtained from different types statistics for all particles at whole range of energies, Fig. \ref{fig:AllStatisticsPerParticle}. The left panels illustrate results for pions. The results for Kaons are given in the middle panels. The right panels presents the results for protons and anti-protons.

\newpage
\subsubsection{Tsallis statistical fits} 
\label{TsallisAA}
Figure \ref{fig:3A} shows a statistical analysis of the transverse momentum spectra $p_{\mathtt{T}}$ of charged particles and their anti-particles by using Tsallis statistics (curves) at energies ranging from $7.7$ to $2760.0~$GeV. The experimental results from A$+$A collisions are given symbols. The open symbols refer to the charged particles while the closed ones refer to anti-particles. It is obvious that Tsallis statistics fit well all studied particles. The fit parameters are compared to those obtained from different types of statistics for all particles at the whole range of energies. The resulting fit parameters are presentedd in Figs. \ref{fig:Genericd} and \ref{fig:AllStatisticsPerParticle}.

\begin{figure}[htb]
\centering{
\includegraphics[width=5cm,angle=-0]{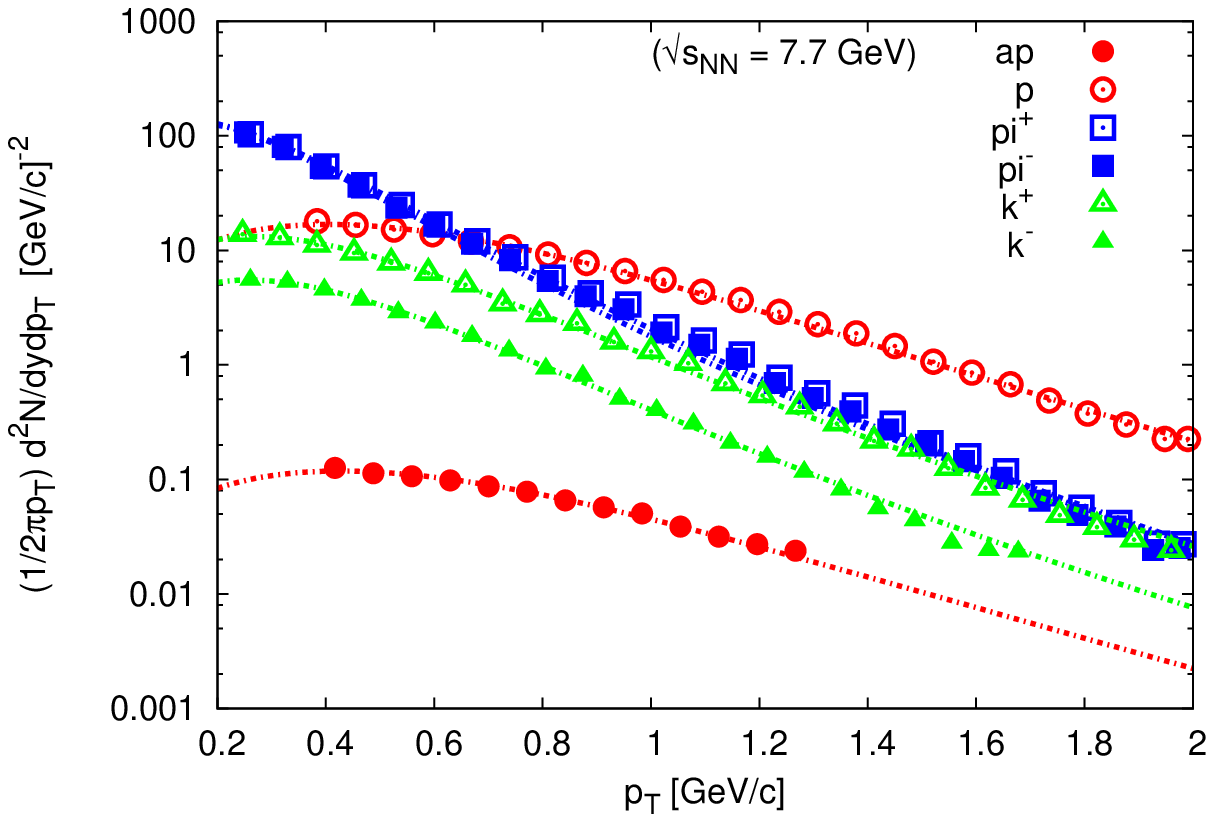}
\includegraphics[width=5cm,angle=-0]{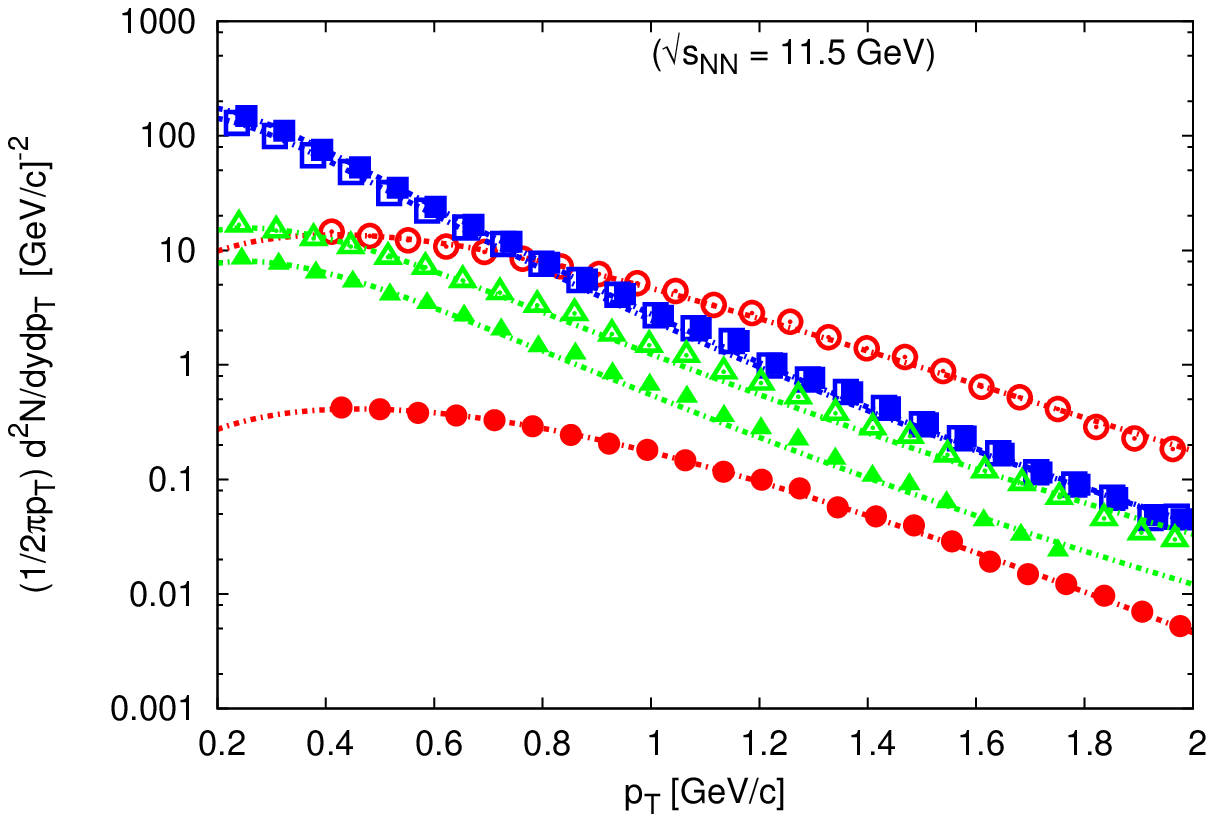}
\includegraphics[width=5cm,angle=-0]{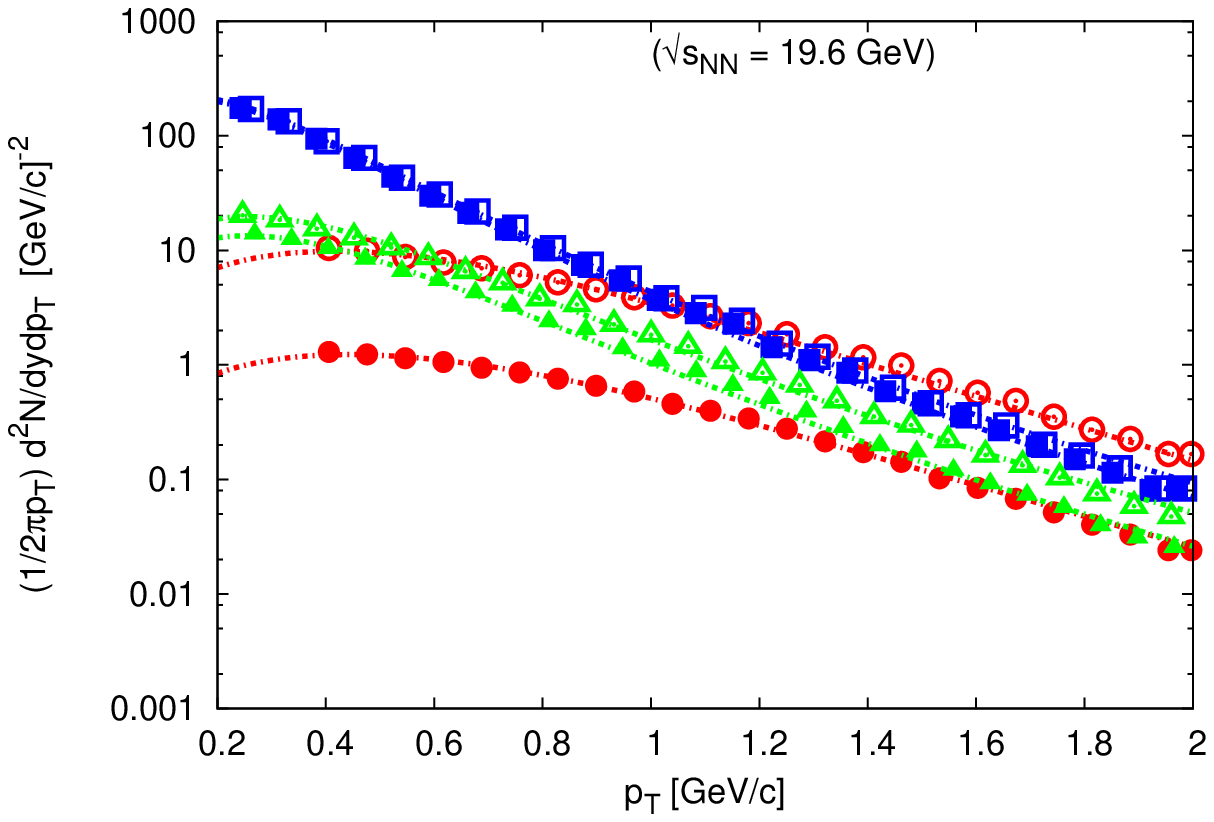}
\includegraphics[width=5cm,angle=-0]{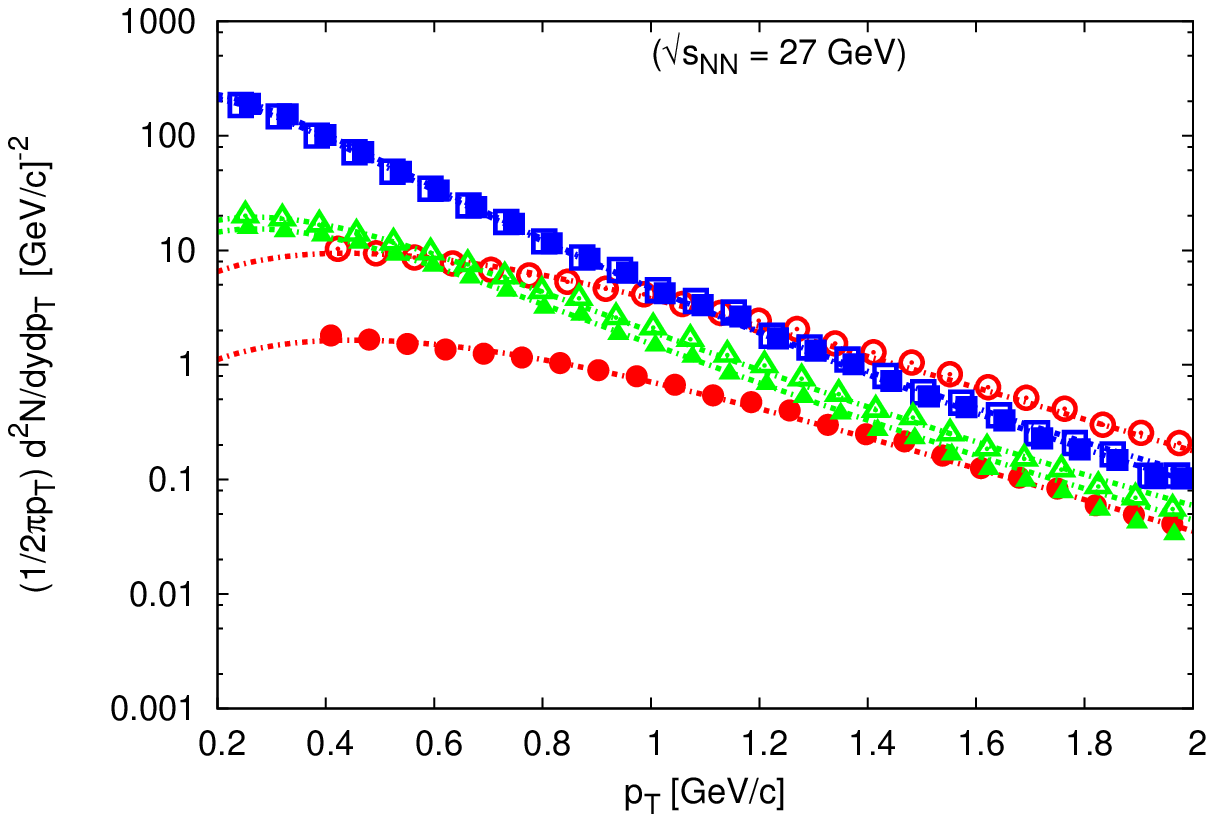}
\includegraphics[width=5cm,angle=-0]{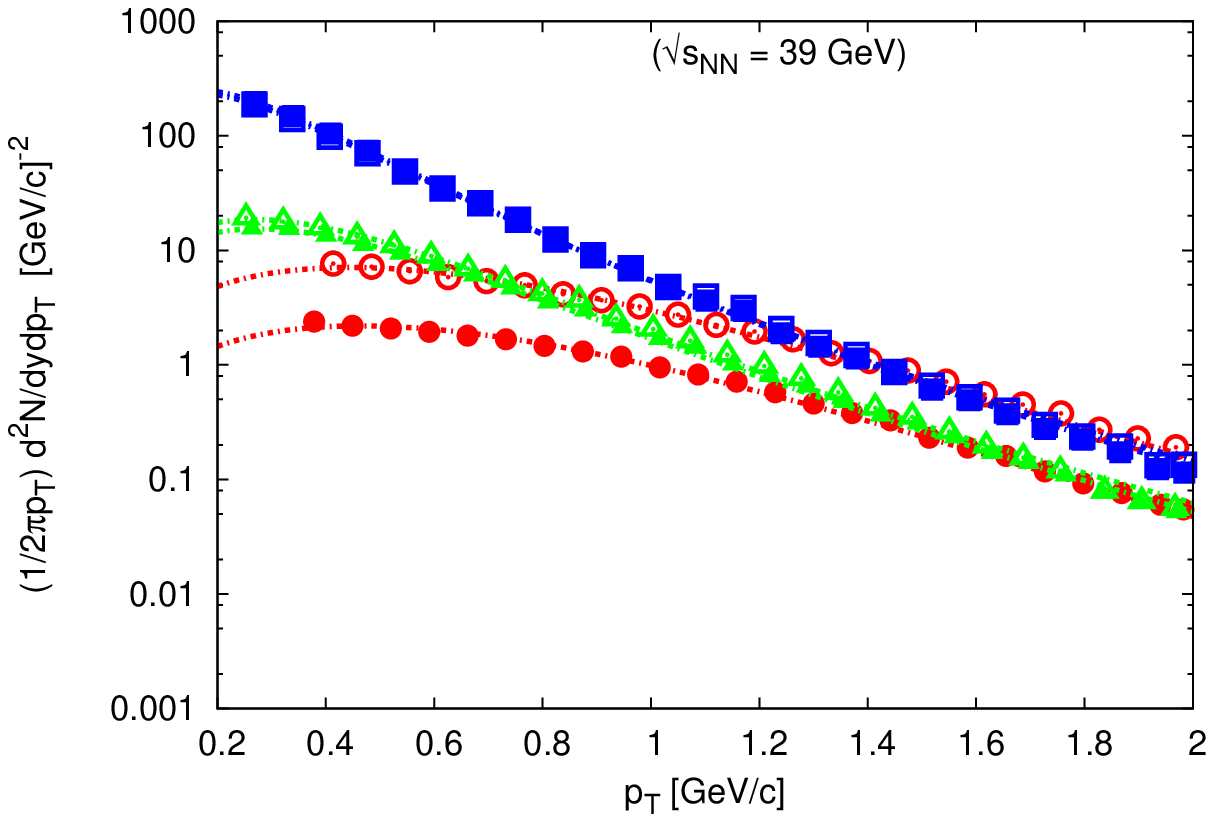}
\includegraphics[width=5cm,angle=-0]{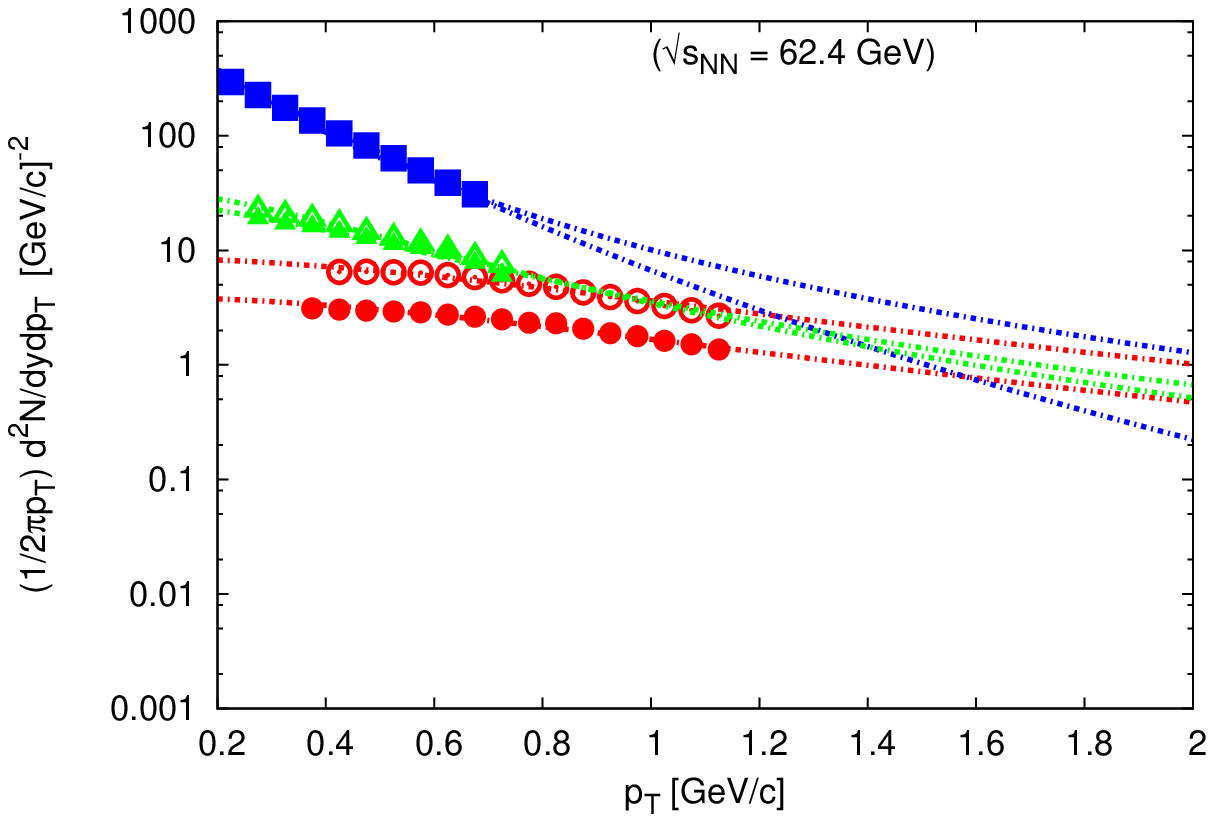}
\includegraphics[width=5cm,angle=-0]{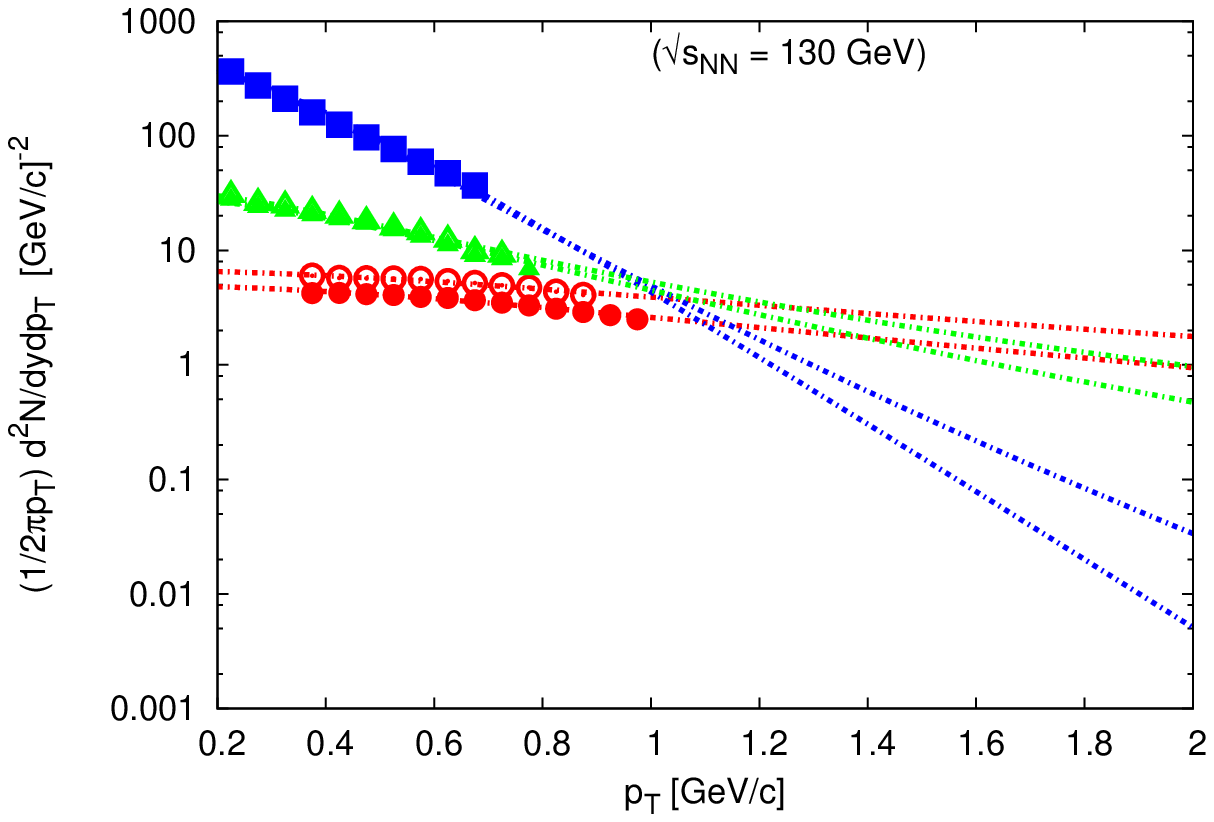}
\includegraphics[width=5cm,angle=-0]{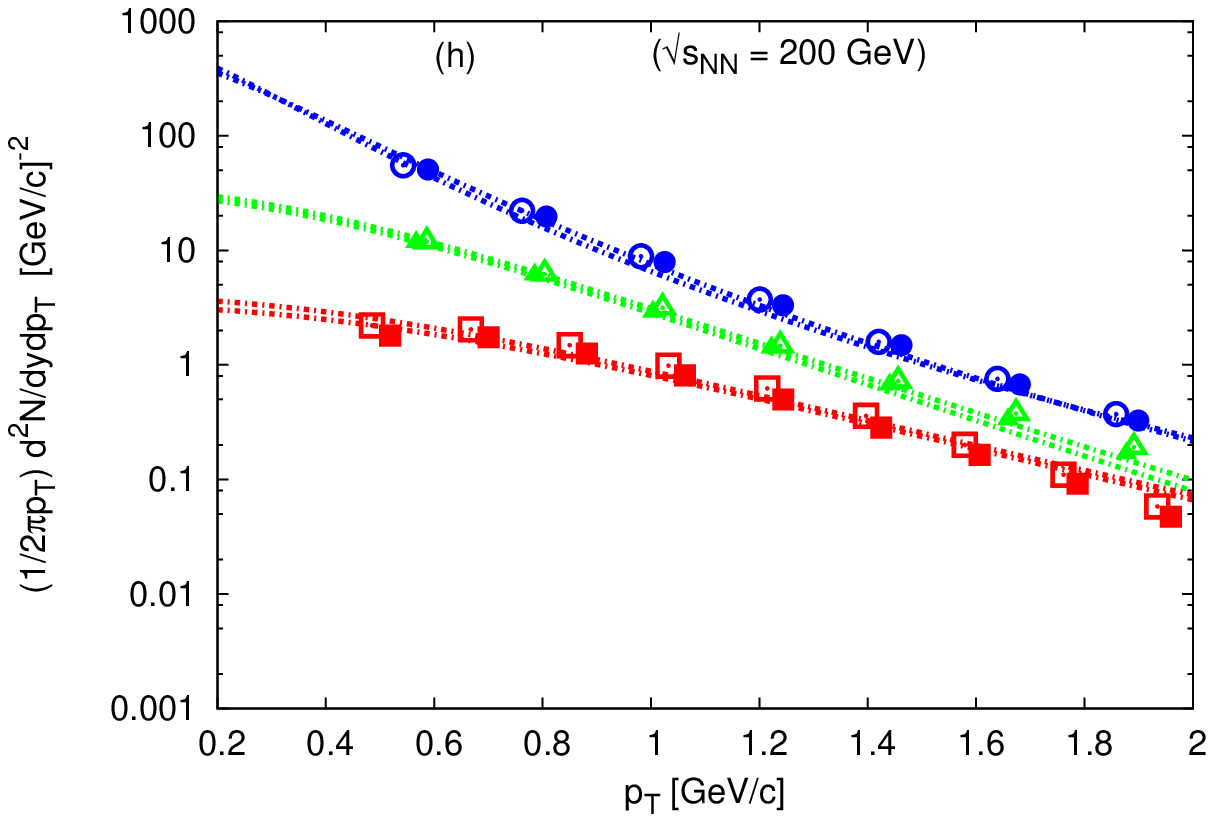}
\includegraphics[width=5cm,angle=-0]{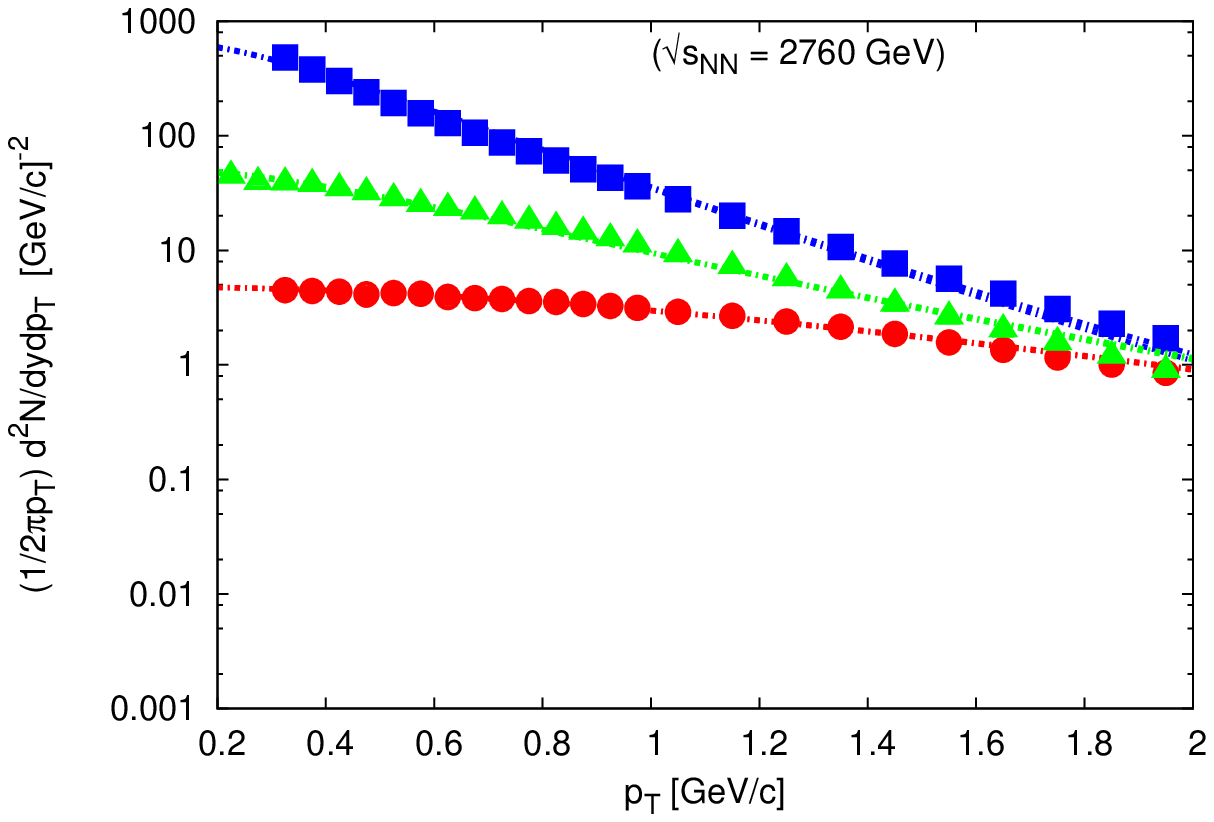}
\caption{(Color online) The transverse momentum spectra $p_{\mathtt{T}}$ for charged particles measured from  A$+$A collisions at a wide range of energies are fitted to Tsallis statistics. Panels (a)-(i) show the impacts of changing energies from $7.7$ to $11.5$ to $19.6$ to $27$ to $39$ to $62.4$ to $130$ to $200$ and to $2760~$GeV, respectively. The corresponding fit parameters are illustrated in Figs. \ref{fig:Genericd} and \ref{fig:AllStatisticsPerParticle} 
\label{fig:3B}
}}
\end{figure} 

\newpage
\subsubsection{Generic axiomatic statistical fits}
\label{GenericAA}
Figure \ref{fig:3A} depicts the transverse momentum spectra $p_{\mathtt{T}}$ for charged particles and their anti-particles measured from A$+$A collisions at energies ranging from $7.7$ to $2760~$GeV (symbols) The open symbols refer to the charged particles while closed symbols to the anti-particles. The fitted spectra by using generic statistics are given as curves. It is found that generic axiomatic statistics reproduce well the transverse momentum distributions for all particles studied. The fit parameters estimated are compared to the ones obtained from different types of statistics for all particles at whole range of energies. The fit parameters are depicted in Figs. \ref{fig:Genericd} and \ref{fig:AllStatisticsPerParticle}.

\begin{figure}[htb]
\centering{
\includegraphics[width=5cm,angle=-0]{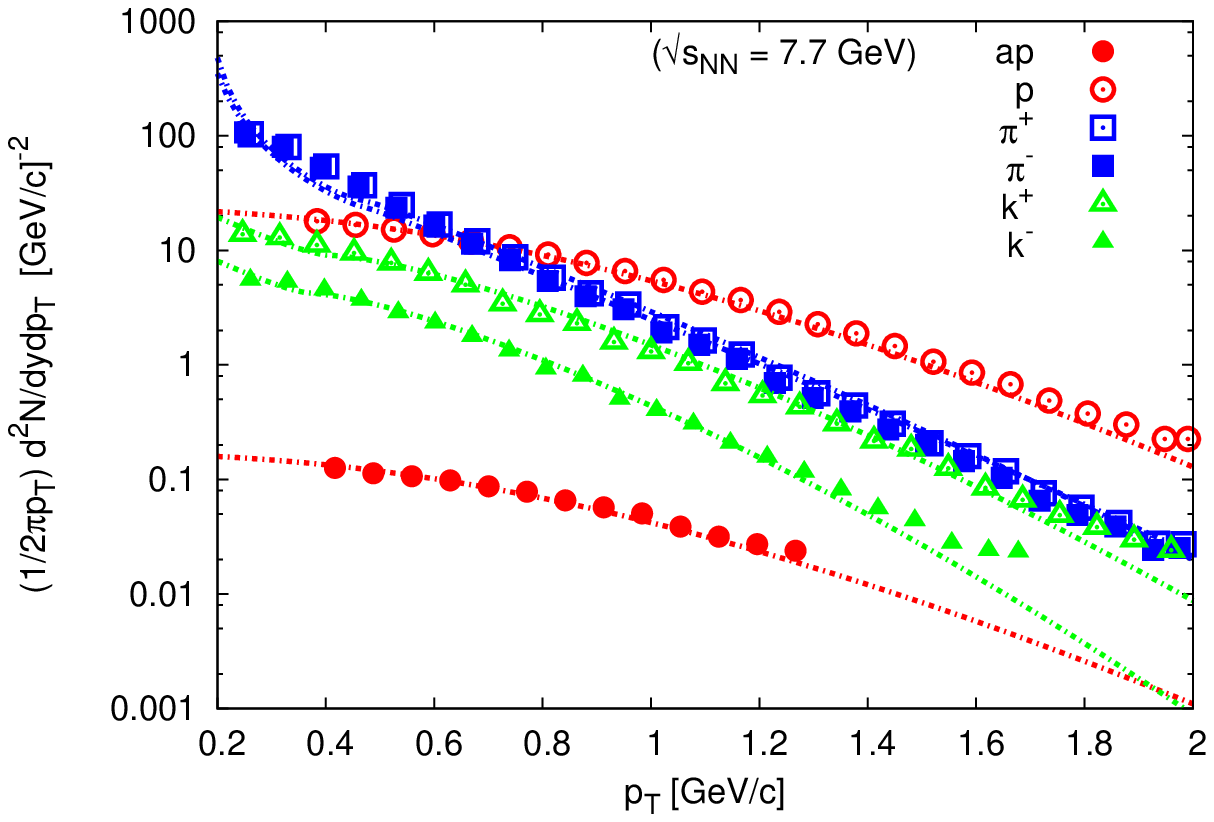}
\includegraphics[width=5cm,angle=-0]{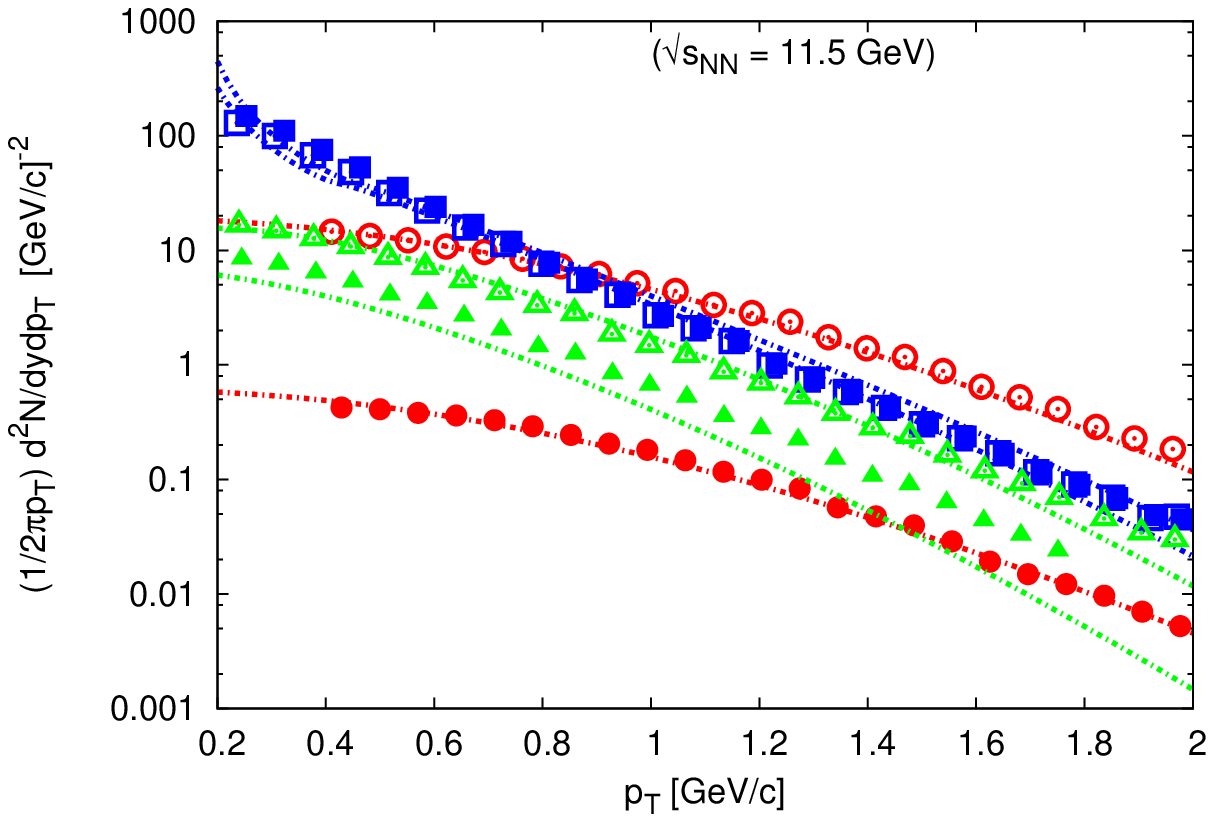}
\includegraphics[width=5cm,angle=-0]{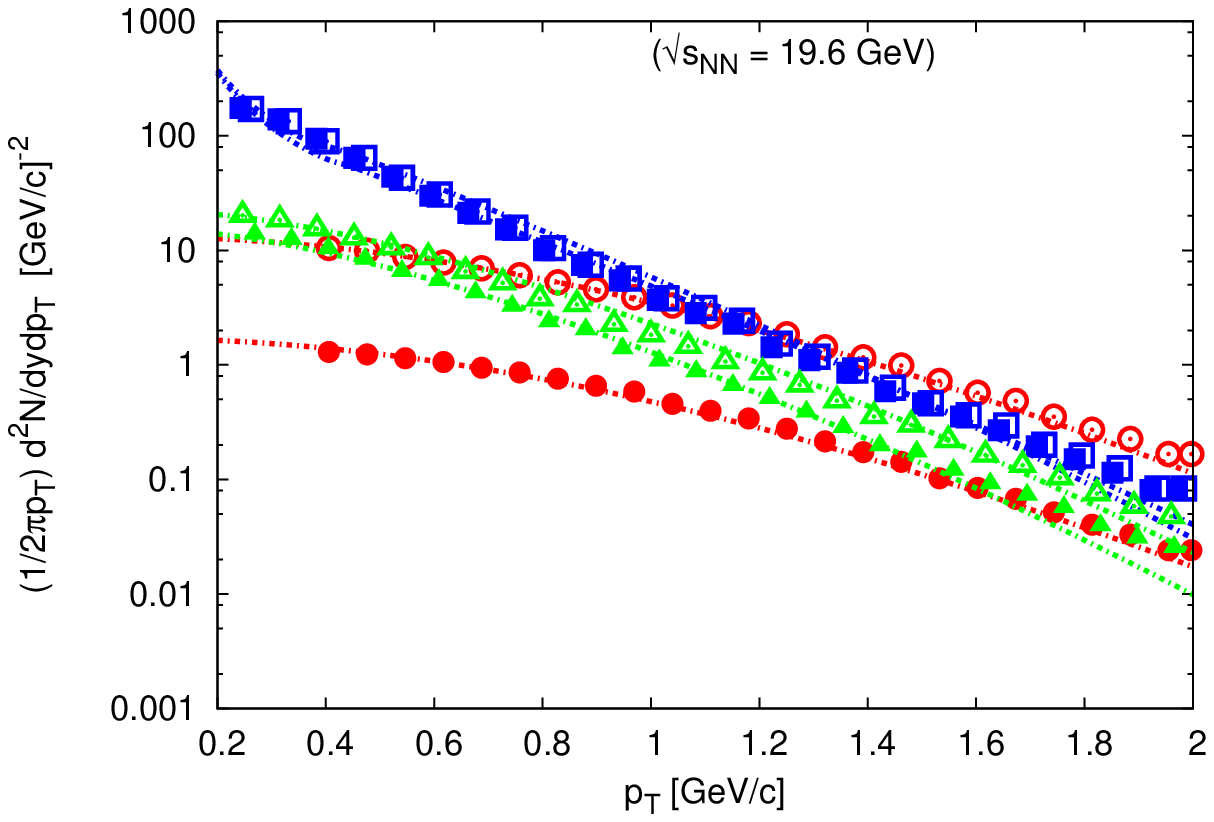}\\
\includegraphics[width=5cm,angle=-0]{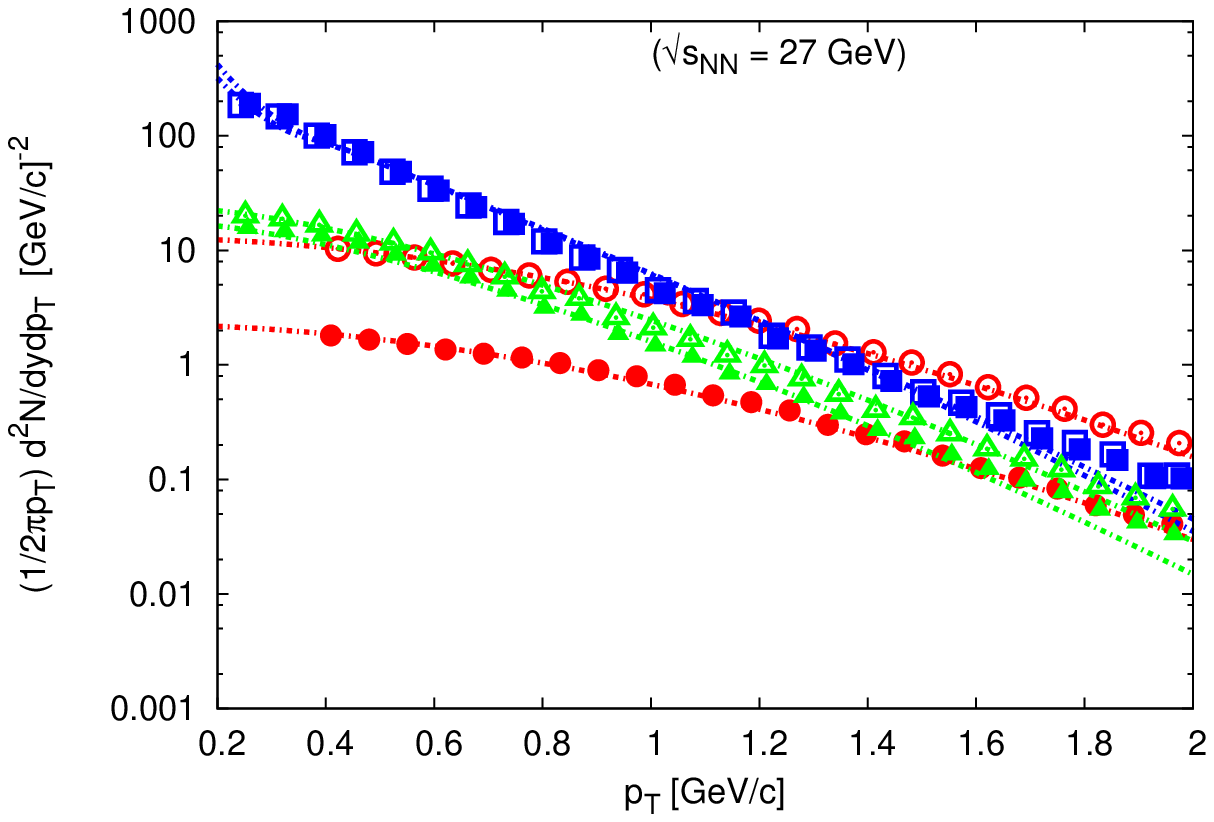}
\includegraphics[width=5cm,angle=-0]{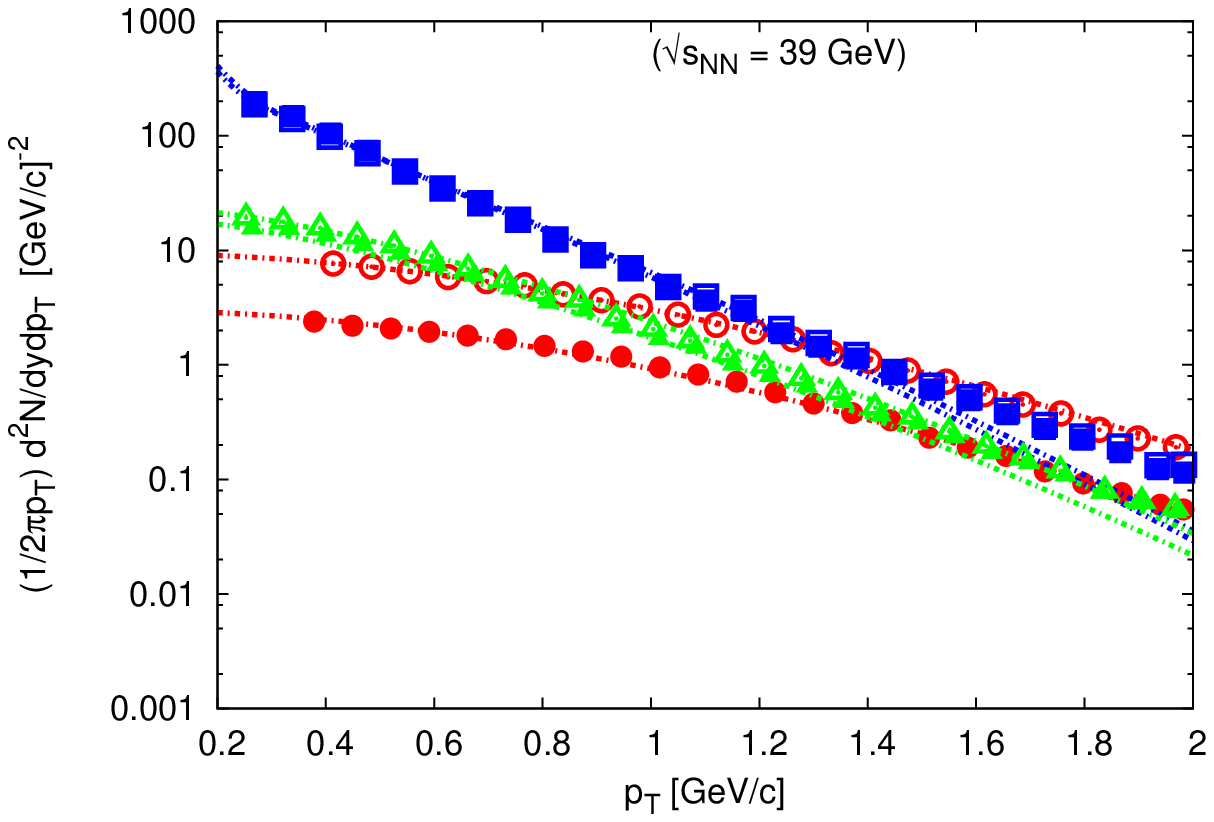}
\includegraphics[width=5cm,angle=-0]{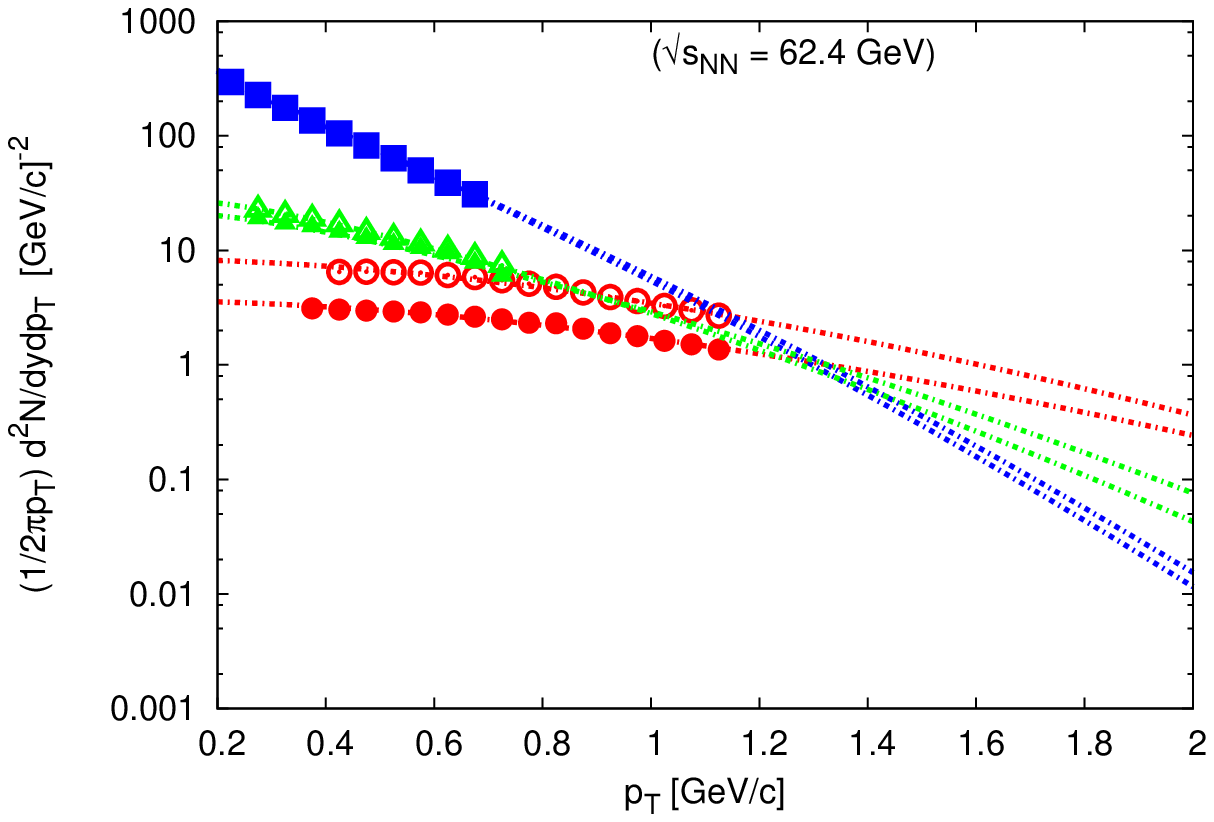}\\
\includegraphics[width=5cm,angle=-0]{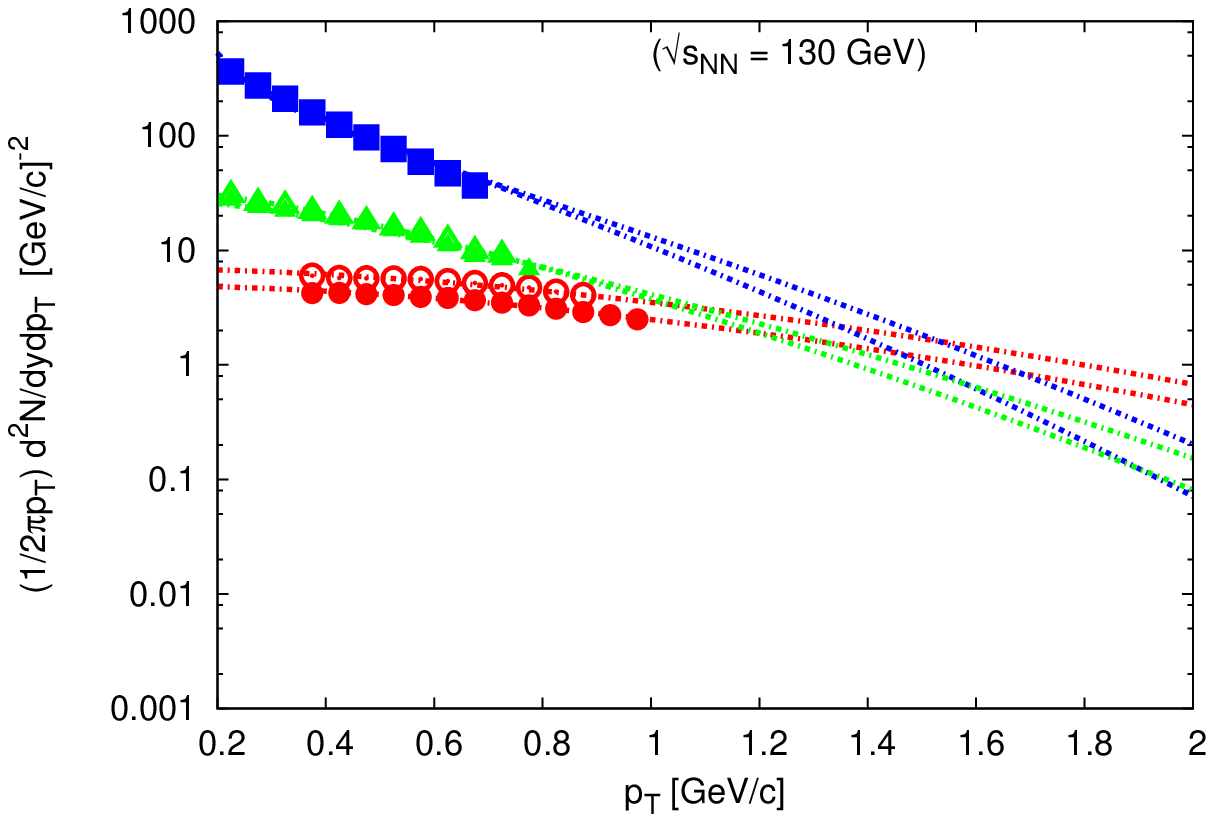}
\includegraphics[width=5cm,angle=-0]{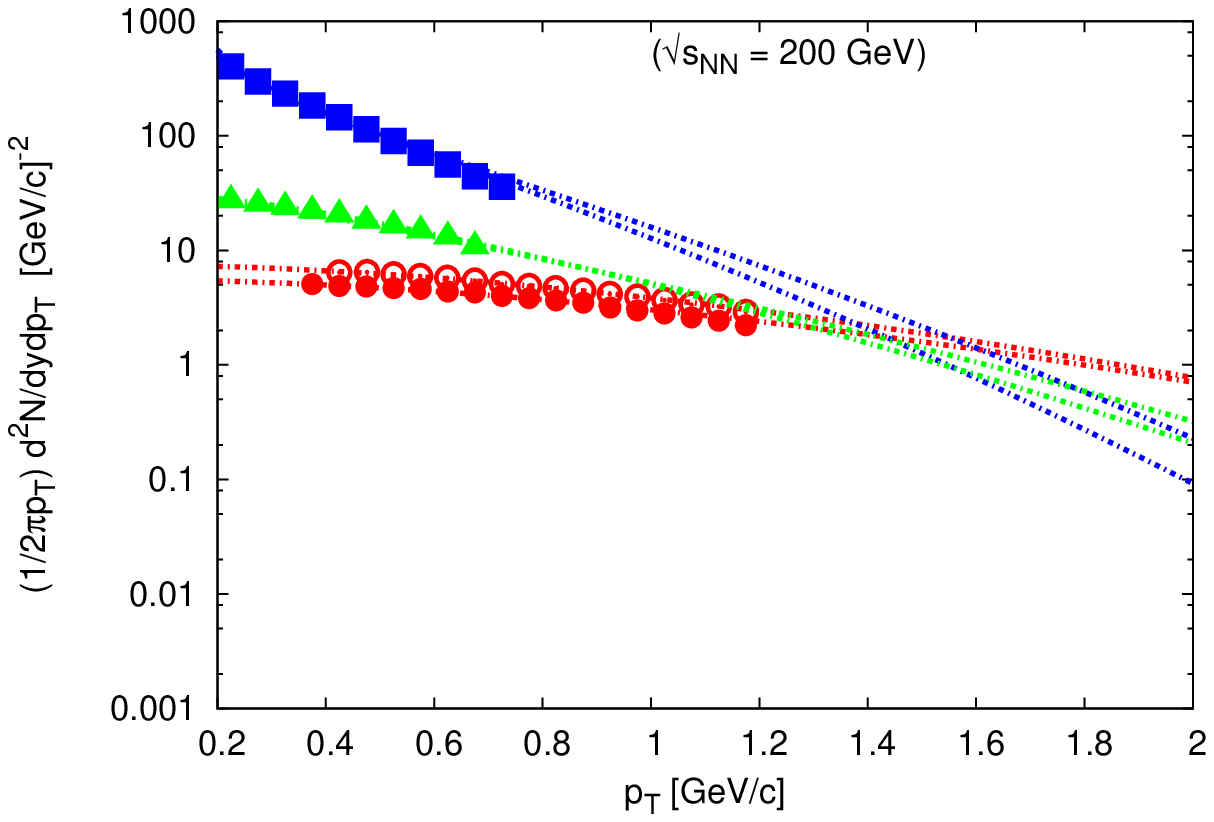}
\includegraphics[width=5cm,angle=-0]{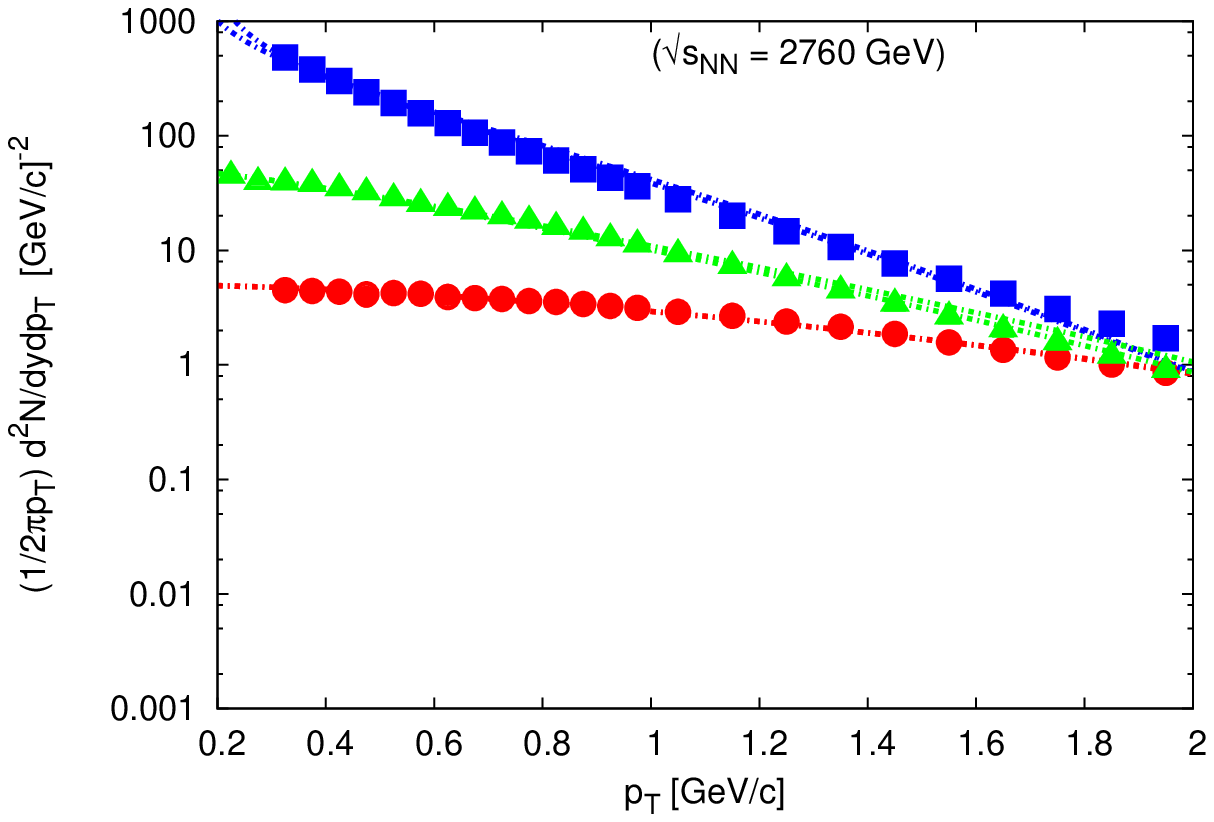}
\caption{(Color online) Transverse momentum spectra $p_{\mathtt{T}}$ for charged particles measured from A$+$A collisions at wide range of energies (symbols) are fitted by using generic axiomatic statistics. Panels (a)-(i) refers to the influence of changing energies from $7.7$ to $11.5$ to $19.6$ to $27$ to $39$ to $62.4$ to $130$ to $200$ and to $2760~$GeV, respectively. The corresponding fit parameters are illustrated in Figs. \ref{fig:Genericd} and \ref{fig:AllStatisticsPerParticle}.  
\label{fig:3}
}}
\end{figure} 

\newpage
\subsection{p$+$p collisions}
\subsubsection{Maxwell-Boltzmann statistical} 
\label{BoltzmannNN}
\begin{figure}[htb]
\centering{
\includegraphics[width=5cm,angle=-0]{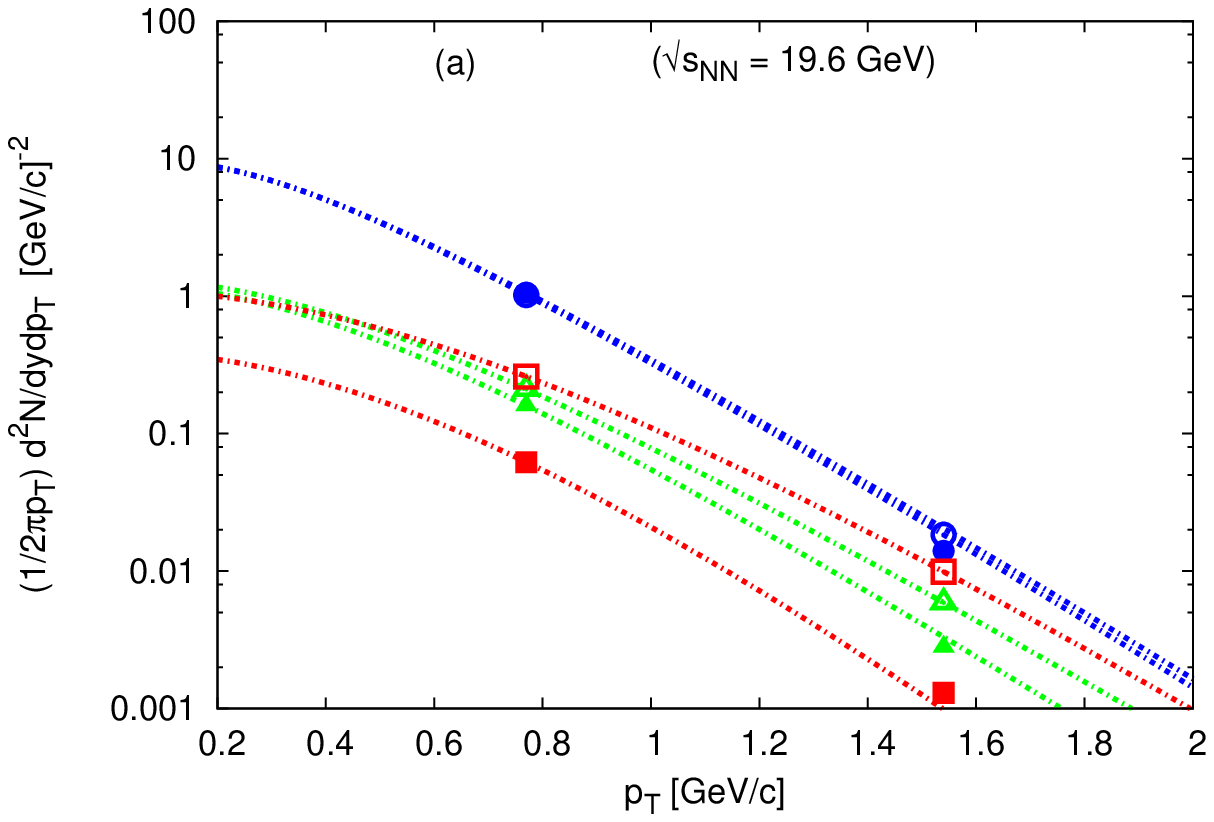}
\includegraphics[width=5cm,angle=-0]{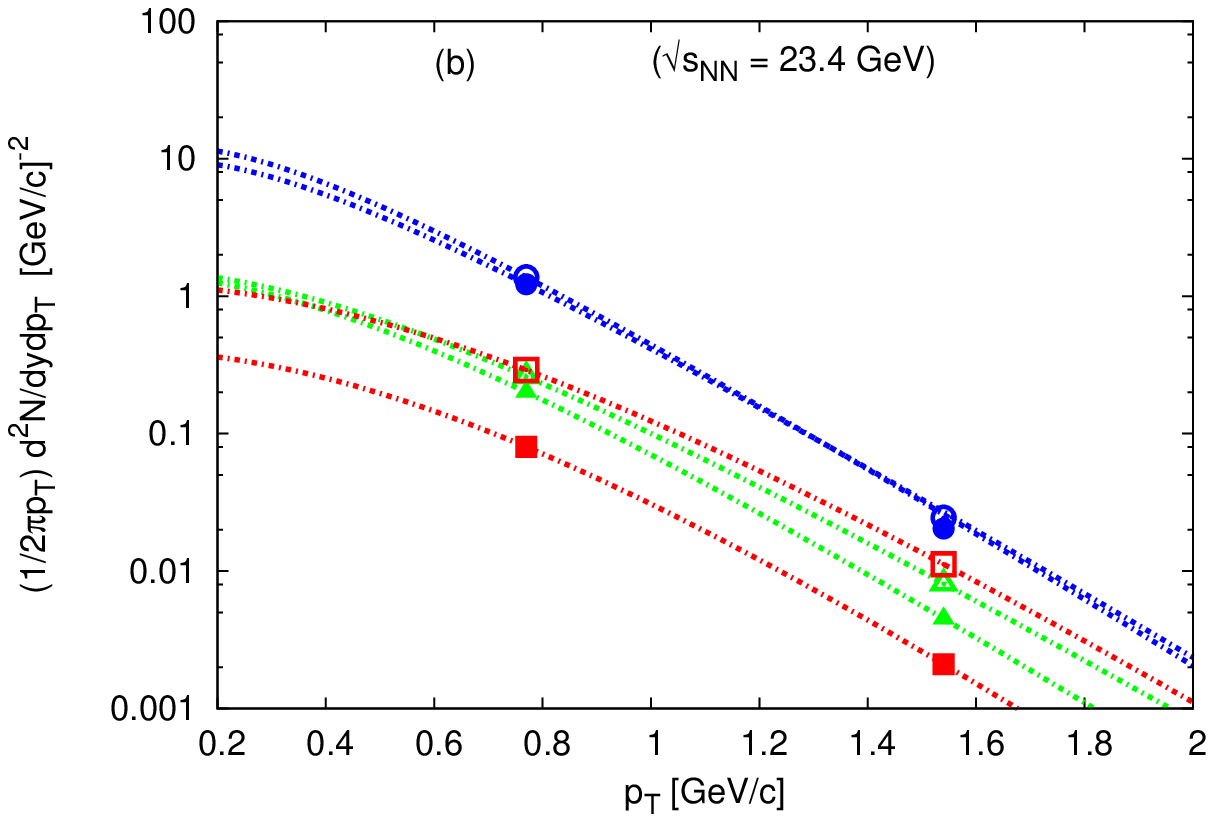}
\includegraphics[width=5cm,angle=-0]{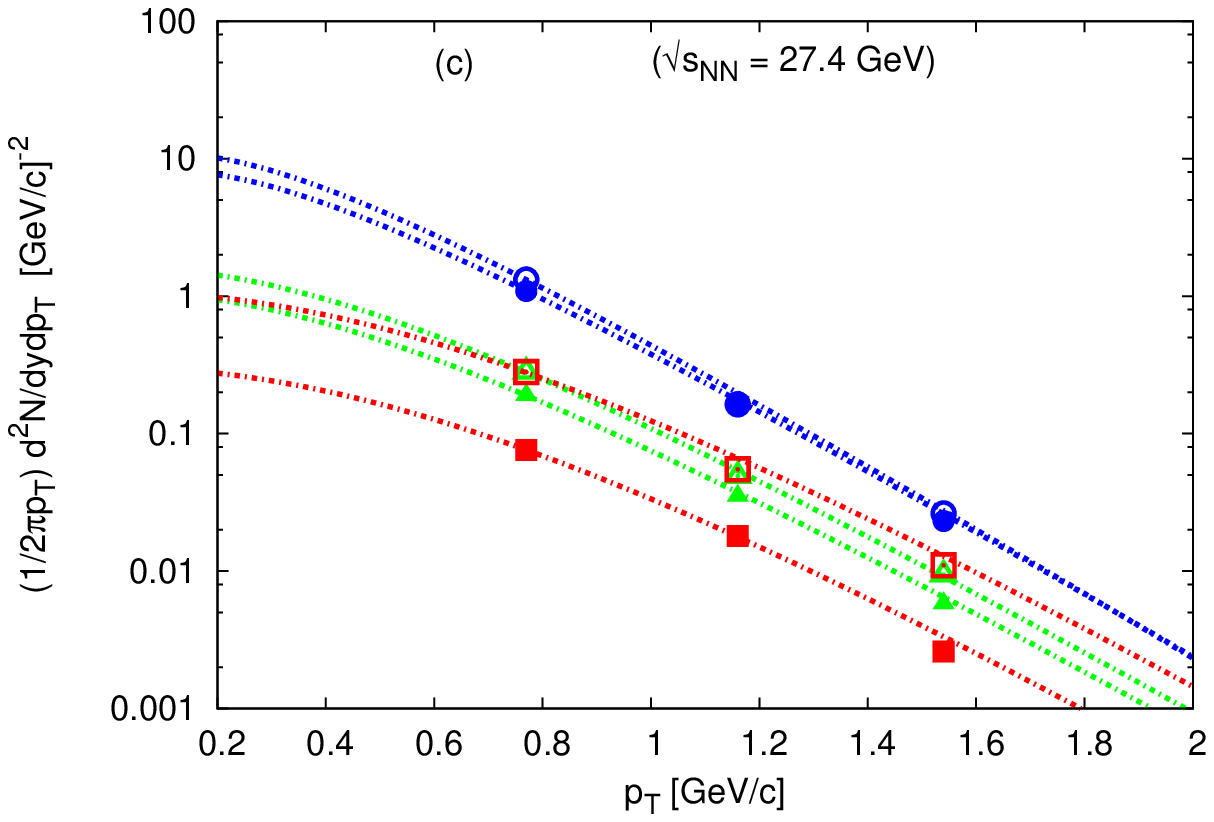}
\includegraphics[width=5cm,angle=-0]{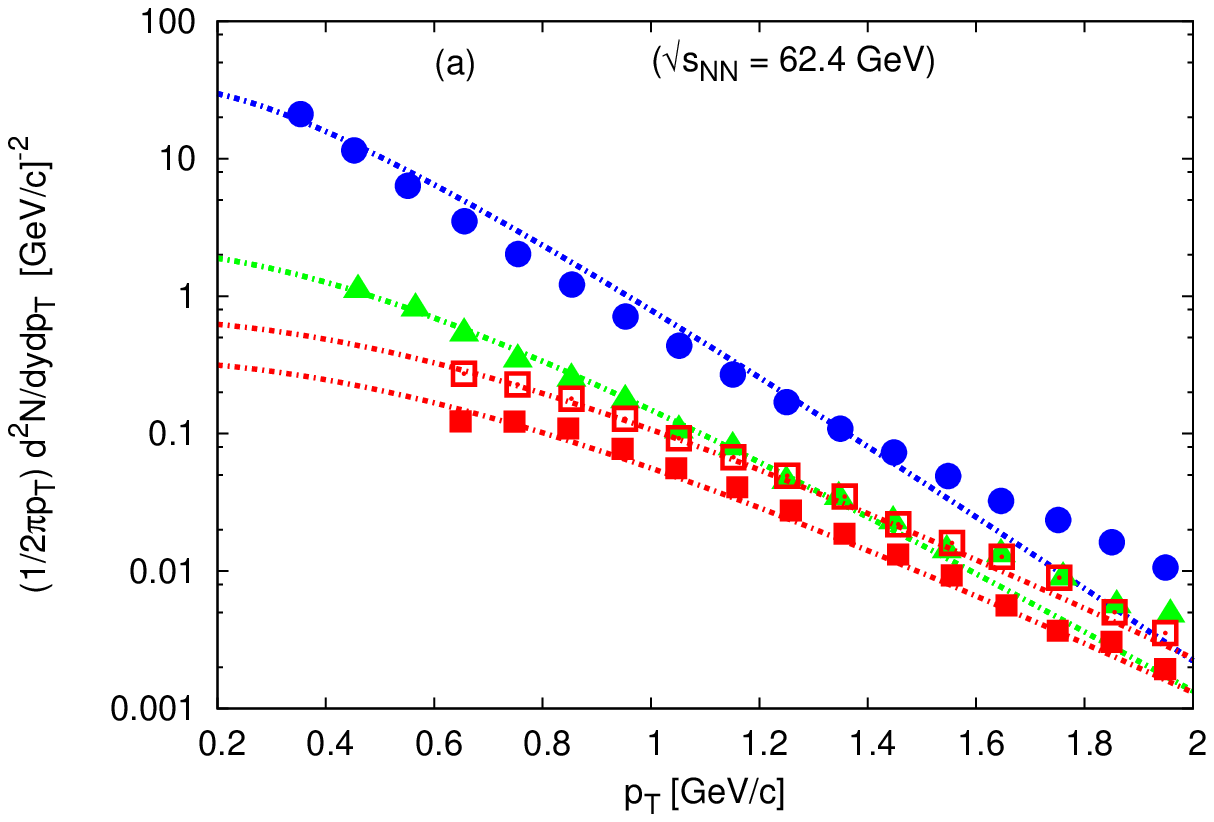}
\includegraphics[width=5cm,angle=-0]{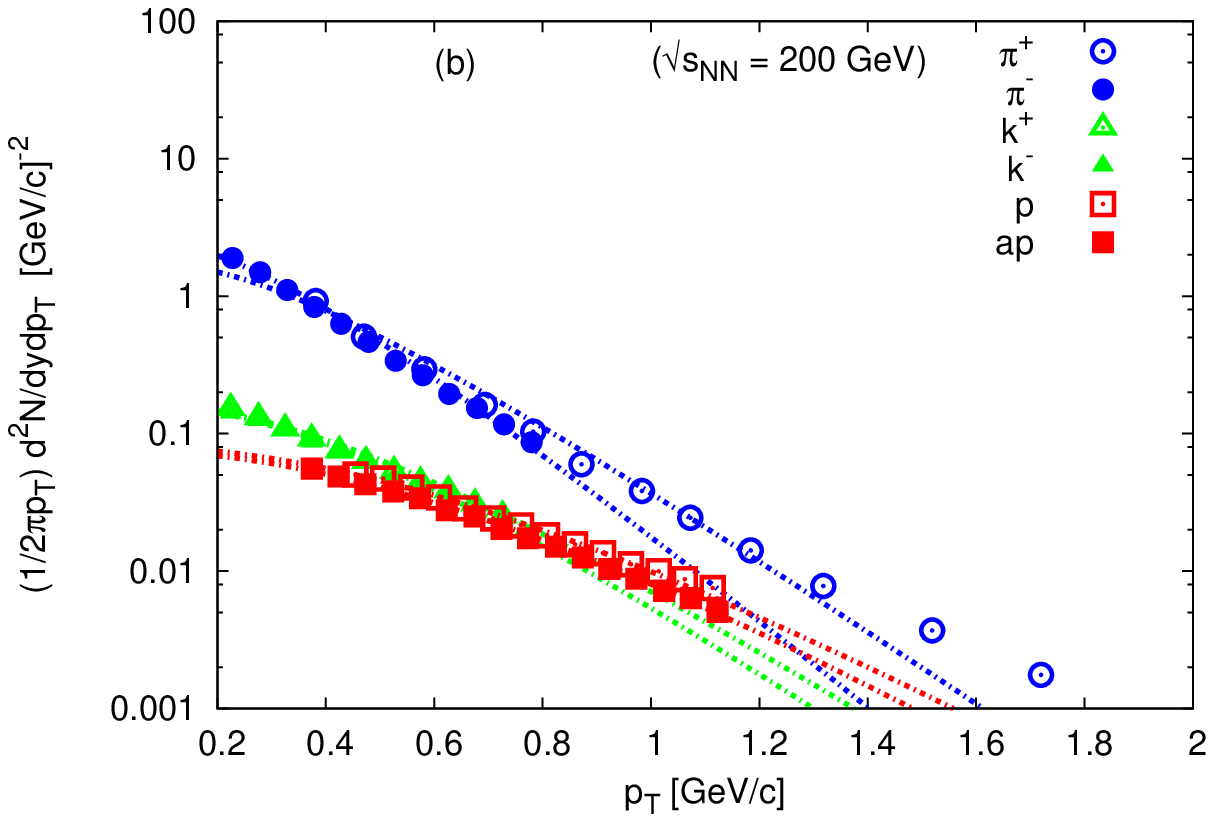}
\includegraphics[width=5cm,angle=-0]{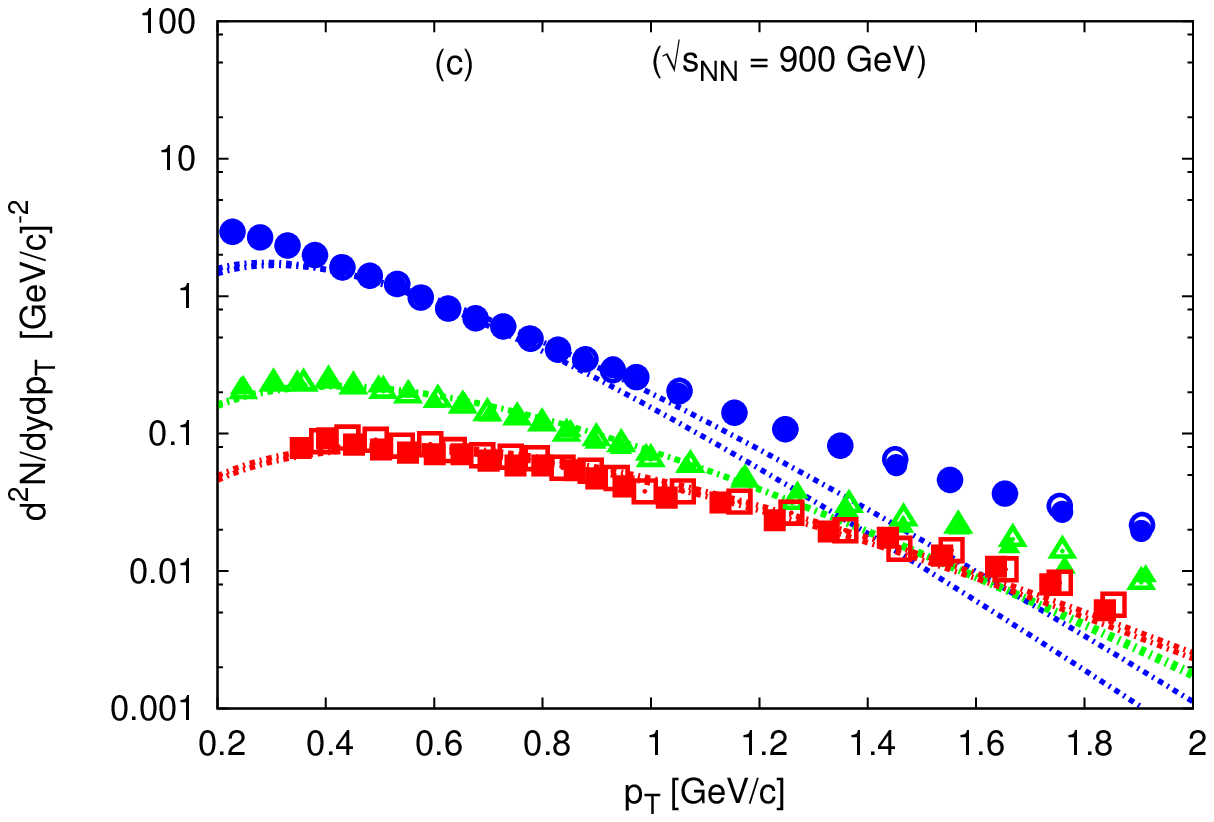}
\includegraphics[width=5cm,angle=-0]{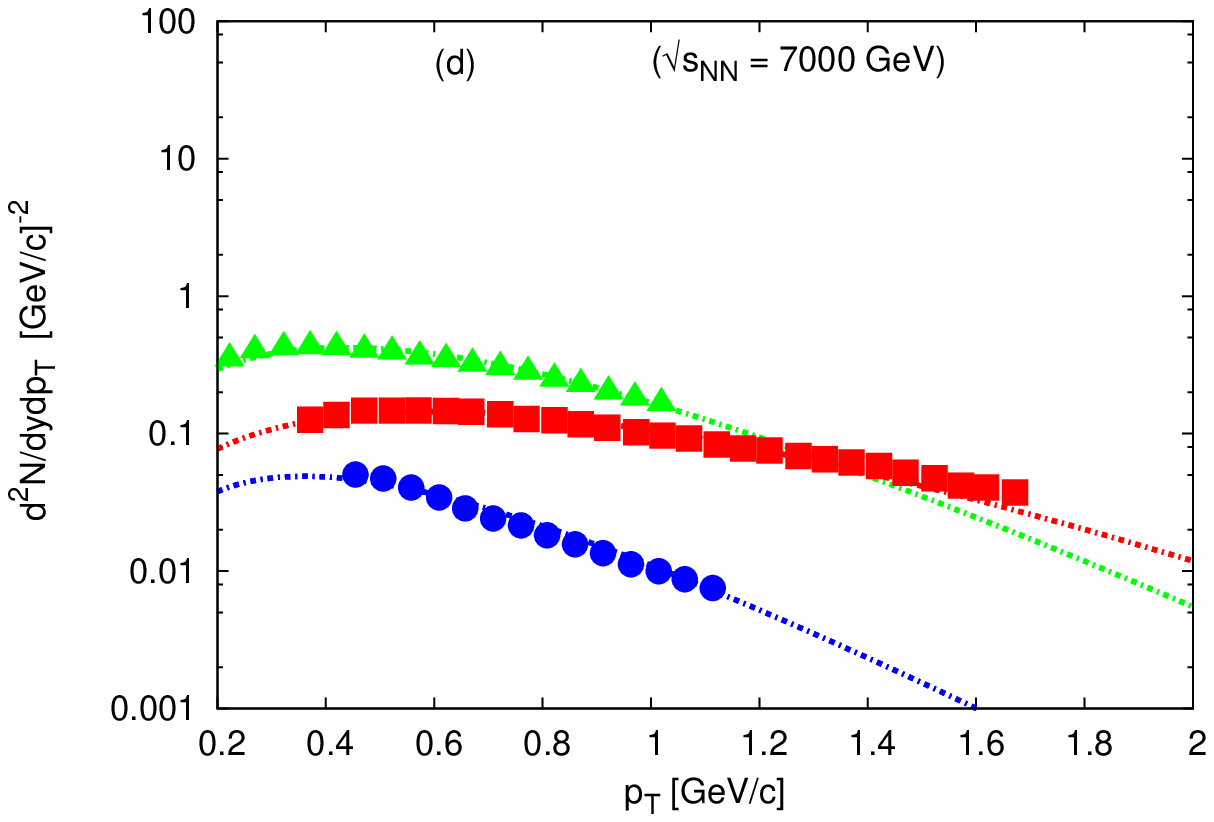}
\caption{(Color online) The transverse momentum spectra $p_{\mathtt{T}}$ for charged particles measured from p$+$p collisions at various energies are fitted by using Maxwell-Boltzmann statistics. Panels (a)-(g) refers to the change in the energies from $19.6$ to $7000~$GeV, respectively. The fit parameters are presented in  Fig. \ref{fig:AllStatisticsPerParticleNN}.
\label{fig:4A}
}}
\end{figure}

The transverse momentum spectra $p_{\mathtt{T}}$ of charged particles and their anti-particles measured in p$+$p collisions at energies ranging from $19.6$ to $7000~$GeV (symbols) are fitted by Maxwell-Boltzmann statistics (curves), Fig. \ref{fig:4A}. Symbols refer to the experimental measurements which are measured in NN collisions. It is apparent that transverse momentum spectra $p_{\mathtt{T}}$ calculated within Maxwell-Boltzmann statistics fit well with the measurements for all studied particles and anti-particles. The fit parameters are depicted in Fig. \ref{fig:GenericAllNN}.

\newpage
\subsubsection{Tsallis statistical fits} 
\label{TsallisNN}
\begin{figure}[htb]
\centering{
\includegraphics[width=5cm,angle=-0]{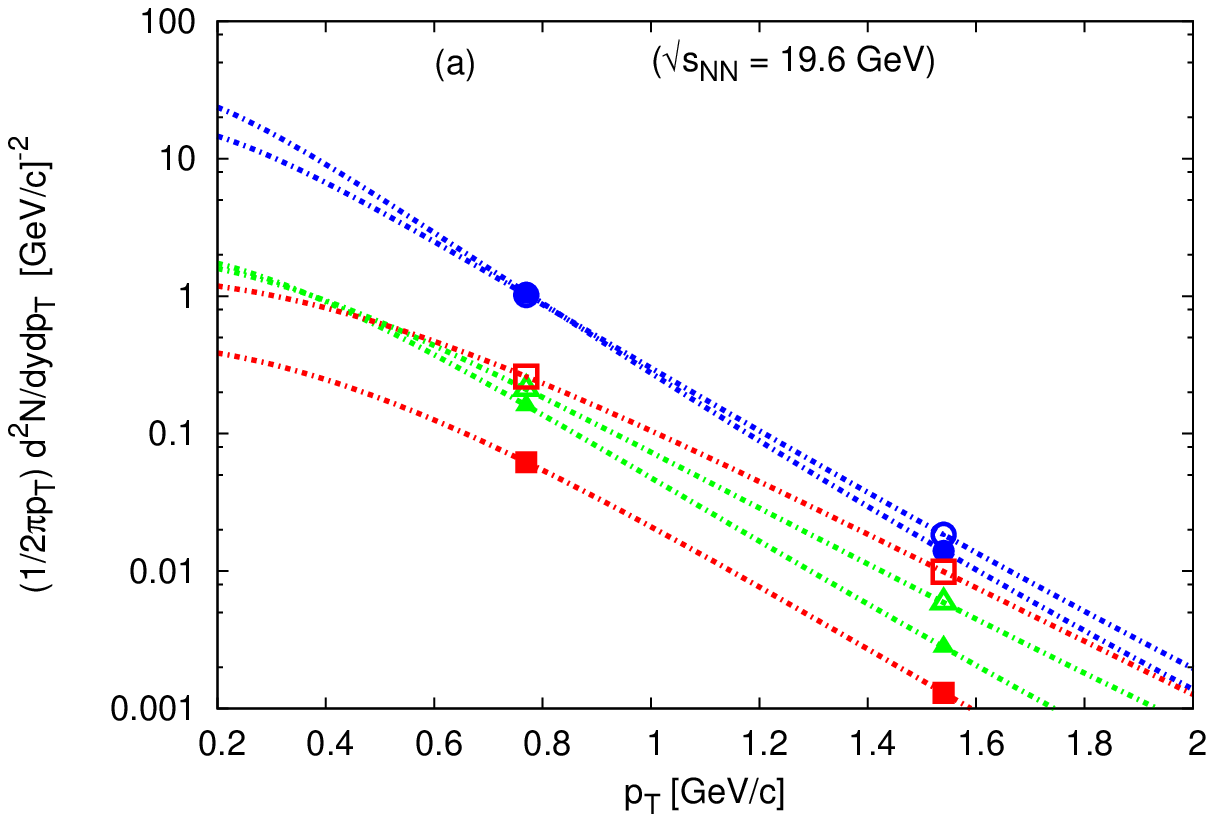}
\includegraphics[width=5cm,angle=-0]{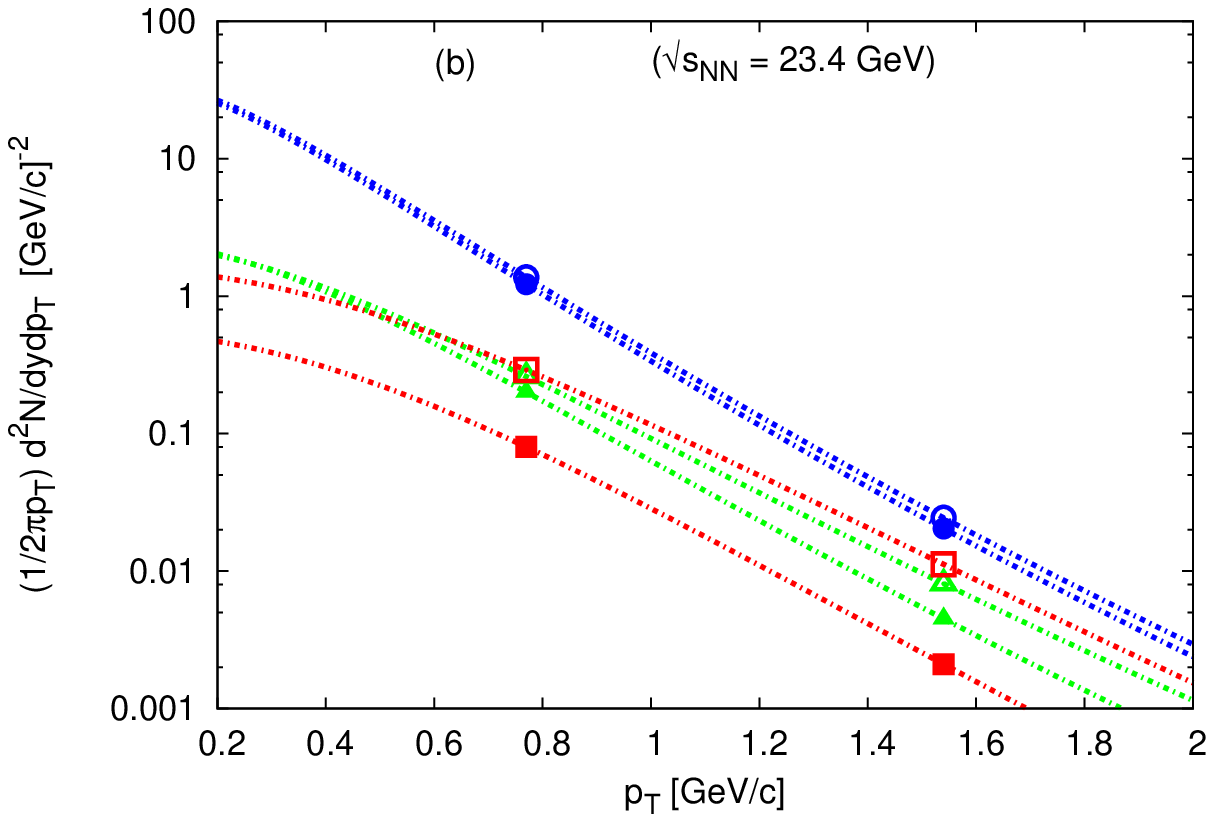}
\includegraphics[width=5cm,angle=-0]{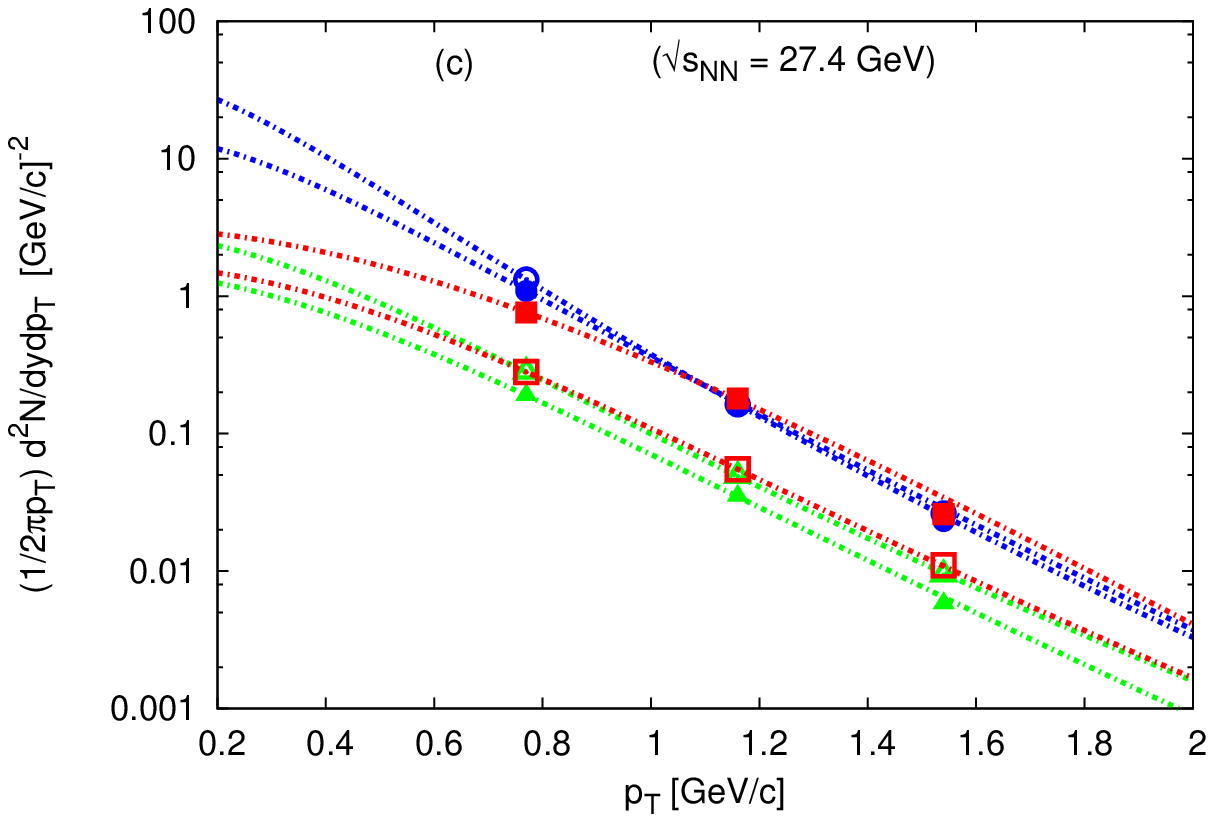}
\includegraphics[width=5cm,angle=-0]{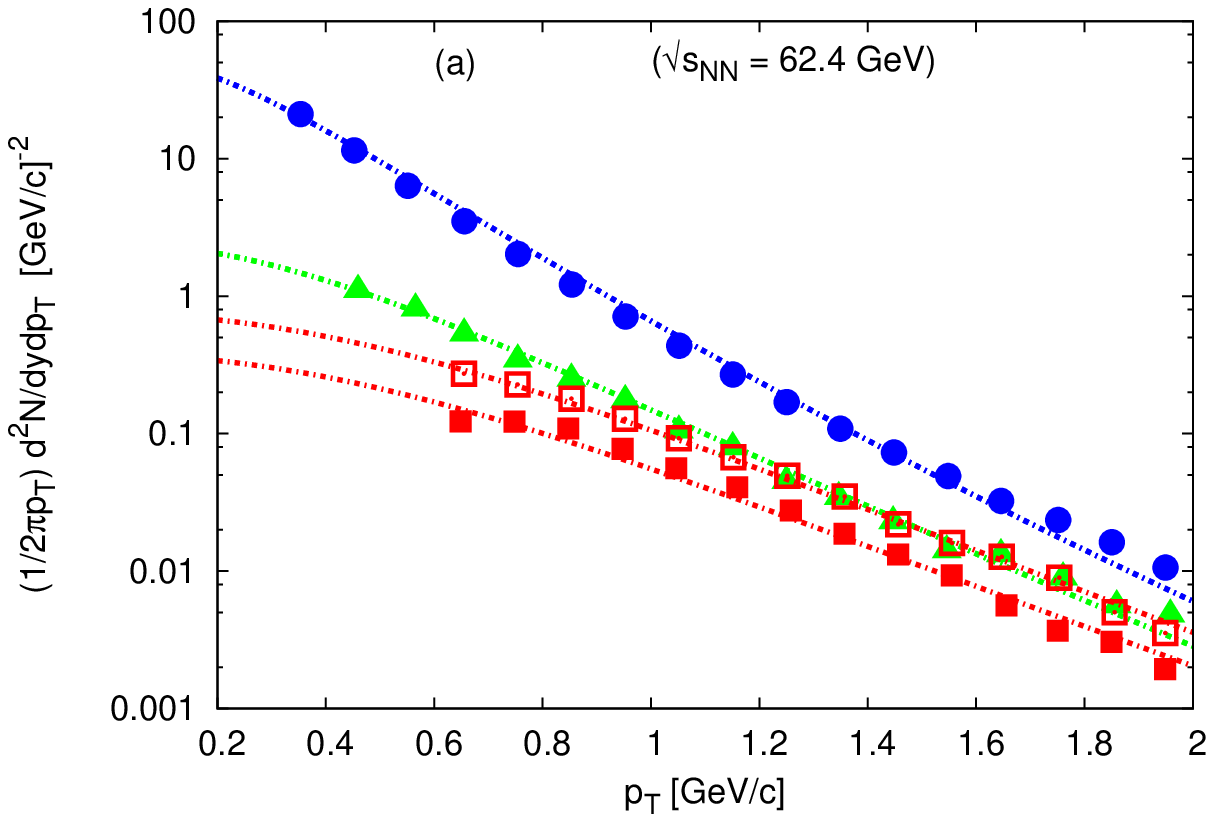}
\includegraphics[width=5cm,angle=-0]{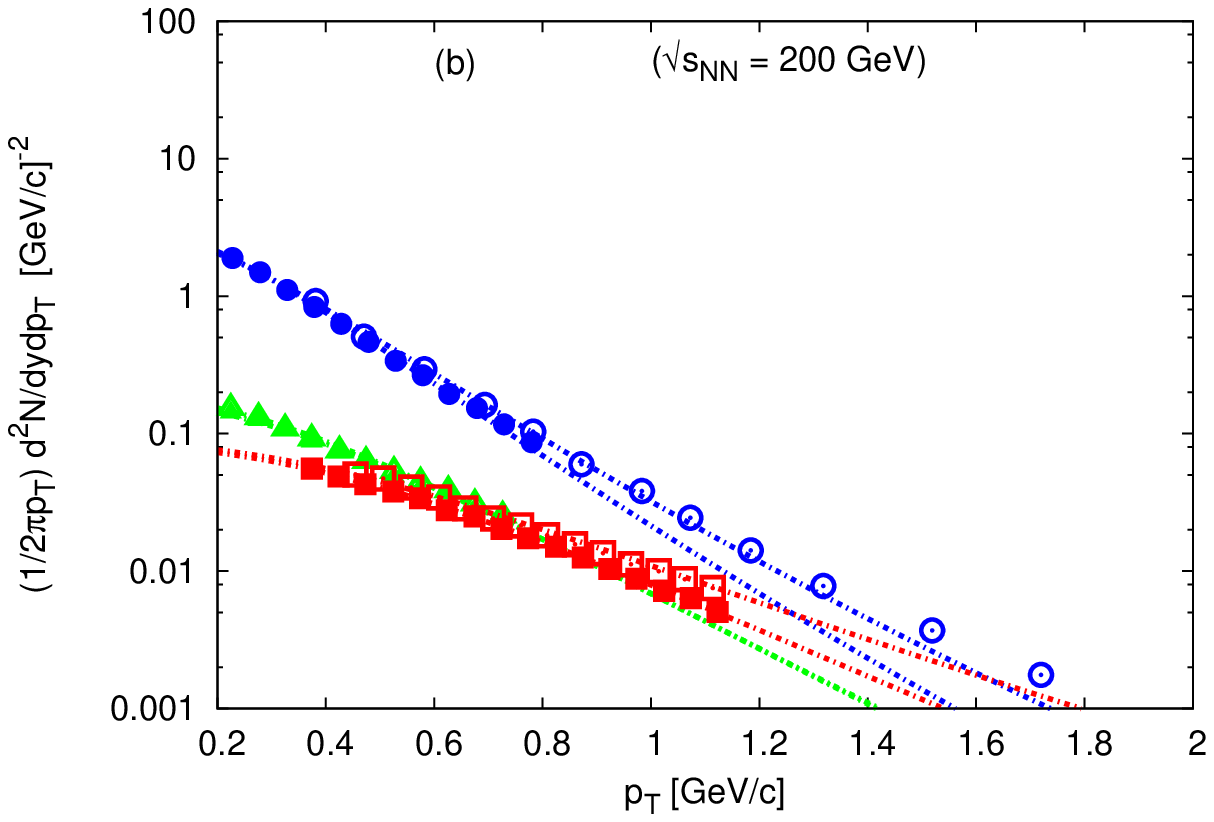}
\includegraphics[width=5cm,angle=-0]{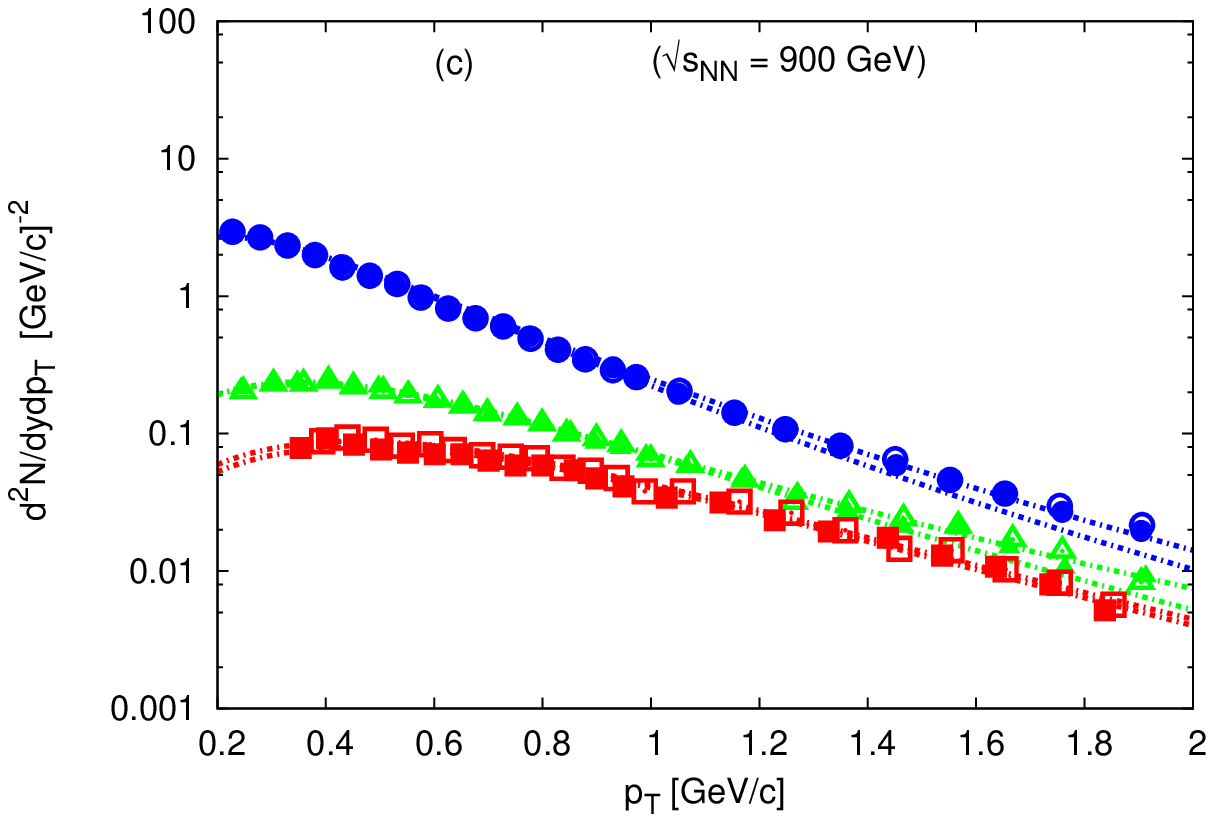}
\includegraphics[width=5cm,angle=-0]{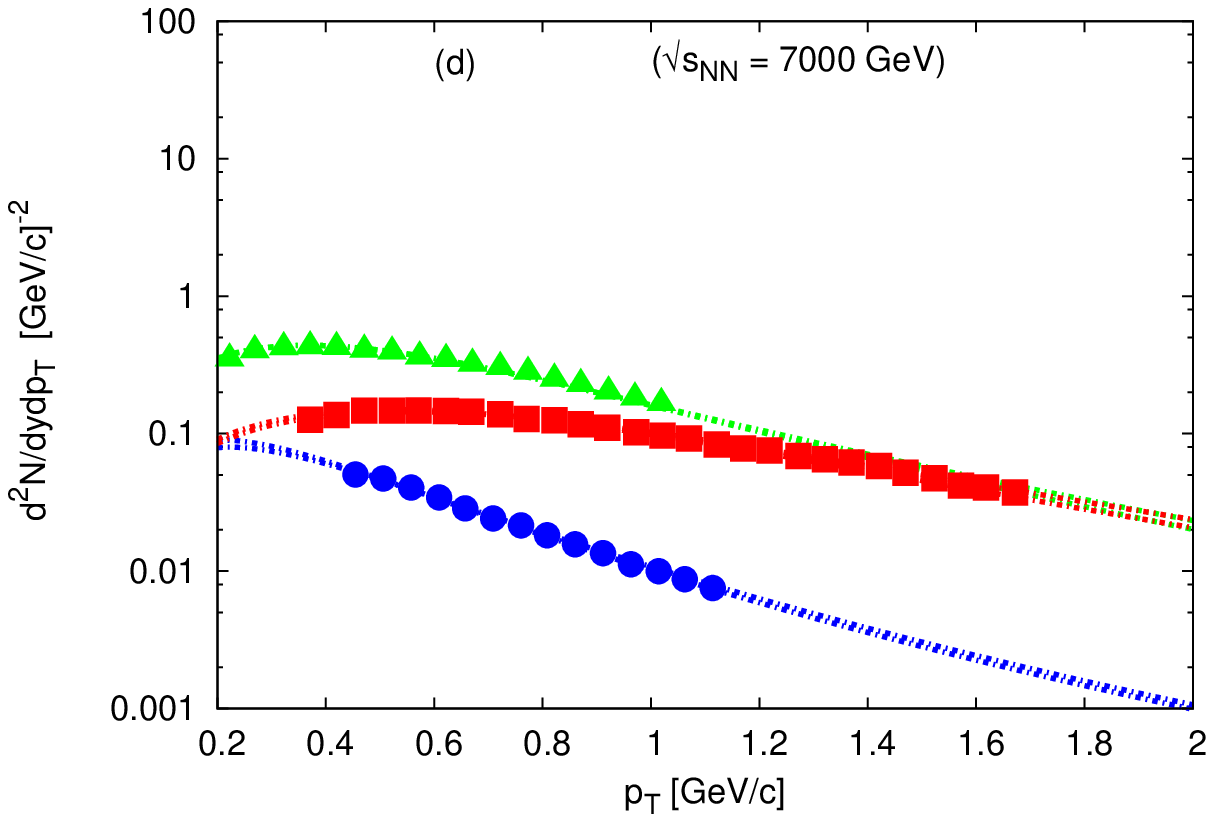}
\caption{(Color online) The same as in Fig. \ref{fig:4A} but by using Tsallis statistics. The corresponding fit parameters are given in Figs. \ref{fig:GenericdNN} and \ref{fig:AllStatisticsPerParticleNN}.
\label{fig:4B}
}}
\end{figure}

For charged particles (open symbols) and their anti-particles (open sysmbols) measured in p$+$p collisions at energies from $19.6$ to $7000~$GeV, the transverse momentum spectra $p_{\mathtt{T}}$ are fitted by using Tsallis statistics (curves) and depicted in Fig. \ref{fig:4B}. The corresponding  fit parameters are depicted in Fig. \ref{fig:GenericAllNN}.

\newpage
\subsubsection{Generic axiomatic statistical fits}
\label{GenericNN}
At energies from $19.6$ to $7000~$GeV, the transverse momentum spectra $p_{\mathtt{T}}$ calculated within generic axiomatic statistics (curves) are fitted to measured $p_{\mathtt{T}}$ from p$+$p collisions for chraged particles (open symbols) and anti-particles (solid symbols). The various fit parameters obtained are depicted in Fig. \ref{fig:GenericAllNN}. With this regard, we recall that the equivalent class $c=0.999506$ remains unchanged, in all cases.

\begin{figure}[htb]
\centering{
\includegraphics[width=5cm,angle=-0]{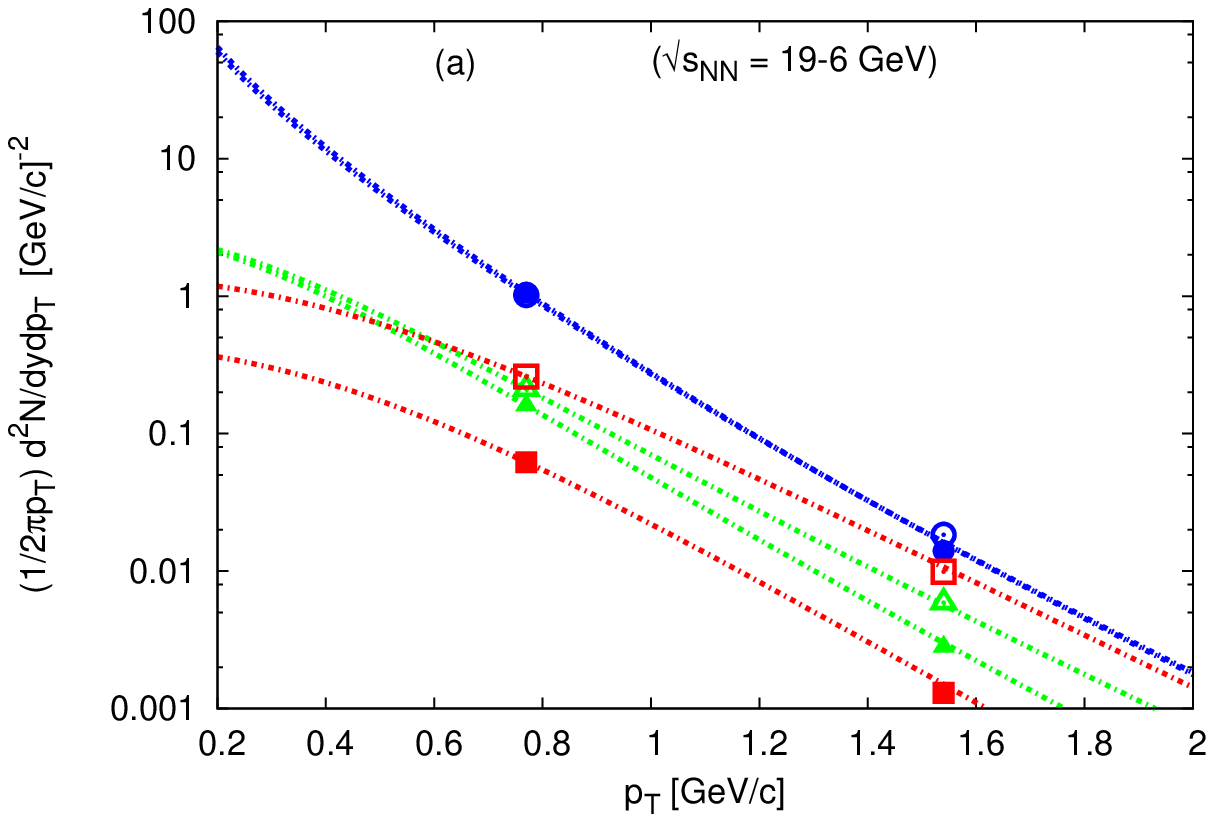}
\includegraphics[width=5cm,angle=-0]{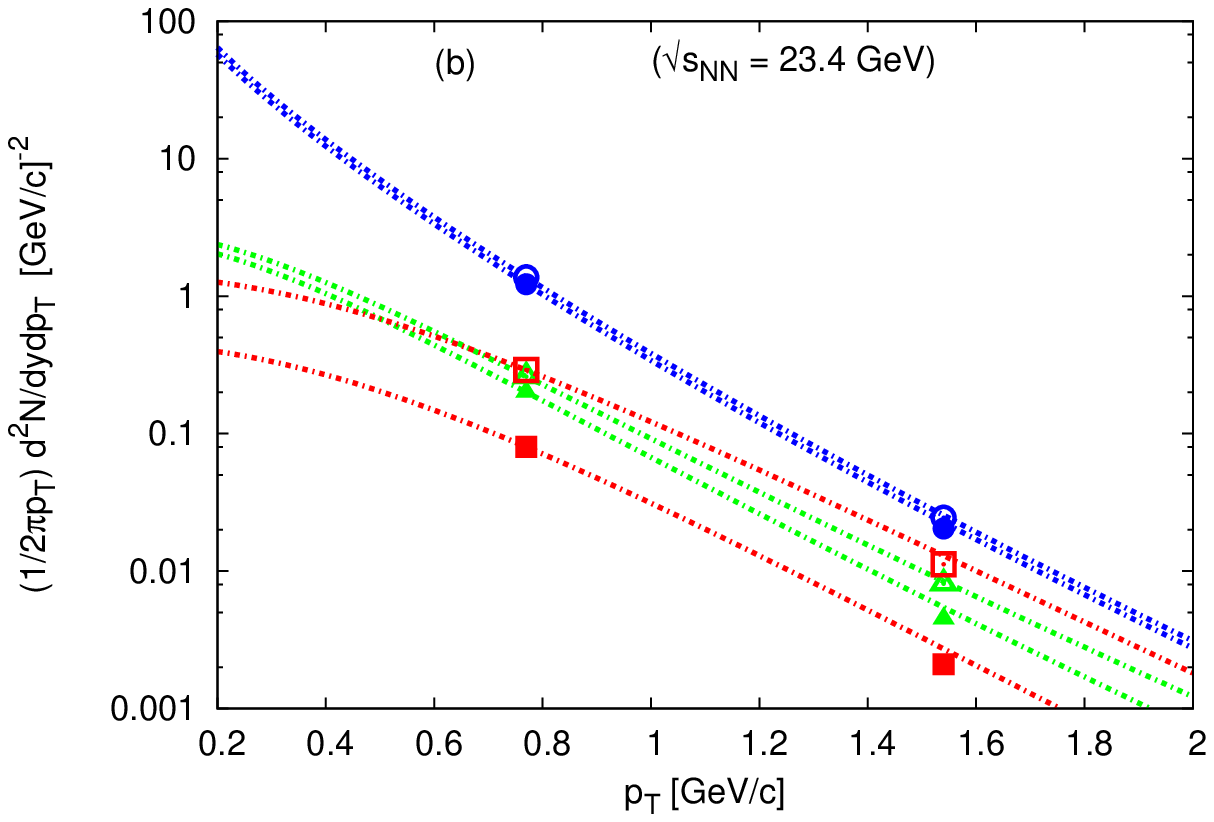}
\includegraphics[width=5cm,angle=-0]{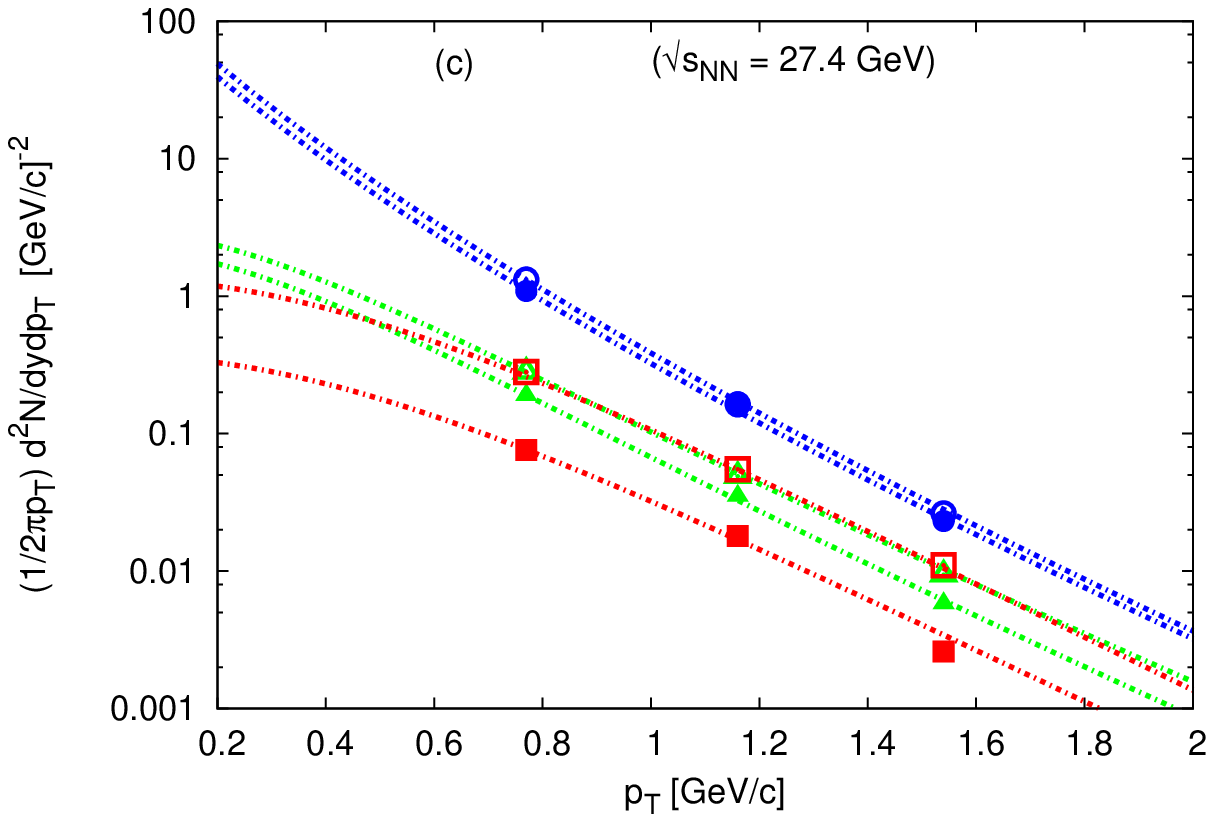}
\includegraphics[width=5cm,angle=-0]{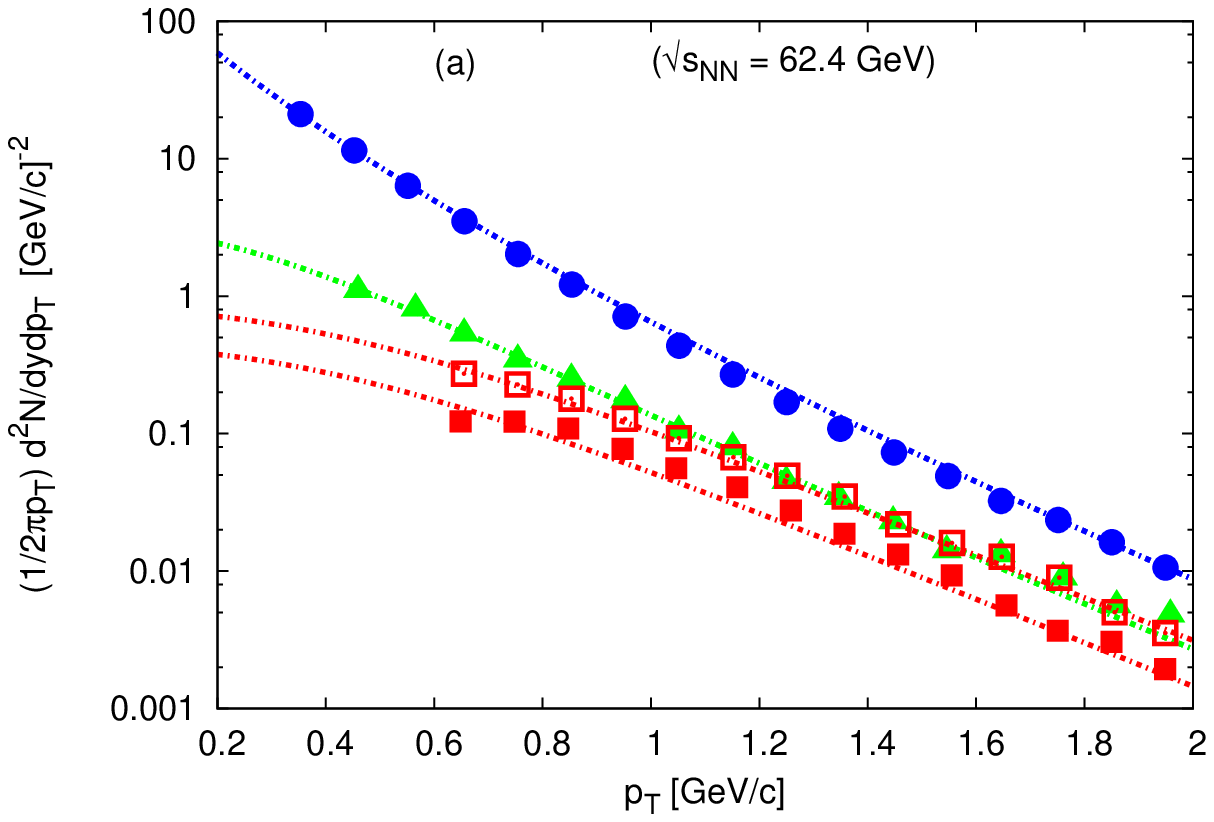}
\includegraphics[width=5cm,angle=-0]{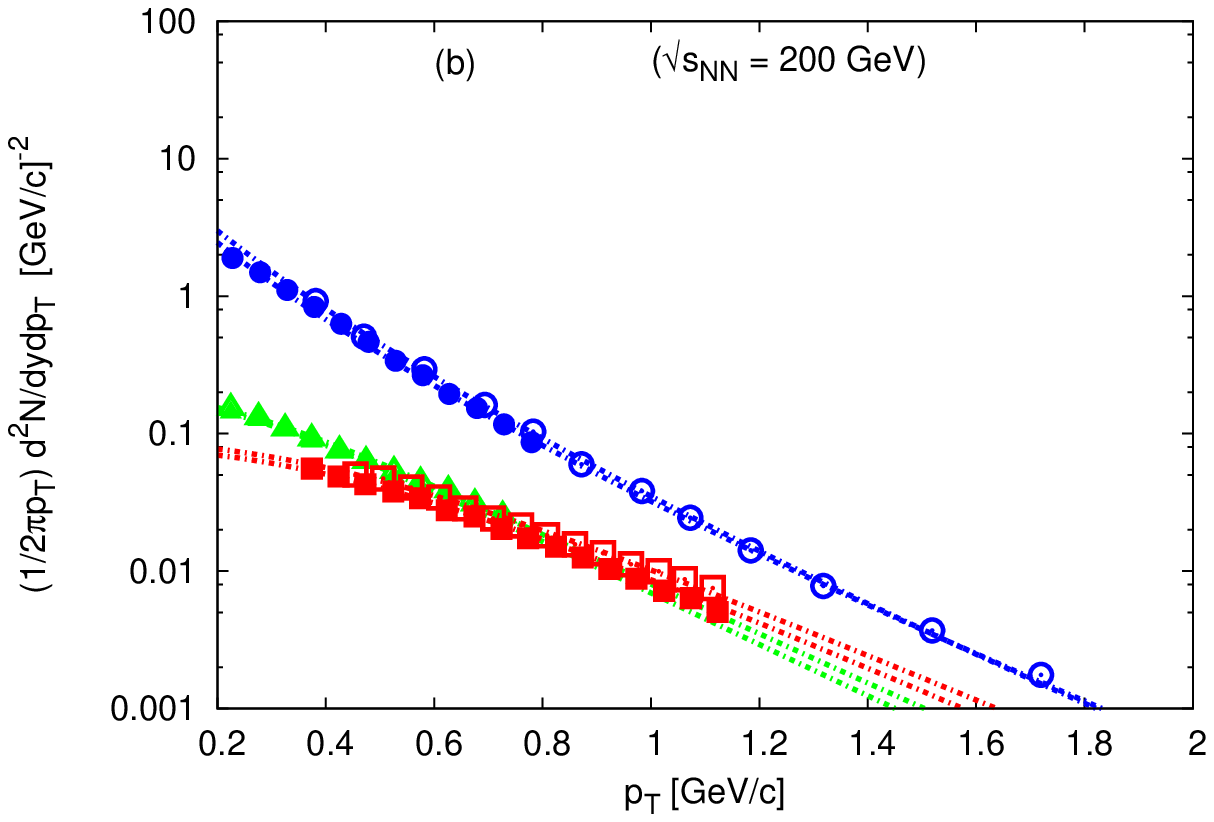}
\includegraphics[width=5cm,angle=-0]{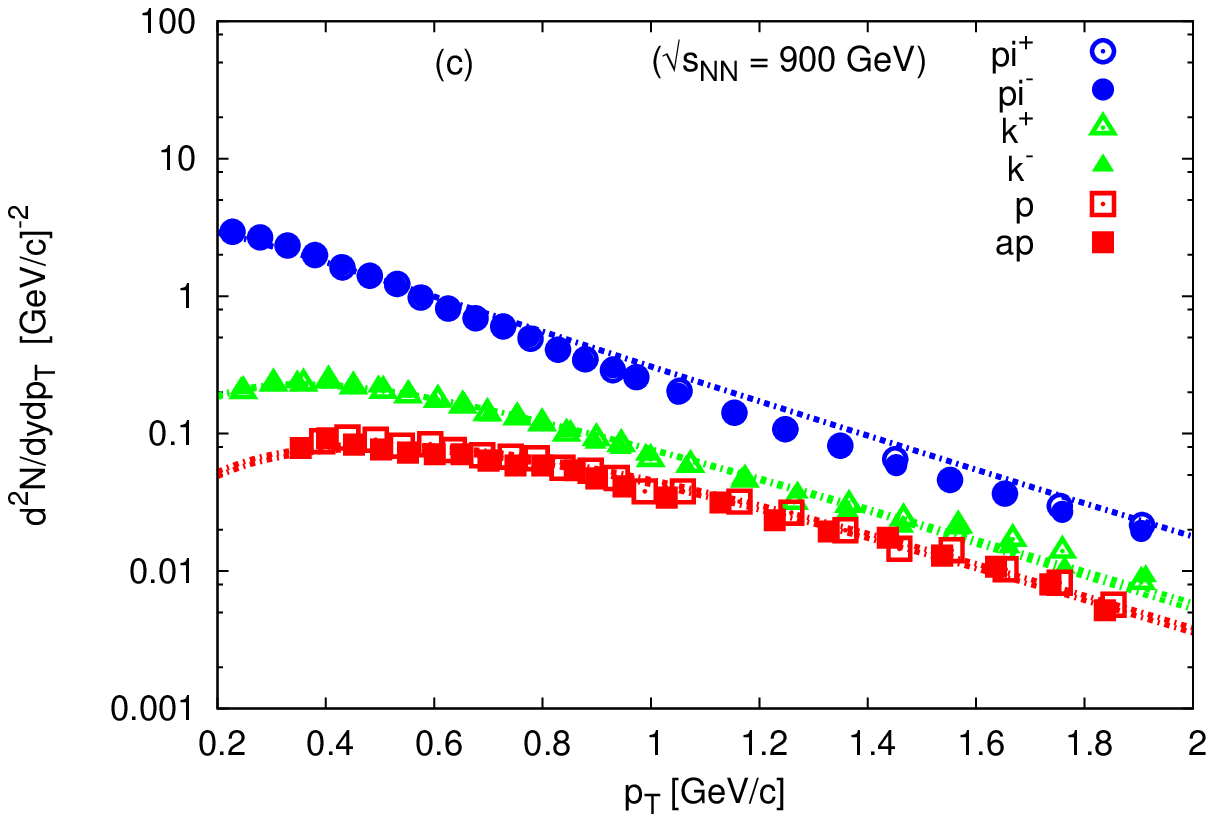}
\includegraphics[width=5cm,angle=-0]{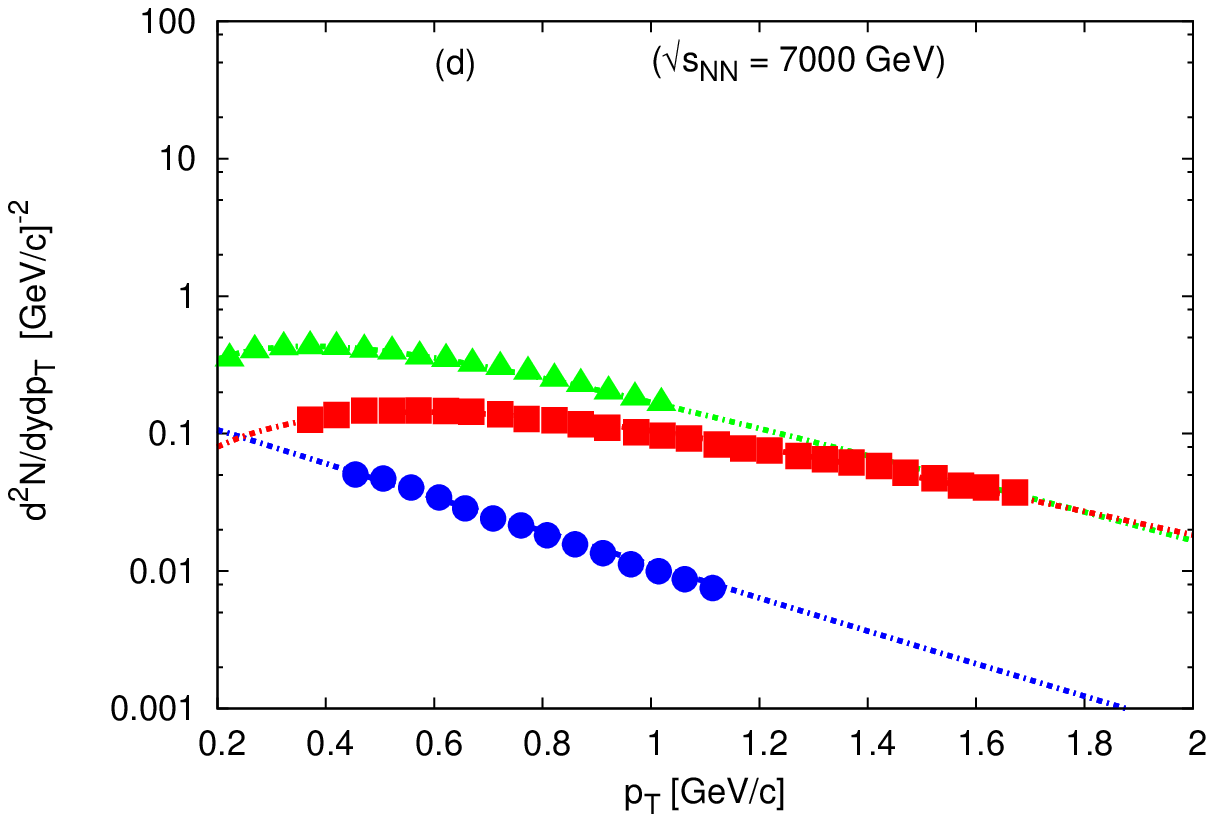}
\caption{(Color online) The same as in Fig. \ref{fig:4A} but by using generic axiomatic statistics. The corresponding fit parameters are graphed in Figs. \ref{fig:GenericdNN} and \ref{fig:AllStatisticsPerParticleNN}.
\label{fig:4}
}}
\end{figure} 

\end{document}